\documentclass[manuscript]{aastex}
\usepackage{longtable}
\title { A survey of Radio Recombination Lines using Ooty Radio Telescope at 328 MHz in the inner Galaxy}
\author {Raju Baddi\altaffilmark{1,2}}  
\affil{Raman Research Institute, C.V. Raman Avenue, Bangalore-80, India.}

\email{baddi@ncra.tifr.res.in}
\begin{document}

\begin{abstract}
A survey of radio recombination lines in the Galactic plane 
with longitude $-32^o < l < +80^o$ and latitude $b<\pm3^o$ using Ooty 
Radio Telescope(ORT) at 328 MHz has been reported. ORT observations were 
made using a New Digital Backend(NDB) augmented to it recently. With NDB 
ORT had a beam of $2^o.3 \times 2^o.2 sec(\delta)$ and a passband of $\sim$1 
MHz in the spectral line mode. The above mentioned Galactic region was 
divided into $\sim 2^o \times 2^o$ patches with the ORT beam pointed to 
the center. The ORT observations form a study of distribution of 
extended low-density warm-ionized medium(ELDWIM) 
in the inner Galaxy using H271$\alpha$ RL. By obtaining kinematical 
distances using $V_{LSR}$ of the H271$\alpha$ RLs the distribution of 
ELDWIM clouds within the inner Galaxy has been deduced for the region 
given above.                 
\end{abstract}


\section{Introduction}
The preliminary unsuccessful attempts to detect RRL from hydrogen were 
made by Egorova and Ryzkov (1960) with the Pulkovo telescope.  Successful 
detection of RRL was made in April 1964 using  an improved radio meter 
with the 22-m radio telescope at Lebedev Physical institute in Puschino.  
Sorochenko and Borodzich detected the hydrogen RRL 
H90$\alpha$ ($\lambda$=3.38cm) towards the omega nebula. Independently at about 
the same time another group(Dravskikh et al 1965) also detected a 
convincing RRL H104$\alpha$. Following this many researchers(Lilley et 
al 1966; Goldberg \& Dupree 1967; Gottesman \& Gordon 1970; ) detected RLs. 
Subsequent attempts(Batty 1976) to detect RRL 
at lower frequencies($<$ 500MHz) seem (Anantharamaiah 1985)to have 
failed due to non availiability/development of radio telescope hardware. 
It was already known(Shaver 1975) in the community that stimulated 
emission($kT/h\nu \times$ spontaneous) would boost RRL at lower 
frequencies. Apparently RRL at frequencies below 500 MHz were first 
observed from the Galactic plane(Anantharamaiah 1985) using Ooty Radio 
Telescope(ORT). A selective survey(Anantharamaiah 1985) of RRL was made 
in the Galactic plane($b \sim 0^o$) with longitude $<60^o$ using ORT at 
325 MHz(H272$\alpha$) towards 53 directions. This observation gave 
numerous RL detections and paved way to a new survey (Roshi \& 
Anantharamaiah 2000; Roshi \& Anantharamaiah 2001)covering the Galactic 
plane systematically. However these positions have been mostly in the 
Galactic plane within a latitude of $b<\pm1^o$. With the introduction 
of a New Digital Backend(NDB) for the ORT in the recent years it has 
been possible to carry forward this survey to higher latitudes within 
less time and better signal to noise ratio(S/N) compared to previous 
observations. Previous observations have used either the entire 
telescope with a beam size of $\sim 2^o \times 6'$(Anantharamaiah 1985; 
Anantharamaiah 1986; Roshi \& Anantharamaiah 2001a) or a couple of 
modules with a beam size of $\sim 2^o \times 2^o$(Roshi \& Anantharamaiah 
2000). In the present observations the NDB could process signals from 
all the 22 modules(Section 2) of the ORT thus reducing the integration 
time required for detection. On the other hand  earlier observations 
have had a larger bandwidth(BW) or more number of  RRL-transitions(Roshi 
\& Anantharamaiah 2000) compared to only one observable transition
(H271$\alpha$) with the recent NDB. \\

ORT observations were made with a New Digital Backend(NDB) augumented 
to it recently(Prabu 2010). In the spectral line mode, with NDB, ORT 
provided a passband of 1.2 MHz and a beam size of $2^o.3 \times 2^o.2 
sec(\delta)$. The Galactic region between $-32^o < l < +80^o$ and 
$b<\pm3^o$ was divided into patches of $2^o \times 2^o$ with ORT pointed 
towards the center of each patch. ORT observed these 165 positions 
distributed in 3 rows and 55 columns for $\sim$3 hrs per pointing. Some 
of the positions were skipped due to existence of earlier observations, 
shortage of telescope time and severe interference at higher declinations. 
Due to technical reasons H271$\alpha$ seemed to be the only appropriate 
RL for ORT with NDB.  Kinematical distances towards H271$\alpha$ line 
emitting regions were obtained using a differential Galactic rotation 
curve(Sofue et al. 2009) which gave a distribution of ELDWIM clouds in 
the inner Galaxy. ORT observations aimed at obtaining the distribution 
of ELDWIM in the inner Galaxy and to obtain new RL detections from above 
and below the Galactic plane at 328 MHz.  \\ 

\section[]{Observations}          
Ooty Radio Telescope(ORT) is situated near the town Ooty, south India at a 
longitude of $283^{o}.3$ and latitude of $11^{o}.4$. ORT(Swarup et al. 1971) 
is an off-axis parabolic cylinder with a length of 530m and width of 30m. 
The telescope is located on a hill which has a natural slope of $11^{o}.4$ 
equal to the geographical latitude of the place. This gives it the feature 
of equatorial mount. The operating frequency of the telescope is centered 
at 326.5 MHz with a maximum BW of 15MHz at the front-end. The 
reflecting surface of the clylinder is made of 1100 stainless steel wires 
running parallel to each other along the entire length of the telescope. 
An array of 1056 half-wave dipoles in front of a $90^{o}$ corner reflector 
forms the primary feed of the telescope. The 1056 dipoles are in groups of 
48. The signals recieved by these groups are added in phase to form 22 group 
outputs, each known as a module. The telescope is divided into northern part 
and southern part. The northern modules are designated as N1 to N11 and the 
southern modules as S1 to S11. The beam width due to each module is $2^o.3$ 
in east-west and $2^o.2 sec(\delta)$ in the north-south, where $\delta$ 
is the declination. This forms the observing mode and beam for the current 
project of RL observations.\\

The RRL observations were made using a new digital backend(Prabu 2010) 
which could be operated in narrow band mode or broad band(10MHz) mode. 
The narrow band mode was the spectral line mode which provided a BW 
of 1.2 MHz. This small BW restricted the observation of only 1 RL at a 
time. The transistion selected was $H271\alpha$ which has a rest frequency 
of 328.5958MHz. ORT has a large front end BW and the NDB's 1.2 MHz 
BW could have accomodated any of the near by RLs. But due to non 
availability of a broad band amplifier for the local oscillator(LO) use 
of ORT's dedicated amplifier with a -3 dB gain BW of 2MHz was employed. 
This restricted the freedom of deviation from the ORT's central LO feed 
frequency of 296.5 MHz. A symbolic block diagram of the instrument  
set up is shown in Figure 1. The observations were carried out using dual 
frequency switching with a shift of $\Delta\nu_{s}$=300 or 400kHz. 
This magnitude of shift ensured that there was no overlap of the 
associated carbon RL C271$\alpha$ with H271$\alpha$ between 2 shifted 
spectra. The $V_{LSR}$ difference between C271$\alpha$ and H271$\alpha$ 
is $\sim$150 km$s^{-1}$. With this arrangement a spectral BW of 
approximately 300 km$s^{-1}$ in $V_{LSR}$ could be covered. At a frequency of 
328 MHz differential frequency roughly transforms into differential 
velocity according to $\Delta V = c\cdot\Delta\nu/\nu$. \\

The exact BW of the spectral line mode was decided by the sampling 
frequency of NDB which was nearly 2.45 MHz, giving a BW of half of 
sampling frequency as decided by Nyquist sampling rate, 1.225MHz. 
The resolution of the observed band would then depend upon the number 
of FFT points performed on the data,

\begin{equation}
\Delta \nu_{res} = \frac{BW}{n_{FFT}}.
\end{equation}  

In the present case $n_{FFT}$ = 256 throughout the observations, so 
$\Delta \nu_{res}$ = 4.785 kHz or $\Delta V_{res}$ = 4.37 km s$^{-1}$. 
This resolution is acceptable for hydrogen 
RL which in the present observations was of primary interest. However 
the same cannot be said for carbon RL. Being heavy and considering 
its origin from cold regions its line width could be completely 
contained within this resolution. So the carbon RL is considerably 
smoothed out.\\

\begin{figure}[ht]
\begin{center}
\includegraphics[width=120mm,height=54mm,angle=0]{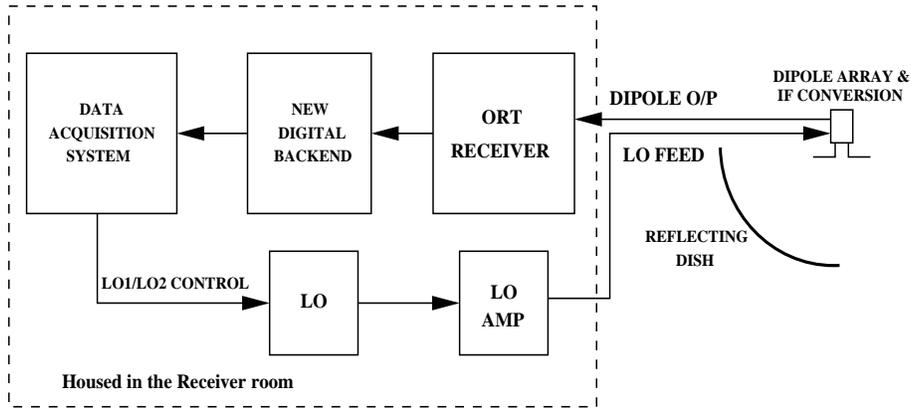}
 \caption{Symbolic block diagram of ORT instrument setup for H271$\alpha$ RL observation.}
\end{center}
\end{figure}

\section{Data Analysis and Calibration}
The frequency switching per second resulted in two sets of power spectra 
corresponding to different settings of LO, LO1 and LO2. Conventionally spectra 
corresponding to LO1 are called $T_{on}$ and the other as 
$T_{off}$. A simple $T_{on}/T_{off} - 1$ eliminated the background continuum 
power simultaneously correcting for the gain variation across the band. 
A folding of $T_{on}/T_{off} - 1$ would average the switched spectra further 
giving a 1/$\sqrt{2}$ improvement in $rms$. Due to presence of interference the 
spectra($T_{on}/T_{off} - 1$) had to be intermittently inspected during 
averaging. This was done using an algorithm(Baddi 2011a) which detected 
interference and clipped it. The clipped portion was replaced by noise 
of equivalent standard deviation corresponding to the spectrum plus a 
baseline connecting the average of a few channel values adjacent to the 
two sides of the interference infected region. ORT data was processed in 
this manner. Further the folded spectra were corrected for baselines by 
polynomial fitting avoiding the regions of astronomical lines. Calibration 
of the spectra was done by performing power measurements on the source and 
cold sky. $T_{on}$ or $T_{off}$ is a measure of continuum power. $T_{on}/T_{off} - 1$ 
gives the line temperature in units of ($T_L/(T_C + T_{e})$). 
To express the line in $^o$K it is necessary to know the value of continuum 
temperature $T_C$. When the telescope is pointed towards a source the power 
level in $T_{on}$ or $T_{off}$ also includes the electronic+spill over 
contribution $T_e$. With this the temperature($T_{onsrc}$) corresponding to 
$T_{on}$ or $T_{off}$ when the telescope is on the source is, 

\begin{equation}
T_{onsrc} = T_C' + T_e = 0.65 T_C + T_e .
\end{equation}   

Where 0.65 is the beam efficiency of ORT. Similarly when the telescope 
is pointed towards the cold sky(sufficiently away from the source towards 
a cold region in the sky keeping the declination constant) we have,

\begin{equation}
T_{offsrc} = 0.65 T_{cold sky} + T_e .
\end{equation}

Using these measurements(which are power levels in dBm) the continuum temperature 
$T_C$ can be obtained as,

\begin{equation}
T_{C} = \frac{123}{0.65} \left[\frac{T_{onsrc}}{T_{offsrc}} - 1 \right]  + T_{cold sky} .
\end{equation}

where $T_{cold sky}$ = 36K and $T_e$ = 100K. This value for $T_e$ also includes the 
spill over contribution. A useful expression for $T_C$ in terms of measured power 
P in dBm is,

\begin{equation}
T_{C} = \frac{123}{0.65} \left[10^{\frac{{P_{onsrc} - P_{offsrc}}}{10}} - 1 \right]  + T_{cold sky} .
\end{equation}

Now calibration is given by,
\begin{equation}
\frac{T_L}{T_C} = \frac{T_{on} - T_{off}}{T_{off}}\left[ \frac{T_C + T_e}{T_C} \right] ~;~ T_L = \frac{T_L}{T_C} T_C.
\end{equation}

The final spectrum is multiplied by $T_C + T_e$ to calibrate the line in $^o$K. 
Measured temperatures towards all the positions have been shown in Figure 2. $T_C$ measurements 
in the plane(b=0$^o$) are in very good agreement with previous observations(Roshi \& Anantharamaiah 2000).

\begin{figure}[ht]
\begin{center}
\includegraphics[width=130mm,height=90mm,angle=0]{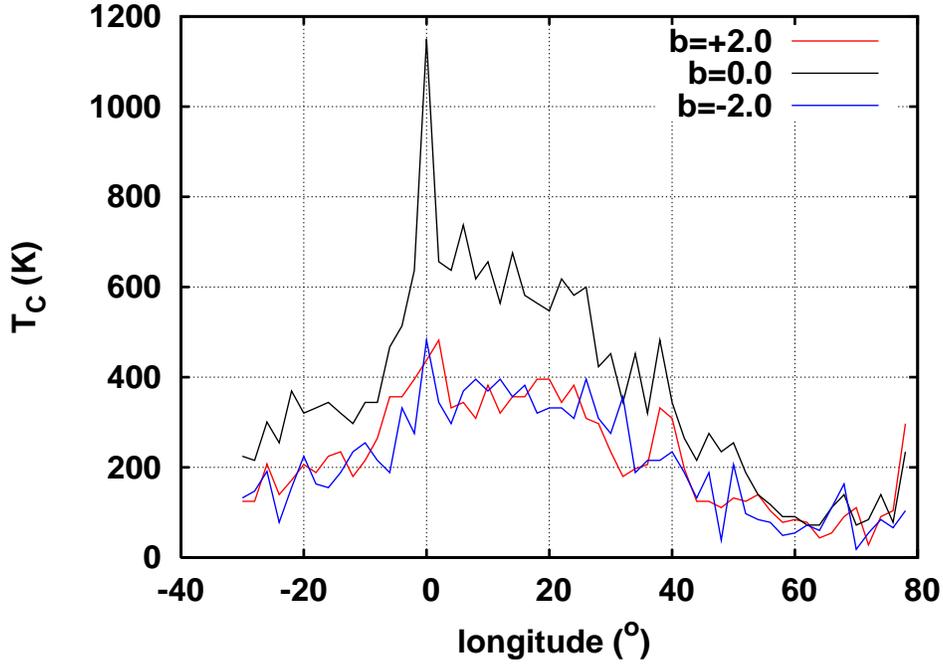}
 \caption{Continuum temperature measured towards all the directions, $-32^o < l < +80^o$ and $b<\pm3^o$ 
at 328 MHz and $2^o$ resolution. The measured temperatures in the Galactic plane(b=0.0) are in very 
good agreement with earlier observations(Roshi \& Anantharamaiah 2000). }
\end{center}
\end{figure}

The final calibrated spectra obtained towards all the positions in the 
Galactic region $-32^o < l < +80^o$ and $b<\pm3^o$ have been displayed 
in Figure 3 to 8. The gaussian parameters fitted to these spectra have 
been given in Table 1.

\section{Distribution of ELDWIM in the inner Galaxy}
Kinematical distances to ELDWIM clouds were obtained from a Galactic rotation 
curve(Sofue et al. 2009) using the $V_{LSR}$ of H271$\alpha$ RLs. Distribution 
of these clouds in the plane(b=$0^o$) of the Galaxy has been shown in Figure 9. 
Due to observed $V_{LSR}$ of clouds above and below the plane there is a similar 
distribution in these regions as well. The width of the lines indicates an 
upper limit on the spread of gas along the line of sight. The FWHM of 
the hydrogen lines mostly lie within 20-60 km$s^{-1}$. The profiles also seem to lack 
pressure broadening(Brocklehurst \& Leeman 1971; Shaver 1975) due 
to absence of extended wings. All the profiles are compatible with a gaussian fit. From 
this, one can deduce that the number density of electrons $n_e$ to have an upper 
cutoff of 10$cm^{-3}$(Baddi 2011b; Baddi 2012) in ELDWIM. At this density and frequency 
the pressure broadening contribution is $\sim$ 10 km$s^{-1}$. The average error on 
the hydrogen line widths is contained within this.  

\section{Acknowledgement}
The author thanks T.Prabu and D.Anish Roshi who helped in doing the observations 
by providing the new system for spectral line observations, related software and 
suggestions. The author thanks the staff of Radio Astronomy Center TIFR who have 
made these observations possible. Ooty Radio Telescope is operated by National 
Center for Radio Astrophysics TIFR. \\

The author thanks the refree for suggestions and comments that improved the presentation 
of the paper. \\

\begin{deluxetable}{cccccccccccccccccccc}

\tablecaption{Gaussian parameters for profiles in Figure 3 to 8, the values in 
the paranthesis are errors. Comments: C-carbon line, H-hydrogen line, 
ND-no data, NLD-no line detection. $D_c$ flags : $f$-far, $n$-near, 
$m$-distance for maximum $V_{LSR}$ possible from rotation curve. Positions 
with errors in $V_{LSR}$ greater than 50\% have been provided with an upper 
limit between bars taking into account the full value of errors, especially 
near $l=0^o$. }
 \startdata
 \multicolumn{2}{c}{Source} &&& $T_{L}$ &&& $V_{LSR}$ &&& $\Delta V_{LSR}$ &&& $T_c$ &&& $D_c$ &&& Comments \\ \cline{1-2}
   $l^{o}$ & $b^{o}$ &&& (mK)  &&& (km$s^{-1}$) &&& (km$s^{-1}$) &&& (K) &&& kpc&&&\\ \hline
   & &&&  &&&  &&&  &&&  &&& &&&\\ 
   78&+2 &&& 83(20) &&& -138(3) &&& 26(7) &&& 297 &&& &&&C \\
     &   &&& 182(13)&&& 0(2) &&& 64(5) &&&  &&& $3.7^f$&&&H \\
   78&0 &&& 78(34) &&& -137(7) &&& 31(15) &&& 234 &&& &&&C\\
     &  &&& 155(29) &&& 2(4) &&& 42(9) &&& &&& $3.5^f$&&&H\\
   78&-2 &&& - &&& - &&& - &&& 104 &&& &&&H\\
   76&+2 &&& - &&& - &&& - &&& 104 &&& &&&NLD\\
   76&0 &&& 45(12) &&& 0(6) &&& 44(13) &&& 78 &&& $4.3^f$&&&H\\
   76&-2 &&& - &&& - &&& - &&& 66 &&& &&&NLD\\
   74&+2 &&& - &&& - &&& - &&& 91 &&& &&&NLD\\
   74&0 &&& - &&& - &&& - &&& 140 &&& &&&NLD\\
   74&-2 &&& - &&& - &&& - &&& 84 &&& &&&NLD\\
   72&+2 &&& - &&& - &&& - &&& 28 &&& &&&NLD\\
   72&0 &&& 40(10) &&& 0(4) &&& 32(9) &&& 84 &&& $5.4^f$&&&H?\\
   72&-2 &&& - &&& - &&& - &&& 55 &&& &&&NLD\\
   70&+2 &&& - &&& - &&& - &&& - &&& &&&ND\\
   70&0 &&& - &&& - &&& - &&& 72 &&& &&&NLD\\
   70&-2 &&& - &&& - &&& - &&& 18 &&& &&&NLD\\
   68&+2 &&& - &&& - &&& - &&& 91 &&& &&&NLD\\
   68&0 &&& - &&& - &&& - &&& 140 &&& &&&NLD\\
   68&-2 &&& - &&& - &&& - &&& 163 &&& &&&NLD\\
   66&+2 &&& - &&& - &&& - &&& 55 &&& &&&NLD\\
   66&0 &&& - &&& - &&& - &&& 111 &&& &&&NLD\\
   66&-2 &&& - &&& - &&& - &&& 111 &&& &&&NLD\\
   64&+2 &&& - &&& - &&& - &&& 44 &&& &&&NLD\\
   64&0 &&& - &&& - &&& - &&& 72 &&& &&&NLD\\
   64&-2 &&& - &&& - &&& - &&& 60 &&& &&&NLD\\ 
   62&+2 &&& - &&& - &&& - &&& 78 &&& &&&NLD\\
   62&0 &&& - &&& - &&& - &&& 72 &&& &&&NLD\\
   62&-2 &&& - &&& - &&& - &&& 72 &&& &&&NLD\\
   60&+2 &&& - &&& - &&& - &&& 84 &&& &&&NLD\\
   60&0 &&& 32(8) &&& -135(5) &&& 40(11) &&& 91 &&& &&&C\\
   & &&& 37(7) &&& 16(5) &&& 56(12) &&&  &&& $1.0^n$&&&H\\
   60&-2 &&& - &&& - &&& - &&& 55 &&& &&&NLD\\
   58&+2 &&& - &&& - &&& - &&& 78 &&& &&&NLD\\
   58&0 &&& - &&& - &&& - &&& 91 &&& &&&ND\\
   58&-2 &&& - &&& - &&& - &&& 49 &&& &&&NLD\\
   56&+2 &&& - &&& - &&& - &&& 104 &&& &&&NLD\\
   56&0 &&& - &&& - &&& - &&& 118 &&& &&&ND\\
   56&-2 &&& - &&& - &&& - &&& 78 &&& &&&NLD\\
   54&+2 &&& - &&& - &&& - &&& 140 &&& &&&NLD\\
   54&0 &&& - &&& - &&& - &&& 140 &&& &&&ND\\
   54&-2 &&& - &&& - &&& - &&& 84 &&& &&&NLD\\
   52&+2 &&& - &&& - &&& - &&& 125 &&& &&&NLD\\
   52&0 &&& - &&& - &&& - &&& 188 &&& &&&ND\\
   52&-2 &&& - &&& - &&& - &&& 97 &&& &&&NLD\\
   50&+2 &&& - &&& - &&& - &&& 132 &&& &&&NLD\\
   50&0 &&& 108(18) &&& 68(3) &&& 41(8) &&& 254 &&& $5.5^m$&&&H\\
   50&-2 &&& - &&& - &&& - &&& 206 &&& &&&NLD\\
   48&+2 &&& - &&& - &&& - &&& 111 &&& &&&NLD\\
   48&0 &&& - &&& - &&& - &&& 234 &&& &&&NLD\\
   48&-2 &&& - &&& - &&& - &&& 38 &&& &&&NLD\\
   46&+2 &&& - &&& - &&& - &&& 125 &&& &&&NLD\\
   46&0 &&& 112(26) &&& 78(3) &&& 28(7) &&& 275 &&& $5.9^m$&&&H\\  
   46&-2 &&& - &&& - &&& - &&& 188 &&& &&&NLD\\
   44&+2 &&& - &&& - &&& - &&& 125 &&& &&&NLD\\
   44&0 &&& - &&& - &&& - &&& 215 &&& &&&NLD\\
   44&-2 &&& - &&& - &&& - &&& 132 &&& &&&NLD\\
   42&+2 &&& - &&& - &&& - &&& 197 &&& &&&NLD\\
   42&0 &&& - &&& - &&& - &&& 265 &&& &&&NLD\\
   42&-2 &&& - &&& - &&& - &&& 188 &&& &&&NLD\\
   40&+2 &&& - &&& - &&& - &&& 308 &&& &&&NLD\\
   40&0 &&& - &&& - &&& - &&& 344 &&& &&&NLD\\
   40&-2 &&& - &&& - &&& - &&& 234 &&& &&&NLD\\
   38&+2 &&& 126(22) &&& -132(3) &&& 34(7) &&& 332 &&& &&&C\\
   & &&& 106(28) &&& 36(3) &&& 21(6) &&&  &&& $2.0^n$&&&H \\
   38&0 &&& - &&& - &&& - &&& 482 &&& &&&NLD\\
   38&-2 &&& - &&& - &&& - &&& 215 &&& &&&NLD\\
   36&+2 &&& 75(15) &&& -132(4) &&& 40(10) &&& 206 &&& &&&C \\
   & &&& 78(17) &&& 95(4) &&& 33(8) &&&  &&& $6.2^n$ &&&H\\
   36&0 &&& 69(12) &&& -107(6) &&& 66(13) &&& 320 &&& &&&C\\
   & &&& 119(15) &&& 65(3) &&& 42(6) &&&  &&& $3.8^n$ &&&H \\
   36&-2 &&& - &&& - &&& - &&& 215 &&& &&&NLD\\
   34&+2 &&& - &&& - &&& - &&& 197 &&& &&&NLD\\
   34&0 &&& 153(22) &&& -121(4) &&& 58(9) &&& 452 &&& &&&C\\
   & &&& 131(18) &&& 59(6) &&& 84(13) &&&  &&& $3.5^n$ &&&H\\
   34&-2 &&& - &&& - &&& - &&& 188 &&& &&&NLD\\
   32&+2 &&& 92(15) &&& 89(2) &&& 27(5) &&& 180 &&& $5.1^n$ &&&H \\
   32&0 &&& 213(18) &&& 90(2) &&& 48(5) &&& 344 &&& $5.2^n$ &&&H\\
   32&-2 &&& 150(50) &&& 83(4) &&& 25(10) &&& 357 &&& $4.8^n$ &&&H\\
   30&+2 &&& 38(11) &&& 81(7) &&& 46(15) &&& 234 &&& $4.7^n$ &&&H\\  
   30&0 &&& 325(35) &&& 91(2) &&& 46(6) &&& 452 &&& $5.2^n$ &&& H\\
   30&-2 &&& 68(17) &&& 87(4) &&& 32(9) &&& 357 &&& $5.0^n$ &&& H\\
   28&+2 &&& 111(26) &&& 83(4) &&& 35(9) &&& 297 &&& $4.8^n$ &&& H\\
   28&0 &&& 296(28) &&& 39(-) &&& 30(-) &&& 423 &&& $2.5^n$ &&& H\\
   & &&& 543(33) &&& 87(-) &&& 35(-) &&& &&& $4.9^n$ &&& H\\
   & &&& 92(40) &&& 104(-) &&& 23(-) &&&  &&& $5.8^n$ &&& H\\
   28&-2 &&& 65(17) &&& 90(5) &&& 37(11) &&& 308 &&& $5.1^n$ &&& H\\
   26&+2 &&& 161(37) &&& 86(3) &&& 30(8) &&& 308 &&& $2.0^n$ &&& H\\
   26.5&0 &&& 92(26) &&& -70(4) &&& 30(10) &&& 500 &&&  &&& C\\
   & &&& 50(30) &&& 21(-) &&& 15(-) &&&  &&& $1.3^n$ &&& H\\
   & &&& 166(24) &&& 67(-) &&& 25(-) &&&  &&& $4.0^n$ &&& H\\
   & &&& 349(21) &&& 98(-) &&& 33(-) &&&  &&& $5.5^n$ &&& H\\
   26&-2 &&& 113(21) &&& 86(4) &&& 43(9) &&& 396 &&& $4.9^n$ &&& H\\
   24&+2 &&& 122(22) &&& 87(4) &&& 41(9) &&& 382 &&& $5.0^n$ &&& H\\
   24&0 &&& 97(23) &&& -105(9) &&& 73(21) &&& 582 &&& &&& C?\\
   & &&& 262(27) &&& 82(3) &&& 56(7) &&&  &&& $4.8^n$ &&& H\\
   24&-2 &&& 75(17) &&& 86(4) &&& 39(10) &&& 308 &&& $5.0^n$ &&& H\\
   22&+2 &&& 132(25) &&& 65(5) &&& 56(12) &&& 344 &&& $4.2^n$ &&& H\\
   22&0 &&& 180(35) &&& 76(4) &&& 45(10) &&& 618 &&& $4.6^n$ &&& H\\
   22&-2 &&& - &&& - &&& - &&& 332 &&& &&& NLD\\
   20&+2 &&& 147(37) &&& 26(3) &&& 27(8) &&& 396 &&& $2.1^n$ &&& H\\
   20&0 &&& 113(52) &&& -115(4) &&& 17(10) &&& 547 &&&  &&& C\\
   & &&& 188(32) &&& 51(4) &&& 46(9) &&&  &&& $3.7^n$ &&& H\\
   20&-2 &&& 43(22) &&& -138(7) &&& 26(16) &&& 332 &&& &&& C\\
   & &&& 87(14) &&& 60(5) &&& 70(13) &&& &&& $4.1^n$ &&& H\\
   18&+2 &&& 72(17) &&& -128(10) &&& 84(23) &&& 396 &&& &&&C\\
   & &&& 227(27) &&& 28(2) &&& 33(5) &&&  &&& $2.4^n$ &&& H\\   
   18&0 &&& 110(33) &&& -115(10) &&& 69(24) &&& 564 &&&  &&& C\\
   & &&& 282(32) &&& 44(4) &&& 71(9) &&&  &&& $3.5^n$ &&& H\\
   18&-2 &&& 66(14) &&& 47(8) &&& 77(19) &&& 320 &&& $3.6^n$ &&& H\\
   16&+2 &&& 102(22) &&& -132(4) &&& 40(10) &&& 357 &&& &&& C\\
   & &&& 179(23) &&& 29(2) &&& 34(5) &&&  &&& $2.7^n$ &&& H\\
   16&0 &&& 162(29) &&& -128(3) &&& 35(7) &&& 582 &&& &&& C\\
   & &&& 337(26) &&& 31(2) &&& 41(4) &&&  &&& $2.8^n$ &&& H\\
   16&-2 &&& 177(31) &&& -151(2) &&& 29(6) &&& 382 &&& &&& C\\
   & &&& 121(26) &&& 32(4) &&& 41(10) &&&  &&& $2.9^n$ &&& H\\
   14&+2 &&& 108(26) &&& 29(4) &&& 34(9) &&& 357 &&& $2.9^n$ &&& H\\
   14&0 &&& 243(27) &&& -122(2) &&& 42(5) &&& 676 &&&  &&& C\\
   & &&& 448(27) &&& 29(1) &&& 41(3) &&&  &&& $2.9^n$ &&& H\\
   14&-2 &&& 122(29) &&& -149(3) &&& 26(7) &&& 357 &&& &&& C\\
   & &&& 94(25) &&& 30(5) &&& 36(11) &&&  &&& $3.0^n$ &&& H\\
   12&+2 &&& 108(23) &&& -150(2) &&& 21(5) &&& 320 &&& &&& C\\
   & &&& 92(19) &&& 21(3) &&& 33(8) &&&  &&& $2.5^n$ &&& H\\
   12&0 &&& 178(28) &&& -126(4) &&& 47(9) &&& 564 &&& &&& C\\
   & &&& 354(29) &&& 32(2) &&& 44(4) &&&  &&& $3.5^n$ &&& H\\
   12&-2 &&& 53(20) &&& -146(11) &&& 58(25) &&& 396 &&& &&& C\\
   & &&& 127(24) &&& 25(4) &&& 40(9) &&&  &&& $2.9^n$ &&& H\\
   10&+2 &&& 69(42) &&& -142(6) &&& 18(13) &&& 382 &&& &&& C\\
   & &&& 102(34) &&& 20(5) &&& 28(11) &&&  &&& $2.8^n$ &&& H \\
   10&0 &&& 172(26) &&& -126(3) &&& 43(8) &&& 656 &&& &&& C\\
   & &&& 381(26) &&& 21(1) &&& 44(3) &&&  &&& $2.9^n$ &&& H\\
   10&-2 &&& 65(21) &&& -143(8) &&& 49(18) &&& 369 &&& &&& C\\
   & &&& 192(25) &&& 23(2) &&& 33(5) &&&  &&& $3.1^n$ &&& H\\
   8&+2 &&& 80(33) &&& 17(4) &&& 21(10) &&& 308 &&& $2.9^n$ &&& H\\
   8&0 &&& 148(19) &&& -139(4) &&& 63(9) &&& 618 &&& &&& C\\
   & &&& 328(25) &&& 19(1) &&& 36(3) &&&  &&& $3.1^n$ &&& H\\
   8&-2 &&& 85(36) &&& -142(4) &&& 20(10) &&& 396 &&& &&& C\\
   & &&& 153(33) &&& 17(2) &&& 23(6) &&&  &&& $2.9^n$ &&& H\\
   6&+2 &&& 44(49) &&& -134(9) &&& 16(21) &&& 344 &&& &&& C\\
   & &&& 70(49) &&& 11(6) &&& 16(13) &&&  &&& $2.6^n$ &&& H\\
   6&0 &&& 268(39) &&& -145(4) &&& 58(10) &&& 738 &&& &&& C\\
   & &&& 441(53) &&& 18(2) &&& 31(4) &&&  &&& $3.6^n$ &&& H\\
   6&-2 &&& 105(24) &&& -148(5) &&& 48(13) &&& 369 &&& &&& C\\
   & &&& 156(35) &&& 14(2) &&& 23(6) &&&  &&& $3.1^n$ &&& H\\
   4&+2 &&& 112(33) &&& 9(4) &&& 30(10) &&& 332 &&& 3.0 &&& H\\
   4&0 &&& 164(34) &&& -141(4) &&& 37(9) &&& 637 &&& &&& C\\
   & &&& 240(39) &&& 9(2) &&& 29(5) &&&  &&& $3.0^n$ &&& H\\
   4&-2 &&& 68(28) &&& -145(3) &&& 16(8) &&& 297 &&& &&& C\\
   & &&& 97(27) &&& 15(2) &&& 18(6) &&&  &&& $4.1^n$ &&& H\\
   2&+2 &&& - &&& - &&& - &&& 482 &&& &&& NLD\\
   2&0 &&& 223(52) &&& -151(2) &&& 22(6) &&& 656 &&& &&& C\\
   & &&& 270(51) &&& 0(2) &&& 22(5) &&&  &&& $|1.5|^n$ &&& H\\
   2&-2 &&& 94(23) &&& 20(6) &&& 48(13) &&& 344 &&& $6.2^n$ &&& H\\
   0&+2 &&& 79(28) &&& -142(7) &&& 42(17) &&& 437 &&& &&& C\\
   & &&& 197(33) &&& 12(3) &&& 31(6) &&&  &&& - &&& H\\
   0&0 &&& 345(74) &&& -149(2) &&& 20(5) &&& 1150 &&& &&& C\\
   & &&& 680(149) &&& 0(2) &&& 24(5) &&&  &&& - &&& H\\
   & &&& 140(57) &&& 33(22) &&& 42(42) &&&  &&& - &&& H\\
   0&-2 &&& 122(25) &&& -145(5) &&& 53(13) &&& 482 &&& &&& C\\
   & &&& 148(30) &&& 5(4) &&& 37(9) &&&  &&& - &&& H\\
   -2&+2 &&& 57(42) &&& -3(7) &&& 18(15) &&& 396 &&& $2.0^n$ &&& H\\
   -2&0 &&& 189(47) &&& -155(4) &&& 29(8) &&& 656 &&& &&& C\\
   & &&& 259 &&& 0(3) &&& 37(7) &&&  &&& $|2|^n$ &&& H\\
   -2&-2 &&& 94(24) &&& -158(3) &&& 25(7) &&& 275 &&& &&& C \\
   & &&& 114(20) &&& 0(3) &&& 35(7) &&&  &&& $|2|^n$ &&& H\\
   -4&+2 &&& 147(60) &&& 0(6) &&& 29(14) &&& 357 &&& $|1.8|^n$ &&& H\\
   -4&0 &&& 188(42) &&& -153(3) &&& 24(6) &&& 514 &&& &&& C\\
   & &&& 233(33) &&& 0(3) &&& 37(6) &&&  &&& $|1.0|^n$ &&& H\\
   -4&-2 &&& 90(28) &&& -154(3) &&& 22(8) &&& 332 &&& &&& C\\
   & &&& 129(27) &&& 0(2) &&& 24(6) &&&  &&& $|1.0|^n$ &&& H\\
   -6&+2 &&& 71(30) &&& -121(7) &&& 36(17) &&& 357 &&& &&& C\\
   & &&& 169(24) &&& 0(4) &&& 54(9) &&&  &&& $|1.0|^n$ &&& H\\
   -6&0 &&& 71(36) &&& -158(9) &&& 36(21) &&& 467 &&& &&& C\\
   & &&& 203(44) &&& 0(3) &&& 24(6) &&&  &&& $|1.0|^n$ &&& H\\
   -6&-2 &&& 100(13) &&& 0(5) &&& 72(11) &&& 188 &&& $|1.2|^n$ &&& H\\
   -8&+2 &&& 147(55) &&& 0(7) &&& 36(15) &&& 265 &&& $|1.4|^n$ &&& H\\
   -8&0 &&& 69(29) &&& -159(4) &&& 20(10) &&& 344 &&& &&& C\\
    & &&& 183(22) &&& -9(2) &&& 33(5) &&&  &&& $2.0^n$ &&& H\\
   -8&-2 &&& 72(23) &&& -119(6) &&& 41(15) &&& 215 &&& &&& C?\\
   & &&& 109(20) &&& -11(5) &&& 51(11) &&&  &&& $2.0^n$ &&& H\\
   -10&+2 &&& 137(46) &&& -10(6) &&& 35(14) &&& 215 &&& $|2.3|^n$ &&& H\\
   -10&0 &&& 106(36) &&& -15(5) &&& 27(11) &&& 344 &&& $2.3^n$ &&& H\\
   -10&-2 &&& 131(36) &&& -10(5) &&& 37(12) &&& 254 &&& $|2.3|^n$ &&& H\\
   -12&+2 &&& 79(15) &&& -13(5) &&& 51(11) &&& 180 &&& $1.7^n$ &&& H\\
   -12&0 &&& - &&& - &&& - &&& 297 &&& &&& NLD\\
   -12&-2 &&& 101(62) &&& -24(19) &&& 64(45) &&& 234 &&& $|4.2|^n$ &&& H?\\
   -14&+2 &&& 113(43) &&& -8(5) &&& 29(13) &&& 234 &&& $|1.6|^n$ &&& H\\
   -14&0 &&& 135(21) &&& -17(4) &&& 49(9) &&& 320 &&& $|2.0|^n$ &&& H\\
   -14&-2 &&& - &&& - &&& - &&& 188 &&& &&& NLD\\
   -16&+2 &&& 125(43) &&& -17(8) &&& 45(18) &&& 224 &&& $1.7^n$ &&& H\\
   -16&0 &&& 154(29) &&& -20(5) &&& 49(11) &&& 344 &&& $2.0^n$ &&& H\\
   -16&-2 &&& - &&& - &&& - &&& 155 &&& &&& NLD\\
   -18&+2 &&& 104(35) &&& -10(5) &&& 32(12) &&& 188 &&& $|1.3|^n$ &&& H\\
   -18&0 &&& 135(45) &&& -154(2) &&& 12(4) &&& 332 &&& &&& C?\\
   & &&& 113(31) &&& -33(3) &&& 24(8) &&&  &&& $2.7^n$ &&& H\\
   -18&-2 &&& - &&& - &&& - &&& 163 &&& &&& ND\\
   -20&+2 &&& - &&& - &&& - &&& 206 &&& &&& NLD\\
   -20&0 &&& 78(33) &&& -40(9) &&& 41(21) &&& 320 &&& $3.0^n$ &&& H\\
   -20&-2 &&& - &&& - &&& - &&& 224 &&& &&& NLD\\
   -22&+2 &&& - &&& - &&& - &&& 171 &&& &&& ND\\
   -22&0 &&& - &&& - &&& - &&& 369 &&& &&& NLD\\
   -22&-2 &&& 50(11) &&& -29(5) &&& 46(11) &&& 155 &&& $2.1^n$ &&& H\\
   -24&+2 &&& - &&& - &&& - &&& 140 &&& &&& NLD\\
   -24&0 &&& - &&& - &&& - &&& 254 &&& &&& NLD\\
   -24&-2 &&& 25(7) &&& -33(6) &&& 44(15) &&& 78 &&& $2.3^n$ &&& H?\\
   -26&+2 &&& - &&& - &&& - &&& 207 &&& &&& NLD\\
   -26&0 &&& 175(45) &&& -56(7) &&& 59(17) &&& 301 &&& $3.6^n$ &&& H\\
   -26&-2 &&& 112(95) &&& -60(16) &&& 32(39) &&& 191 &&& $3.8^n$ &&& H\\
   & &&& 78(95) &&& -11(22) &&& 33(56) &&&  &&& $2.2^n$ &&& H\\
   -28&+2 &&& - &&& - &&& - &&& 125 &&& &&& NLD\\
   -28&0 &&& 108(29) &&& -55(7) &&& 56(17) &&& 215 &&& $3.5^n$ &&& H\\
   -28&-2 &&& 90(33) &&& -77(5) &&& 30(13) &&& 147 &&& $4.6^n$ &&& H\\
   -30&+2 &&& - &&& - &&& - &&& 125 &&&  &&& NLD\\
   -30&0 &&& 139(28) &&& -50(2) &&& 23(5) &&& 225 &&& $3.0^n$ &&& H\\
   -30&-2 &&& - &&& - &&& - &&& 132 &&& &&& NLD\\
\enddata
\end{deluxetable}

\begin{figure}[ht]
\begin{center}
\includegraphics[width=32mm,height=24mm,angle=0]{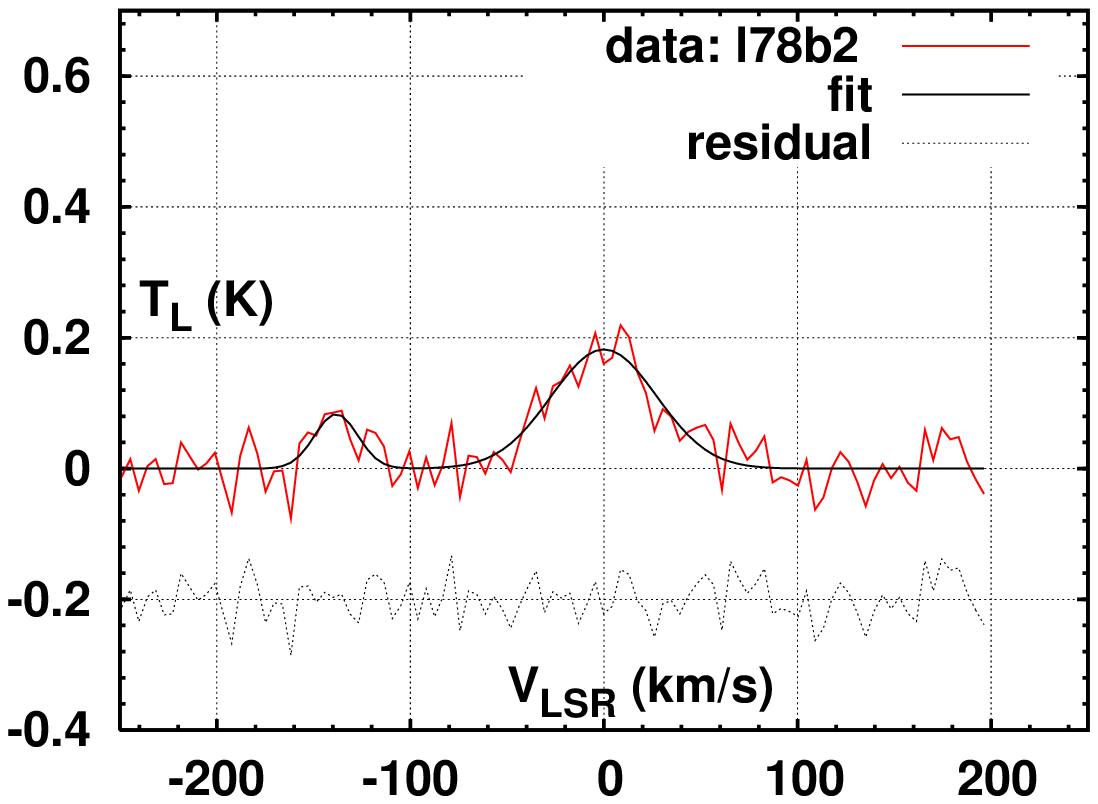}
\includegraphics[width=32mm,height=24mm,angle=0]{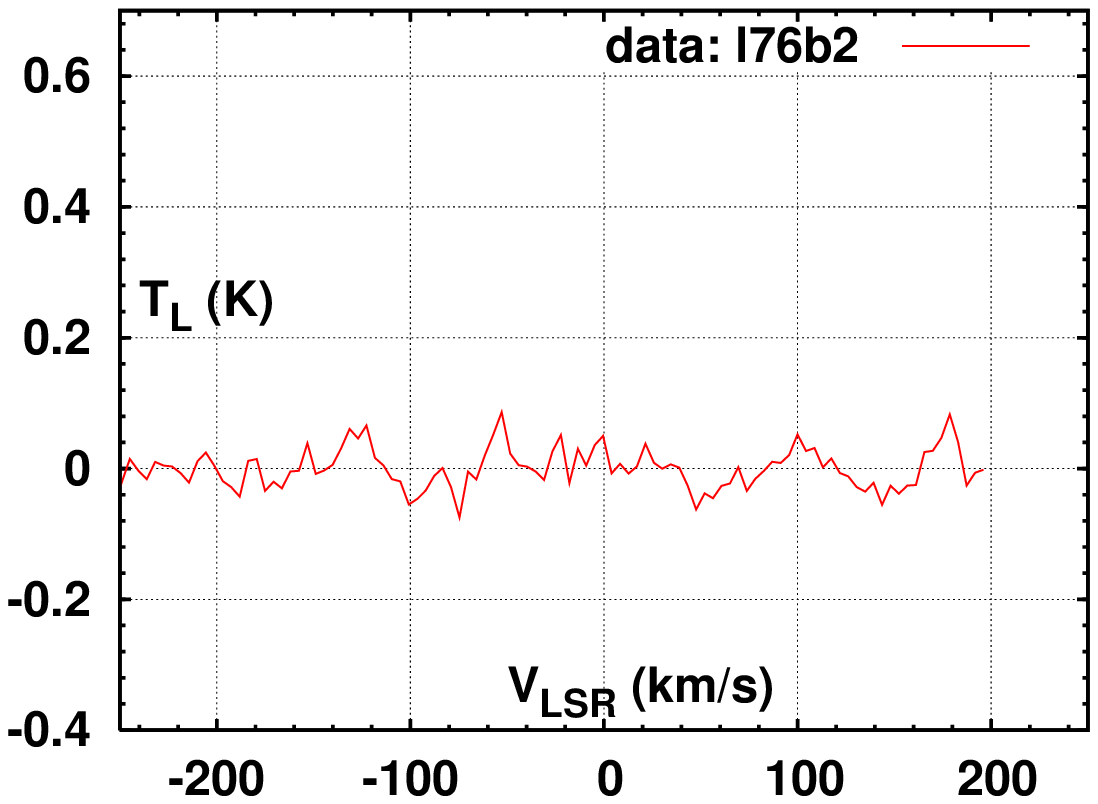}
\includegraphics[width=32mm,height=24mm,angle=0]{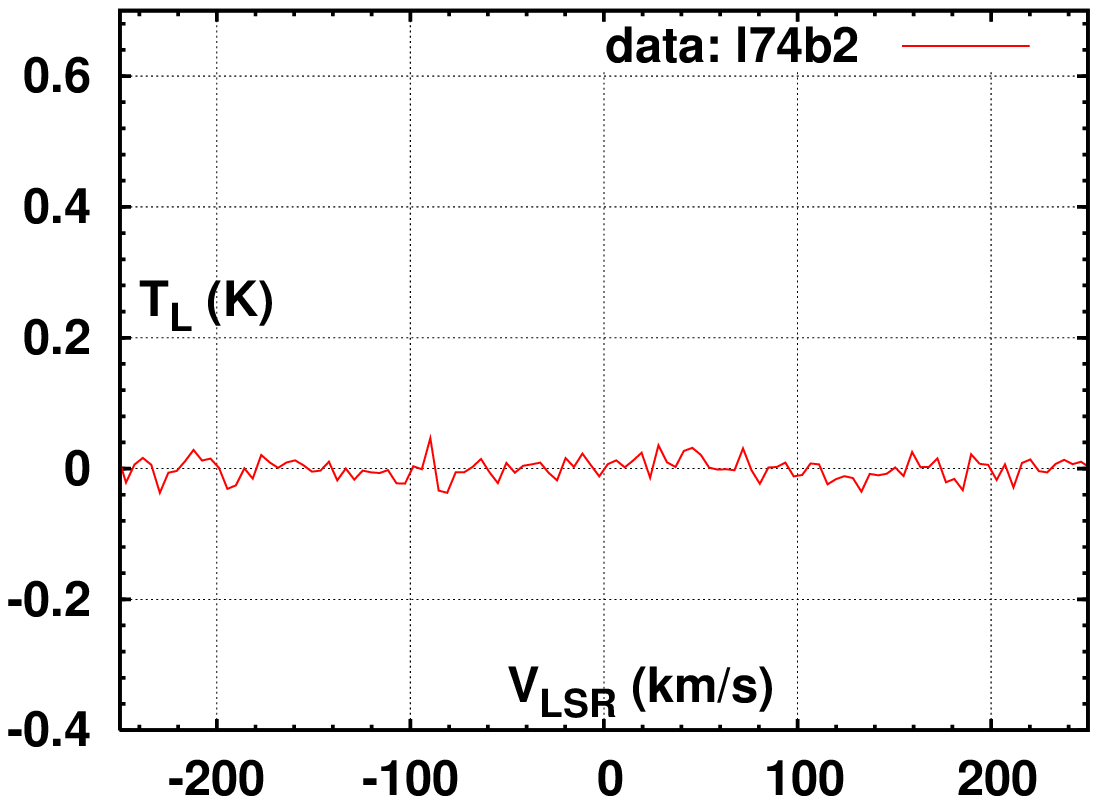}
\includegraphics[width=32mm,height=24mm,angle=0]{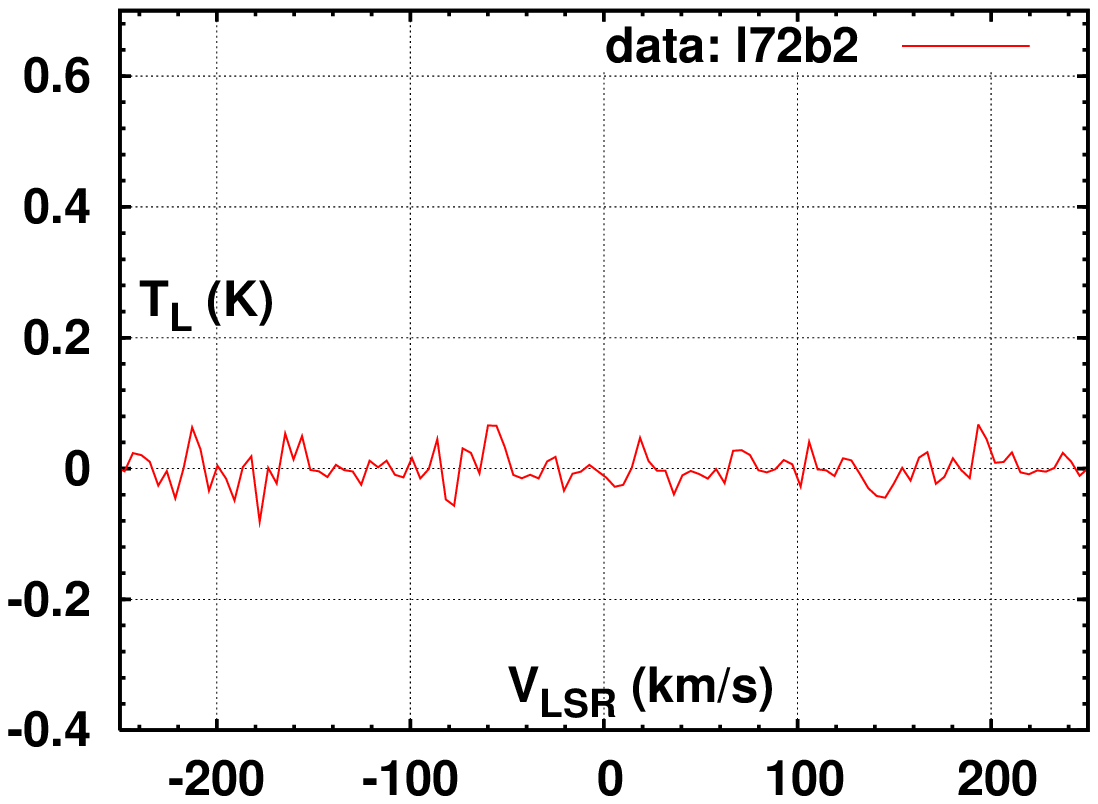}
\includegraphics[width=32mm,height=24mm,angle=0]{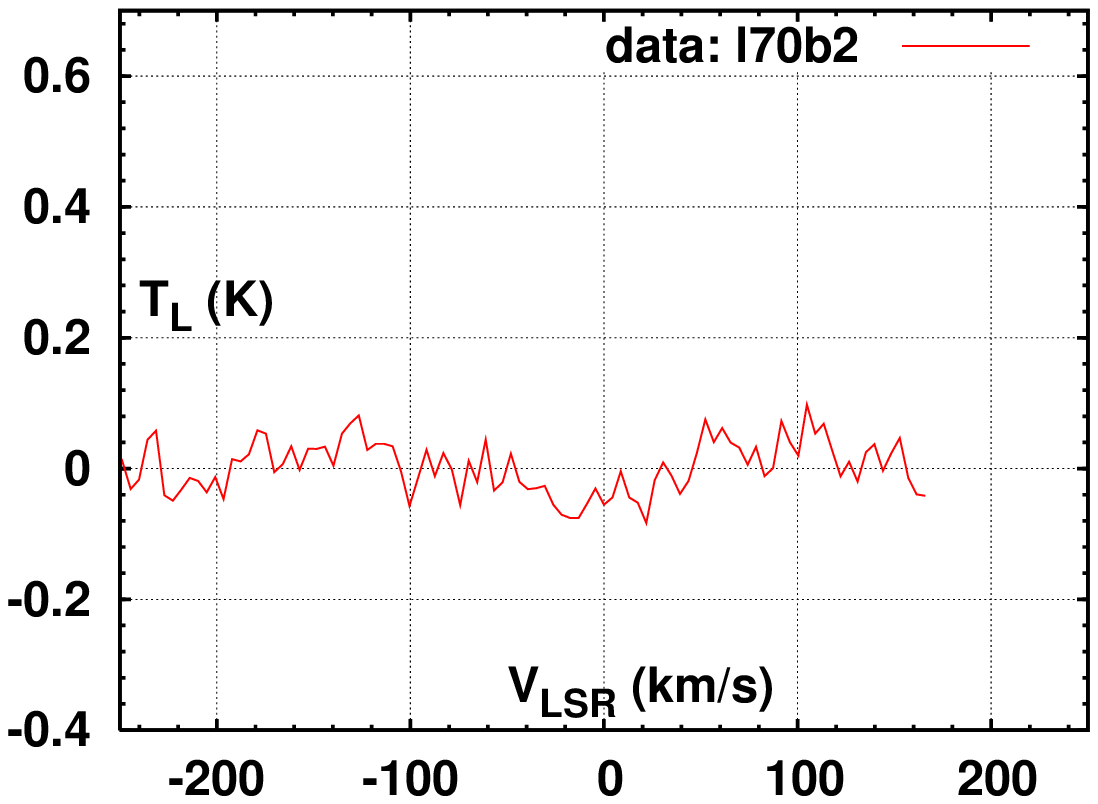}
\includegraphics[width=32mm,height=24mm,angle=0]{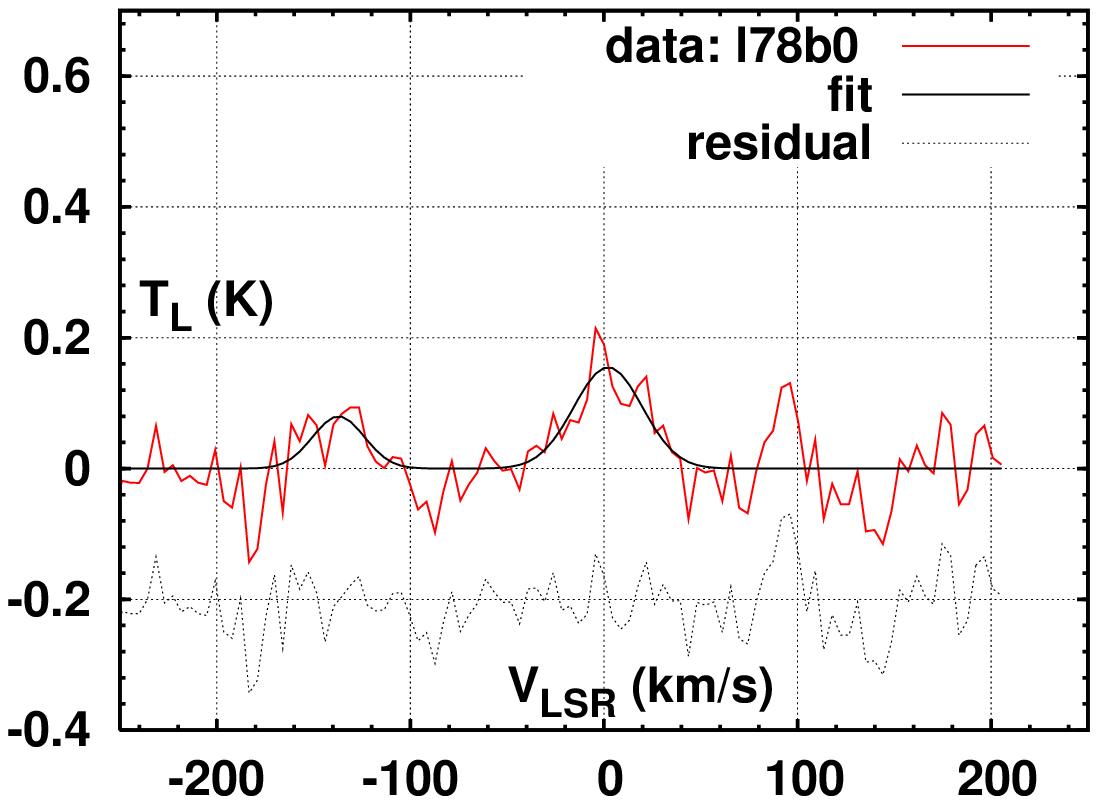}
\includegraphics[width=32mm,height=24mm,angle=0]{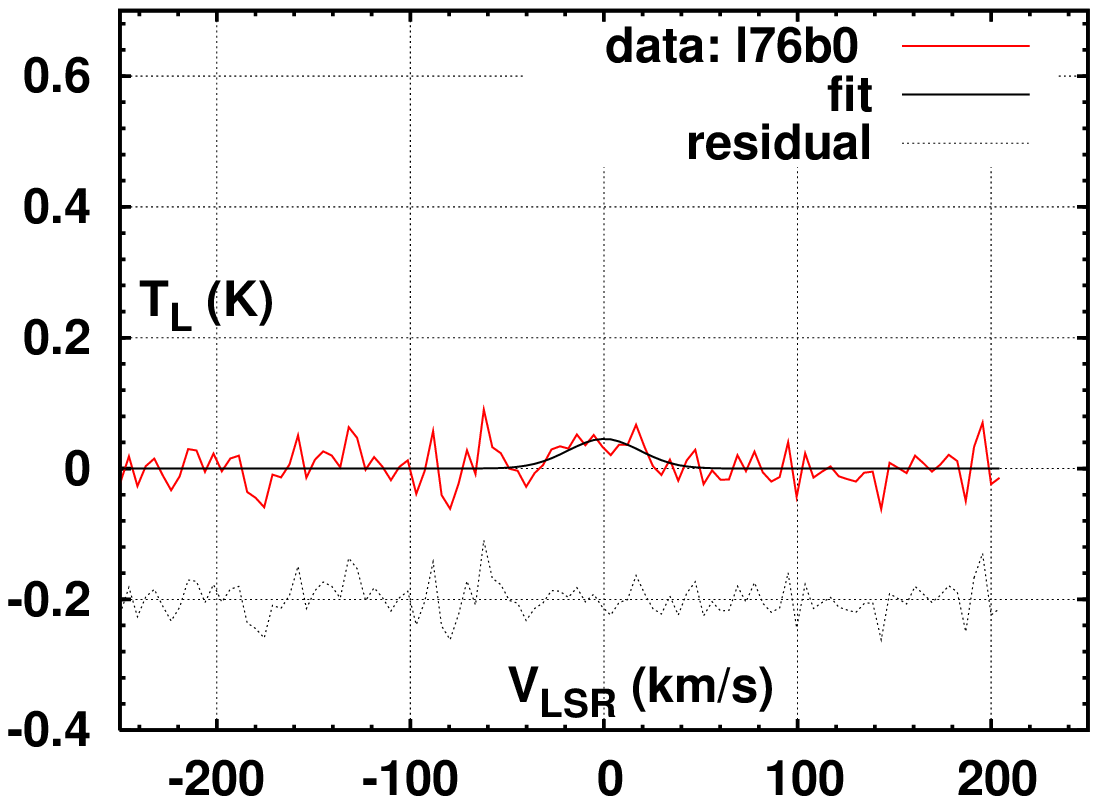}
\includegraphics[width=32mm,height=24mm,angle=0]{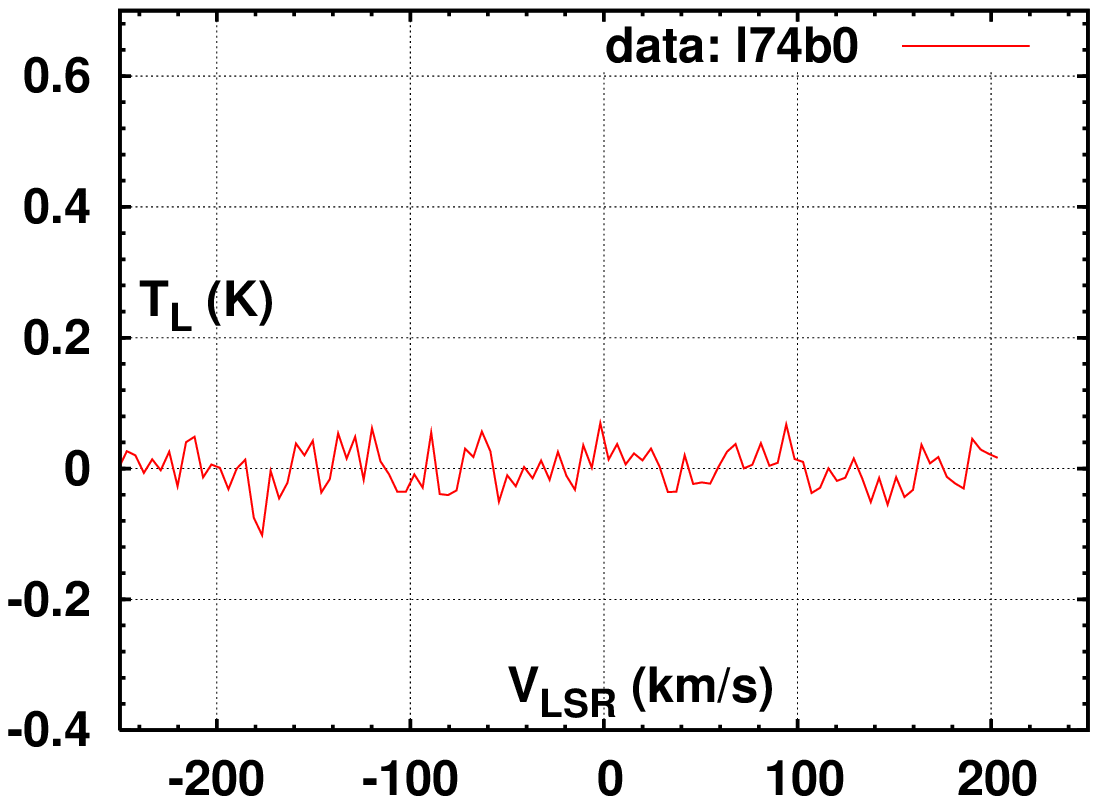}
\includegraphics[width=32mm,height=24mm,angle=0]{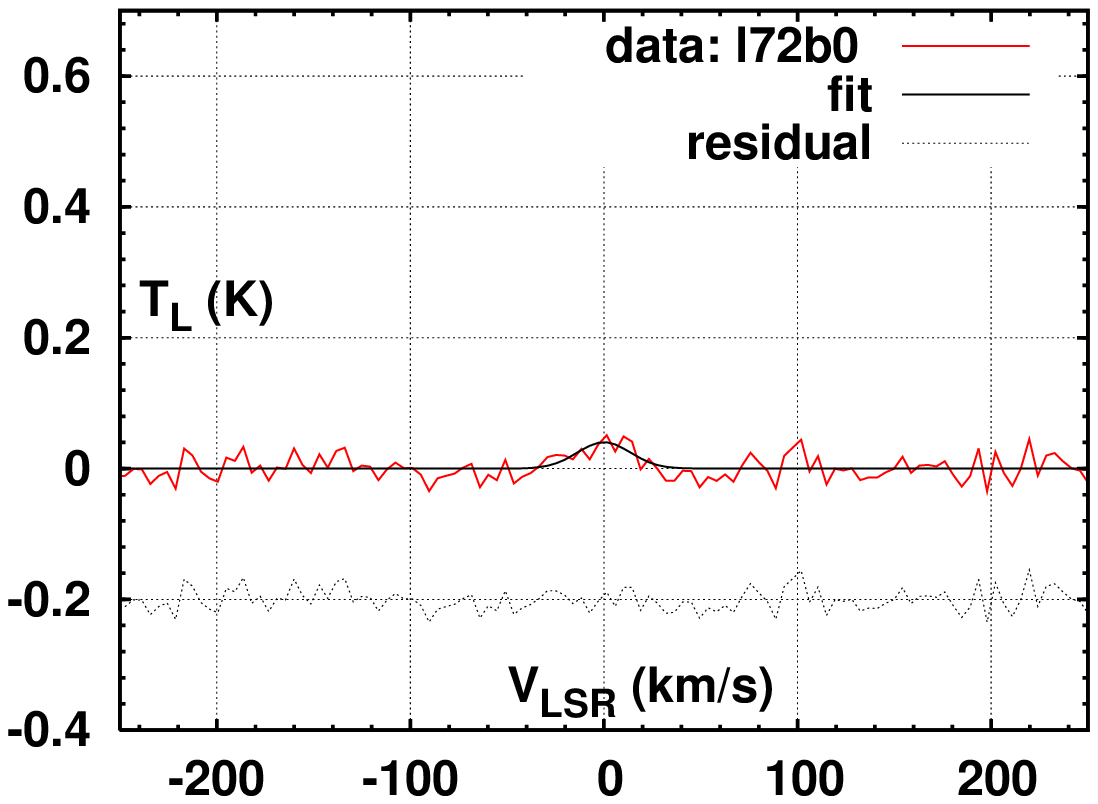}
\includegraphics[width=32mm,height=24mm,angle=0]{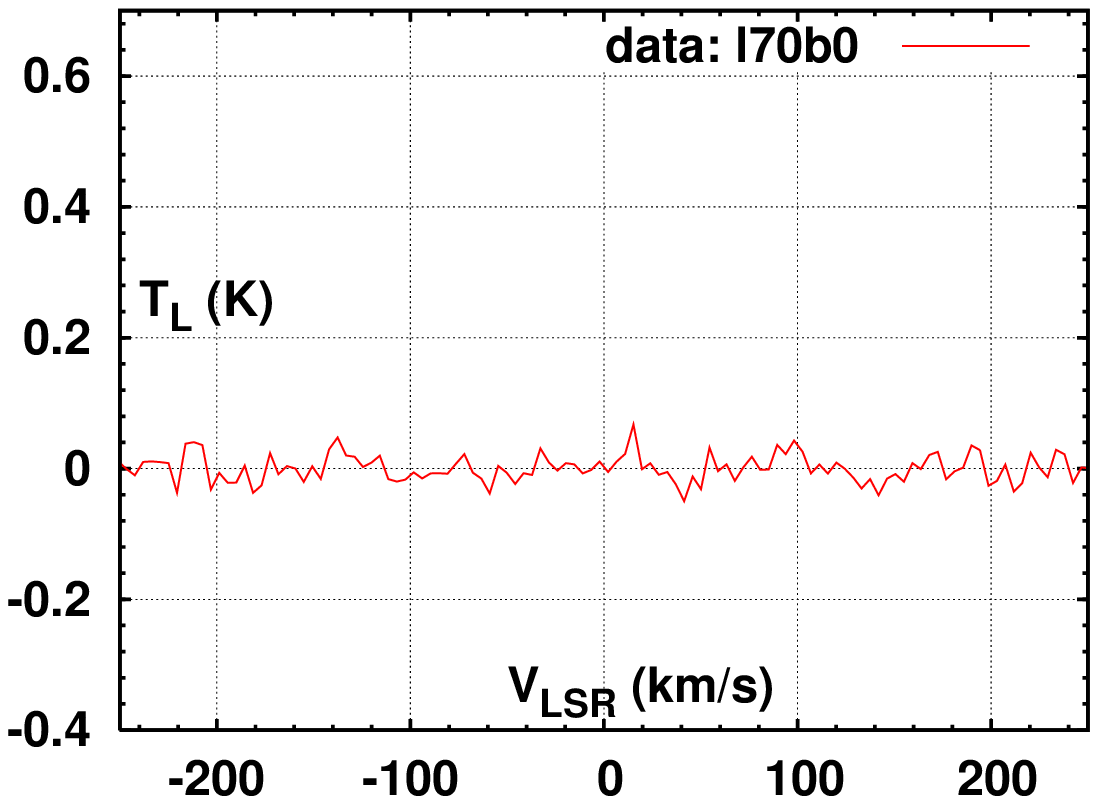}
\includegraphics[width=32mm,height=24mm,angle=0]{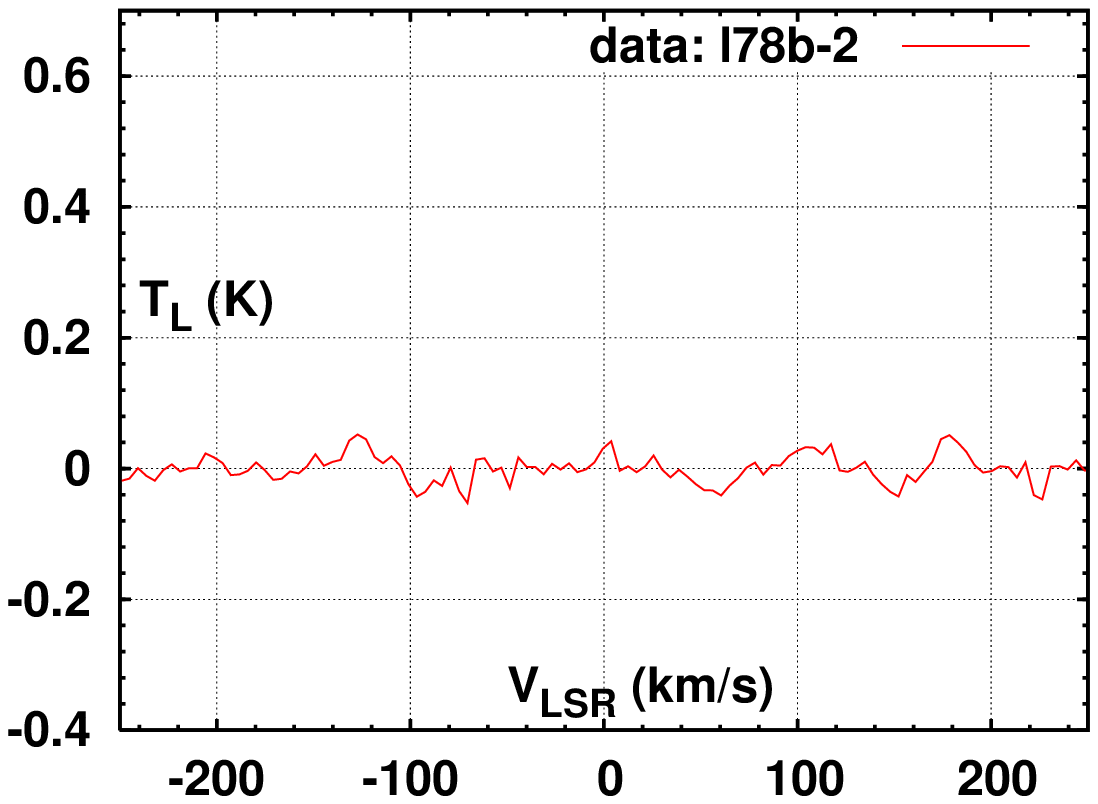}
\includegraphics[width=32mm,height=24mm,angle=0]{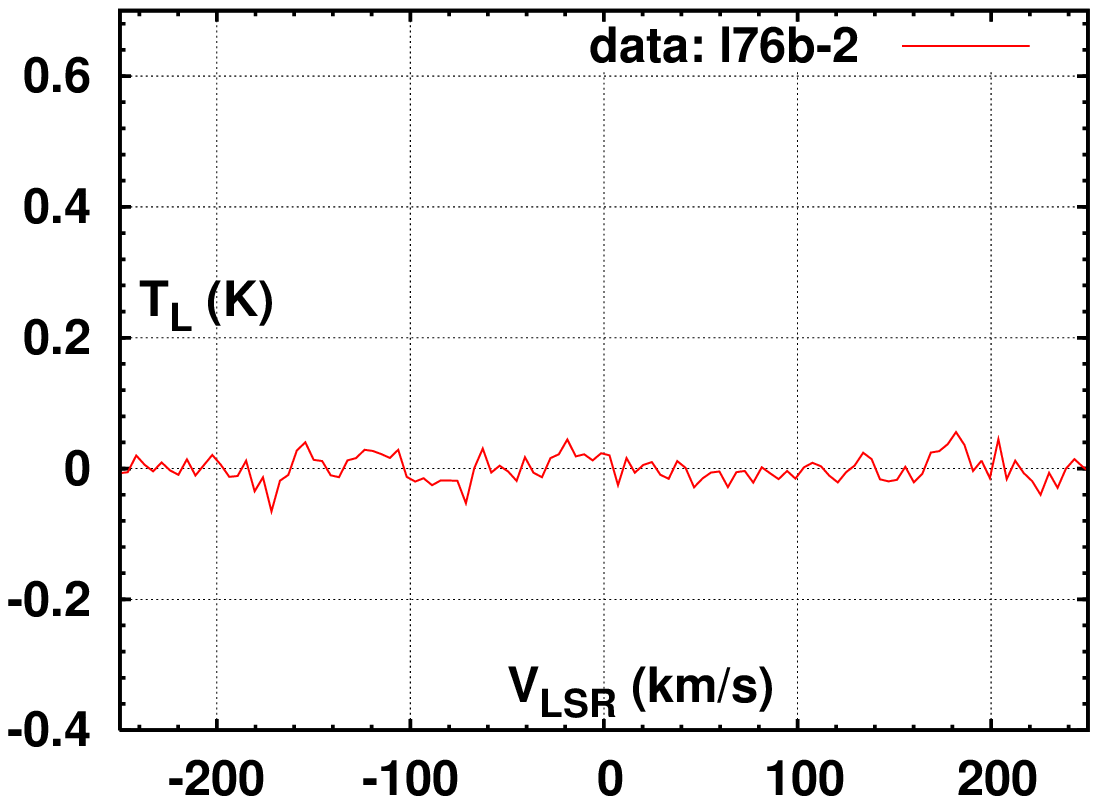}
\includegraphics[width=32mm,height=24mm,angle=0]{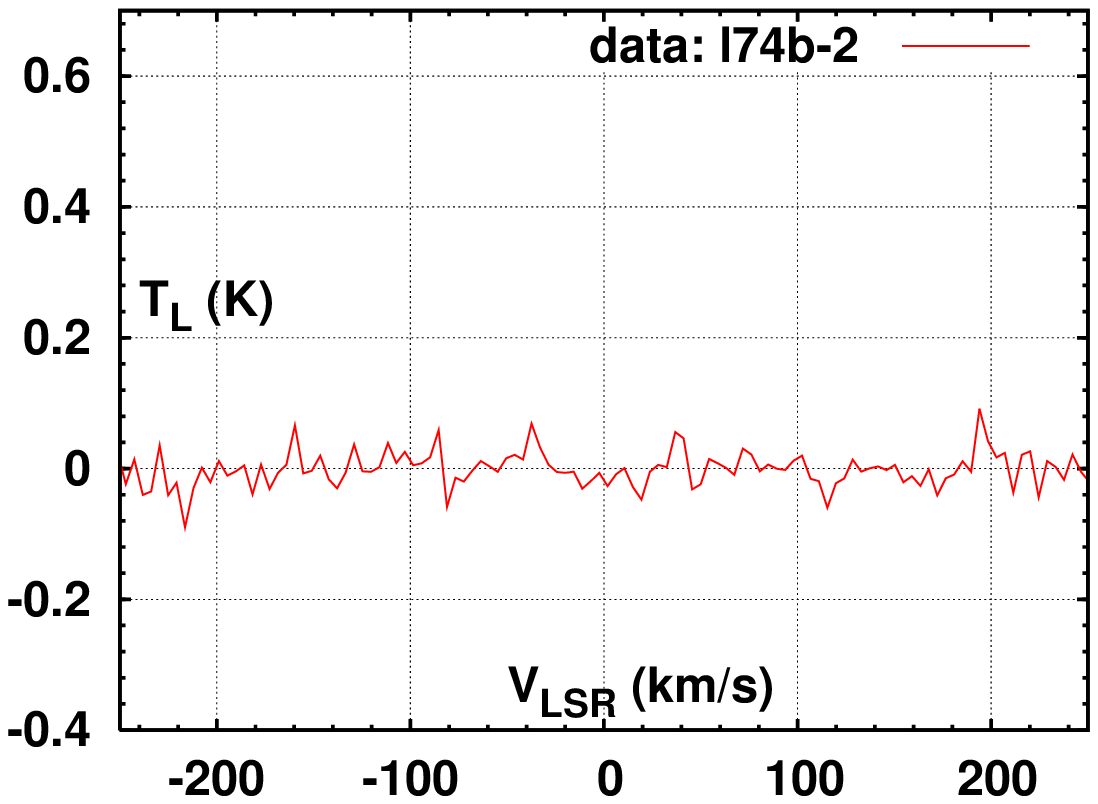}
\includegraphics[width=32mm,height=24mm,angle=0]{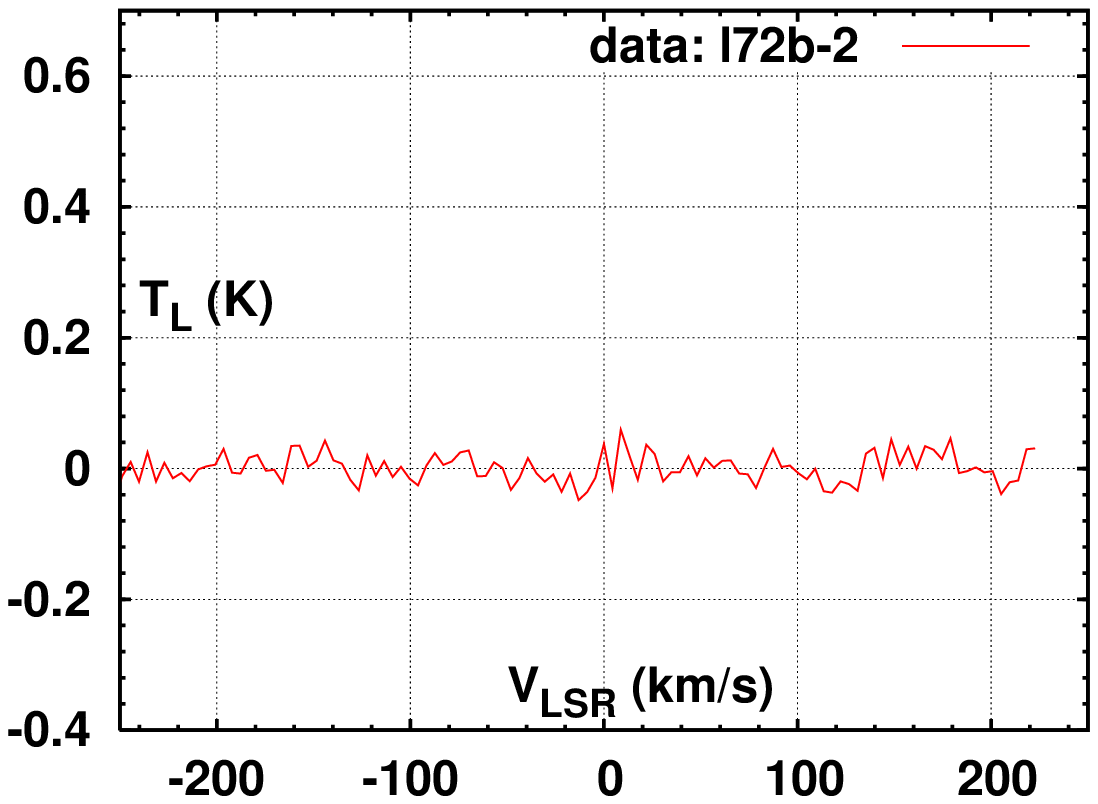}
\includegraphics[width=32mm,height=24mm,angle=0]{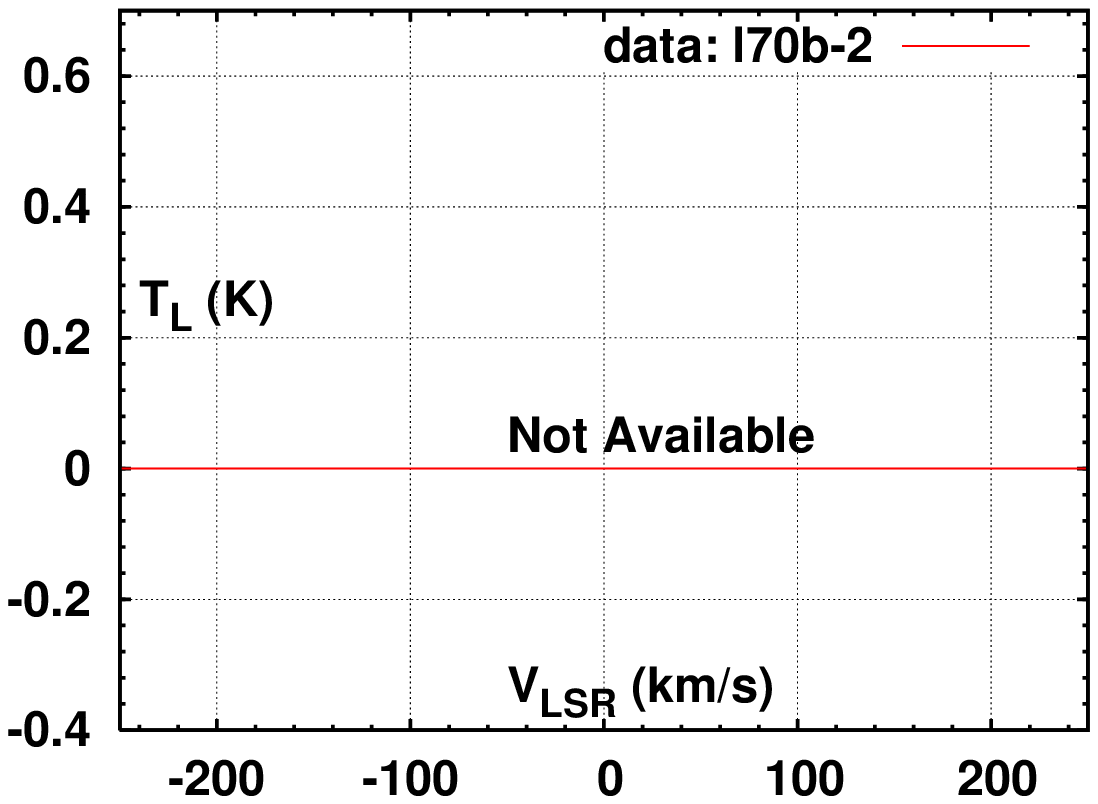}\vspace{1cm}
\includegraphics[width=32mm,height=24mm,angle=0]{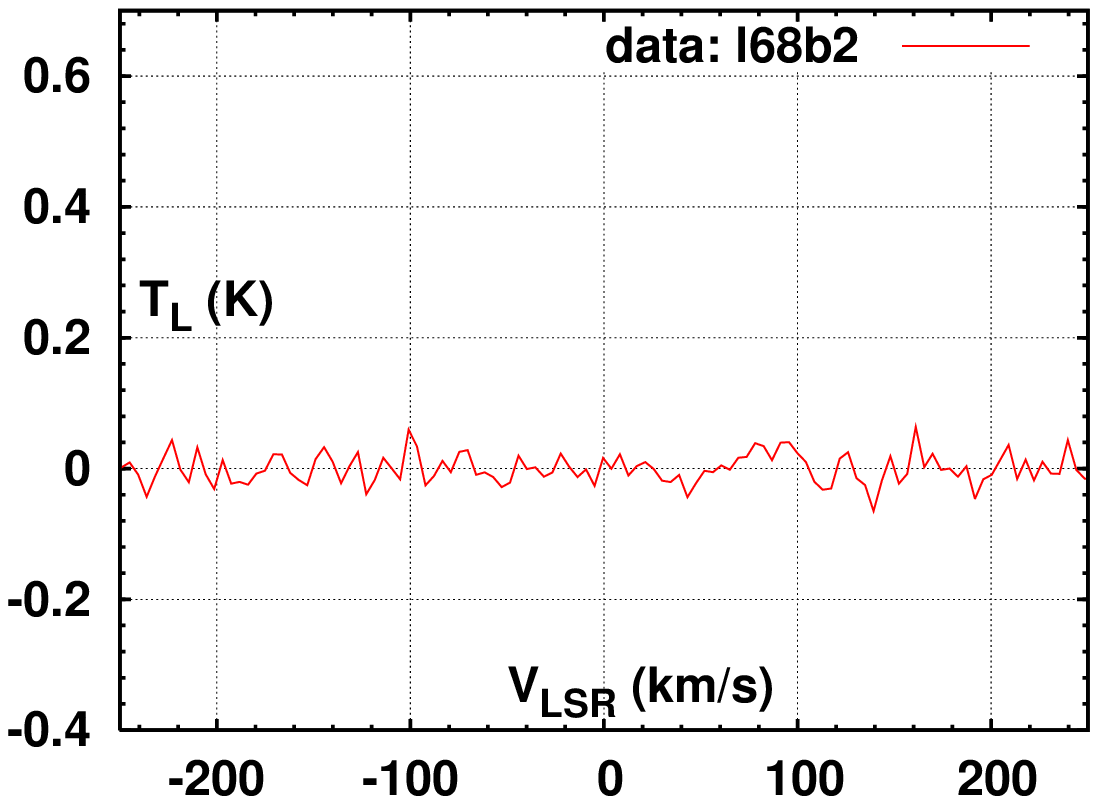}
\includegraphics[width=32mm,height=24mm,angle=0]{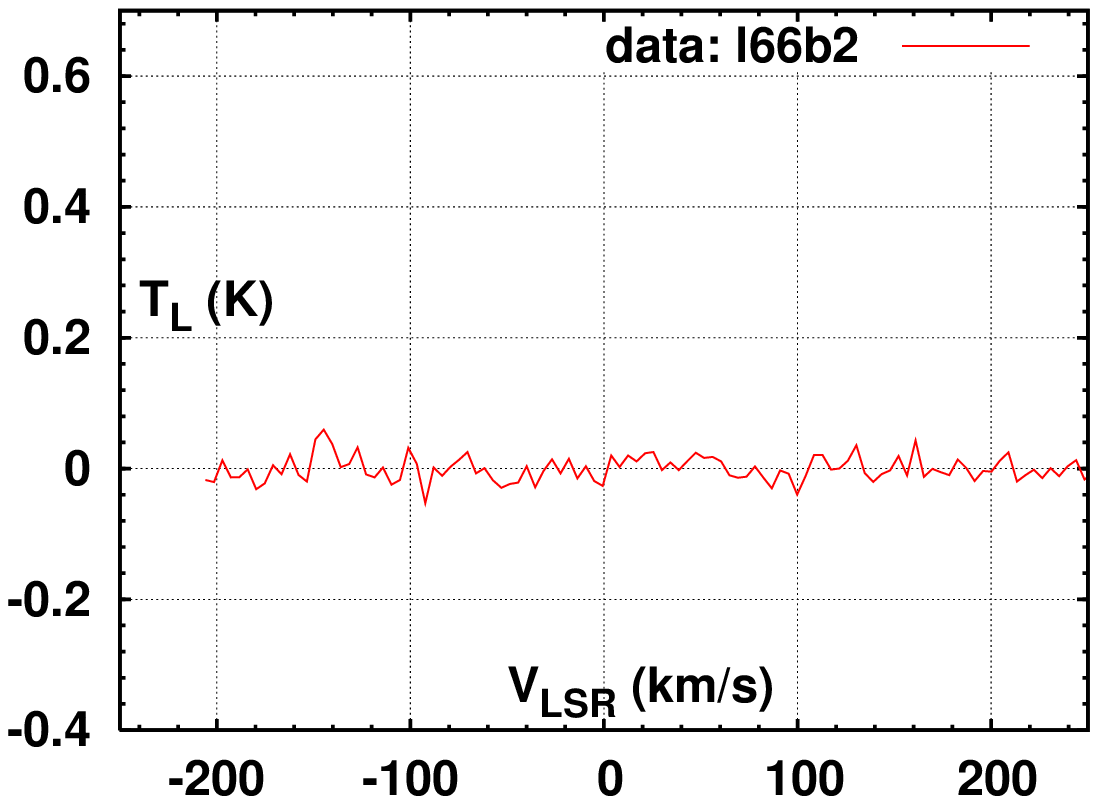}
\includegraphics[width=32mm,height=24mm,angle=0]{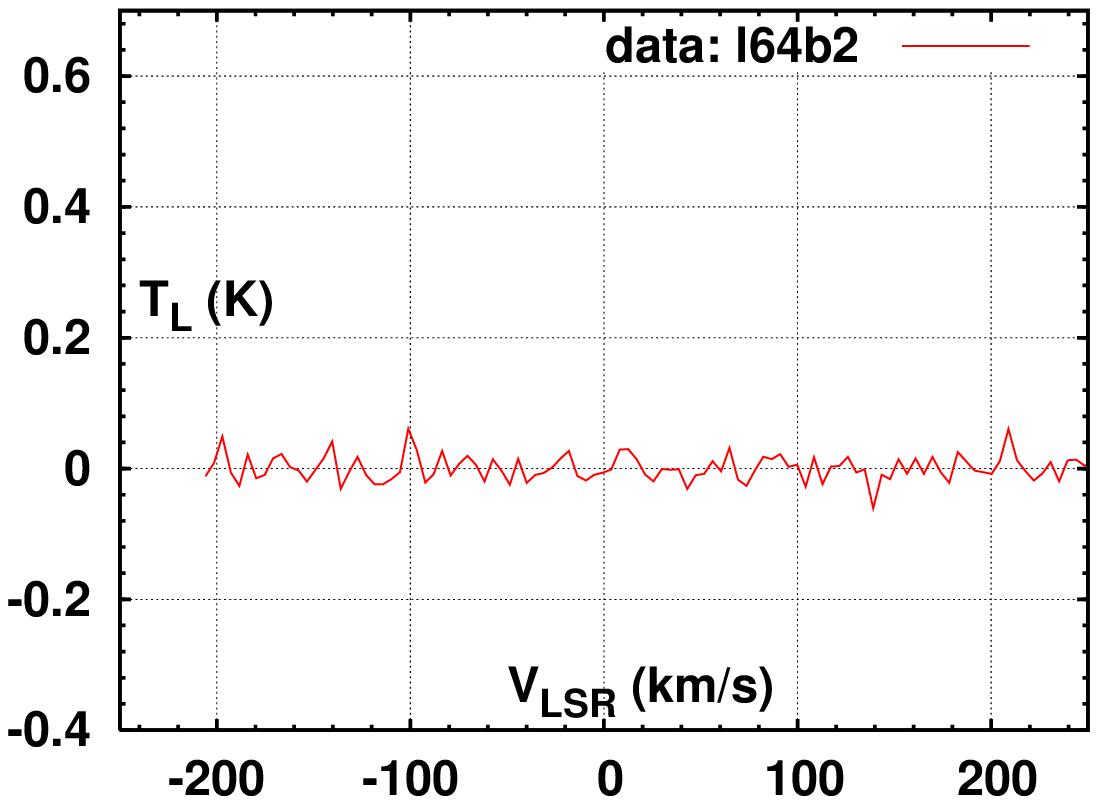}
\includegraphics[width=32mm,height=24mm,angle=0]{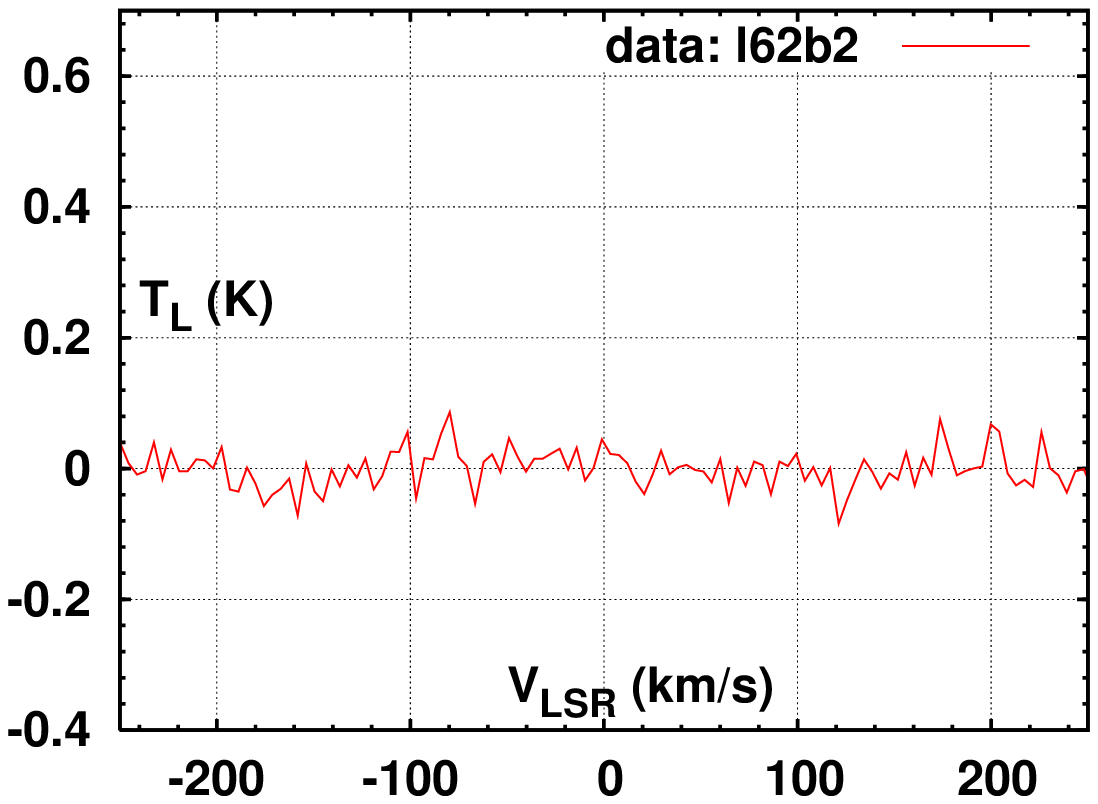}
\includegraphics[width=32mm,height=24mm,angle=0]{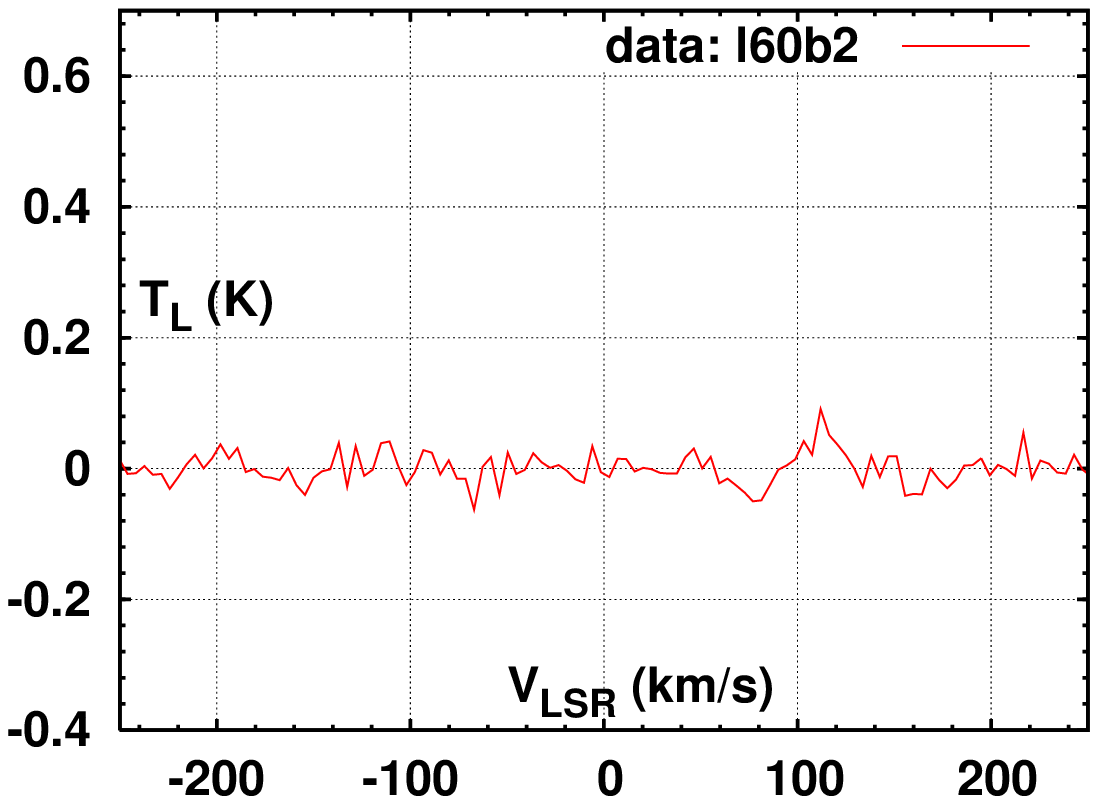}
\includegraphics[width=32mm,height=24mm,angle=0]{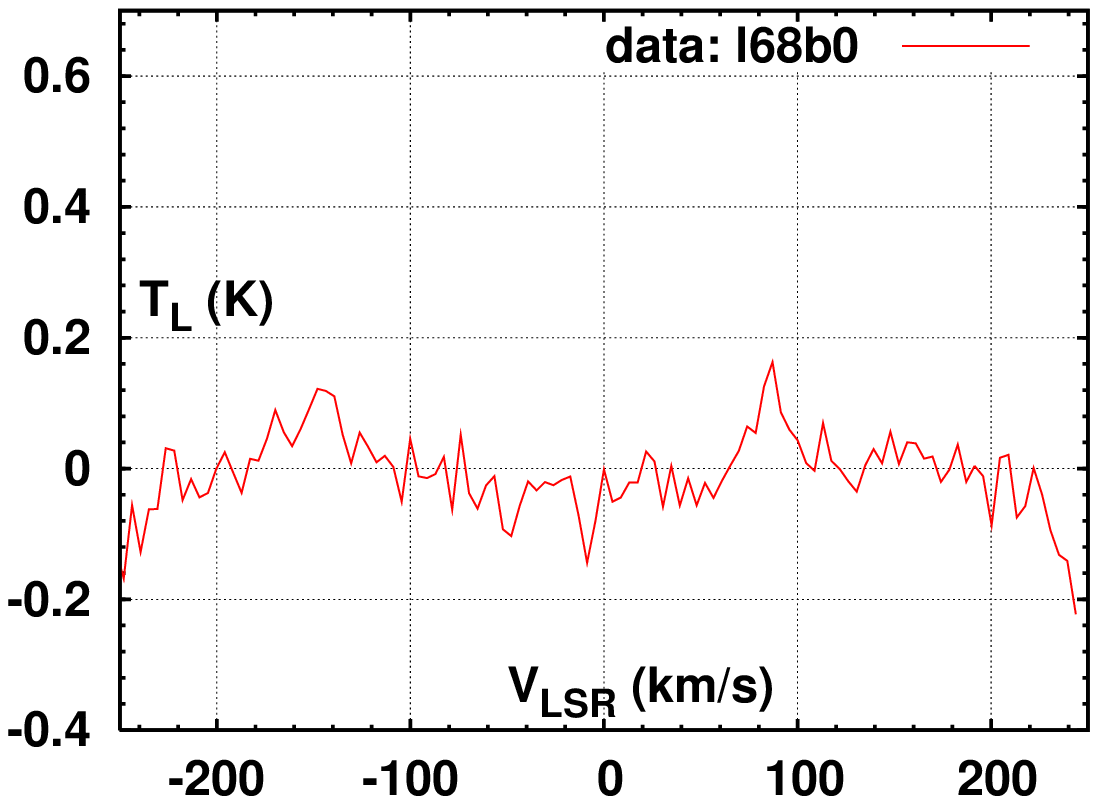}
\includegraphics[width=32mm,height=24mm,angle=0]{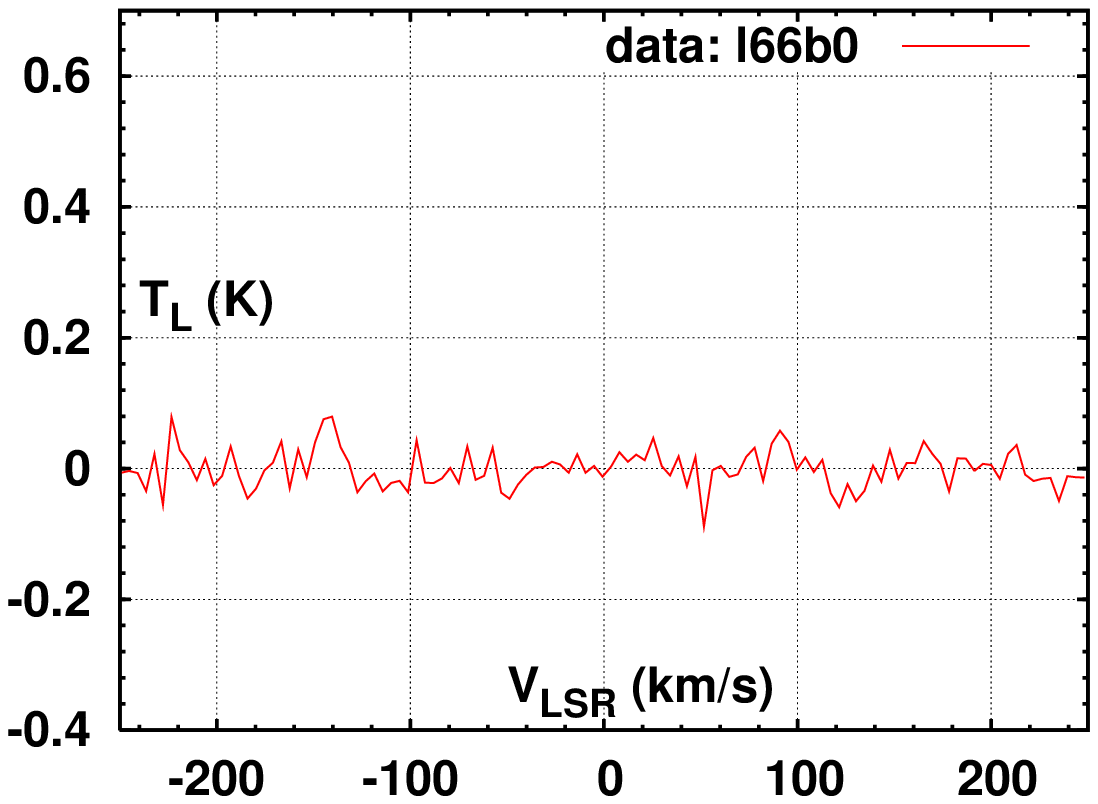}
\includegraphics[width=32mm,height=24mm,angle=0]{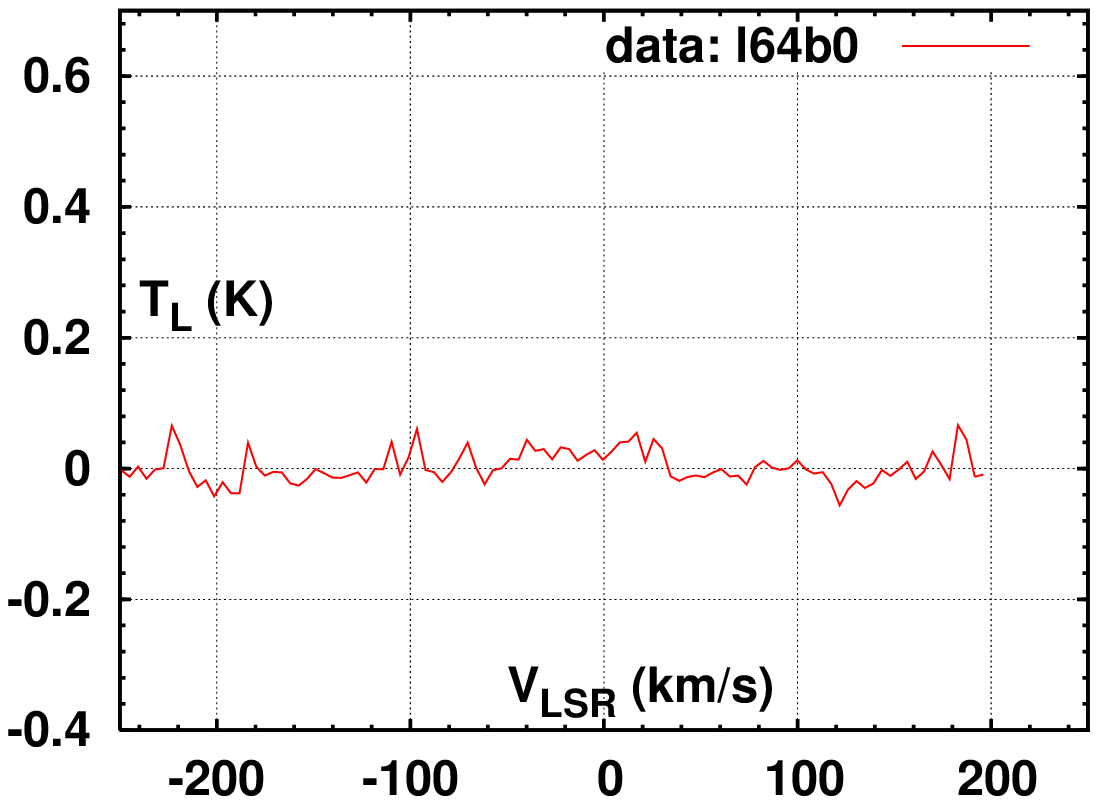}
\includegraphics[width=32mm,height=24mm,angle=0]{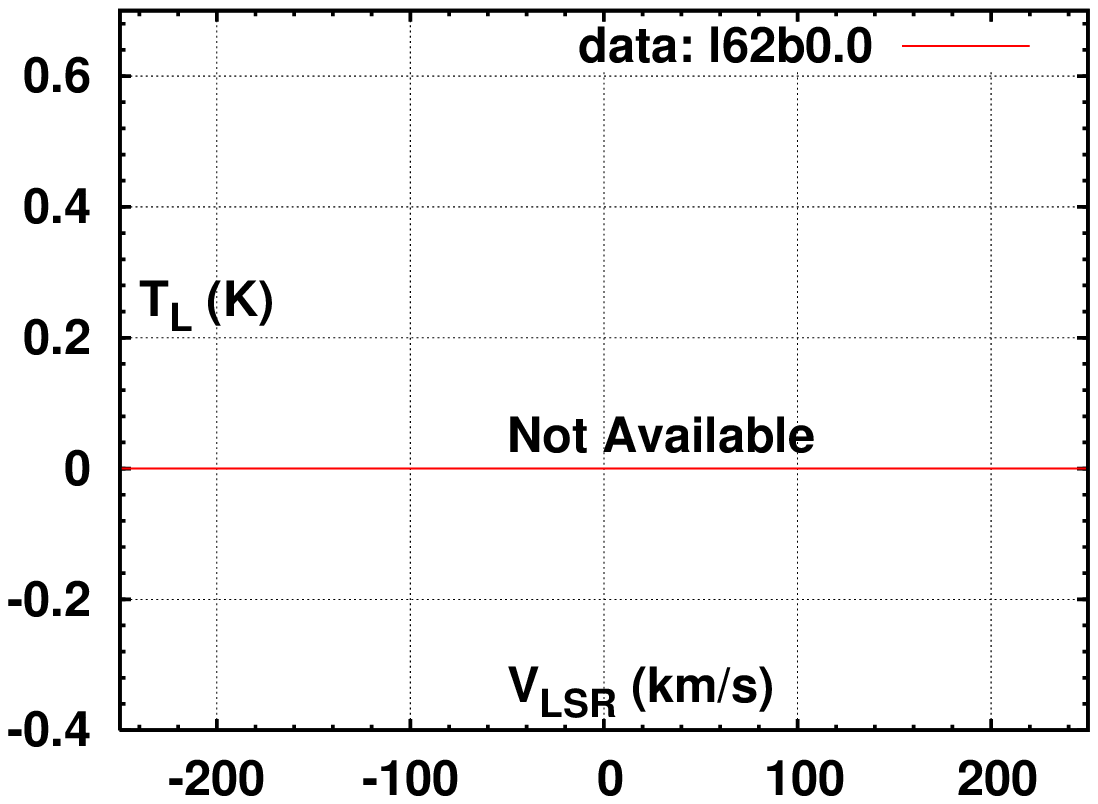}
\includegraphics[width=32mm,height=24mm,angle=0]{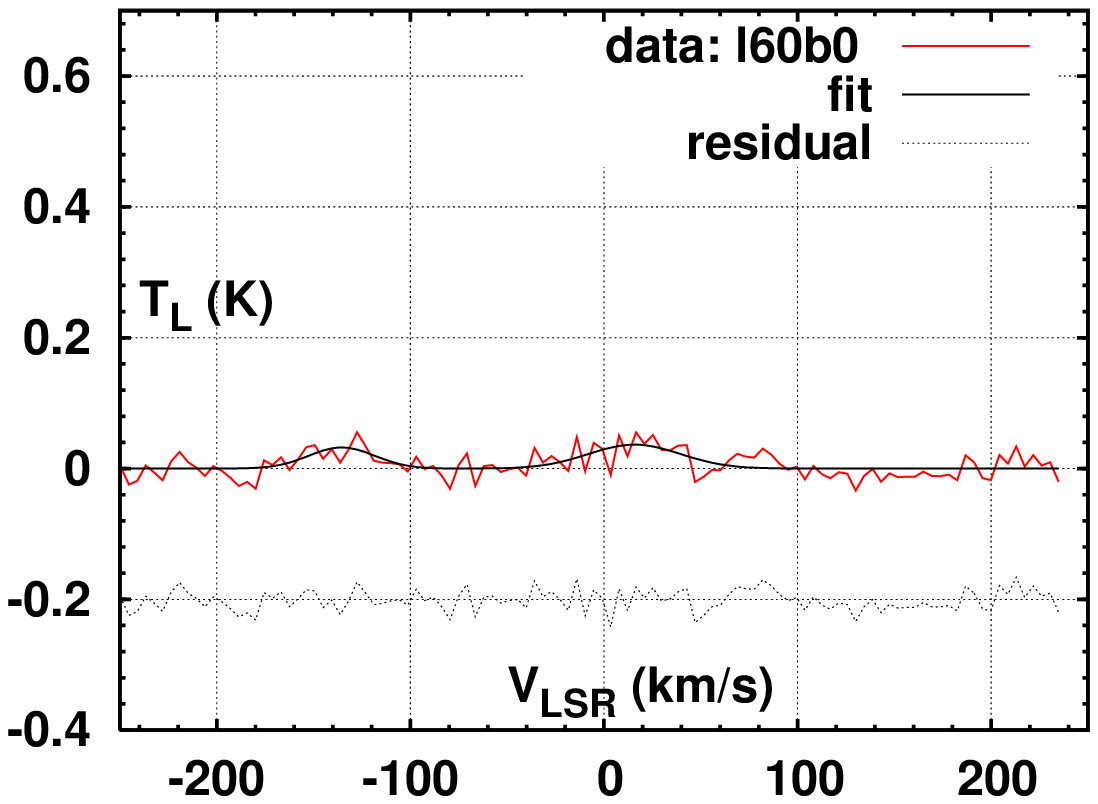}
\includegraphics[width=32mm,height=24mm,angle=0]{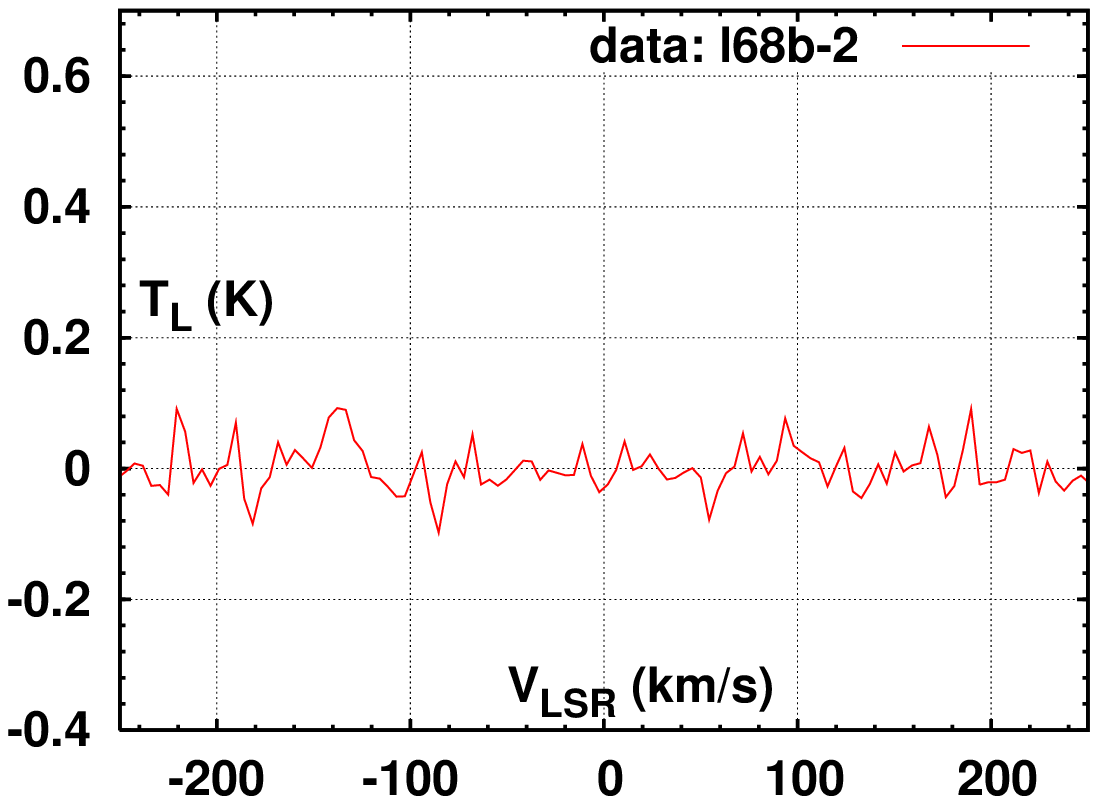}
\includegraphics[width=32mm,height=24mm,angle=0]{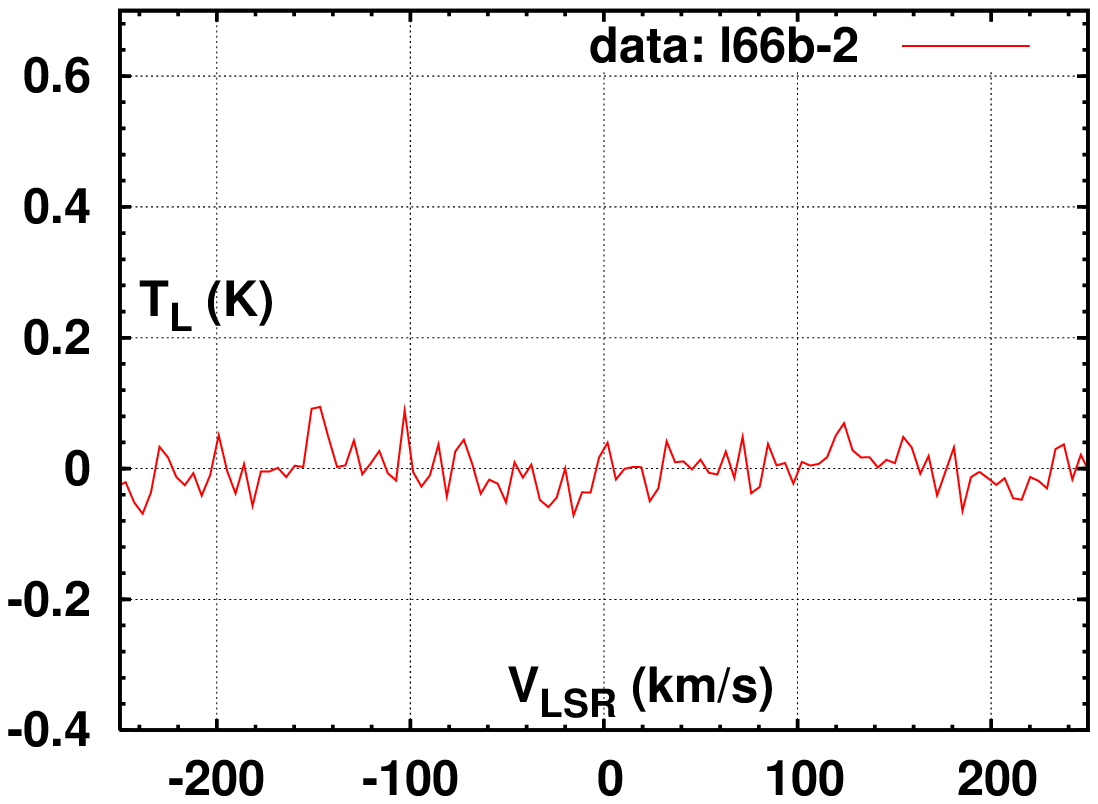}
\includegraphics[width=32mm,height=24mm,angle=0]{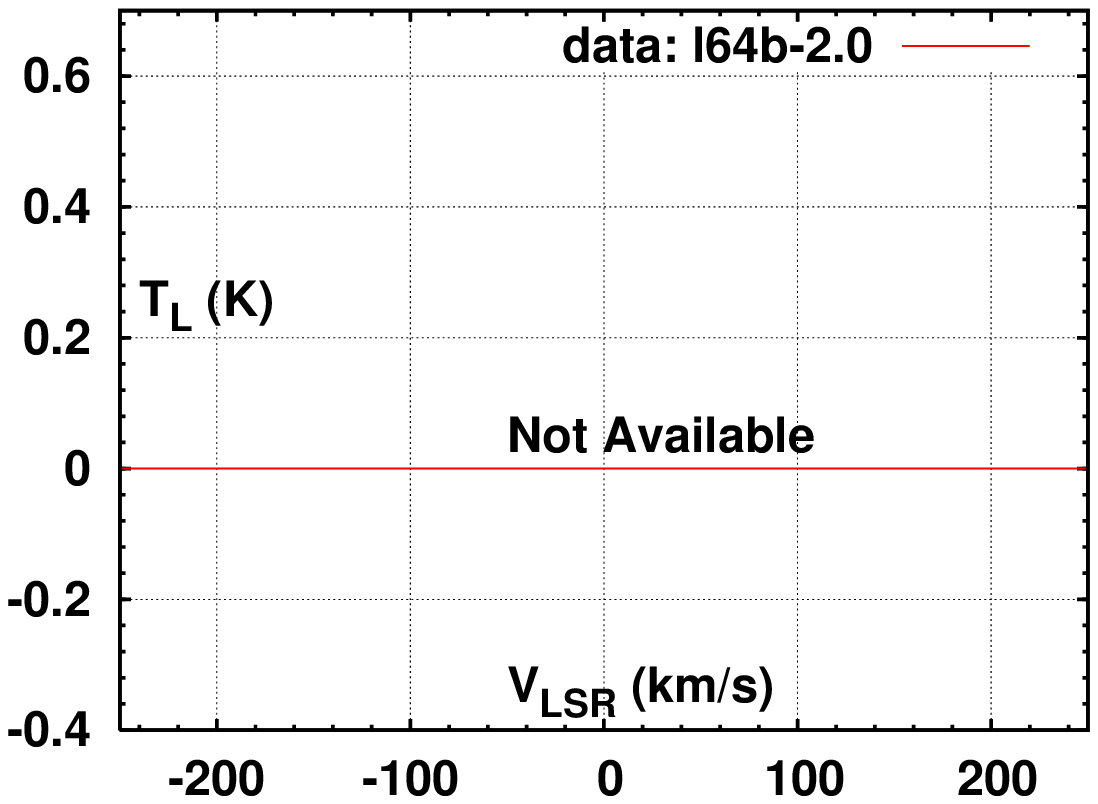}
\includegraphics[width=32mm,height=24mm,angle=0]{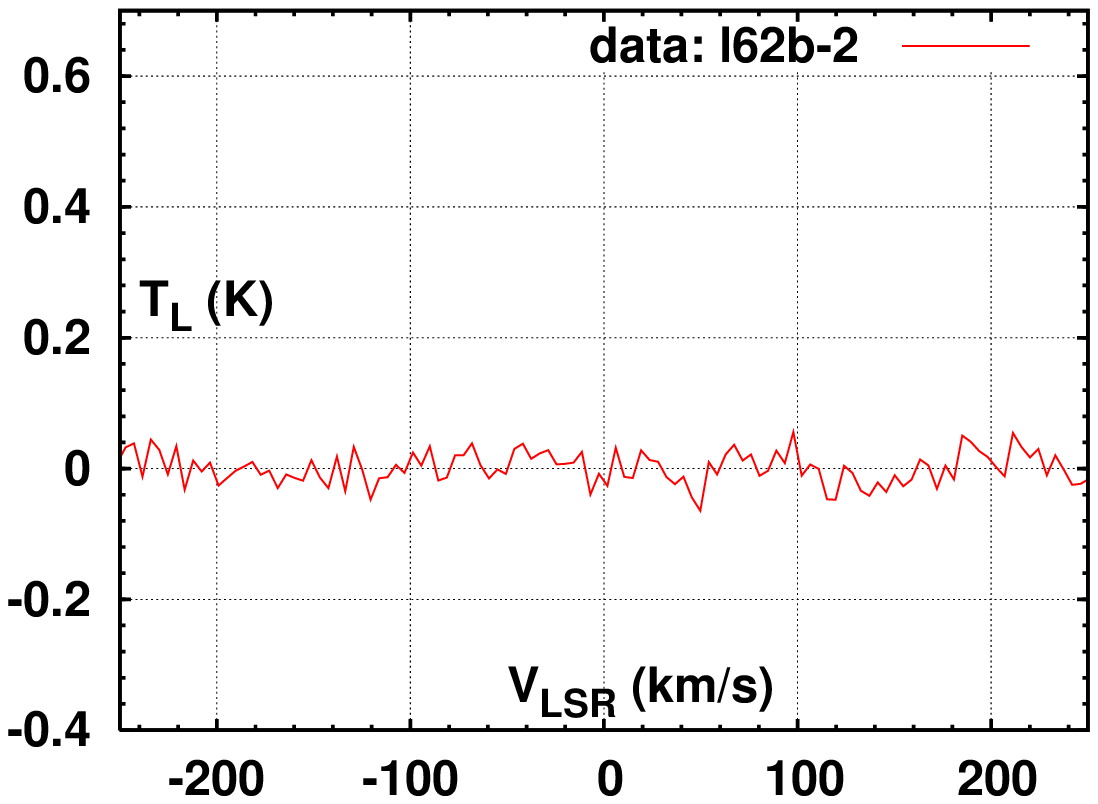}
\includegraphics[width=32mm,height=24mm,angle=0]{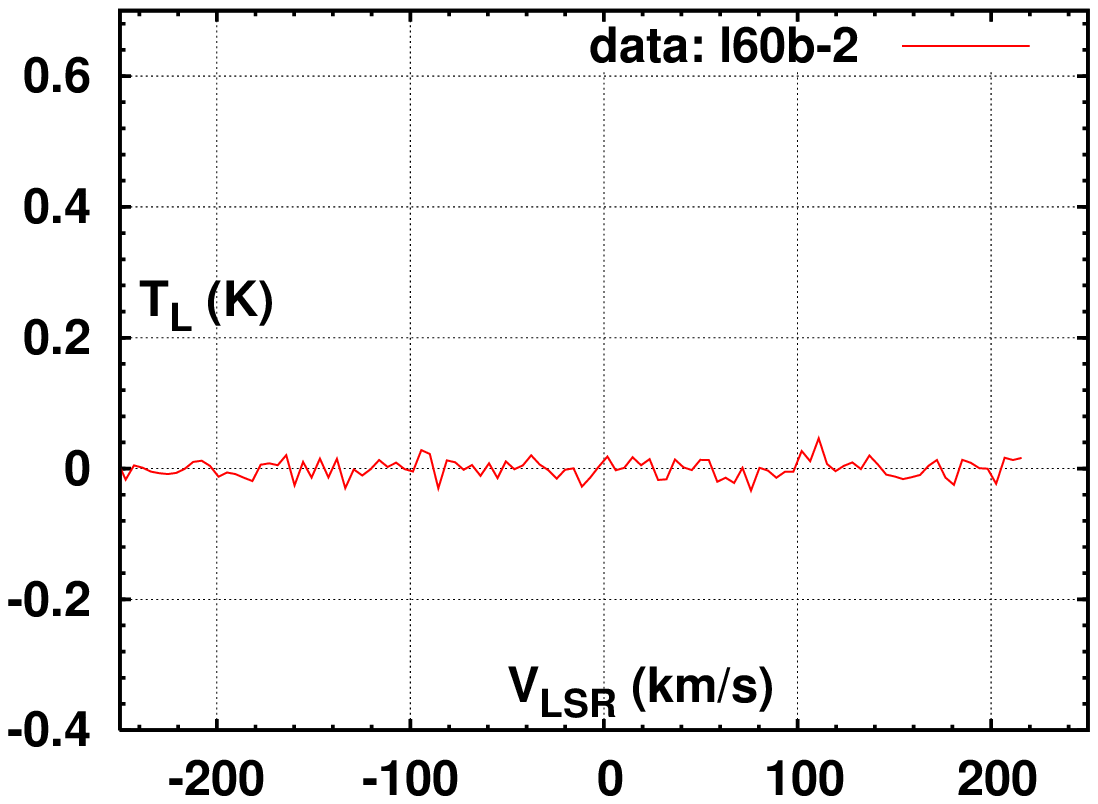}

 \caption{ORT H271$\alpha$ RL observation.}
\end{center}
\end{figure}

\begin{figure}[ht]
\begin{center}
\includegraphics[width=32mm,height=24mm,angle=0]{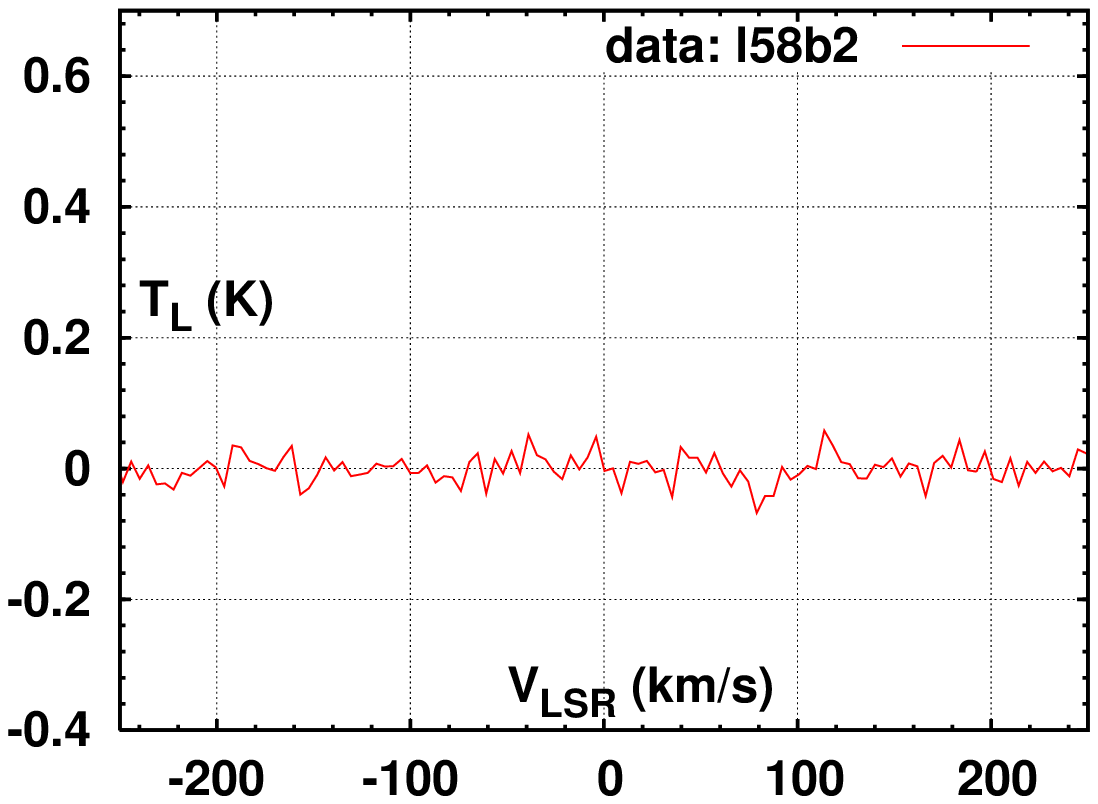}
\includegraphics[width=32mm,height=24mm,angle=0]{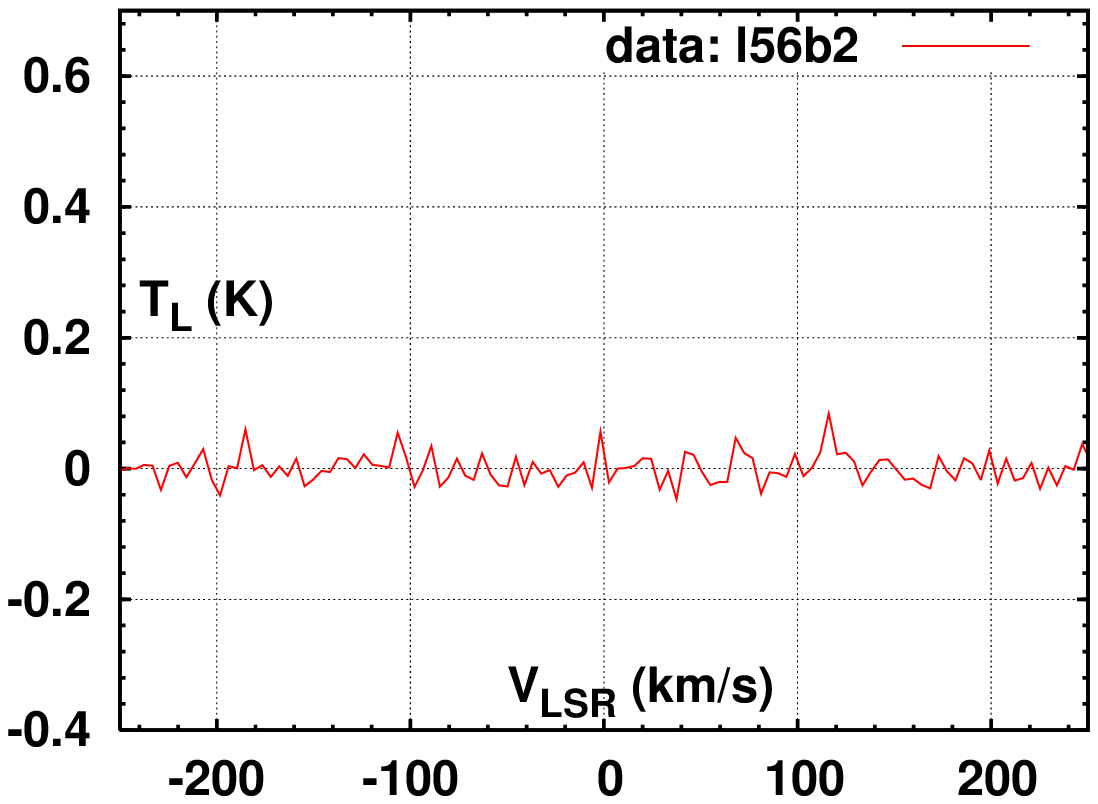}
\includegraphics[width=32mm,height=24mm,angle=0]{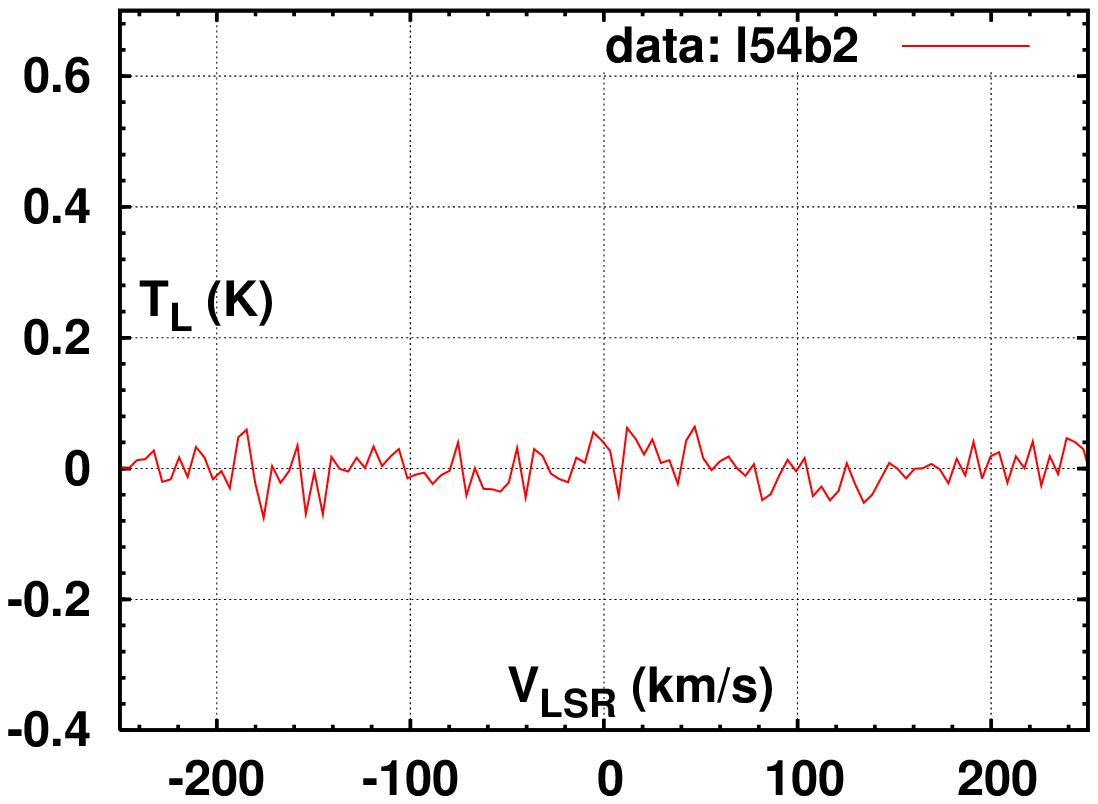}
\includegraphics[width=32mm,height=24mm,angle=0]{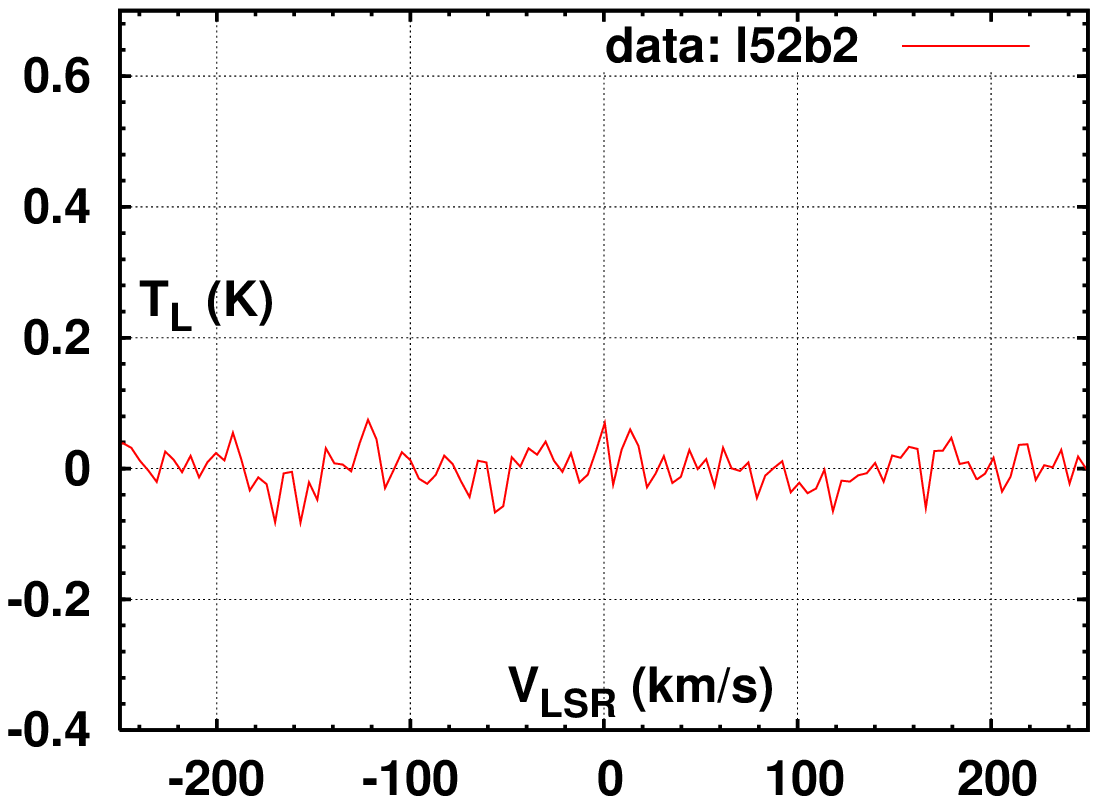}
\includegraphics[width=32mm,height=24mm,angle=0]{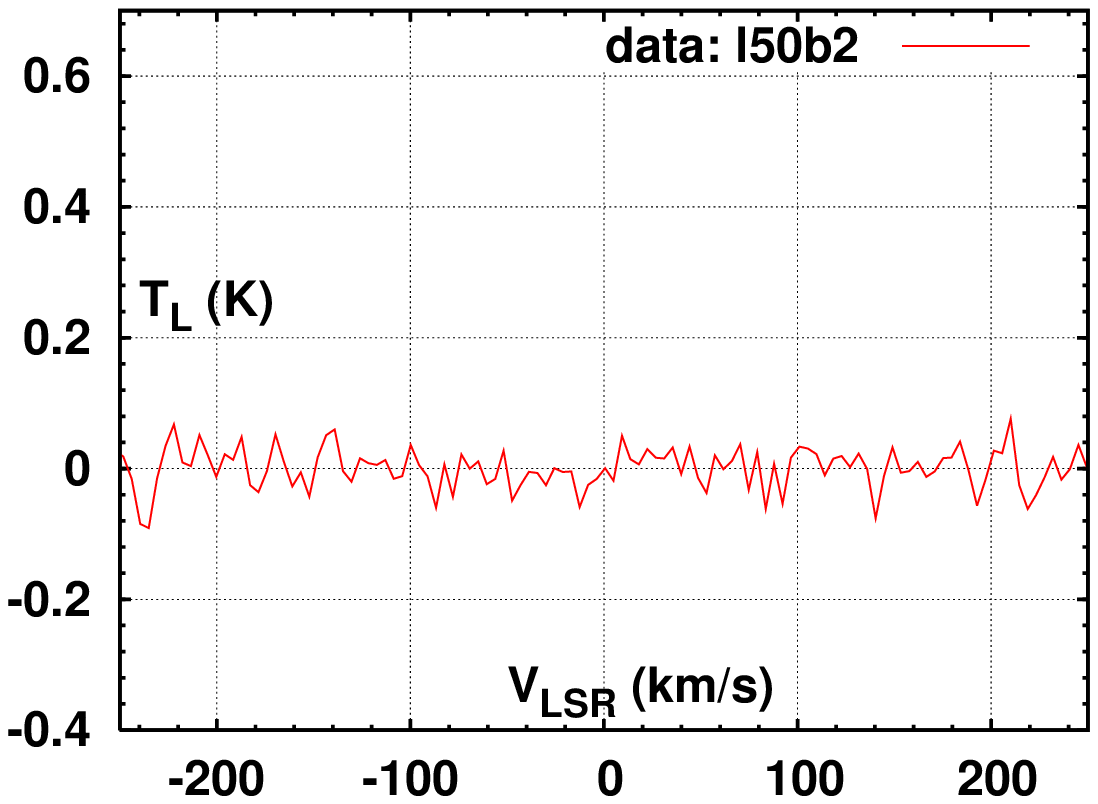}
\includegraphics[width=32mm,height=24mm,angle=0]{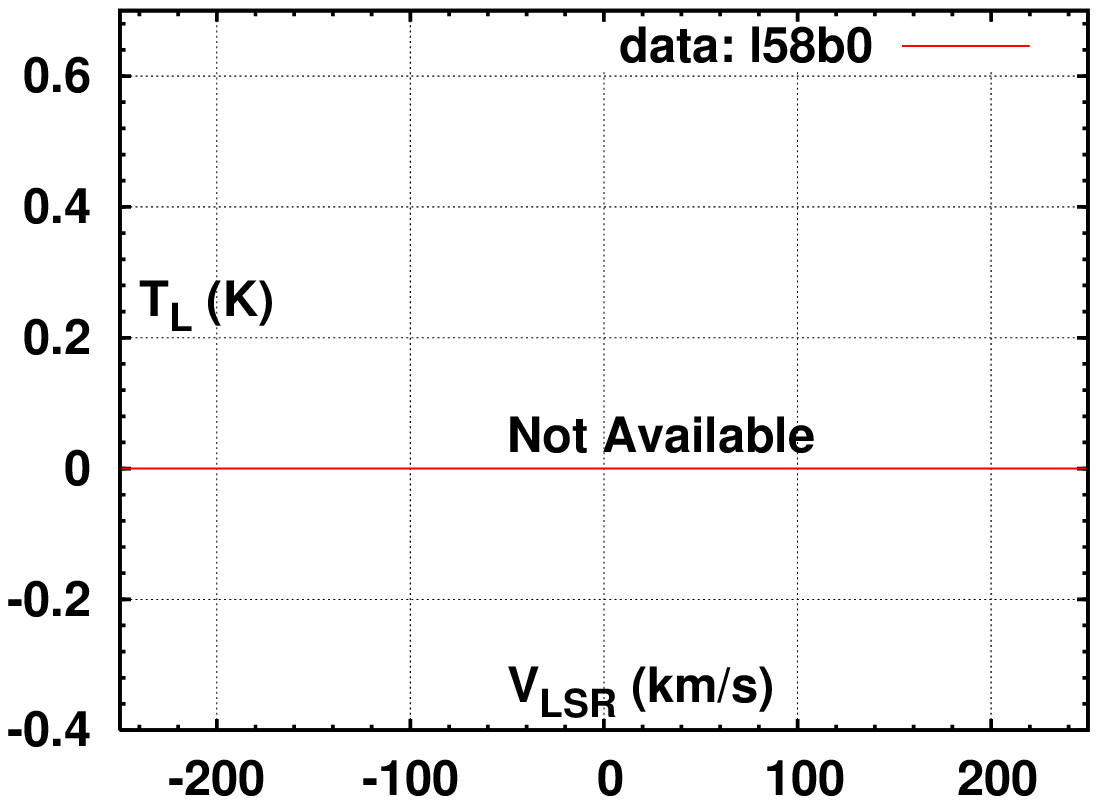}
\includegraphics[width=32mm,height=24mm,angle=0]{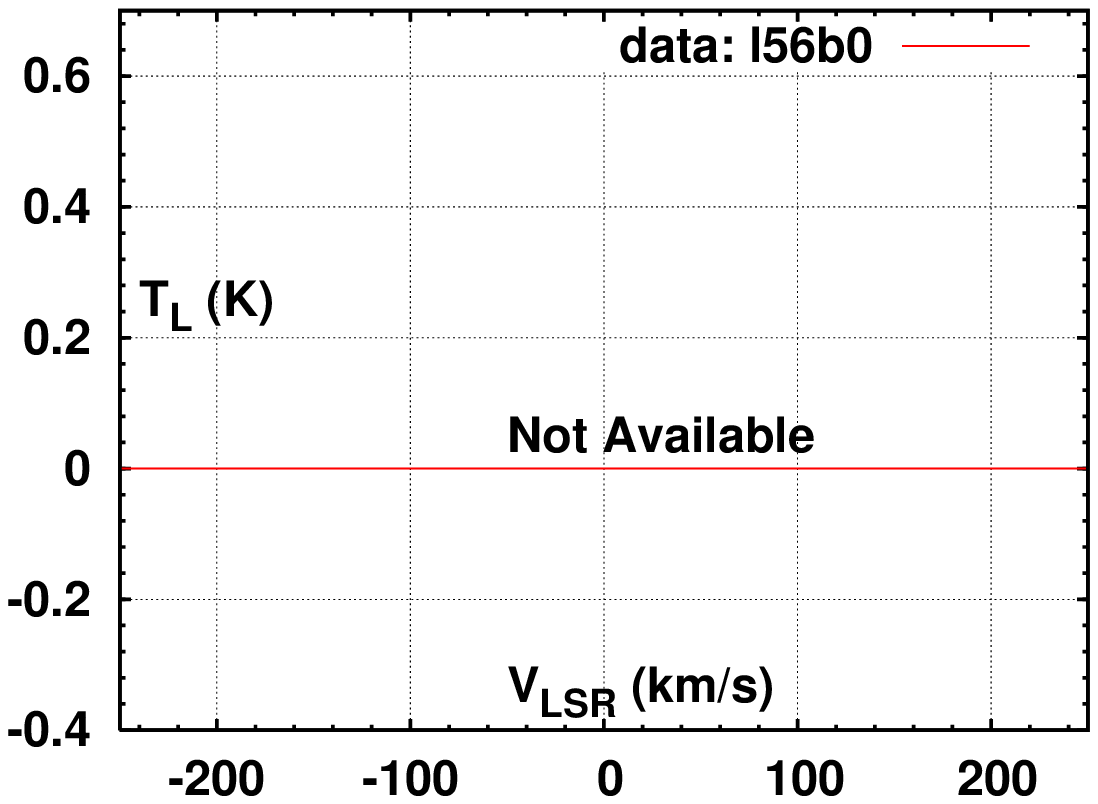}
\includegraphics[width=32mm,height=24mm,angle=0]{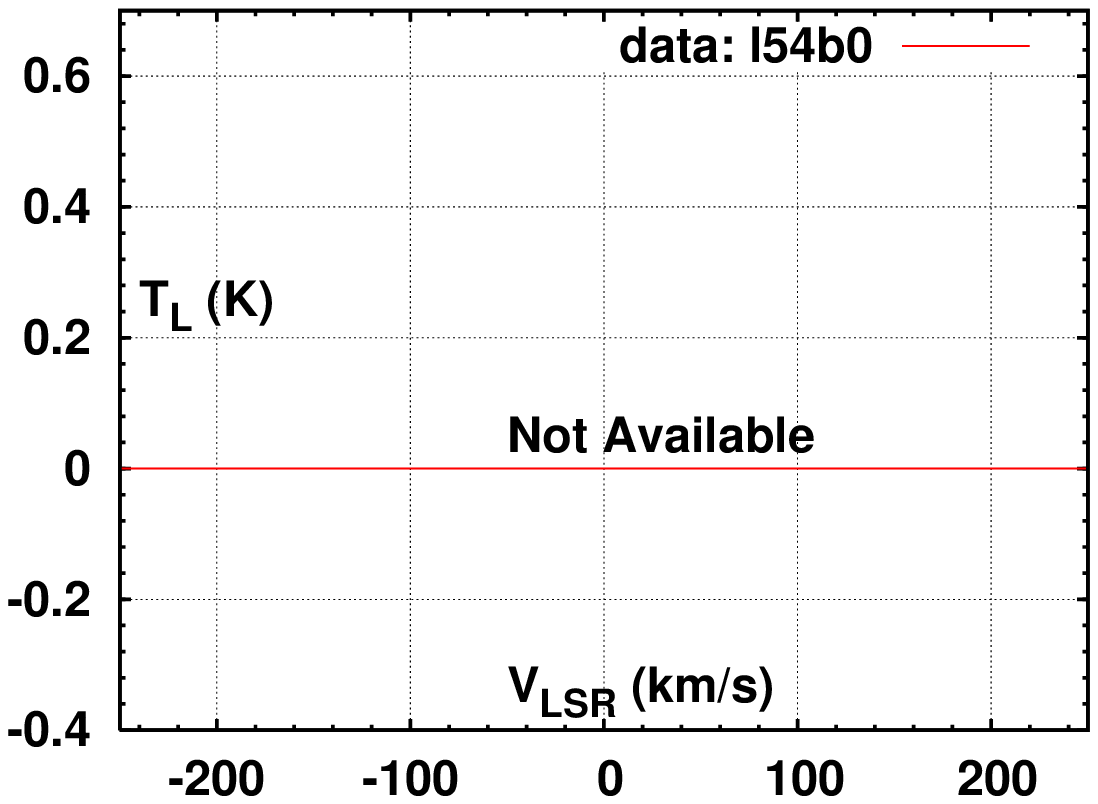}
\includegraphics[width=32mm,height=24mm,angle=0]{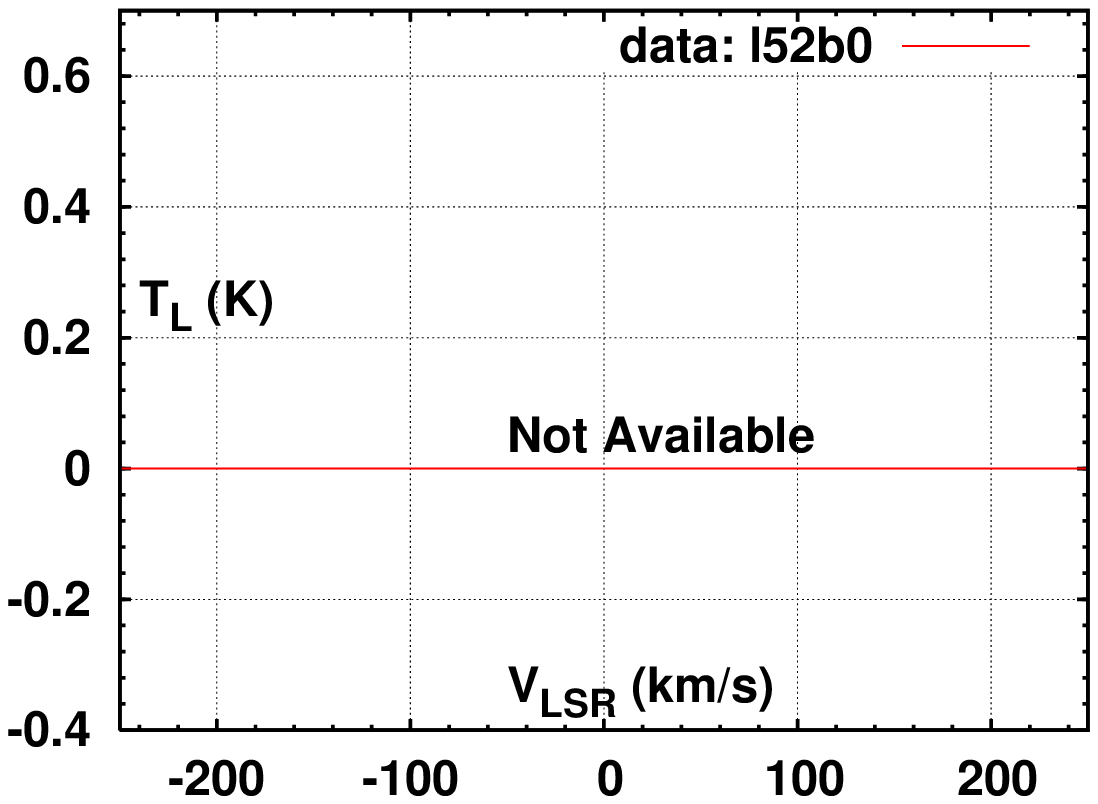}
\includegraphics[width=32mm,height=24mm,angle=0]{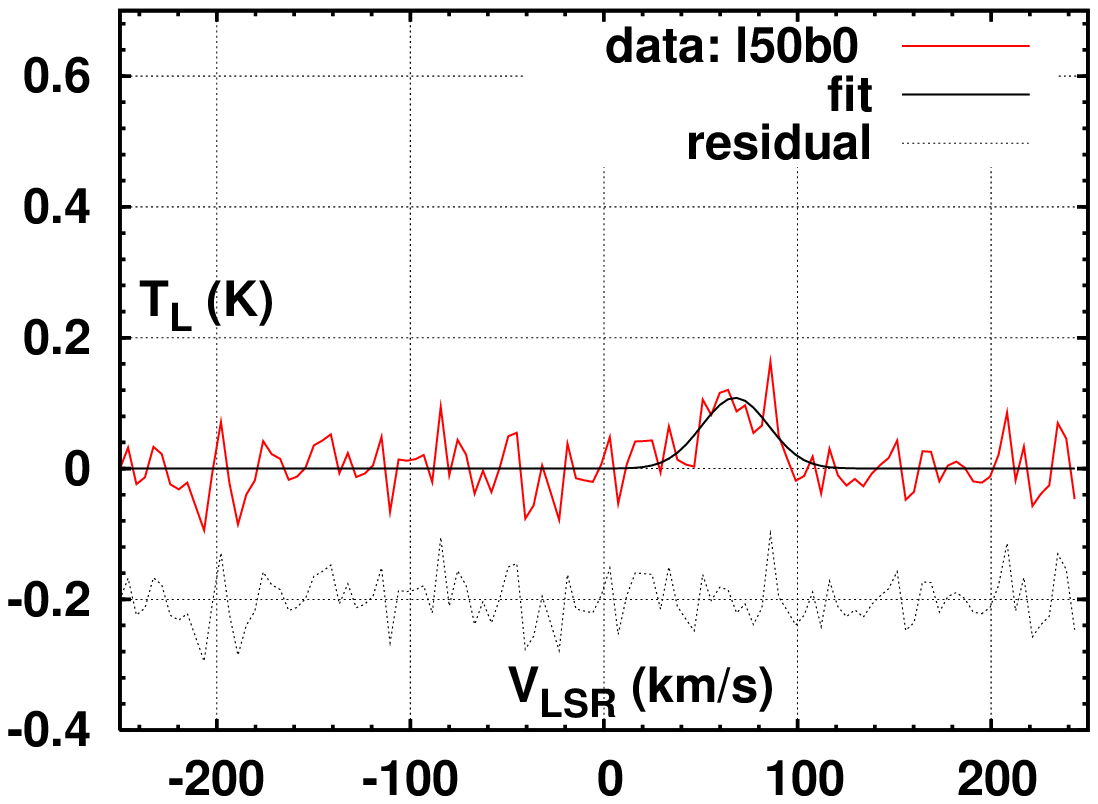}
\includegraphics[width=32mm,height=24mm,angle=0]{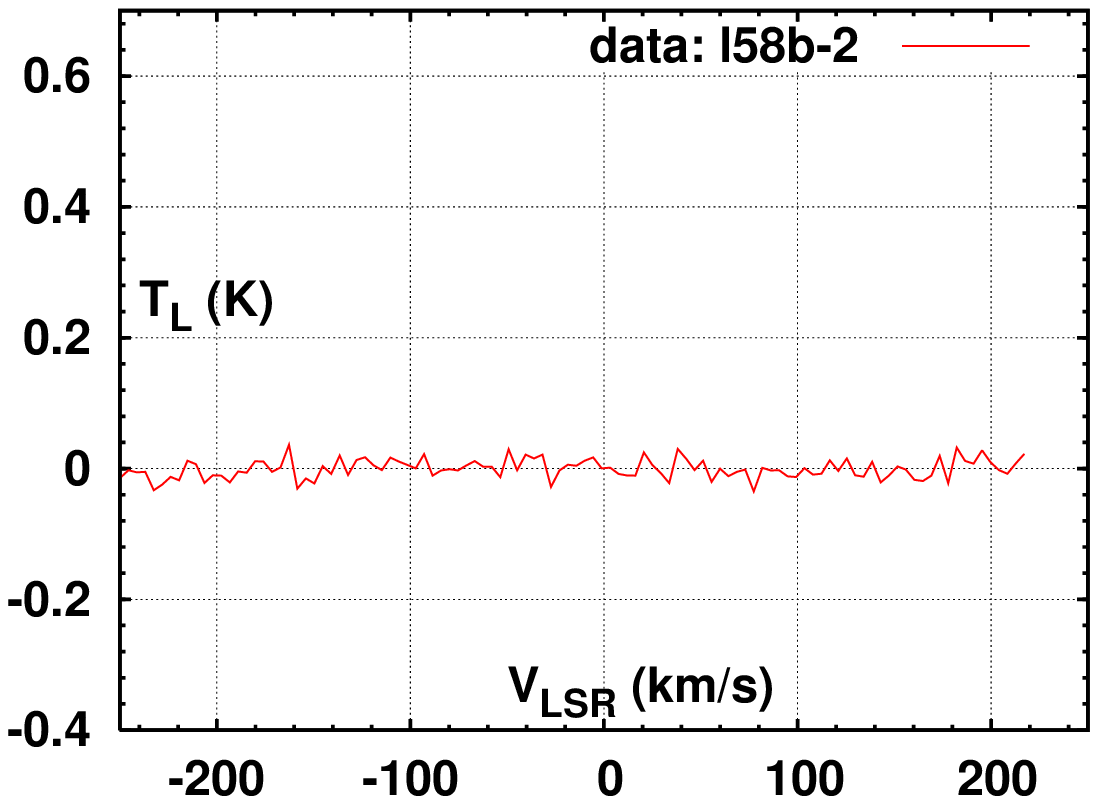}
\includegraphics[width=32mm,height=24mm,angle=0]{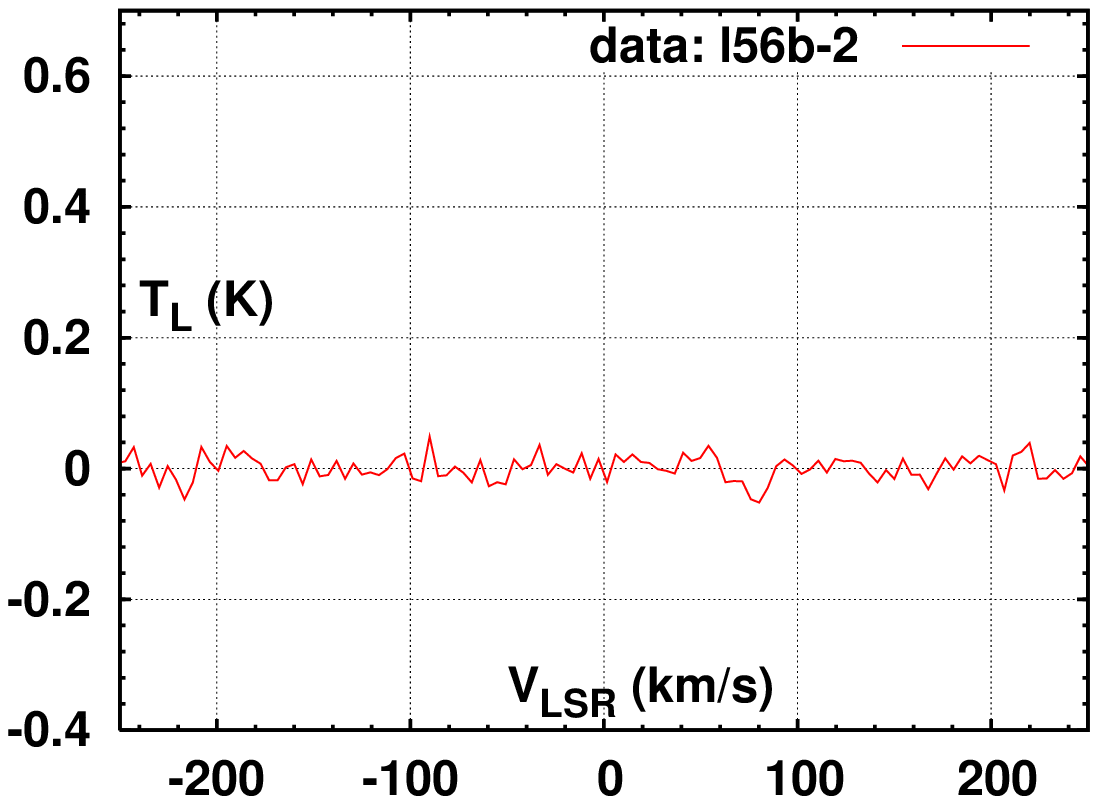}
\includegraphics[width=32mm,height=24mm,angle=0]{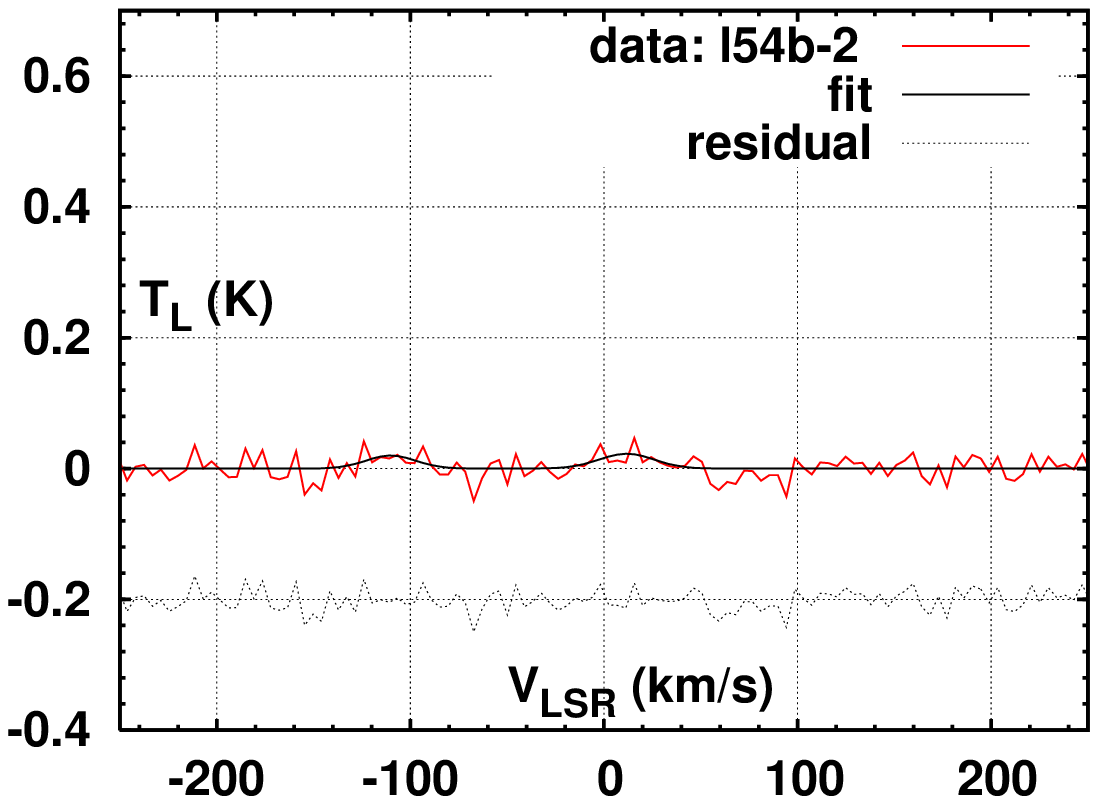}
\includegraphics[width=32mm,height=24mm,angle=0]{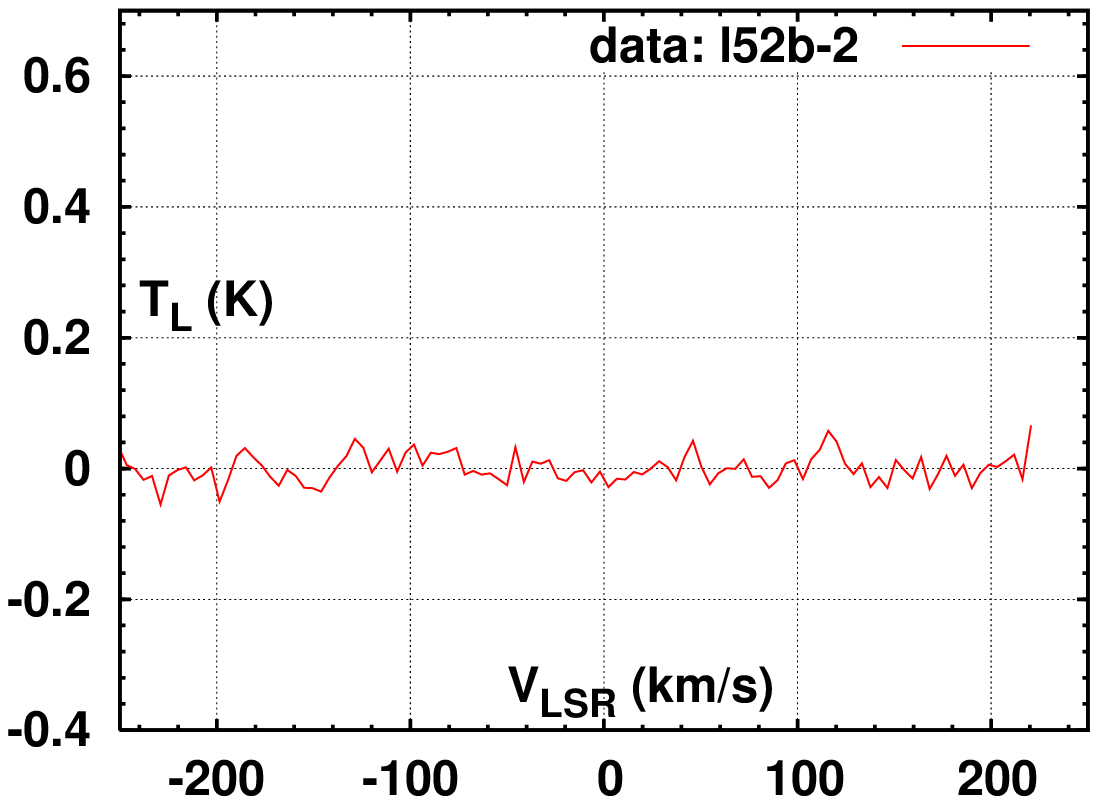}
\includegraphics[width=32mm,height=24mm,angle=0]{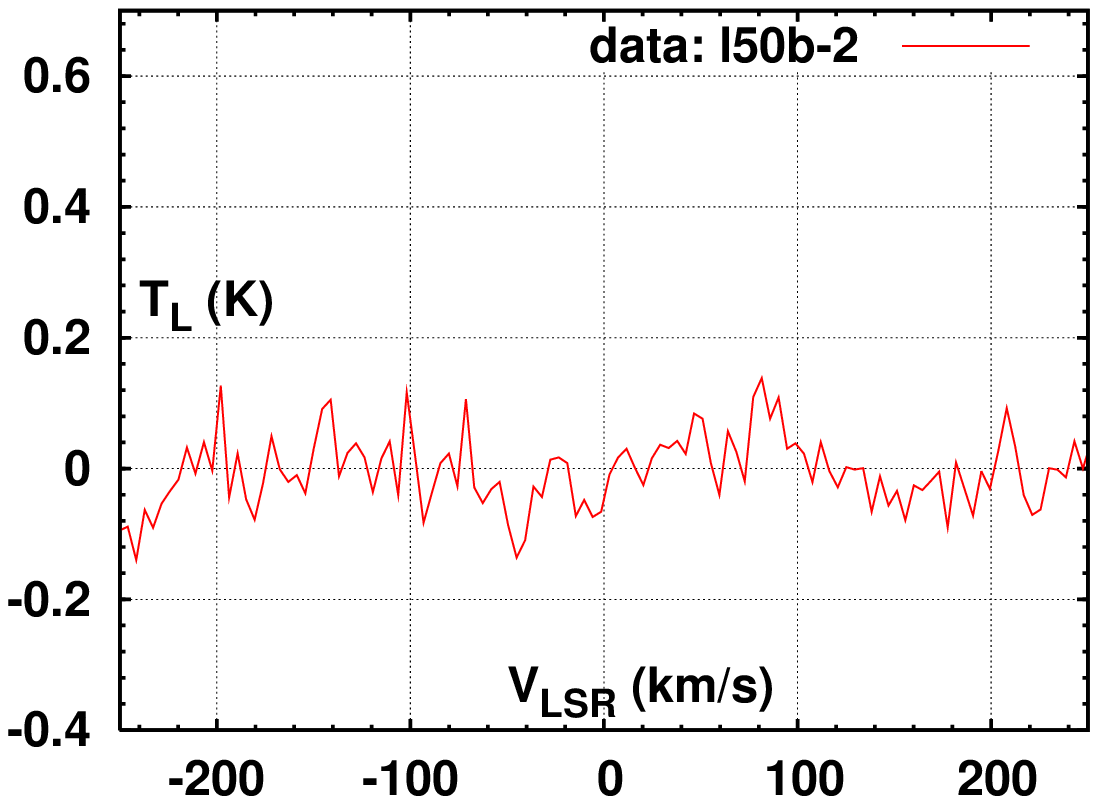}\vspace{1cm}
\includegraphics[width=32mm,height=24mm,angle=0]{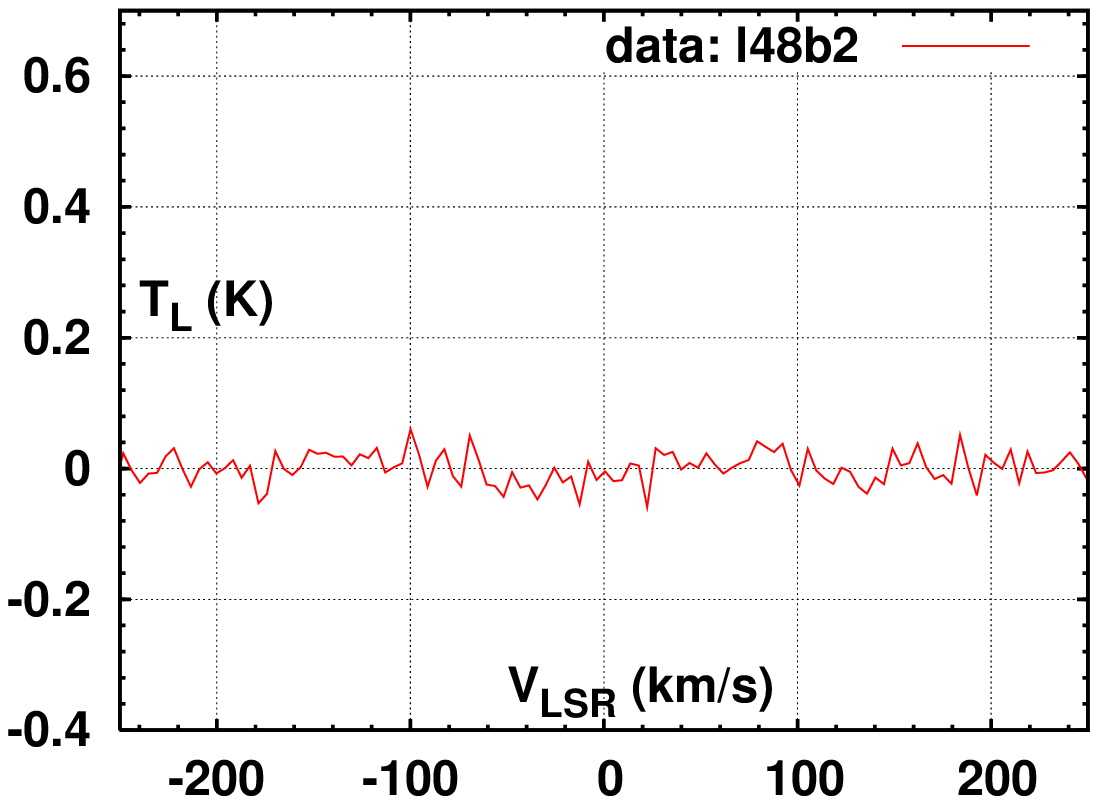}
\includegraphics[width=32mm,height=24mm,angle=0]{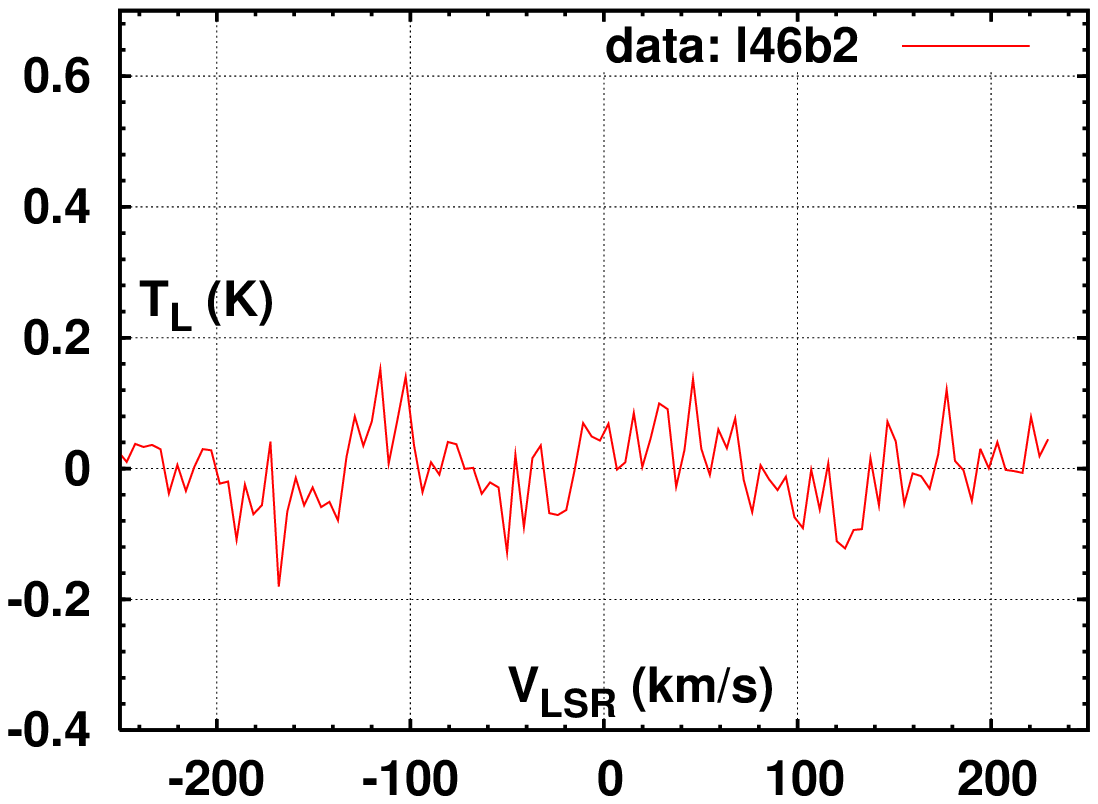}
\includegraphics[width=32mm,height=24mm,angle=0]{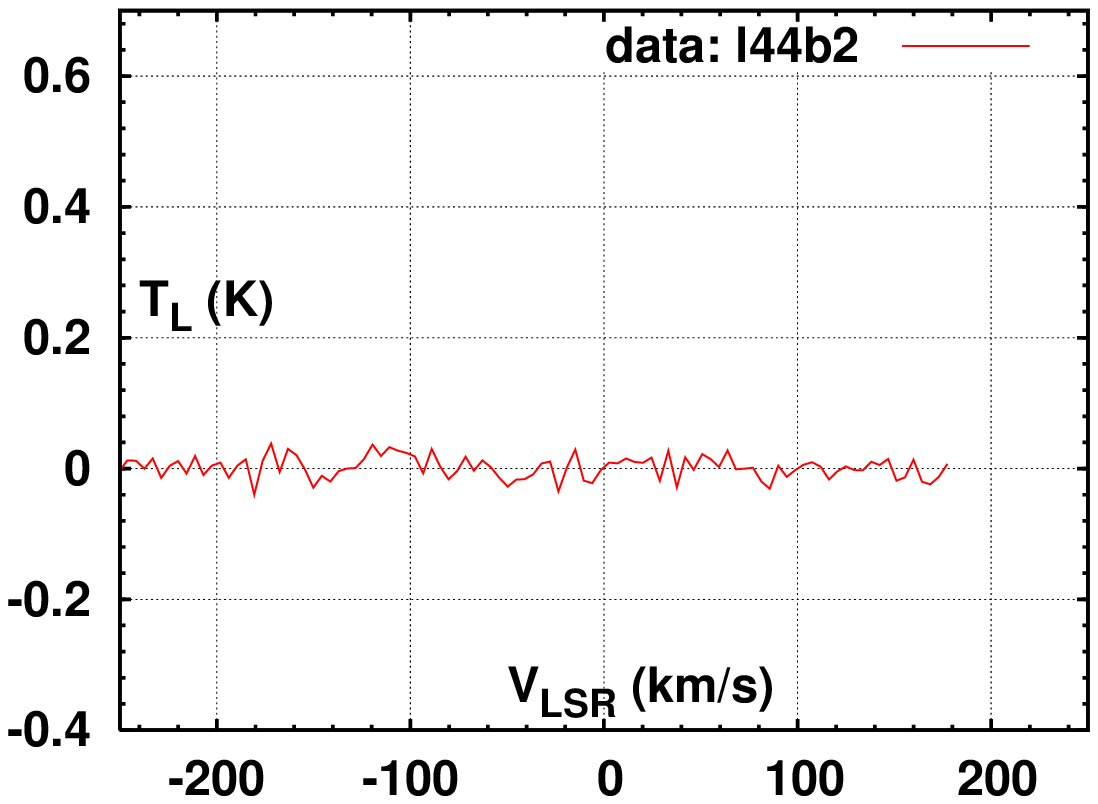}
\includegraphics[width=32mm,height=24mm,angle=0]{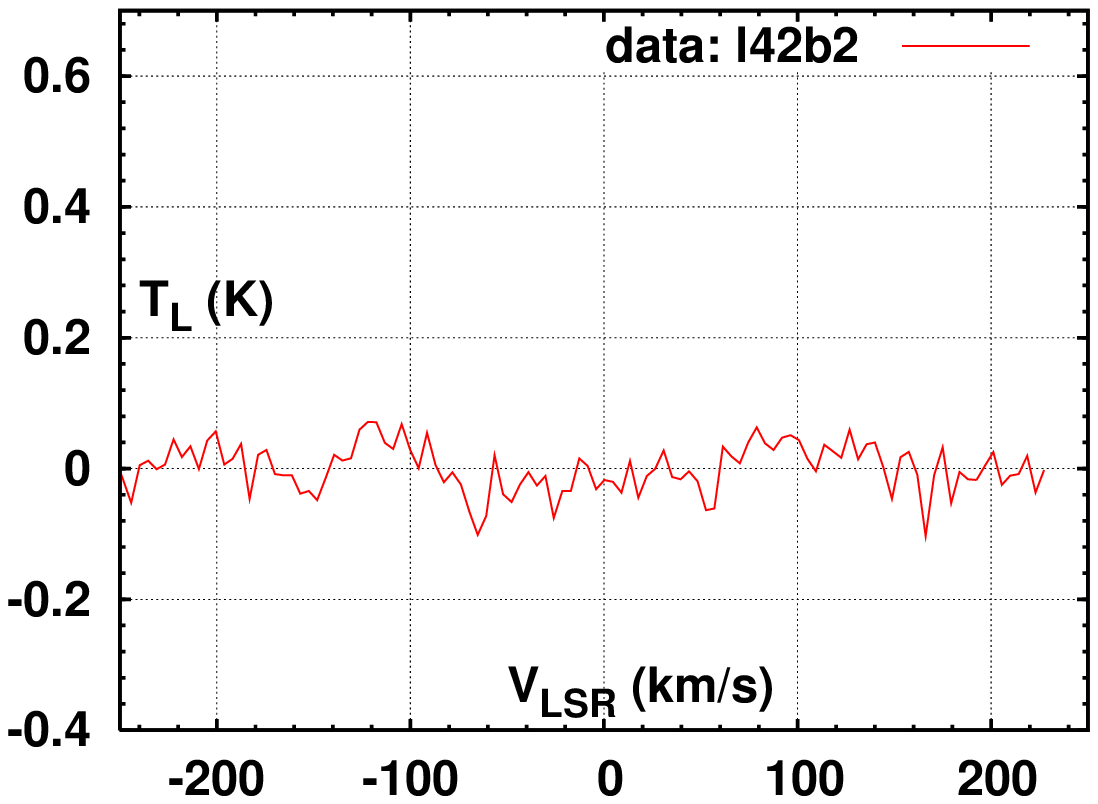}
\includegraphics[width=32mm,height=24mm,angle=0]{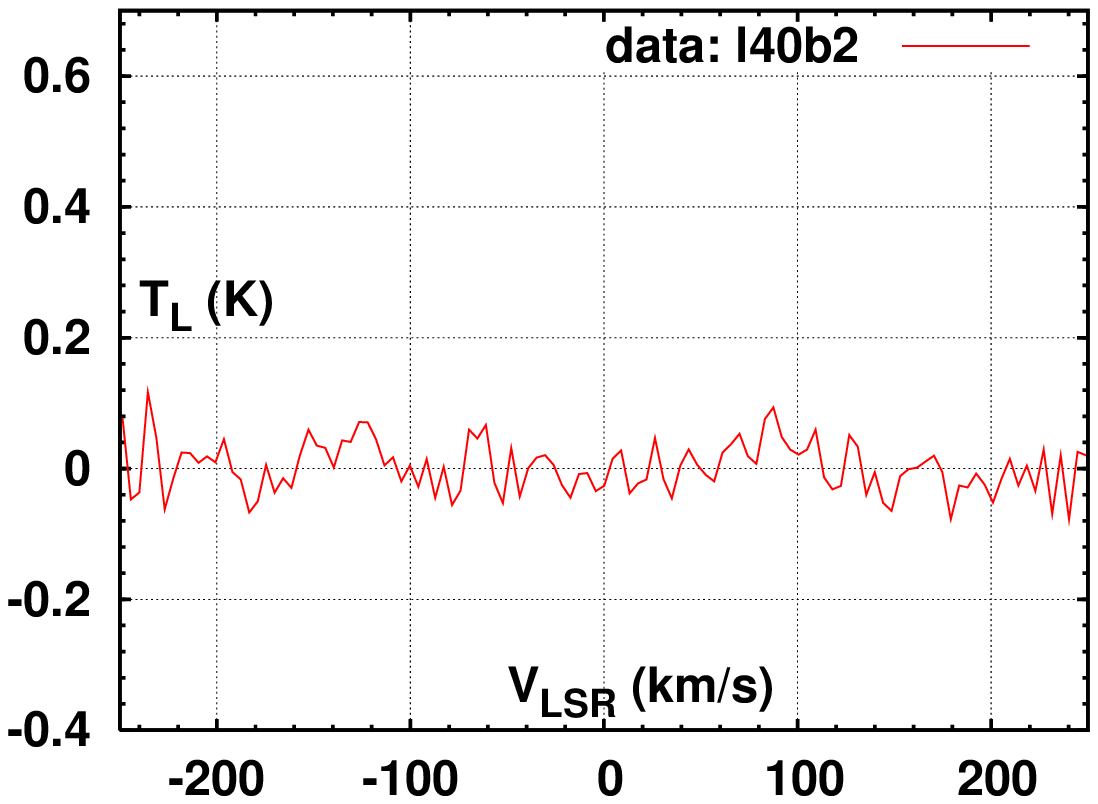}
\includegraphics[width=32mm,height=24mm,angle=0]{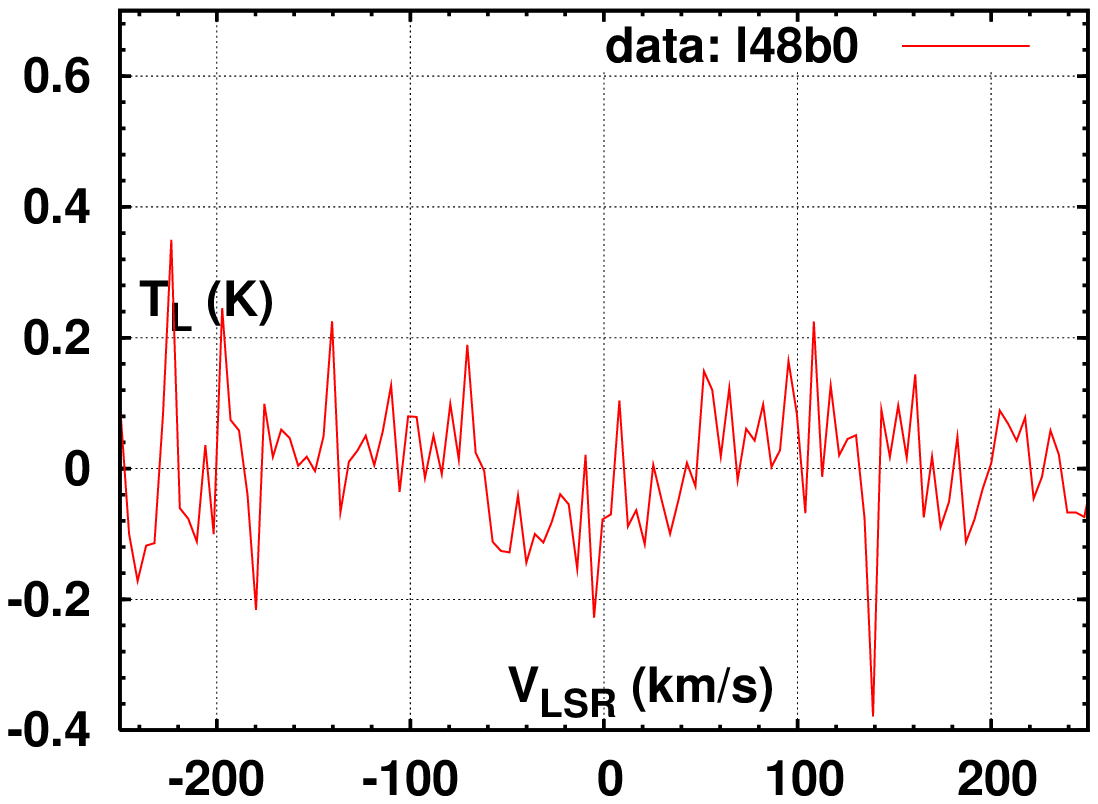}
\includegraphics[width=32mm,height=24mm,angle=0]{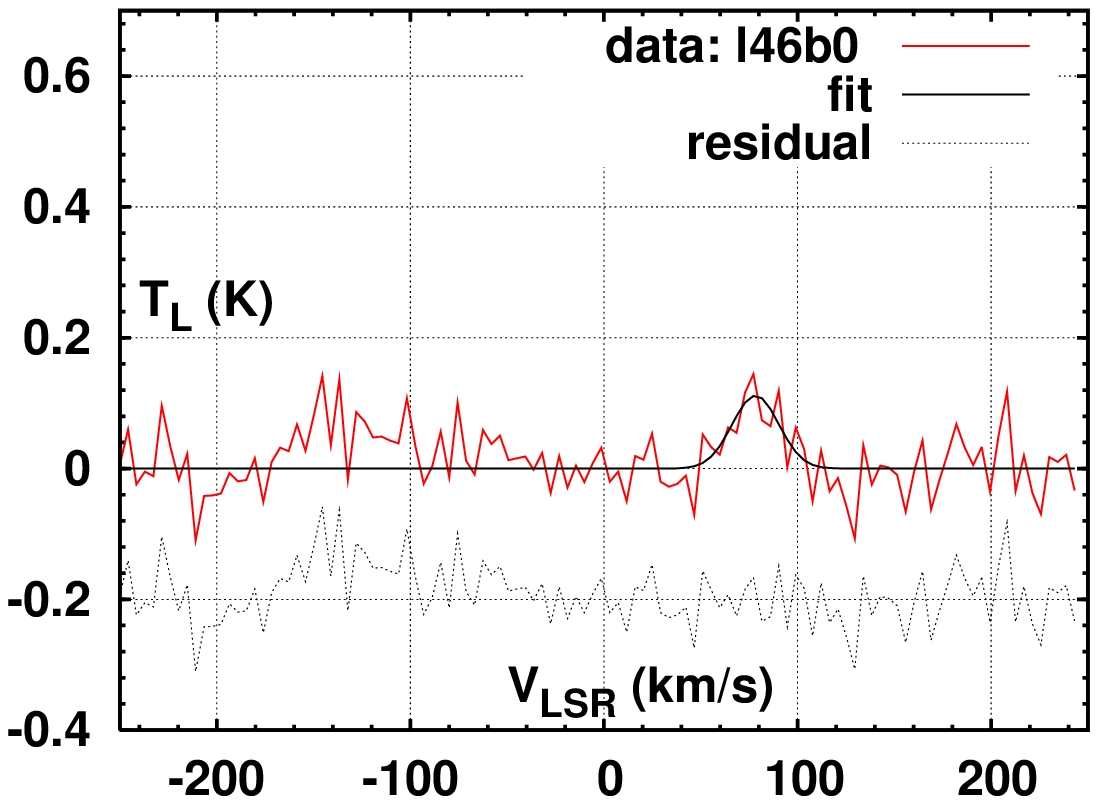}
\includegraphics[width=32mm,height=24mm,angle=0]{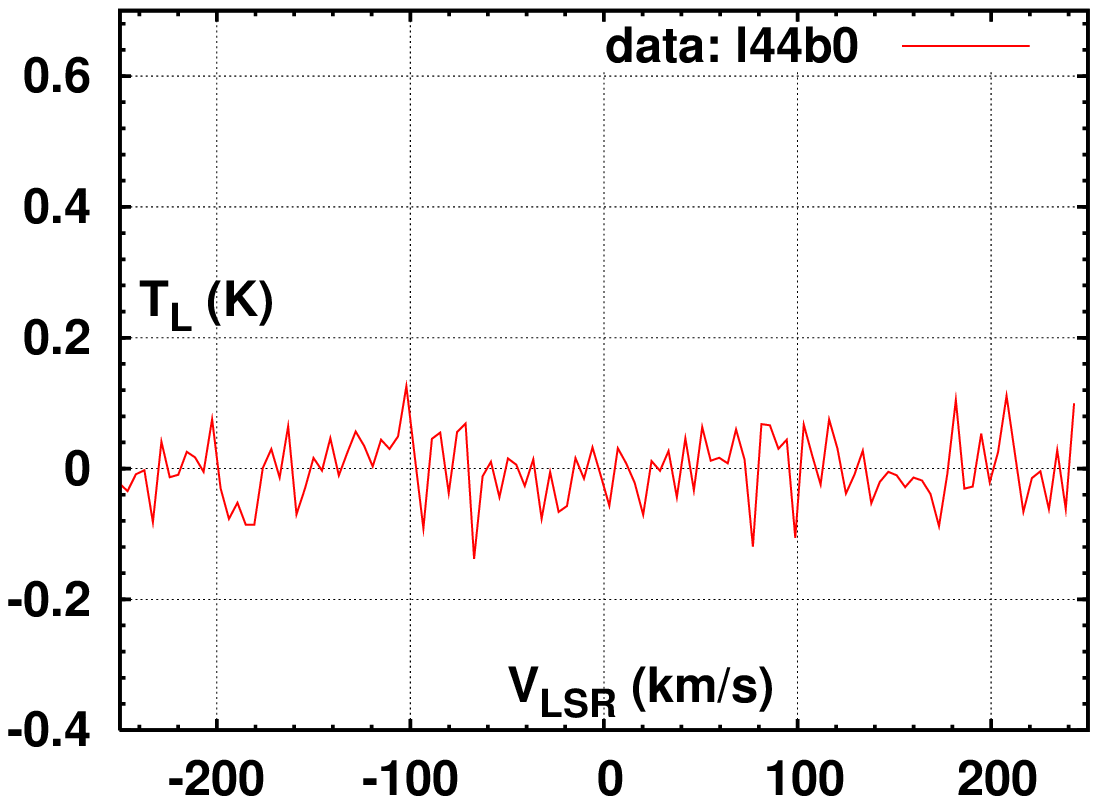}
\includegraphics[width=32mm,height=24mm,angle=0]{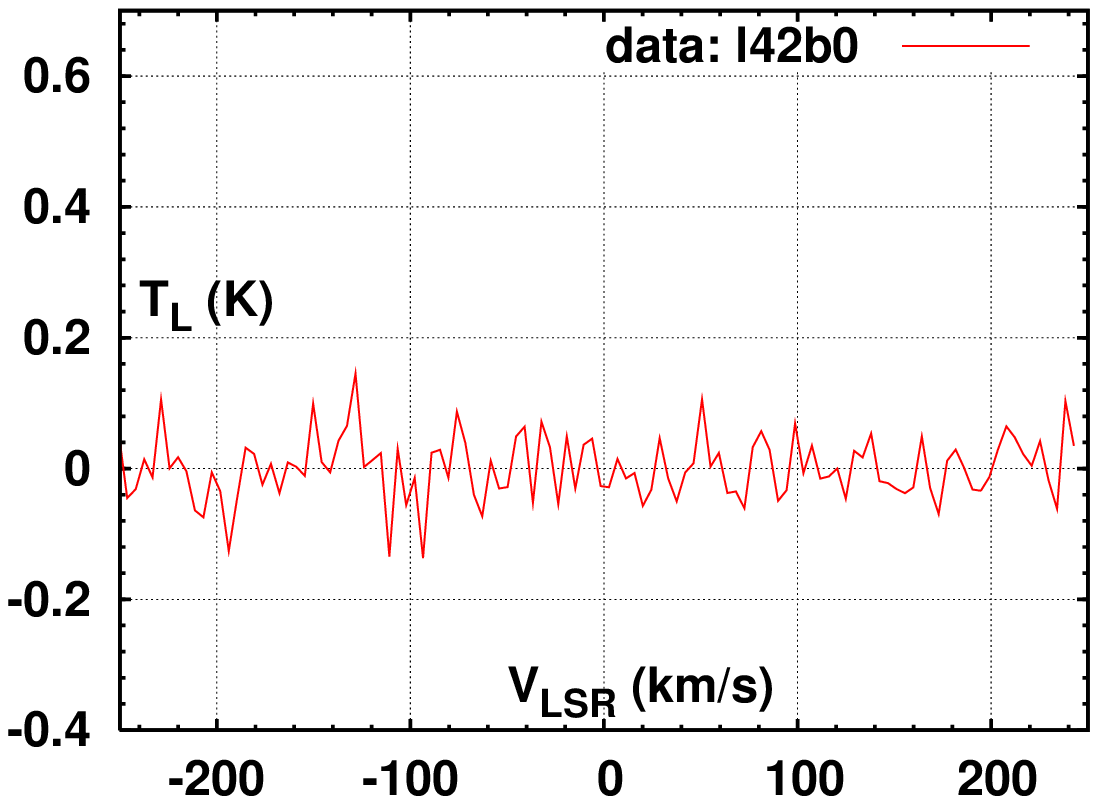}
\includegraphics[width=32mm,height=24mm,angle=0]{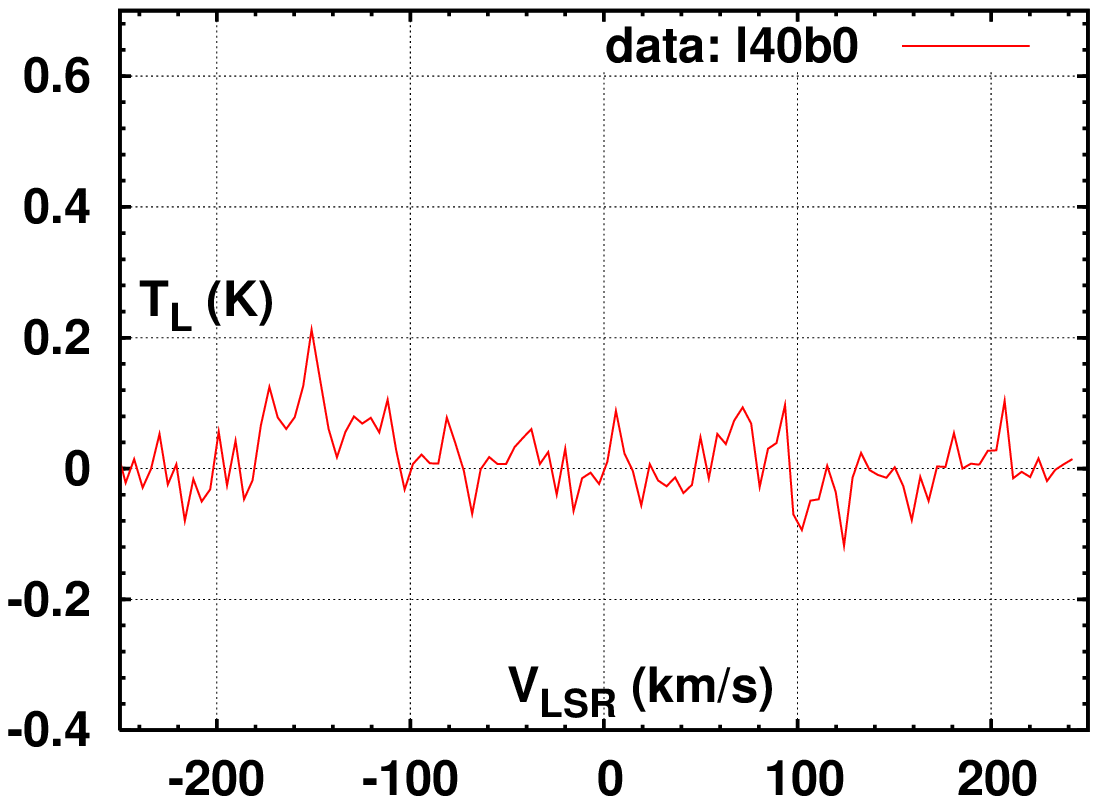}
\includegraphics[width=32mm,height=24mm,angle=0]{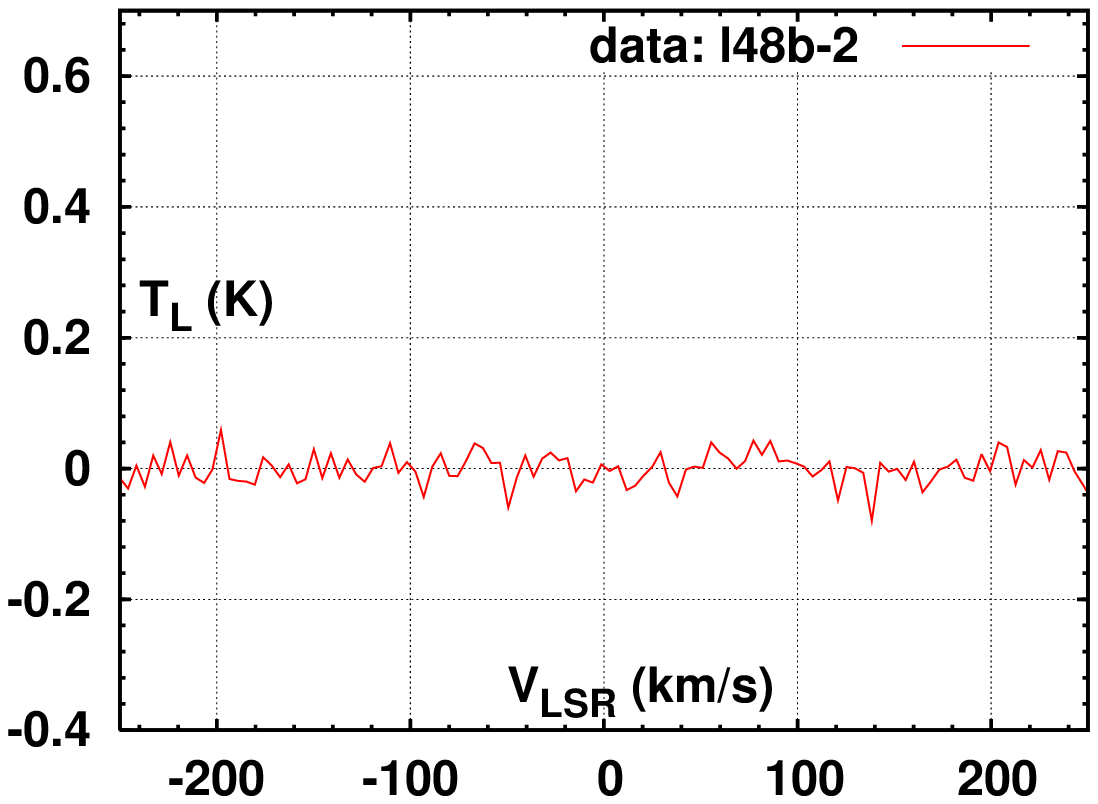}
\includegraphics[width=32mm,height=24mm,angle=0]{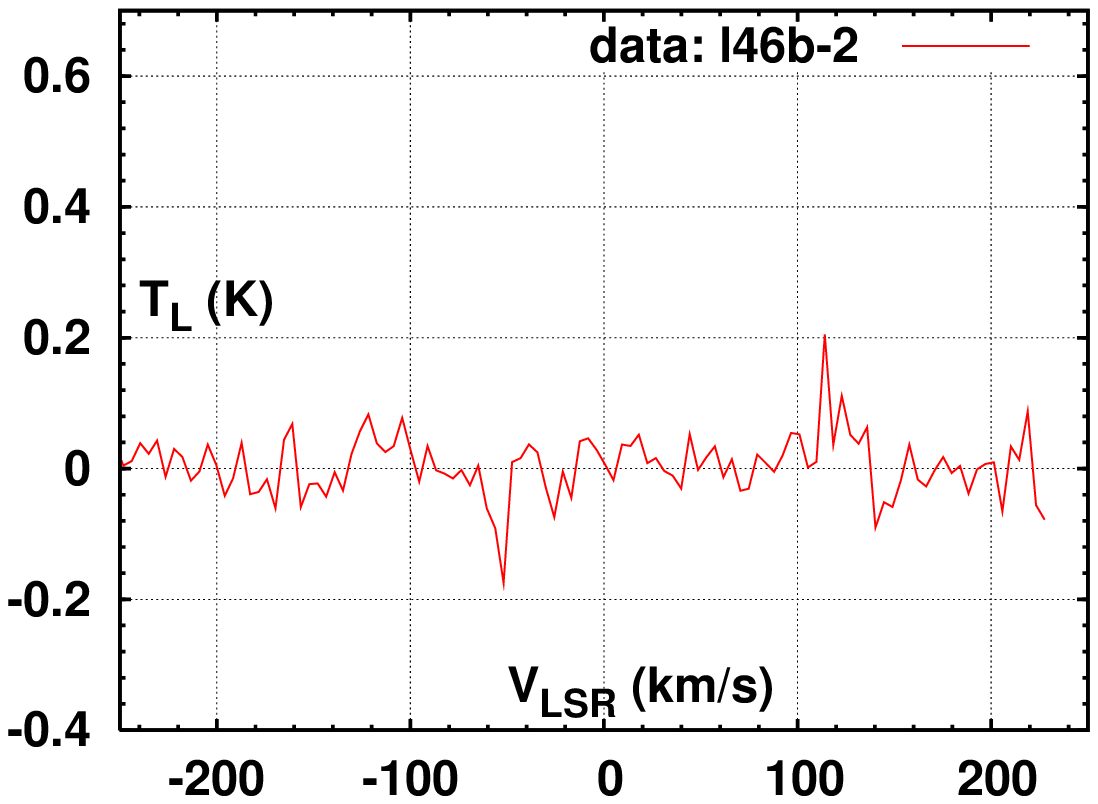}
\includegraphics[width=32mm,height=24mm,angle=0]{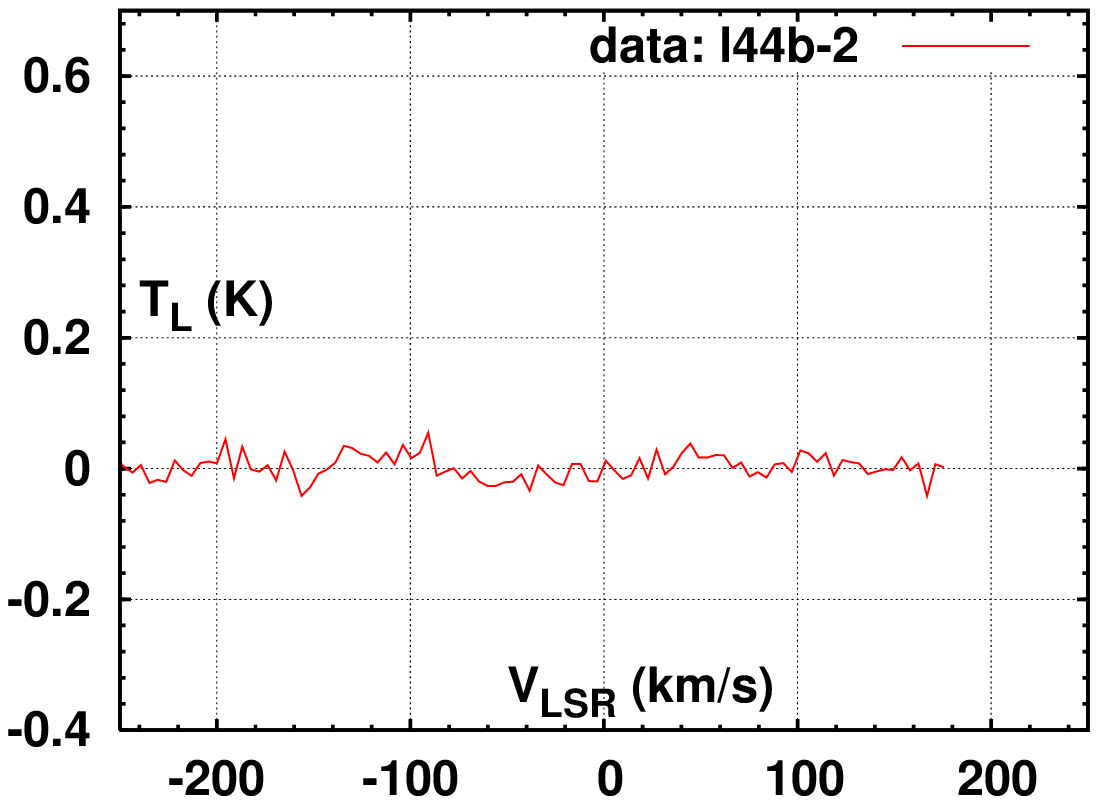}
\includegraphics[width=32mm,height=24mm,angle=0]{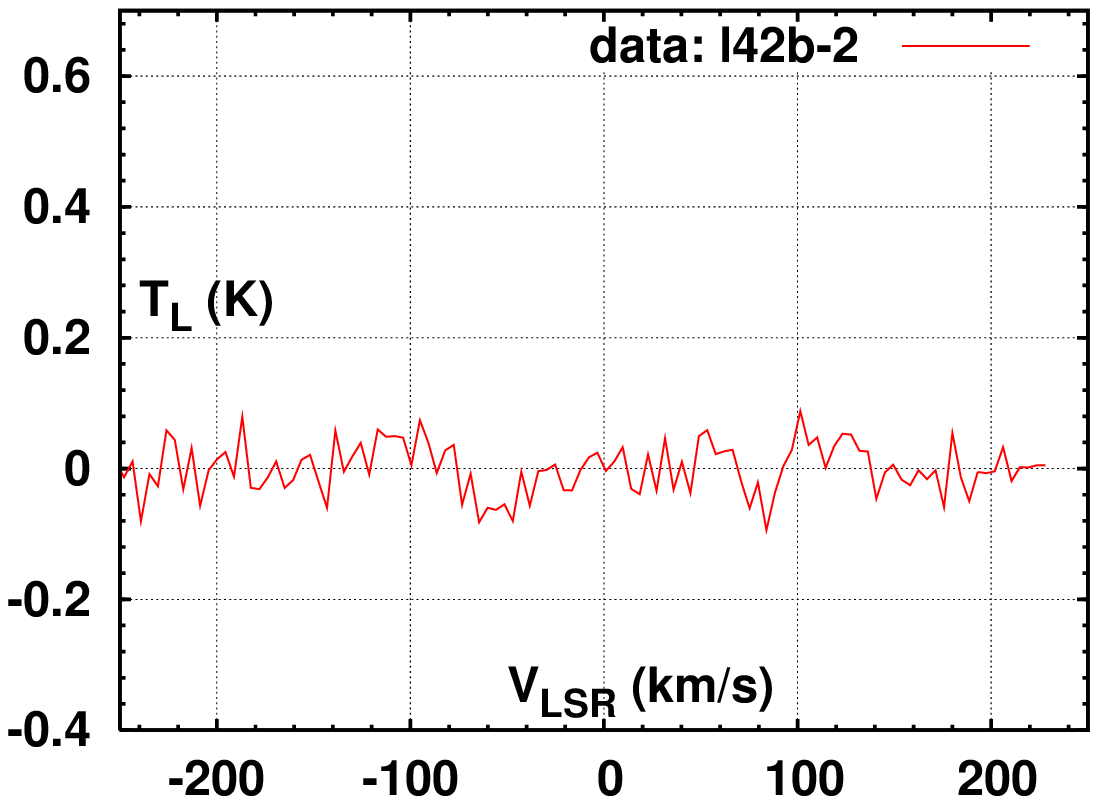}
\includegraphics[width=32mm,height=24mm,angle=0]{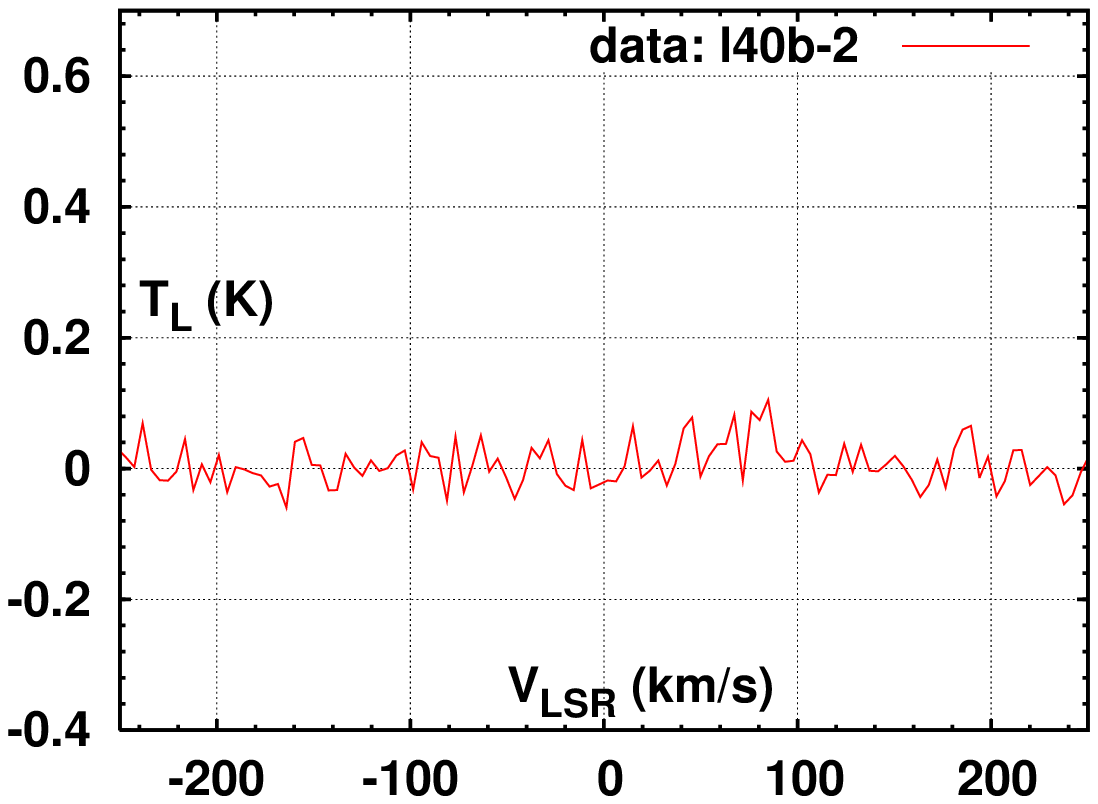}

 \caption{ORT H271$\alpha$ RL observation.}
\end{center}
\end{figure}

\begin{figure}[ht]
\begin{center}
\includegraphics[width=32mm,height=24mm,angle=0]{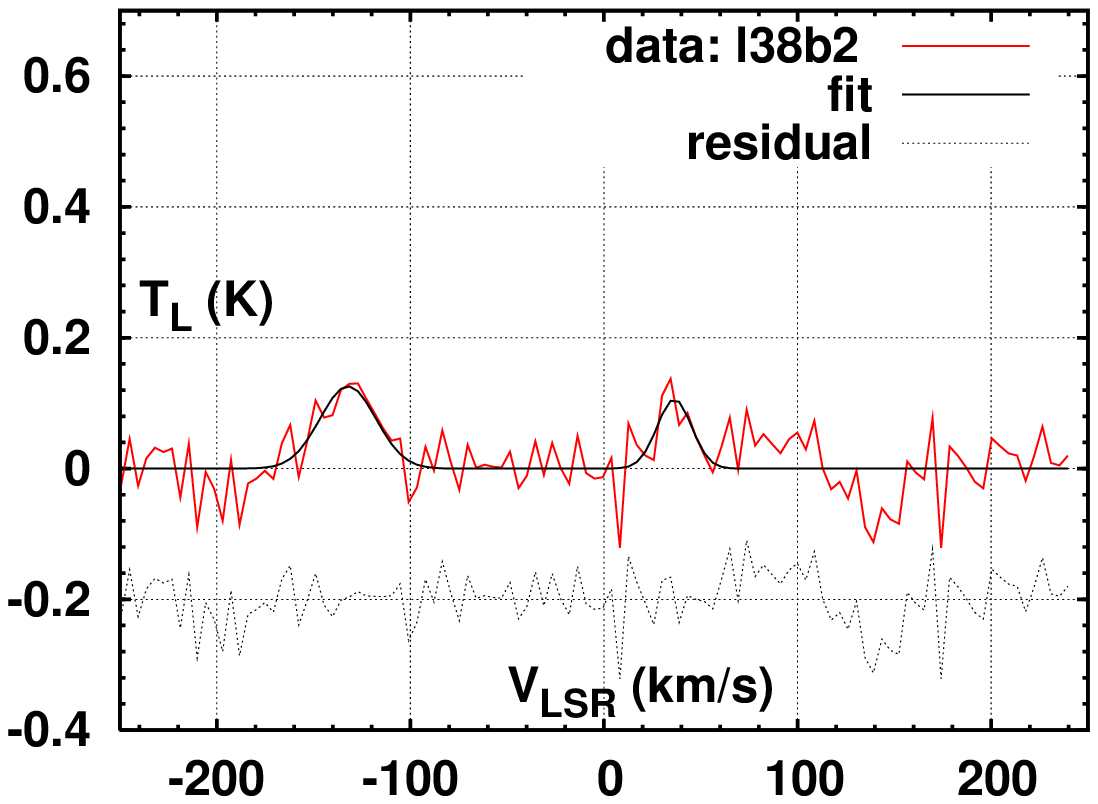}
\includegraphics[width=32mm,height=24mm,angle=0]{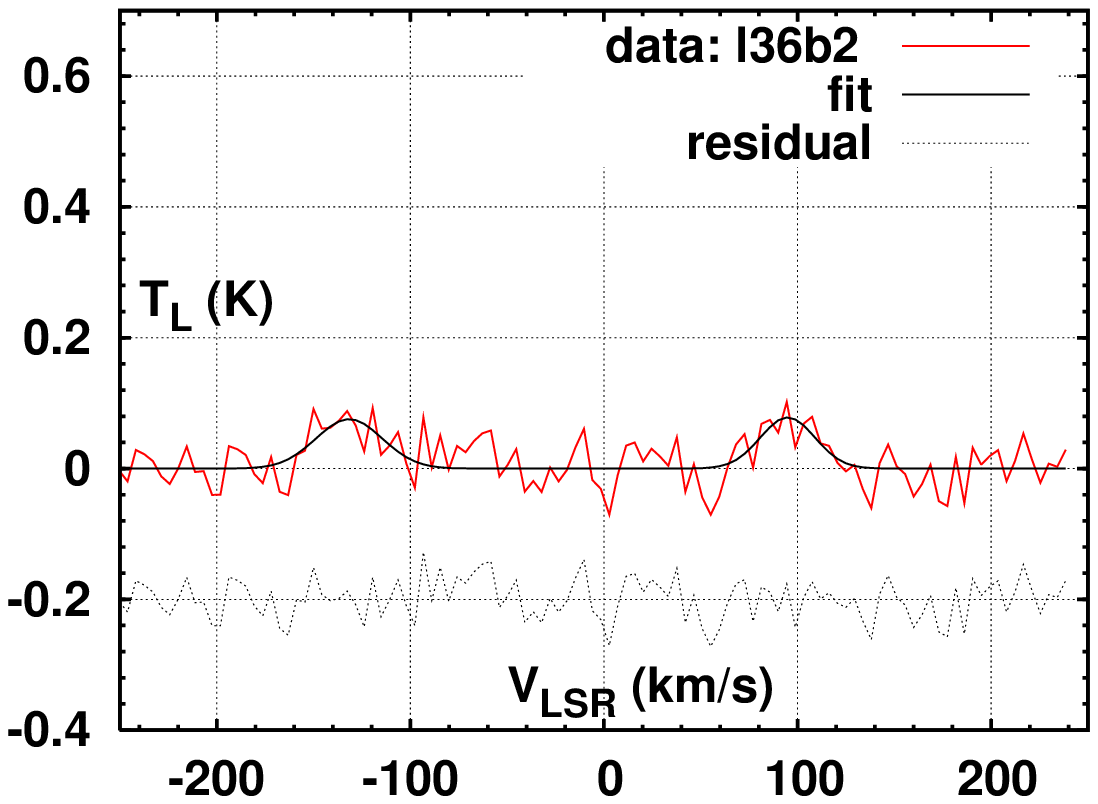}
\includegraphics[width=32mm,height=24mm,angle=0]{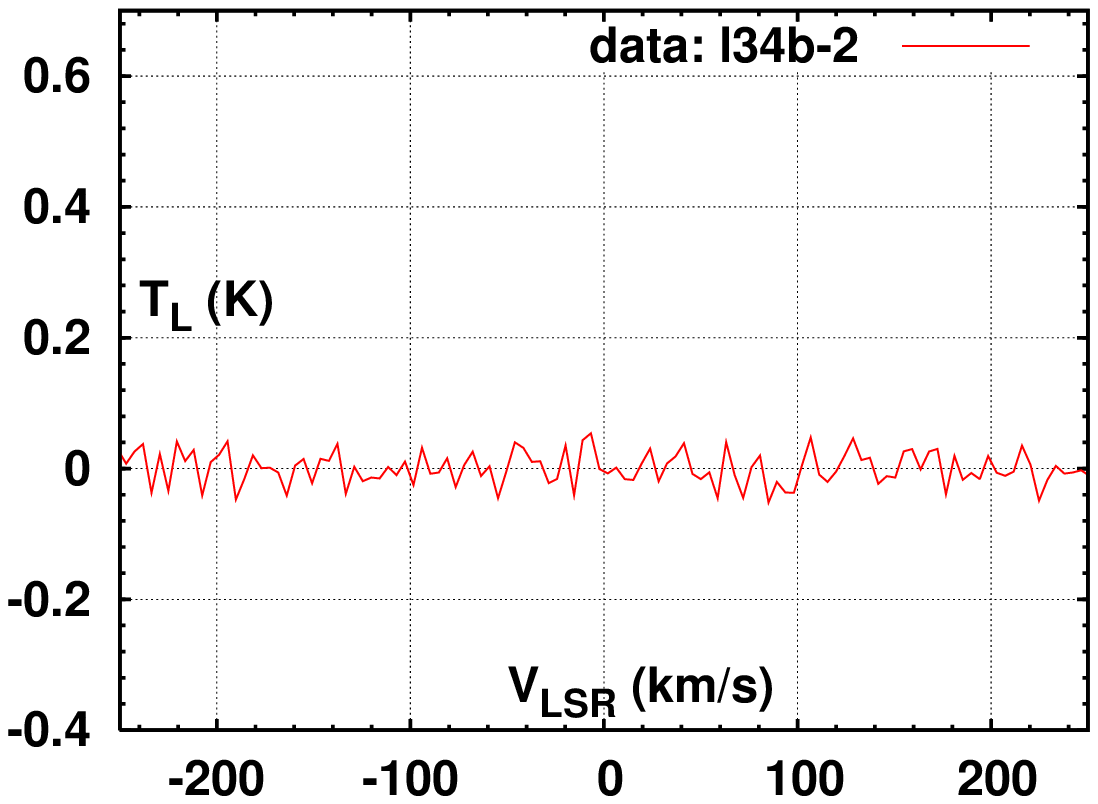}
\includegraphics[width=32mm,height=24mm,angle=0]{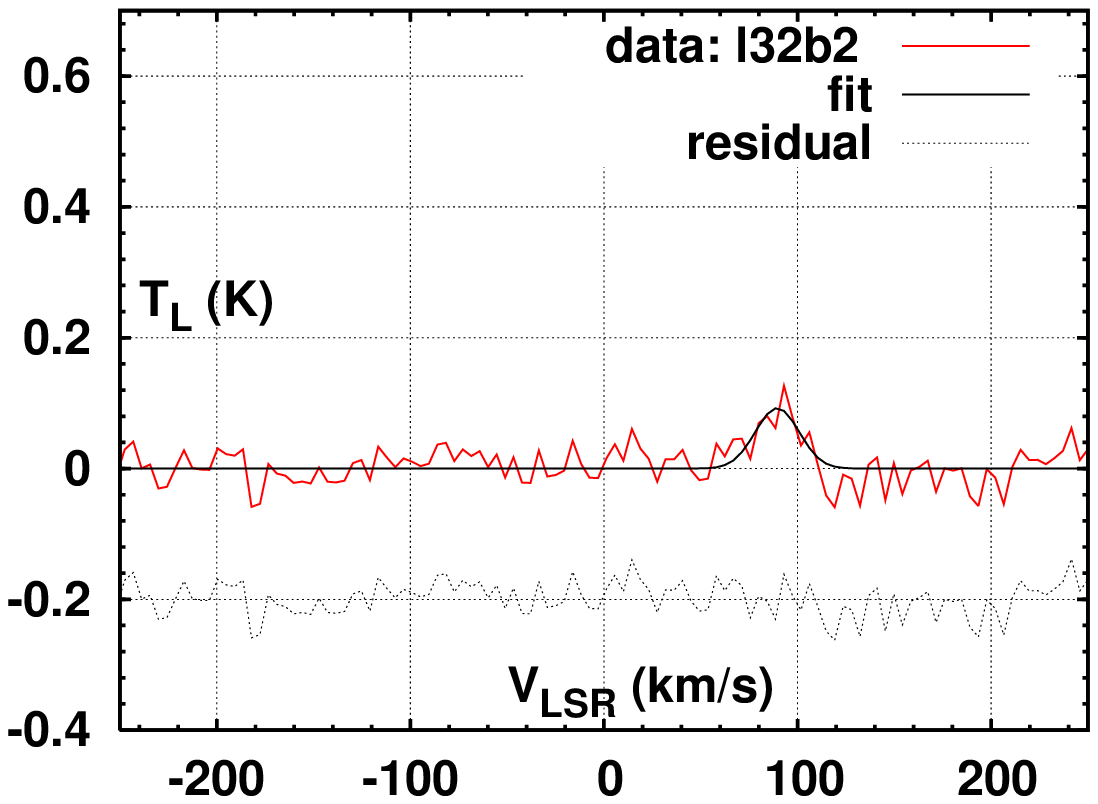}
\includegraphics[width=32mm,height=24mm,angle=0]{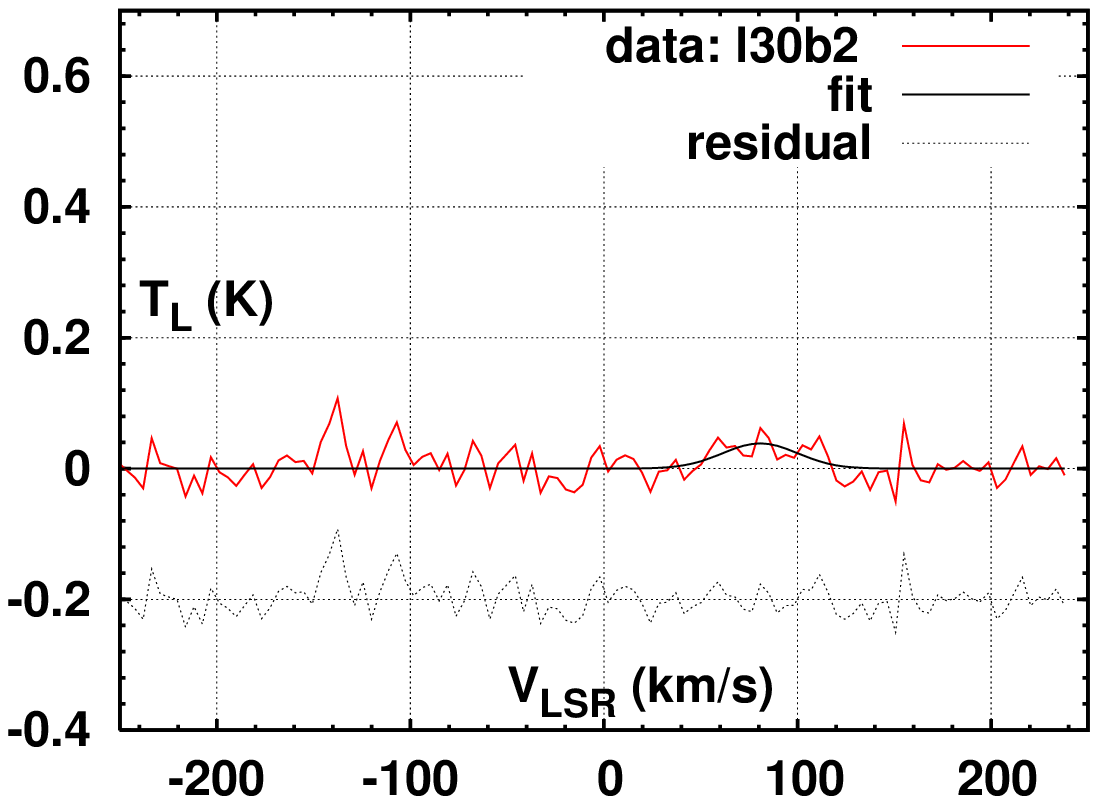}
\includegraphics[width=32mm,height=24mm,angle=0]{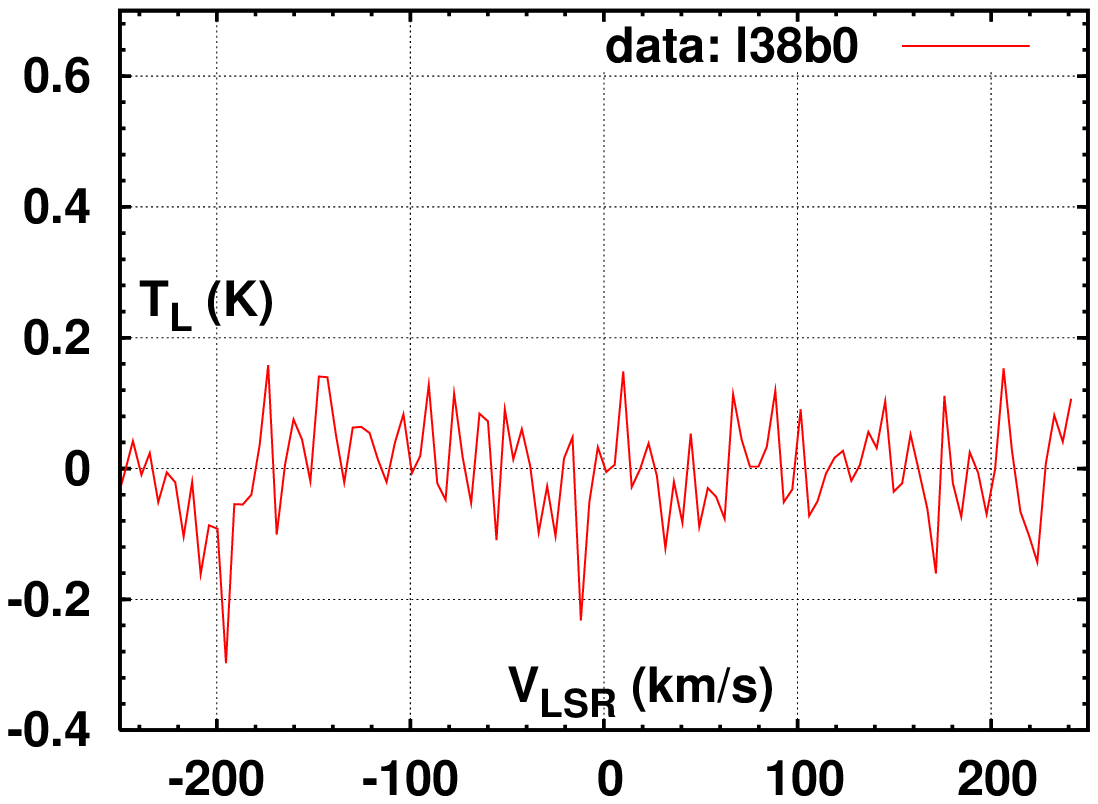}
\includegraphics[width=32mm,height=24mm,angle=0]{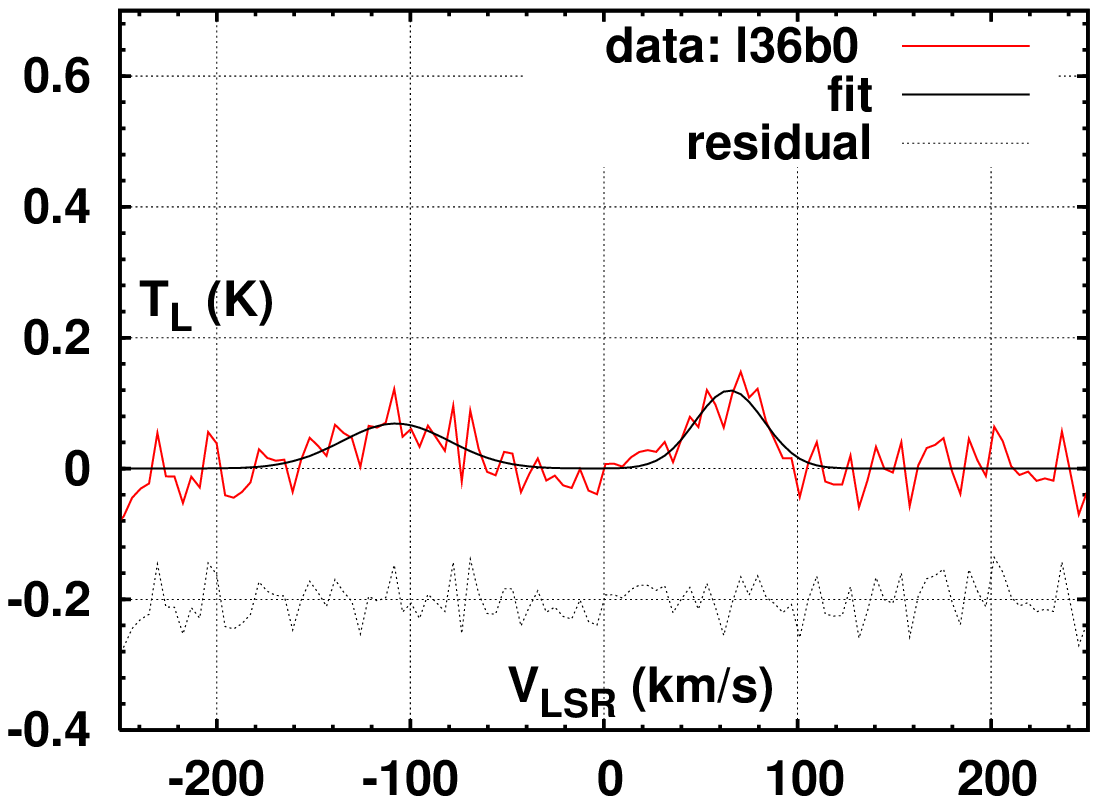}
\includegraphics[width=32mm,height=24mm,angle=0]{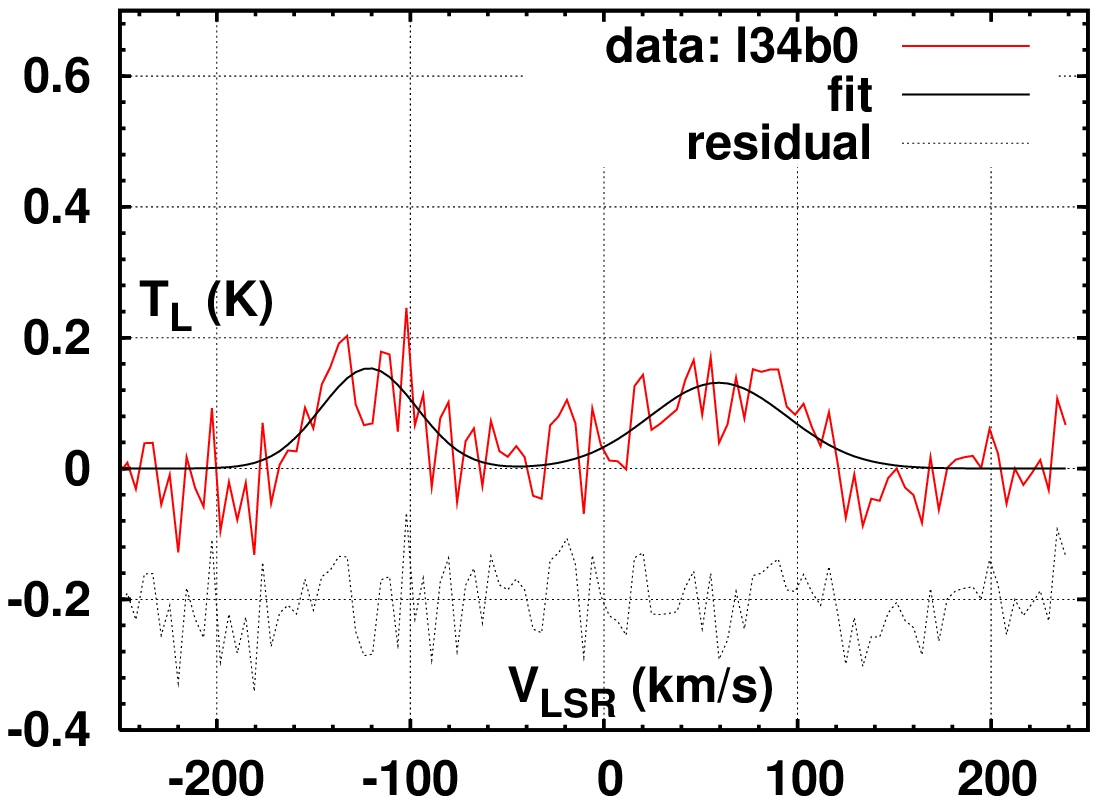}
\includegraphics[width=32mm,height=24mm,angle=0]{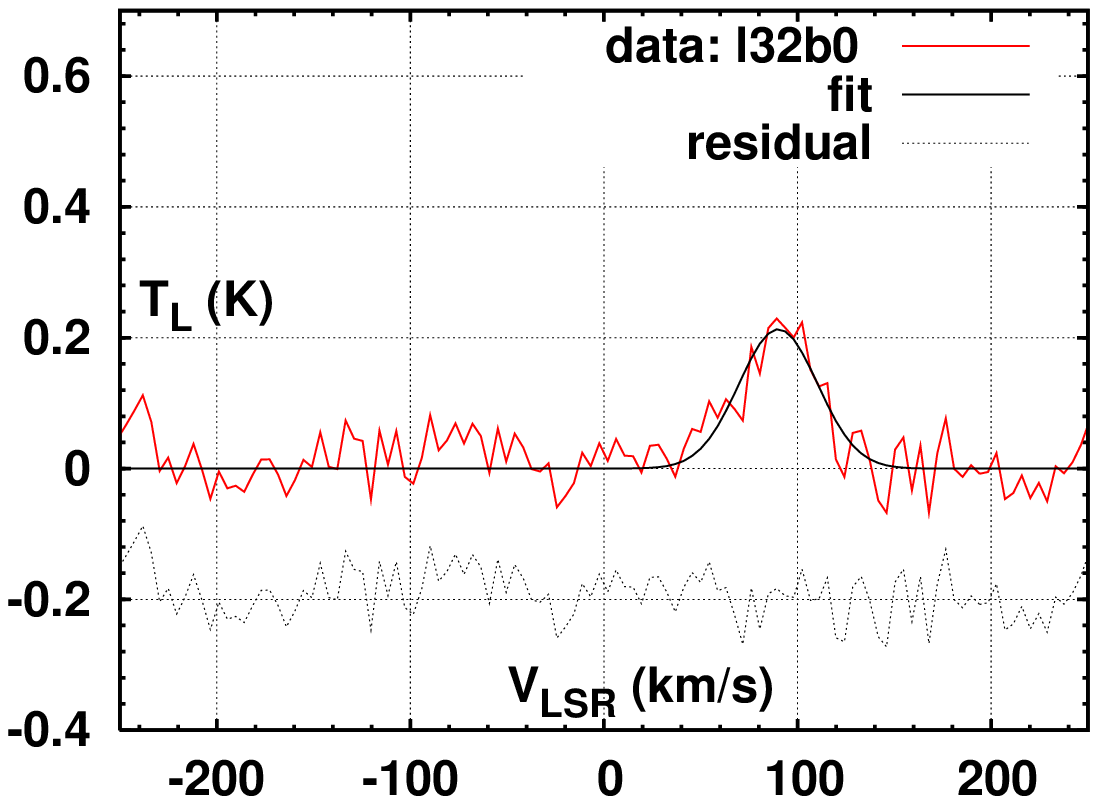}
\includegraphics[width=32mm,height=24mm,angle=0]{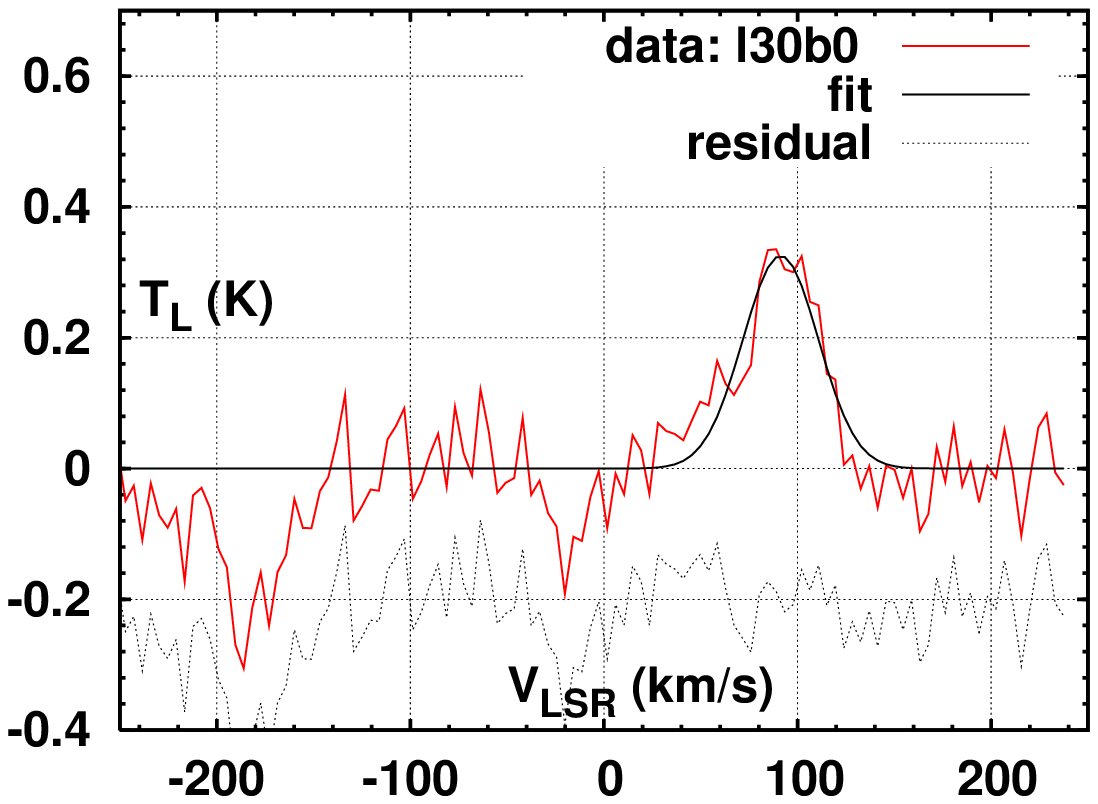}
\includegraphics[width=32mm,height=24mm,angle=0]{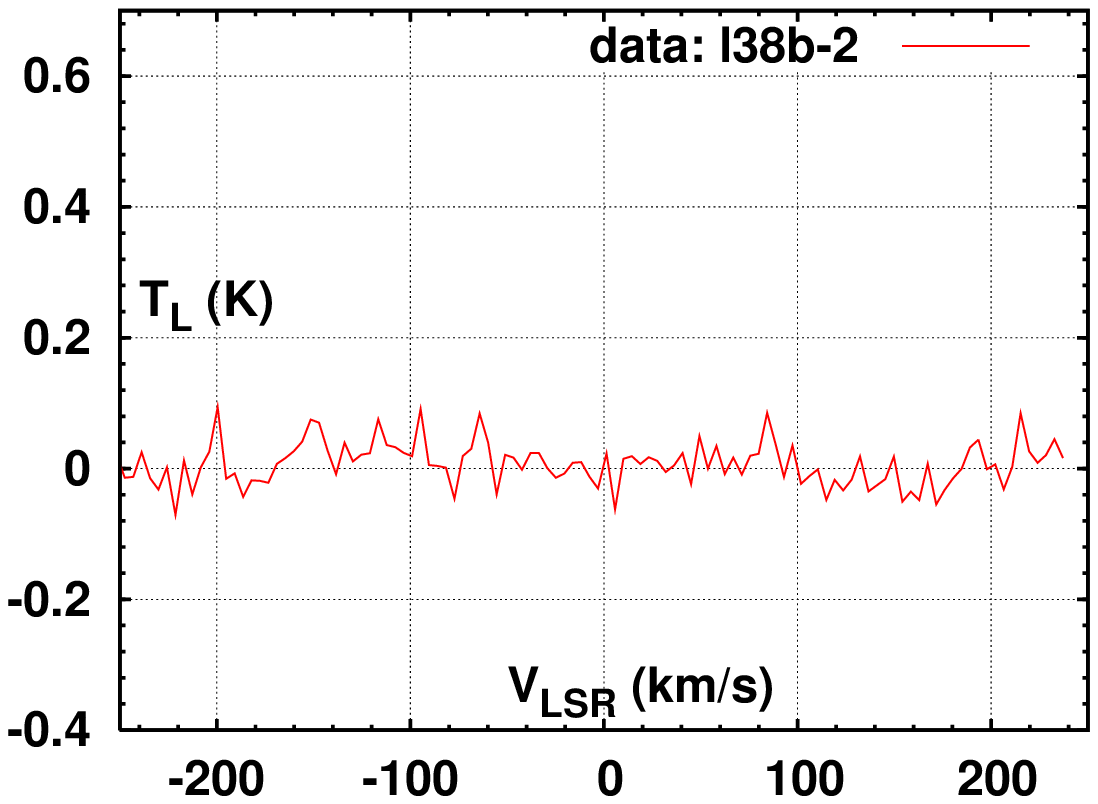}
\includegraphics[width=32mm,height=24mm,angle=0]{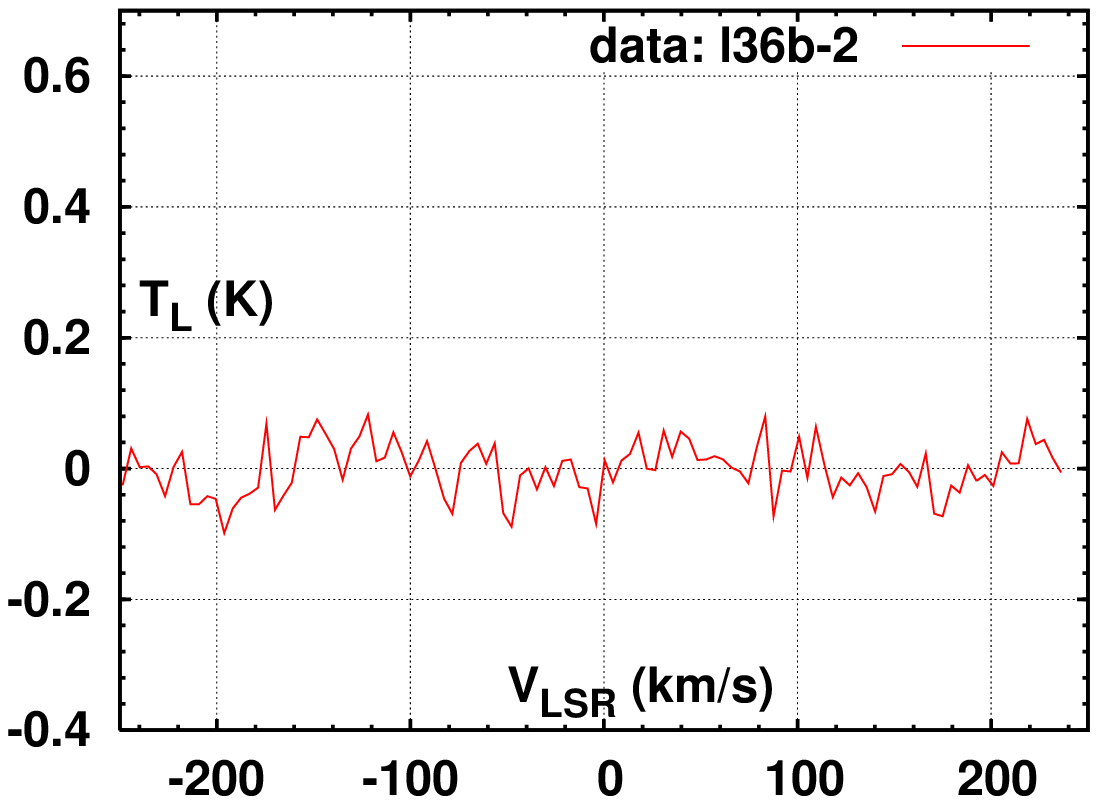}
\includegraphics[width=32mm,height=24mm,angle=0]{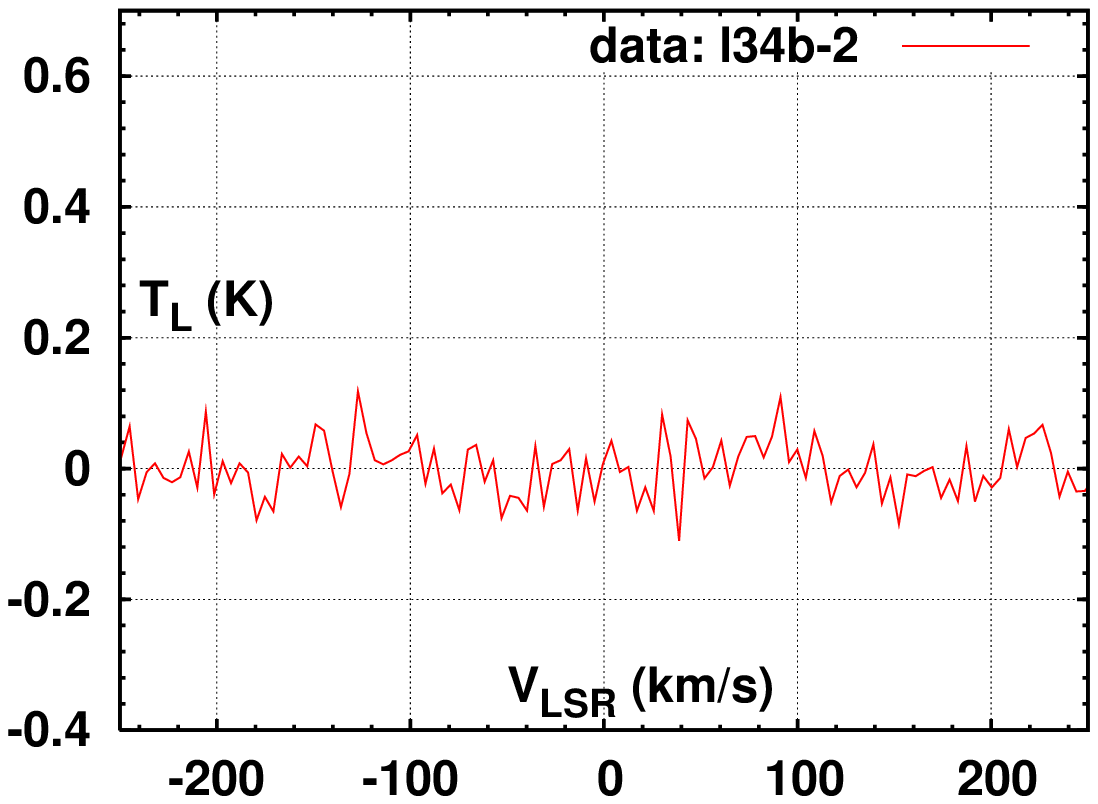}
\includegraphics[width=32mm,height=24mm,angle=0]{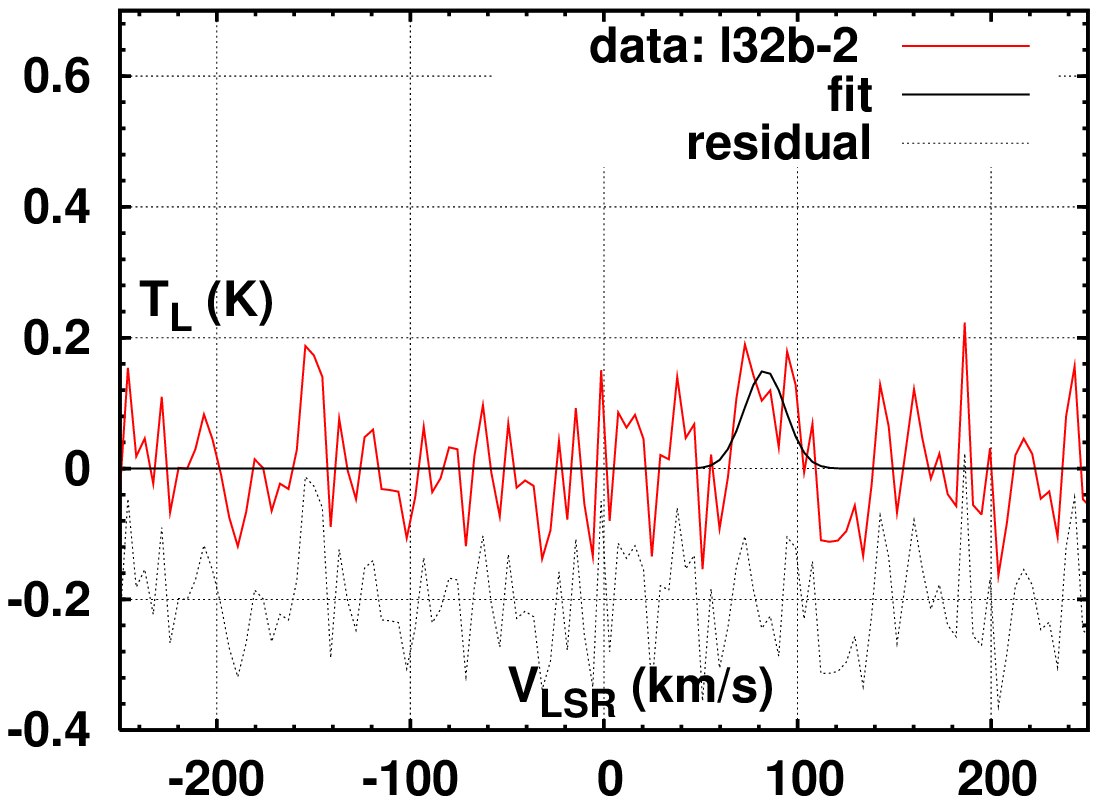}
\includegraphics[width=32mm,height=24mm,angle=0]{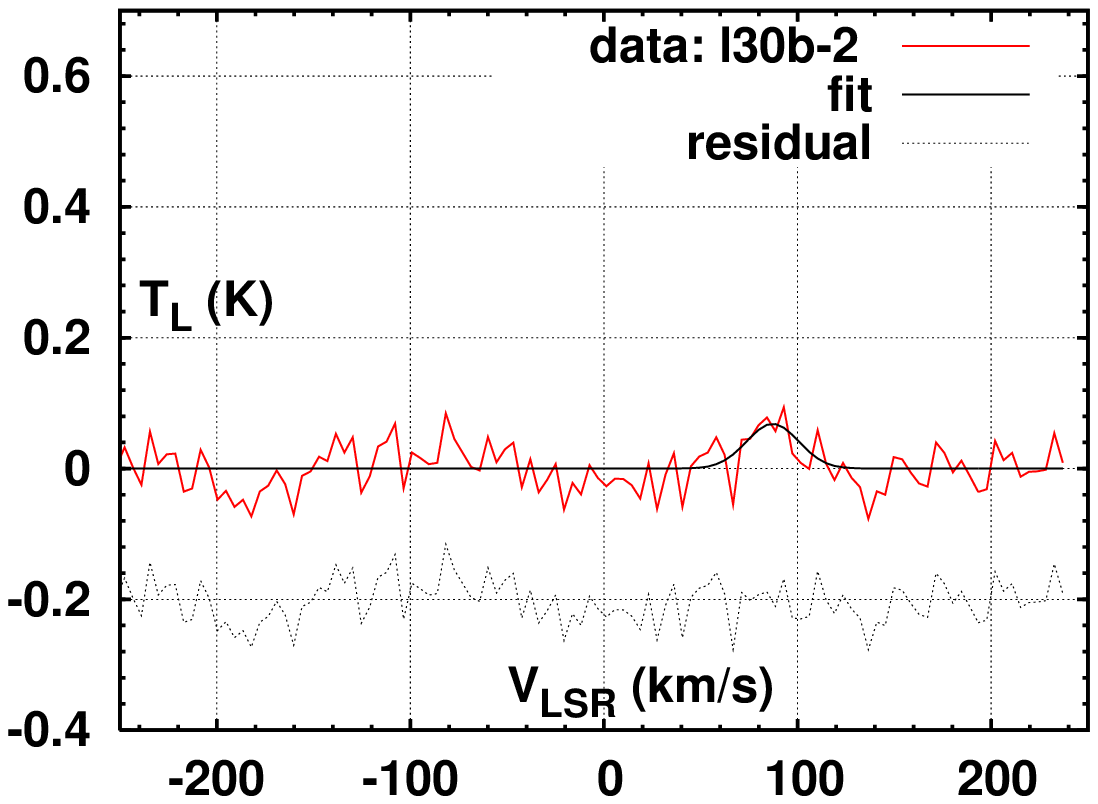}\vspace{1cm}
\includegraphics[width=32mm,height=24mm,angle=0]{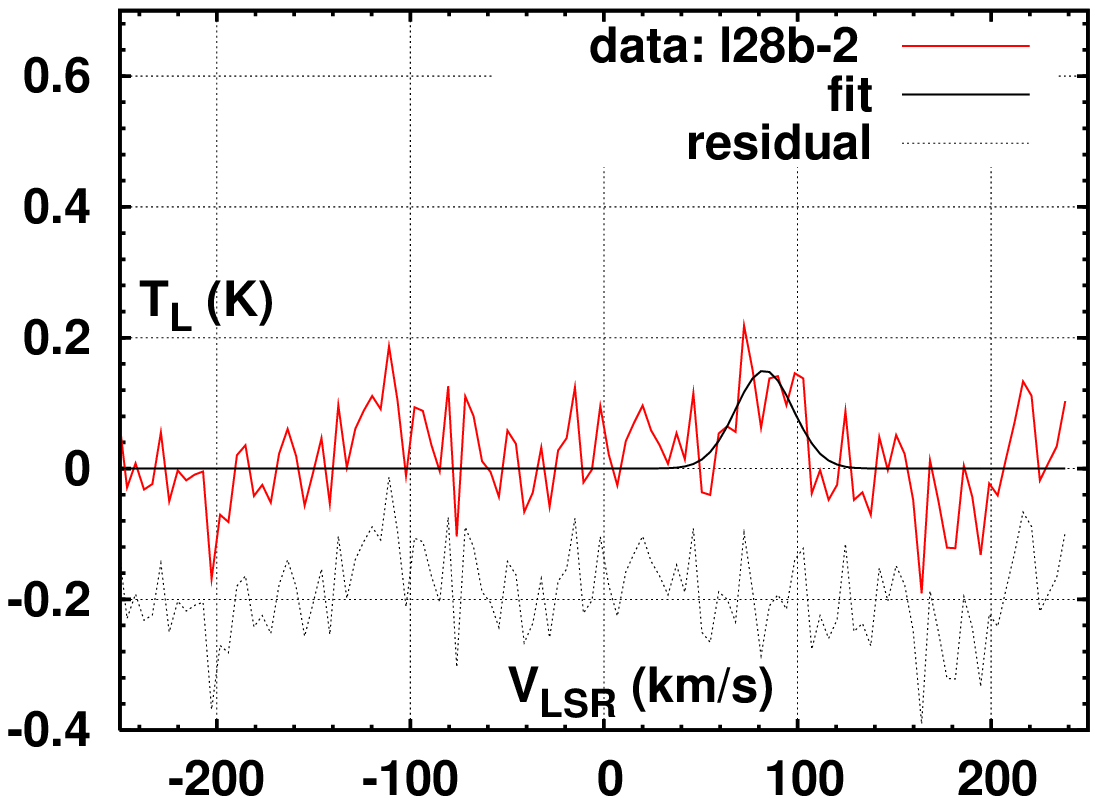}
\includegraphics[width=32mm,height=24mm,angle=0]{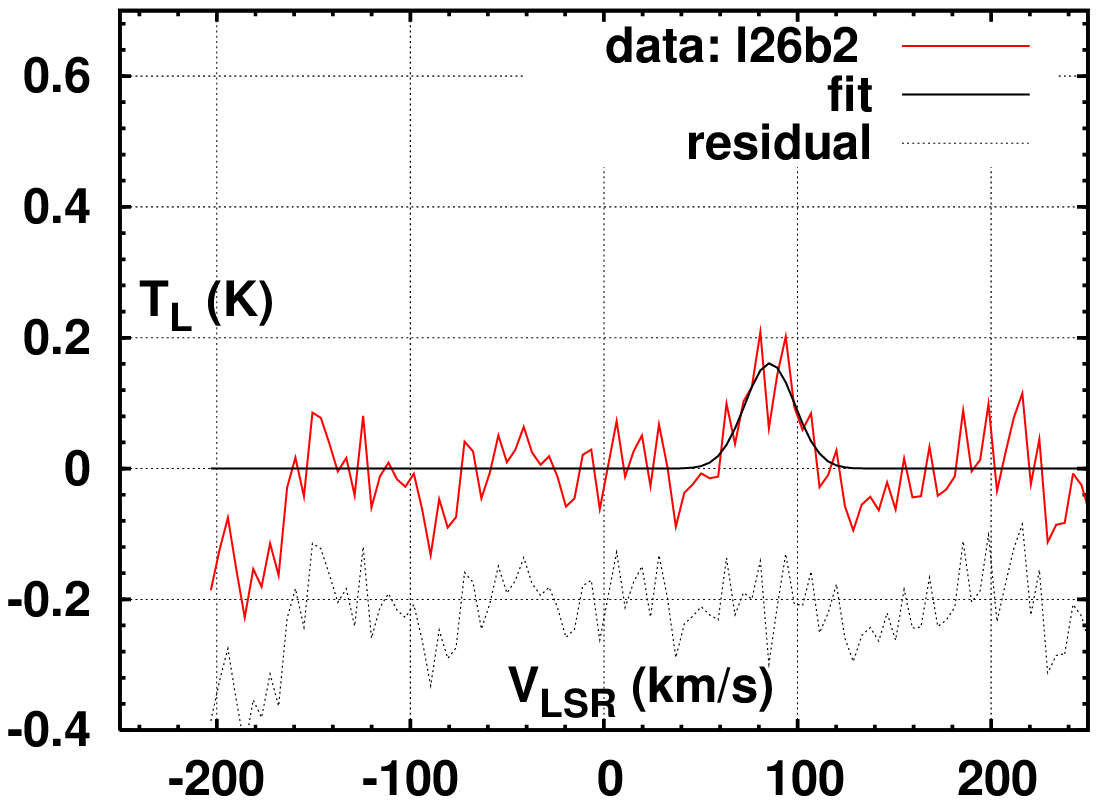}
\includegraphics[width=32mm,height=24mm,angle=0]{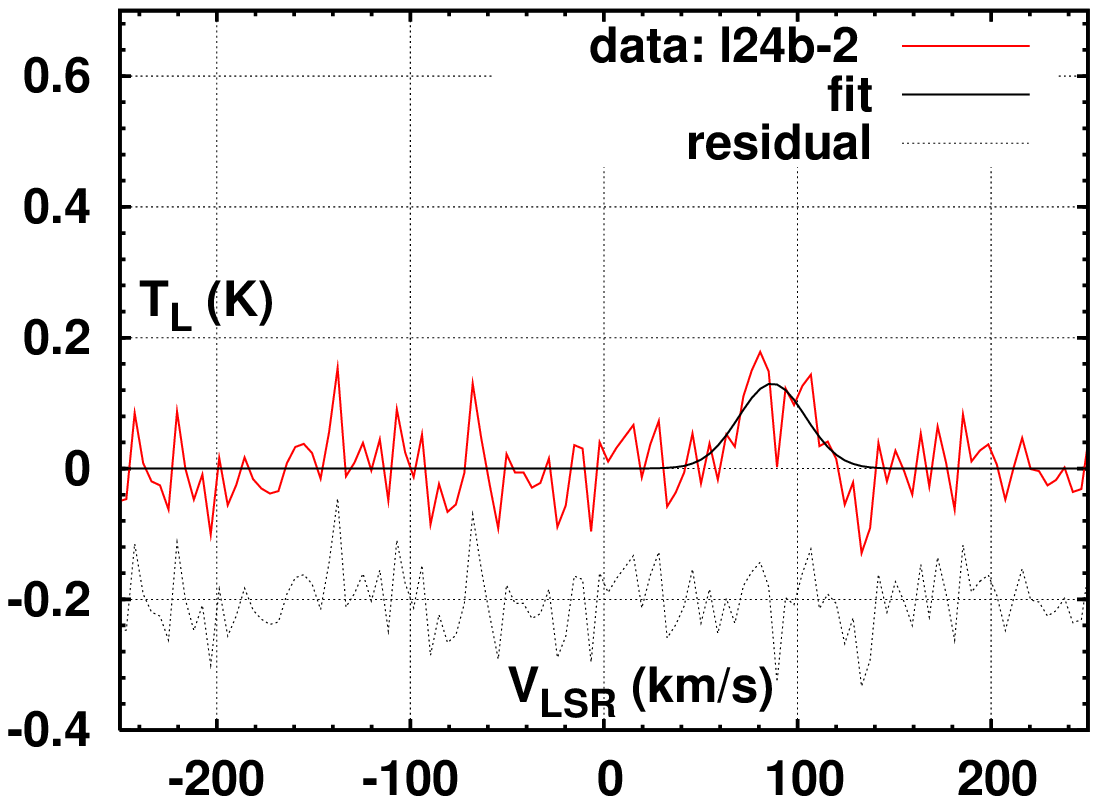}
\includegraphics[width=32mm,height=24mm,angle=0]{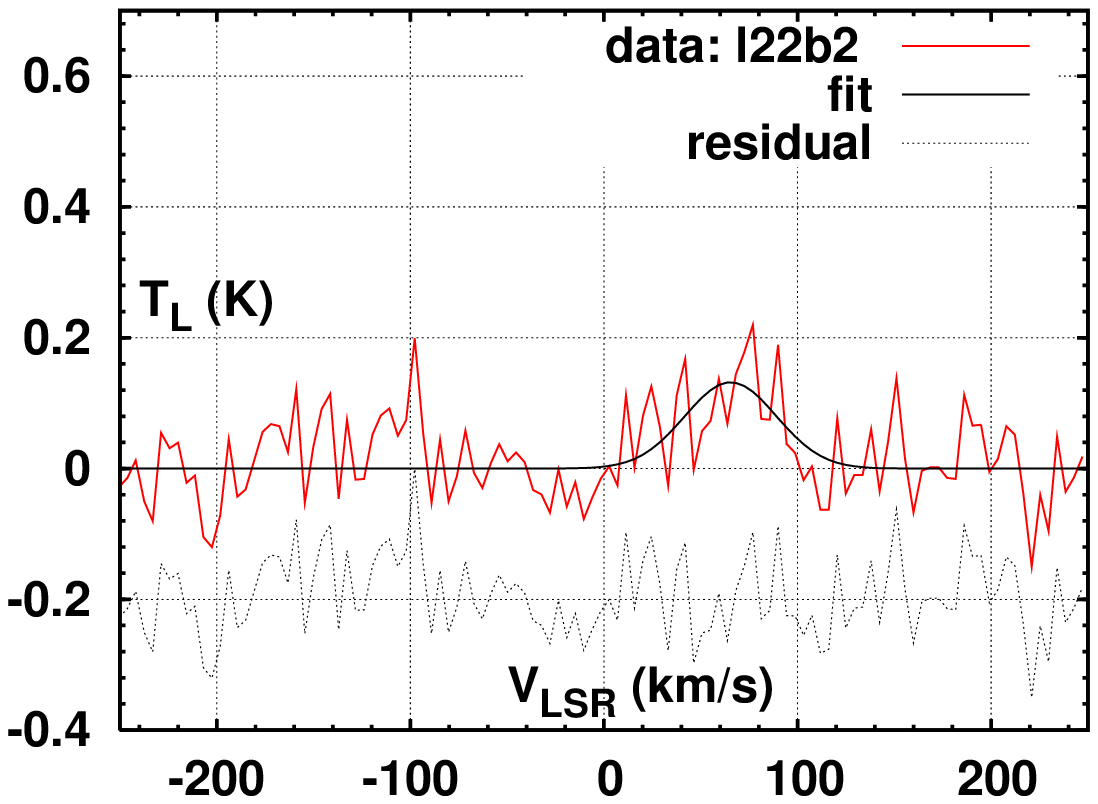}
\includegraphics[width=32mm,height=24mm,angle=0]{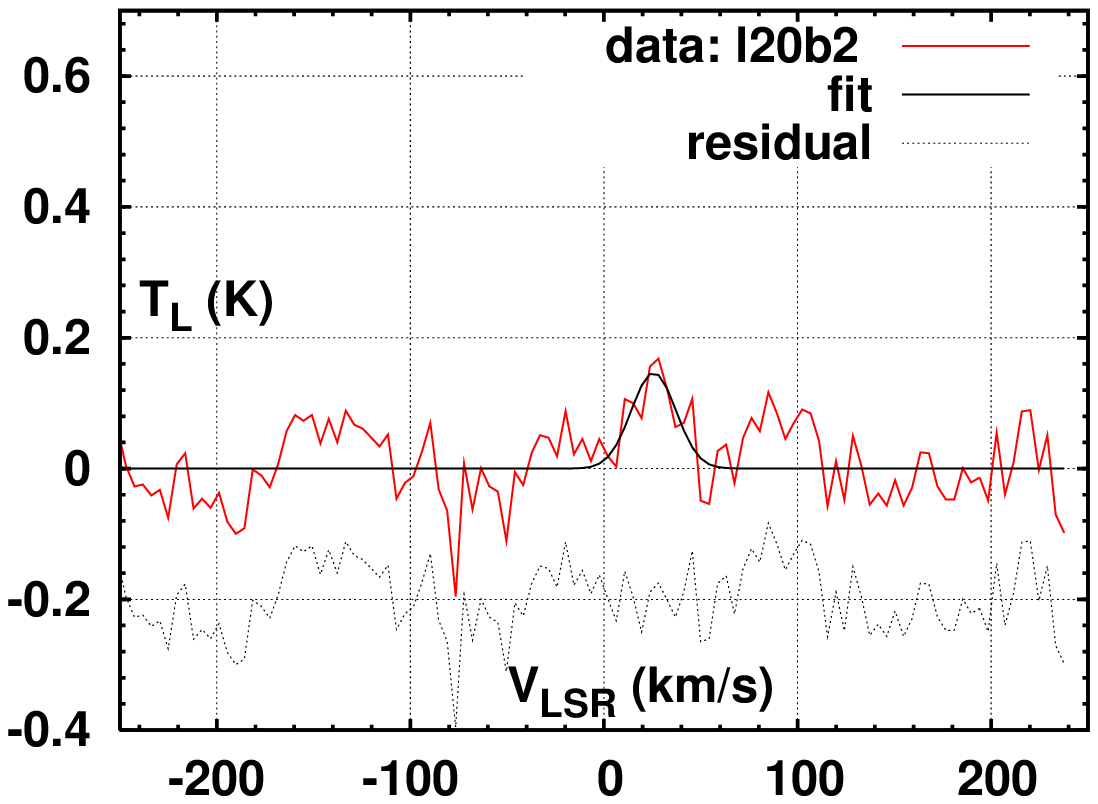}
\includegraphics[width=32mm,height=24mm,angle=0]{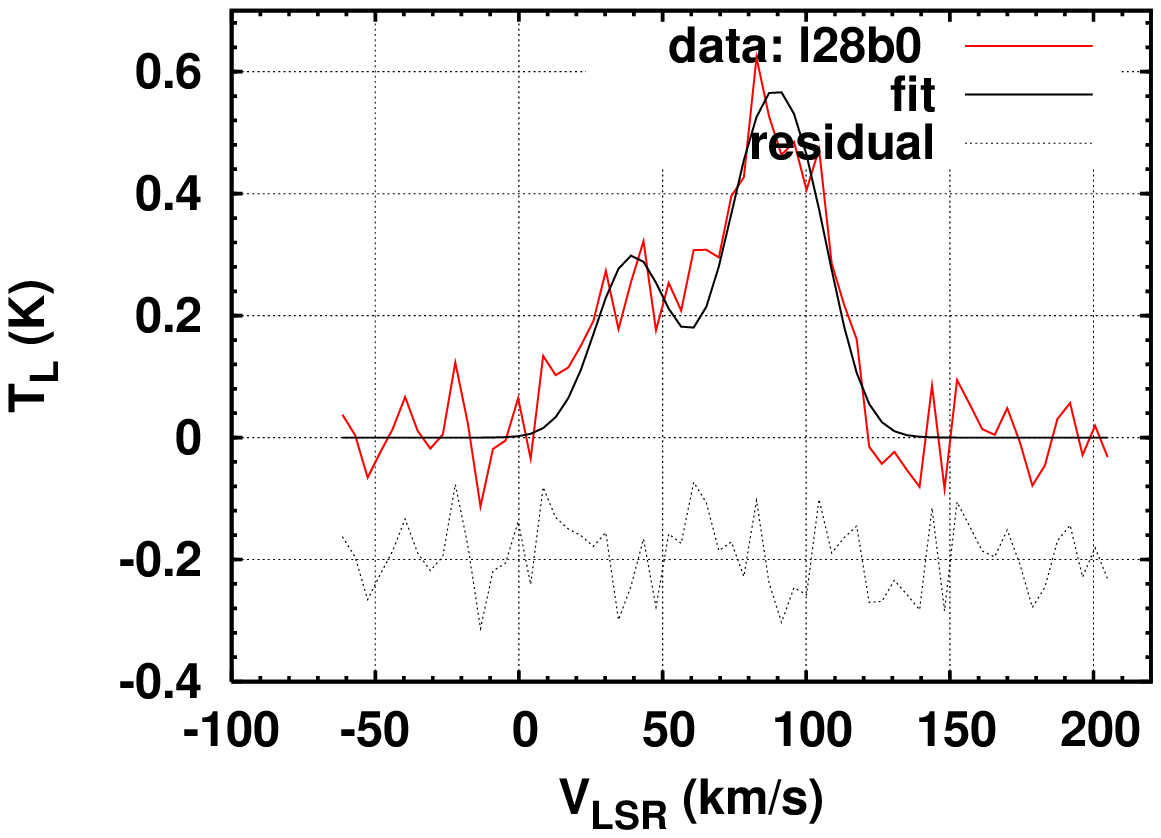}
\includegraphics[width=32mm,height=24mm,angle=0]{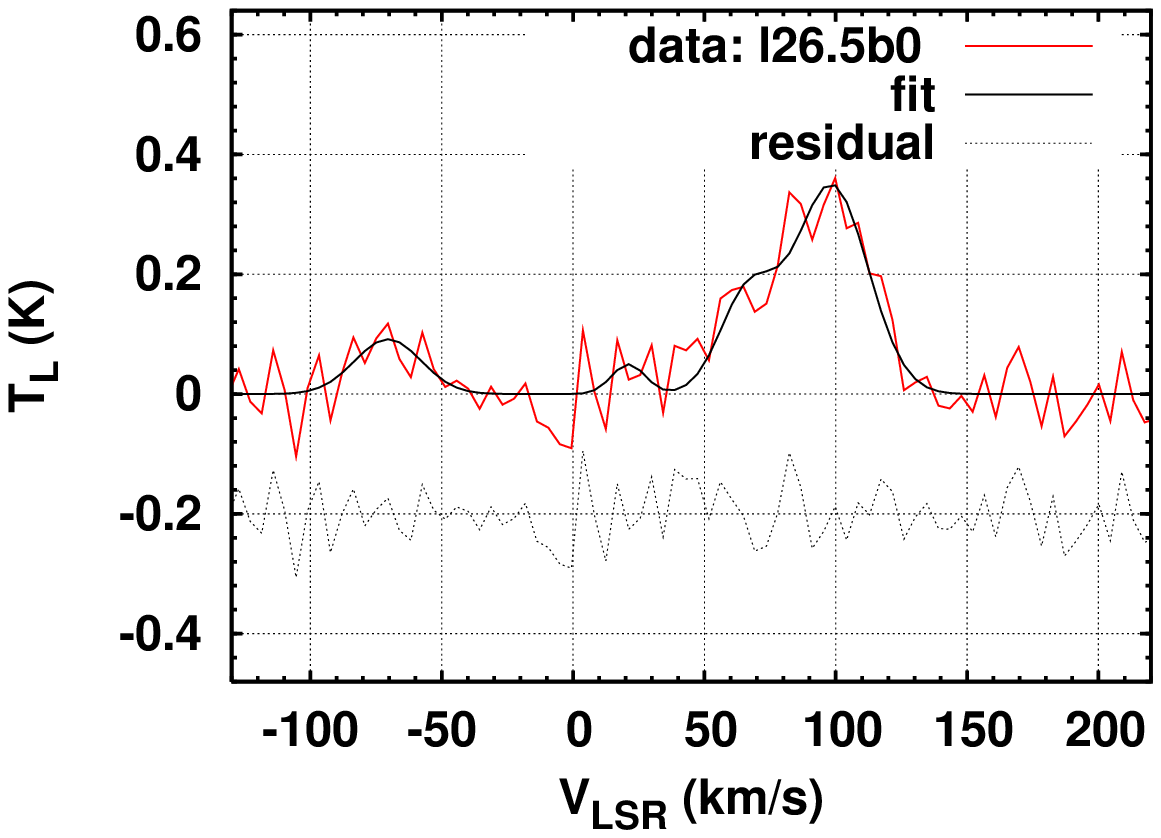}
\includegraphics[width=32mm,height=24mm,angle=0]{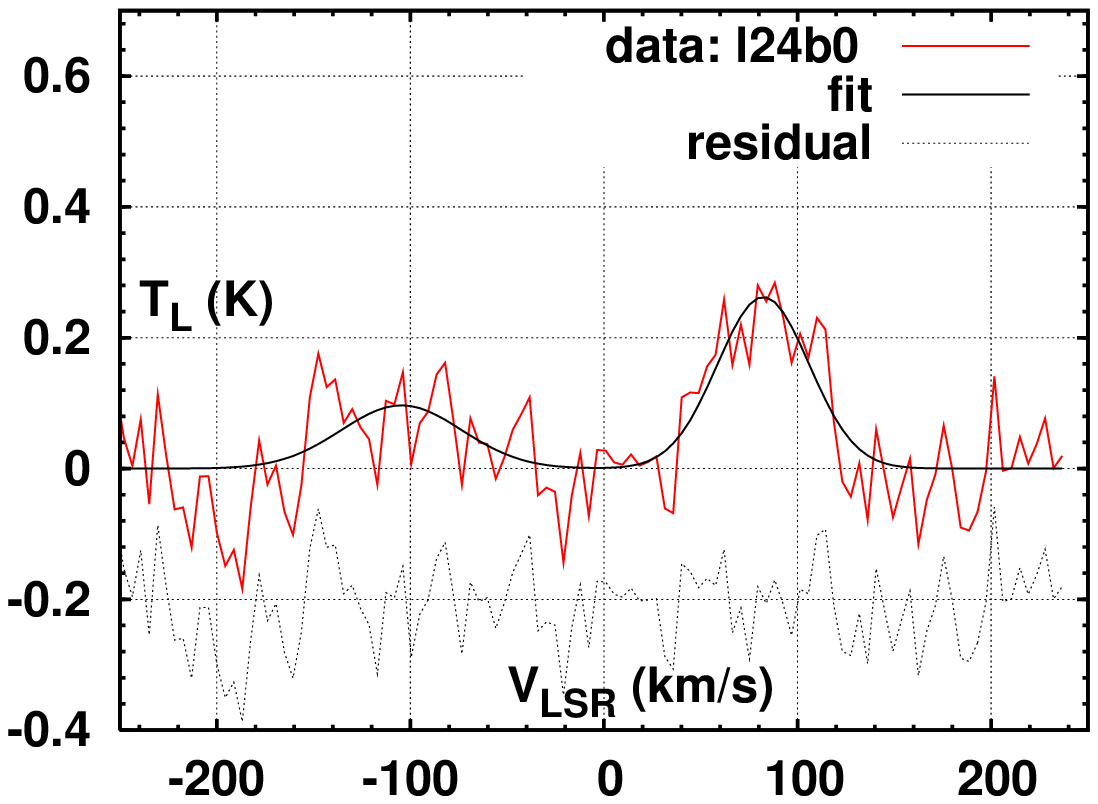}
\includegraphics[width=32mm,height=24mm,angle=0]{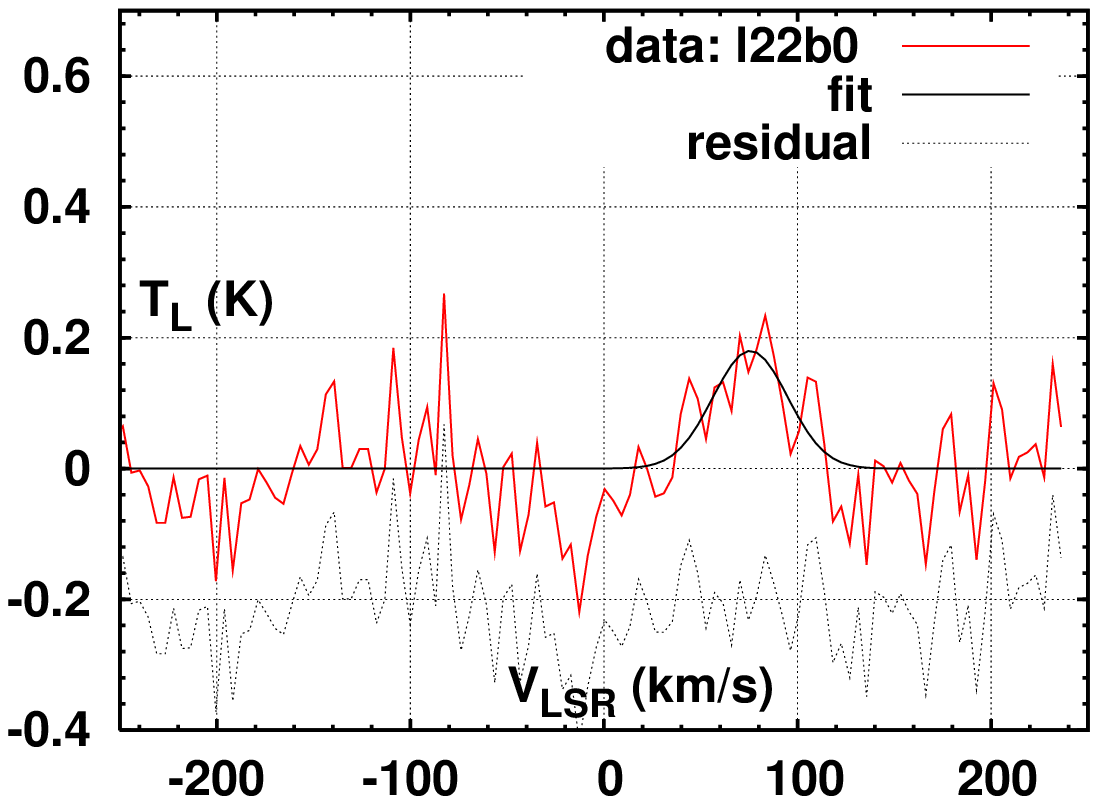}
\includegraphics[width=32mm,height=24mm,angle=0]{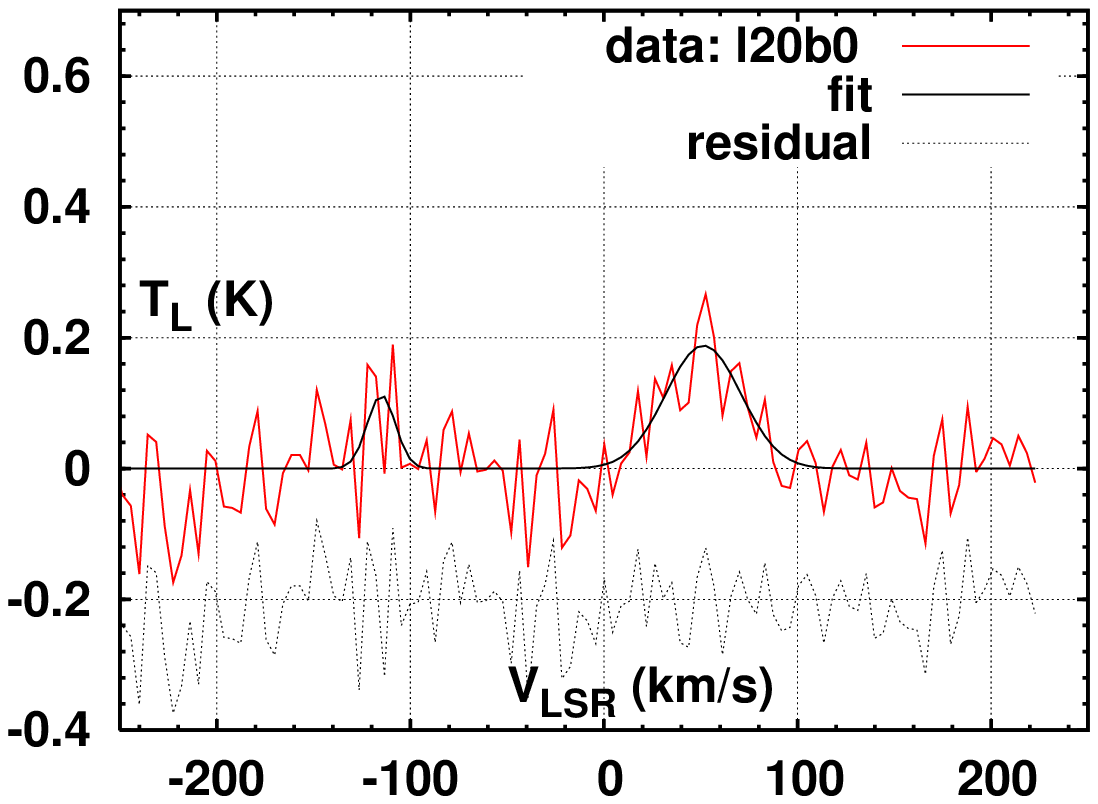}
\includegraphics[width=32mm,height=24mm,angle=0]{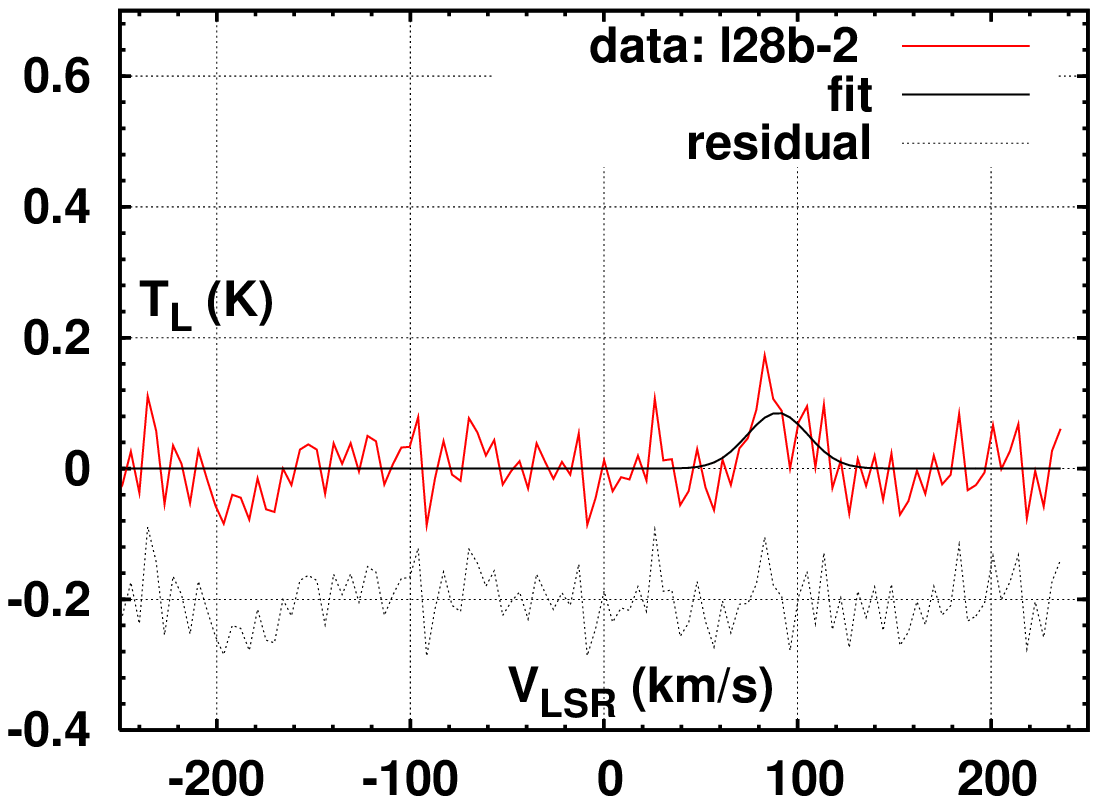}
\includegraphics[width=32mm,height=24mm,angle=0]{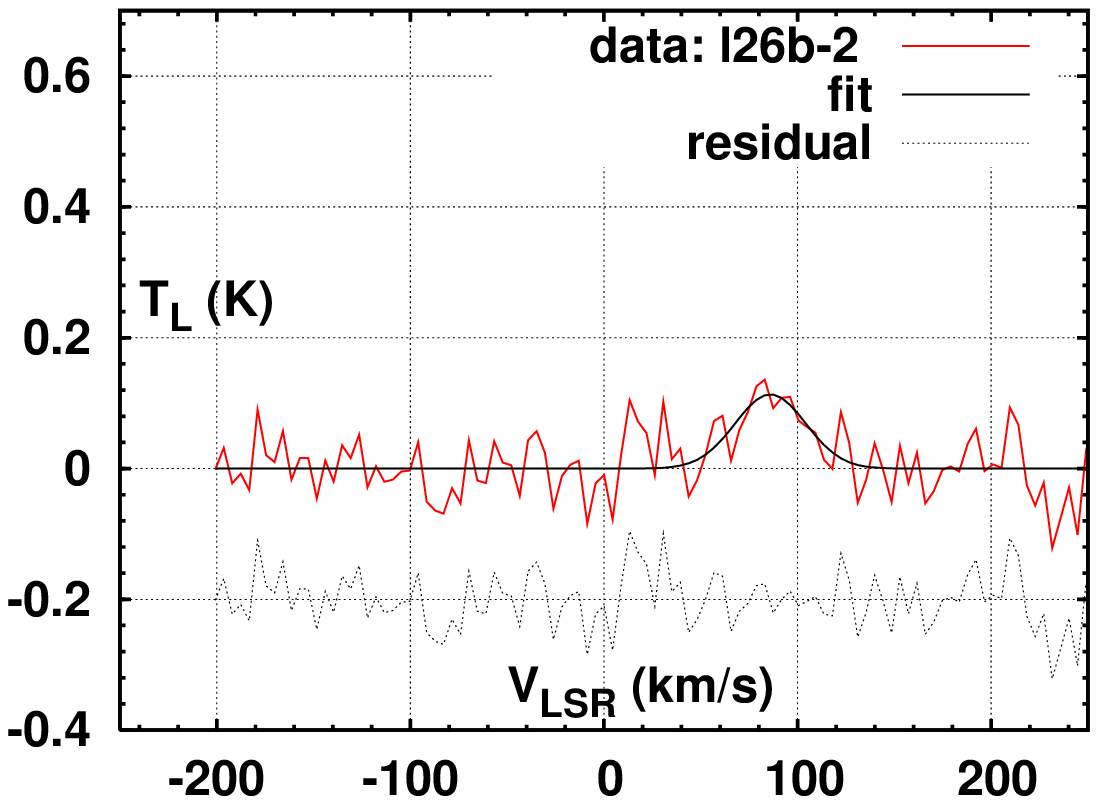}
\includegraphics[width=32mm,height=24mm,angle=0]{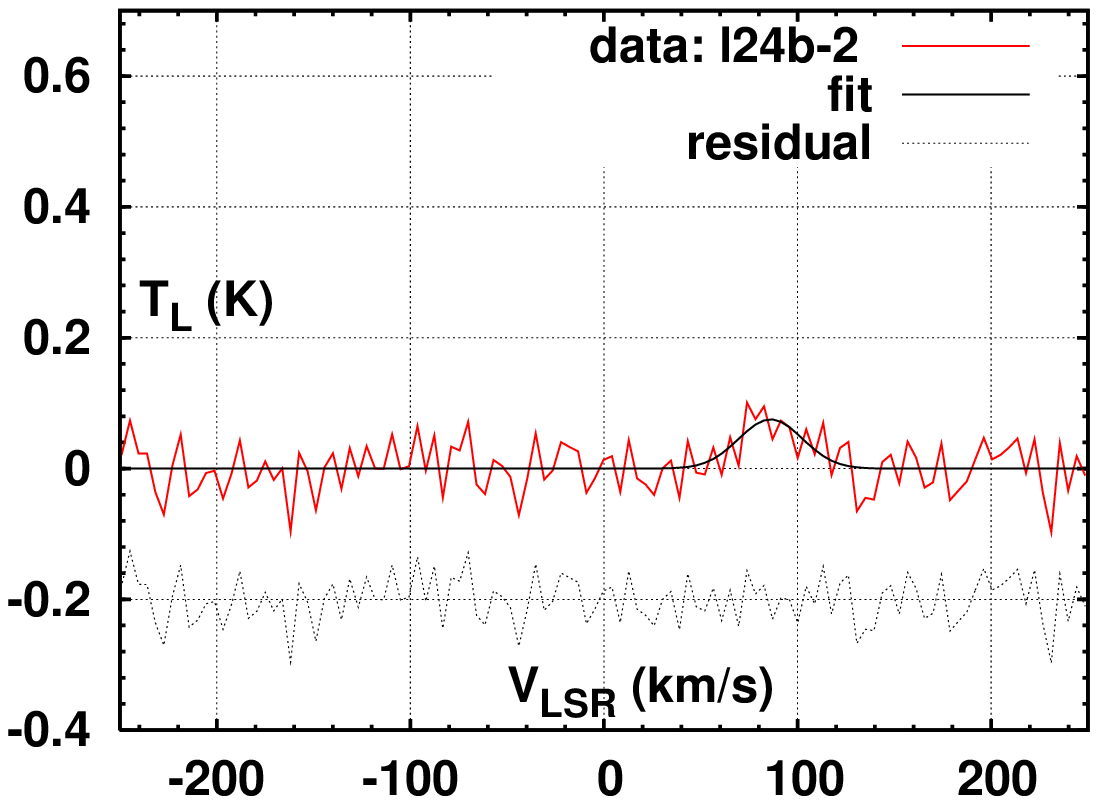}
\includegraphics[width=32mm,height=24mm,angle=0]{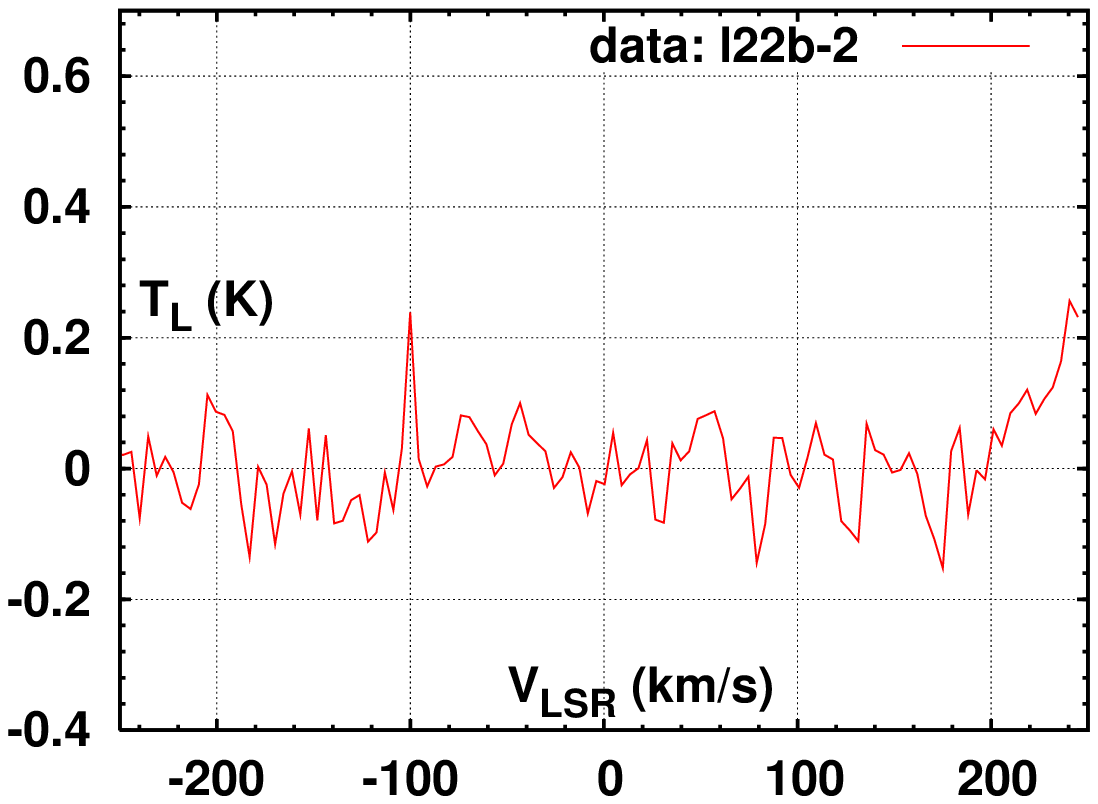}
\includegraphics[width=32mm,height=24mm,angle=0]{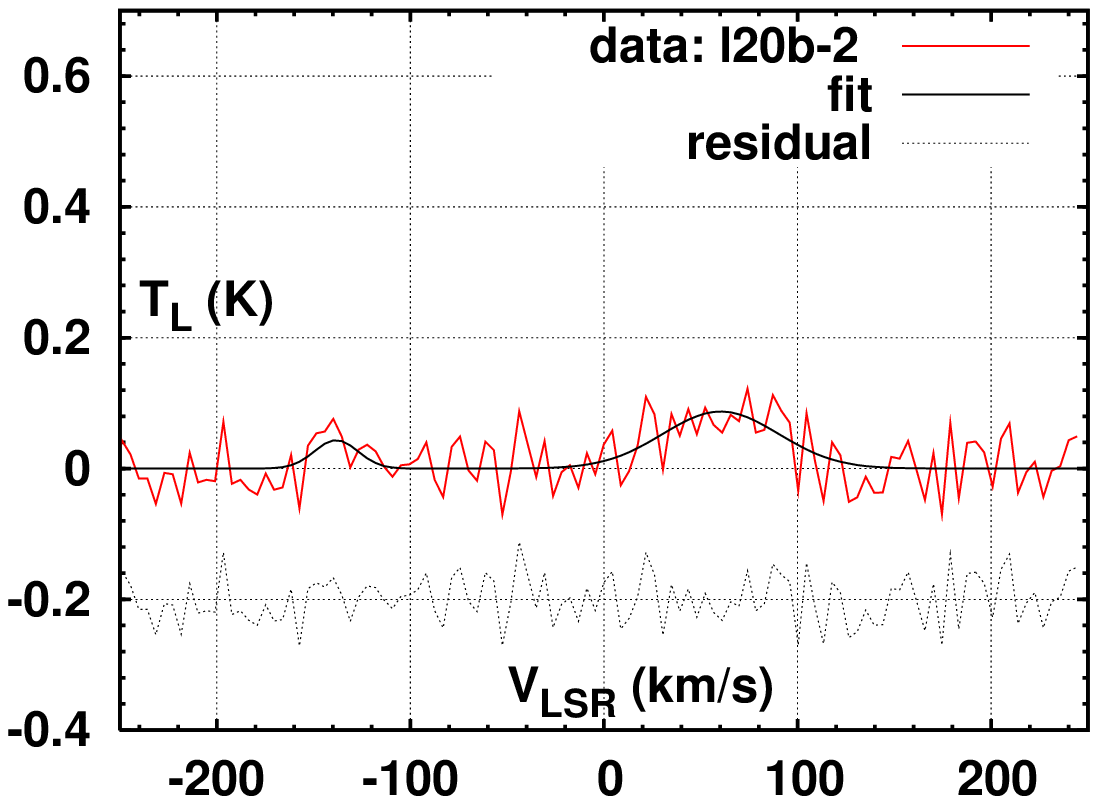}

 \caption{ORT  H271$\alpha$ RL observation.}
\end{center}
\end{figure}

\begin{figure}[ht]
\begin{center}
\includegraphics[width=32mm,height=24mm,angle=0]{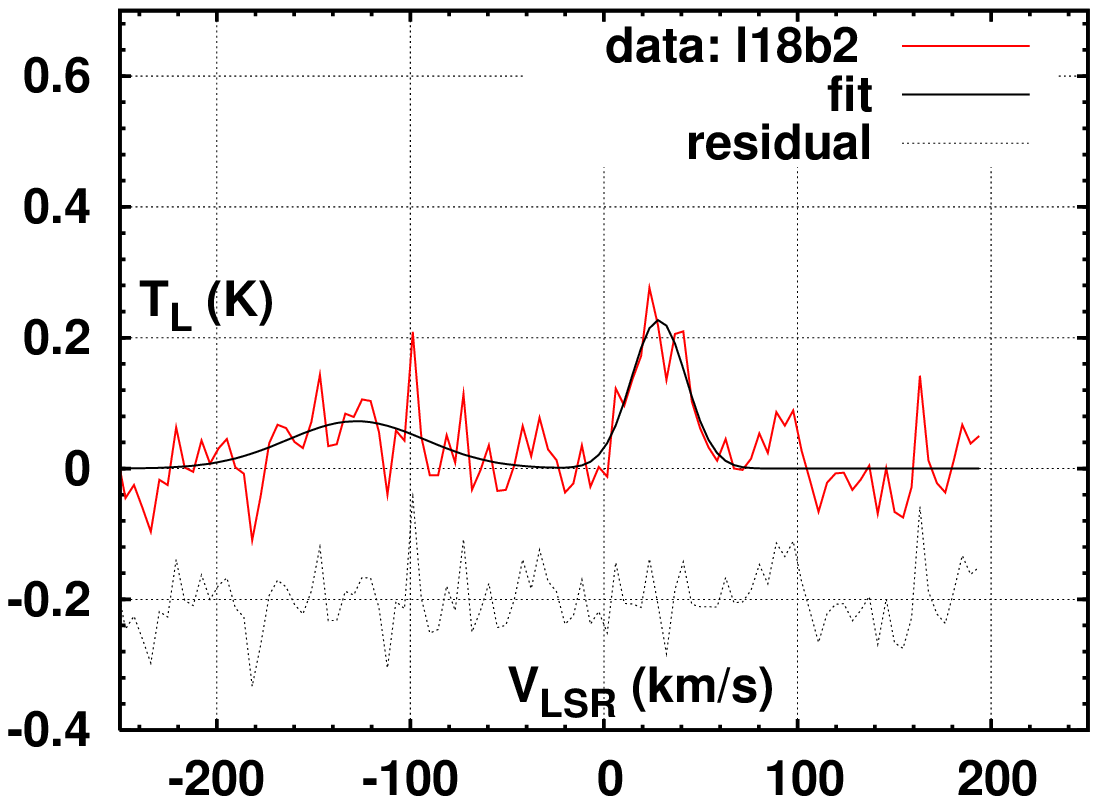}
\includegraphics[width=32mm,height=24mm,angle=0]{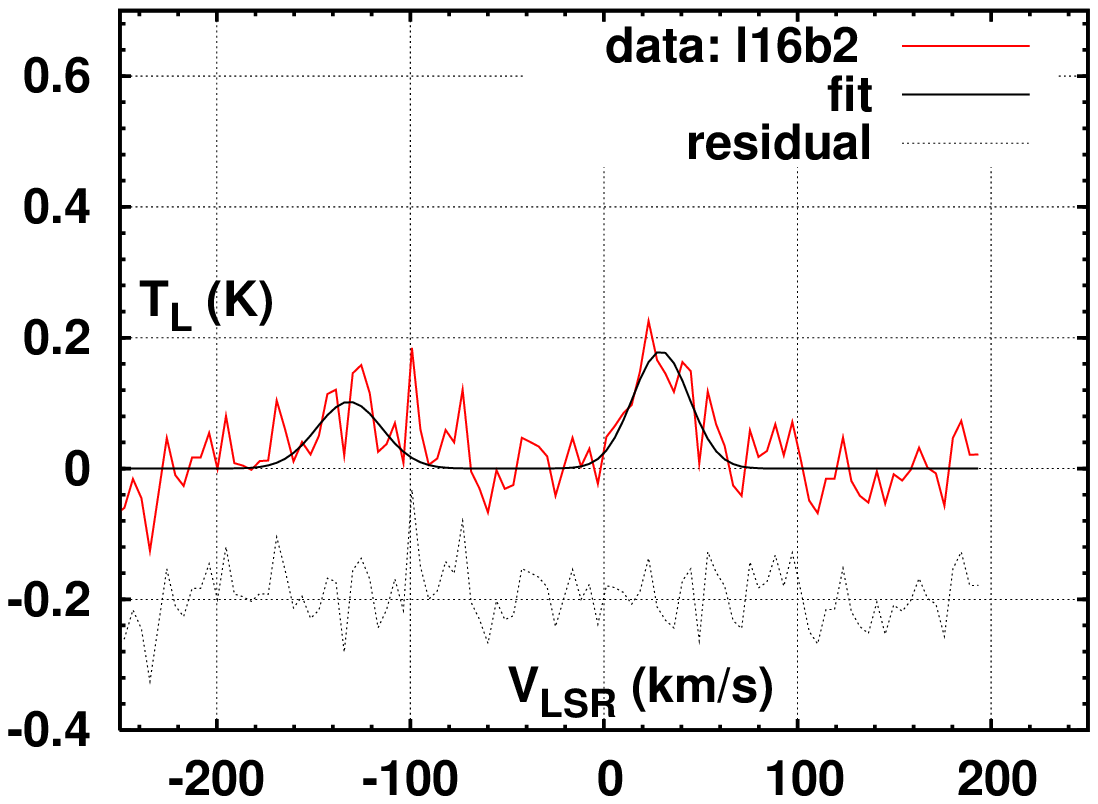}
\includegraphics[width=32mm,height=24mm,angle=0]{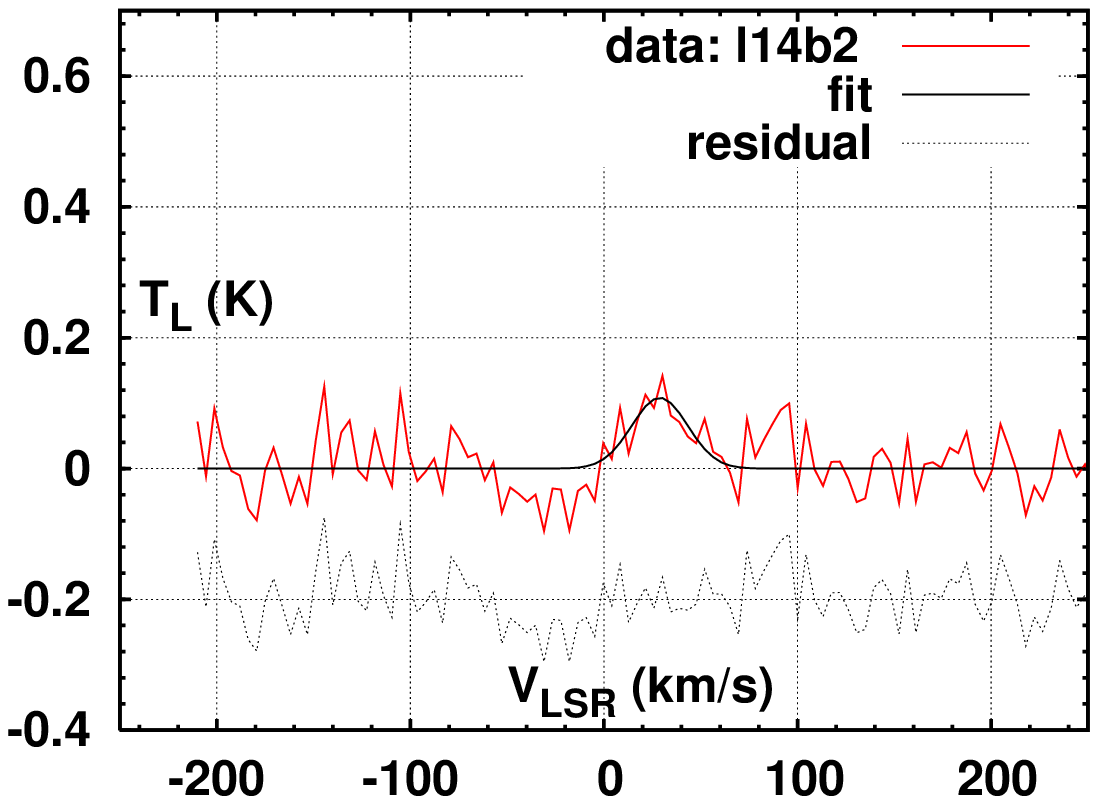}
\includegraphics[width=32mm,height=24mm,angle=0]{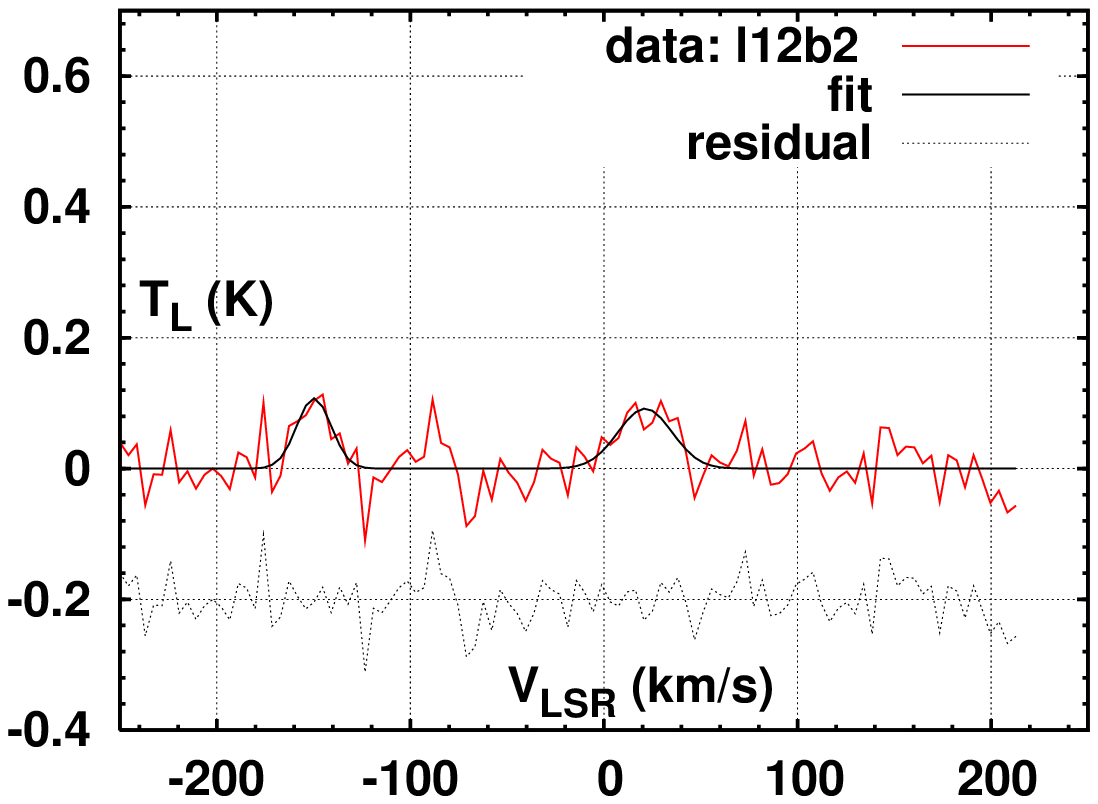}
\includegraphics[width=32mm,height=24mm,angle=0]{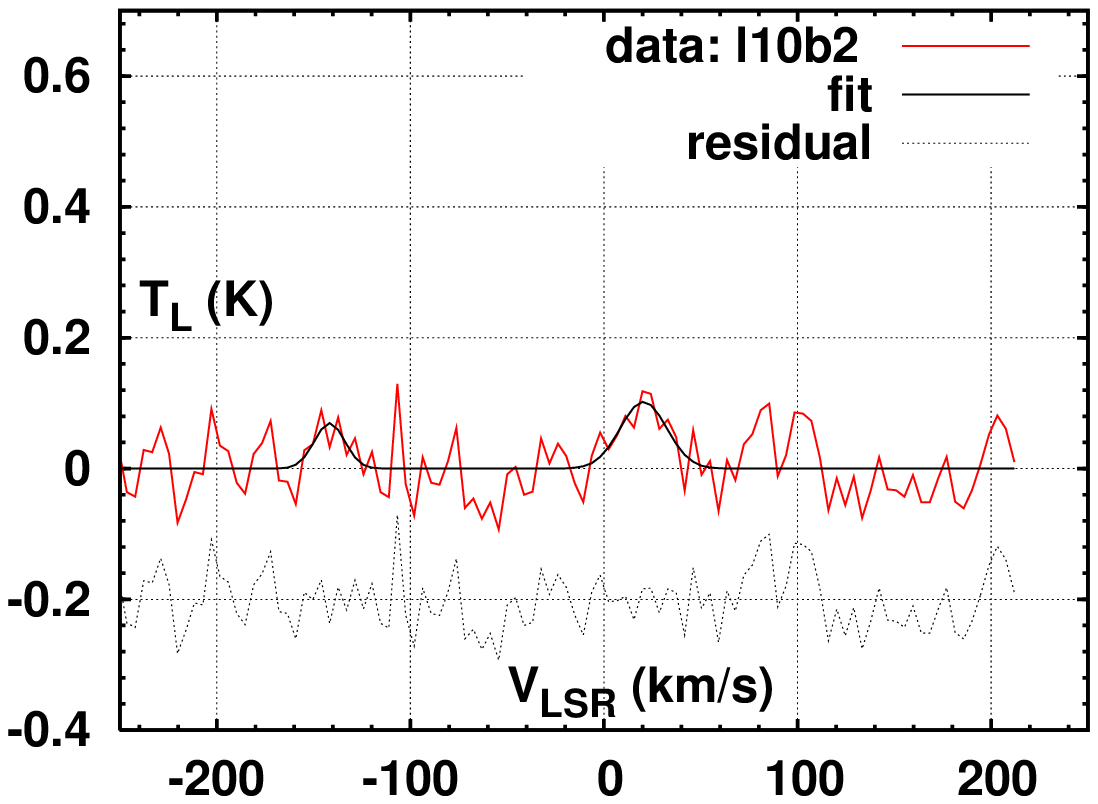}
\includegraphics[width=32mm,height=24mm,angle=0]{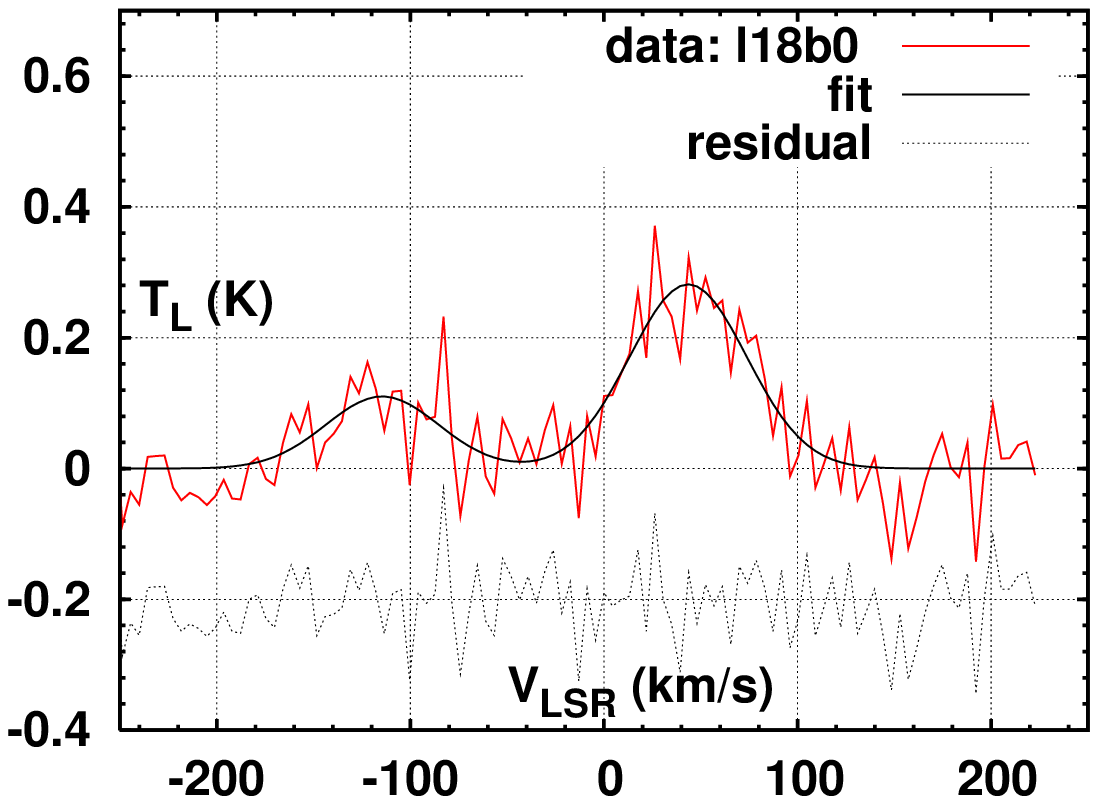}
\includegraphics[width=32mm,height=24mm,angle=0]{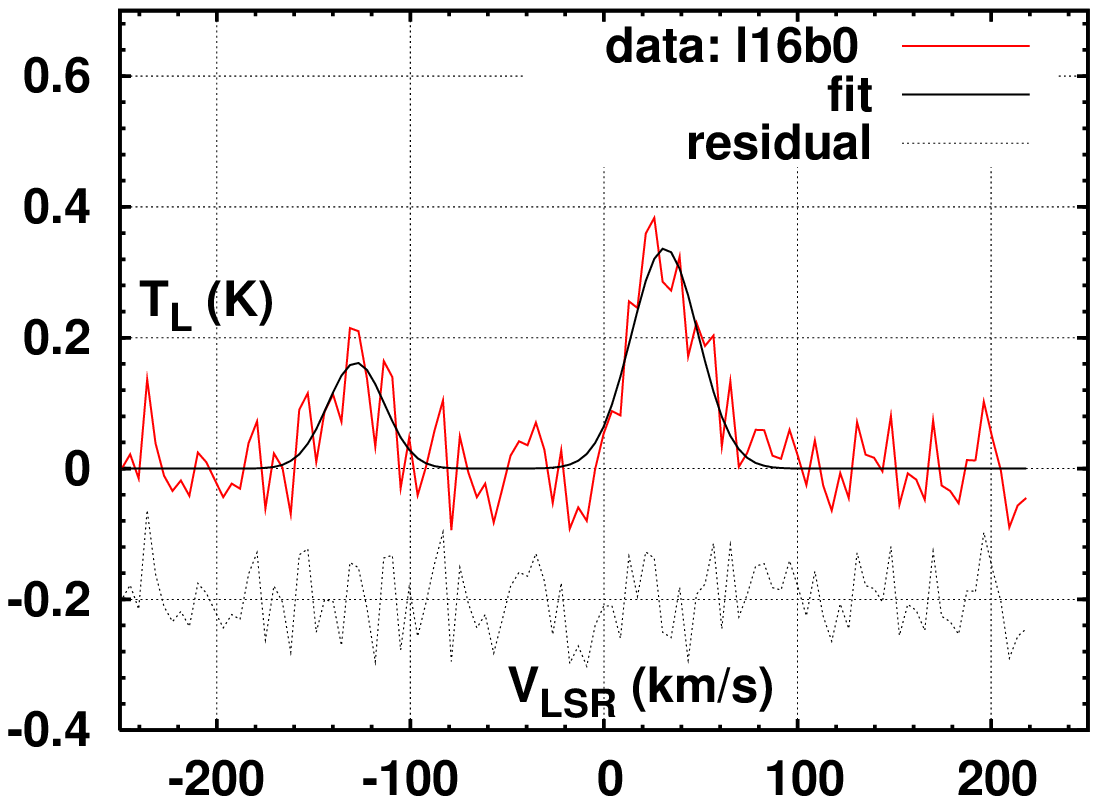}
\includegraphics[width=32mm,height=24mm,angle=0]{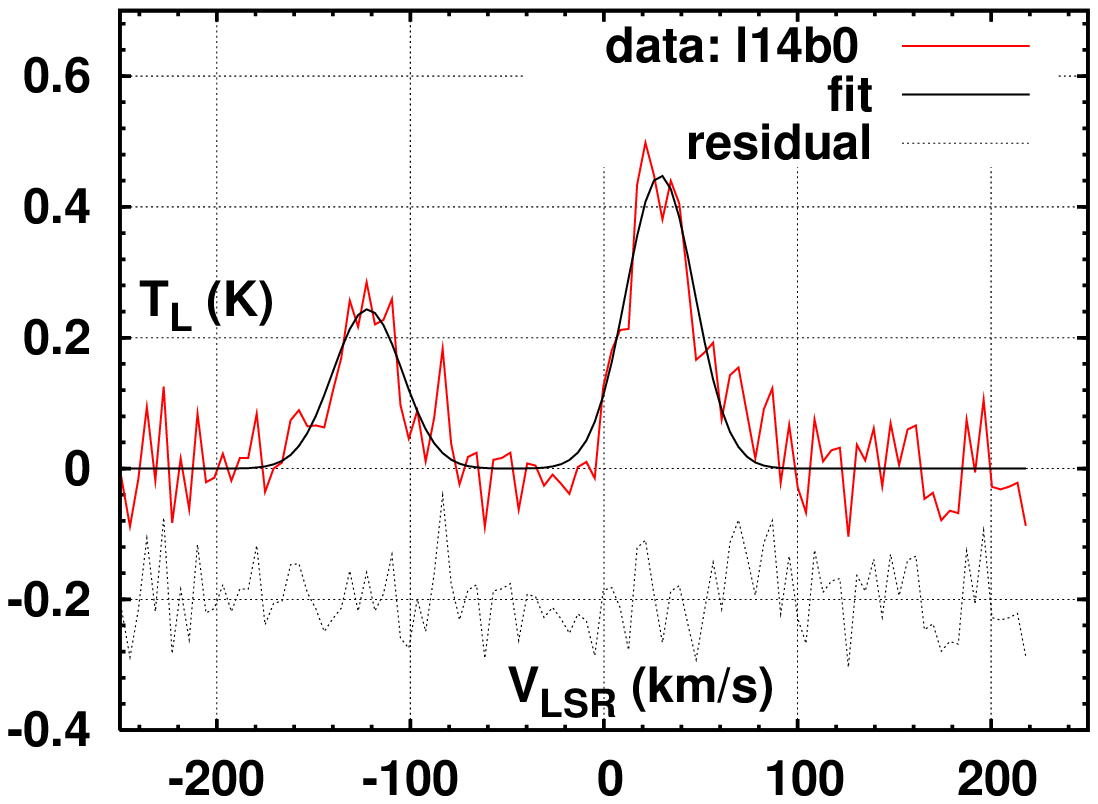}
\includegraphics[width=32mm,height=24mm,angle=0]{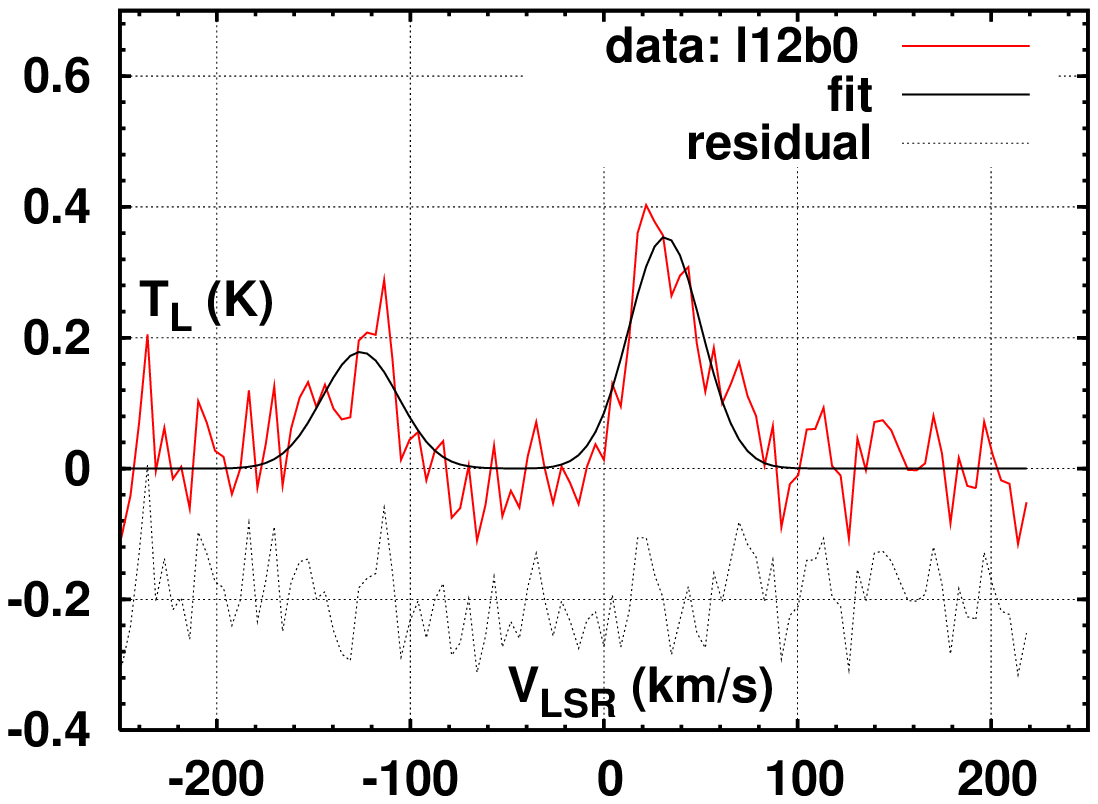}
\includegraphics[width=32mm,height=24mm,angle=0]{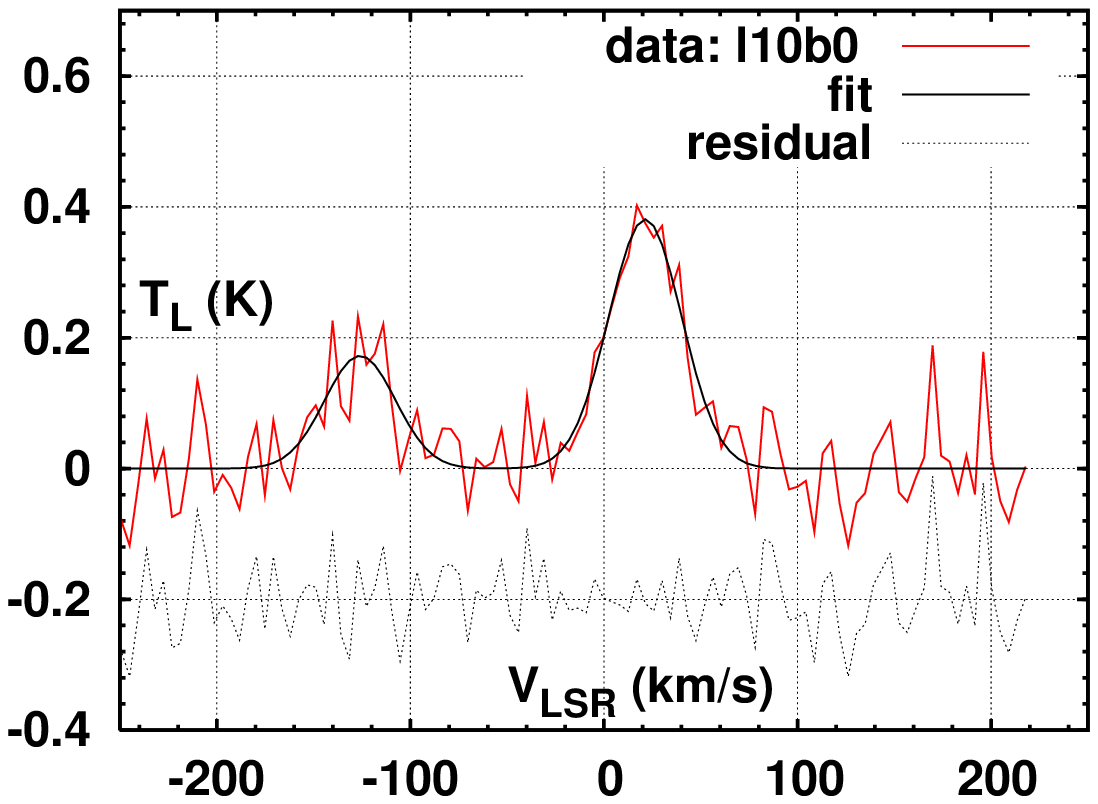}
\includegraphics[width=32mm,height=24mm,angle=0]{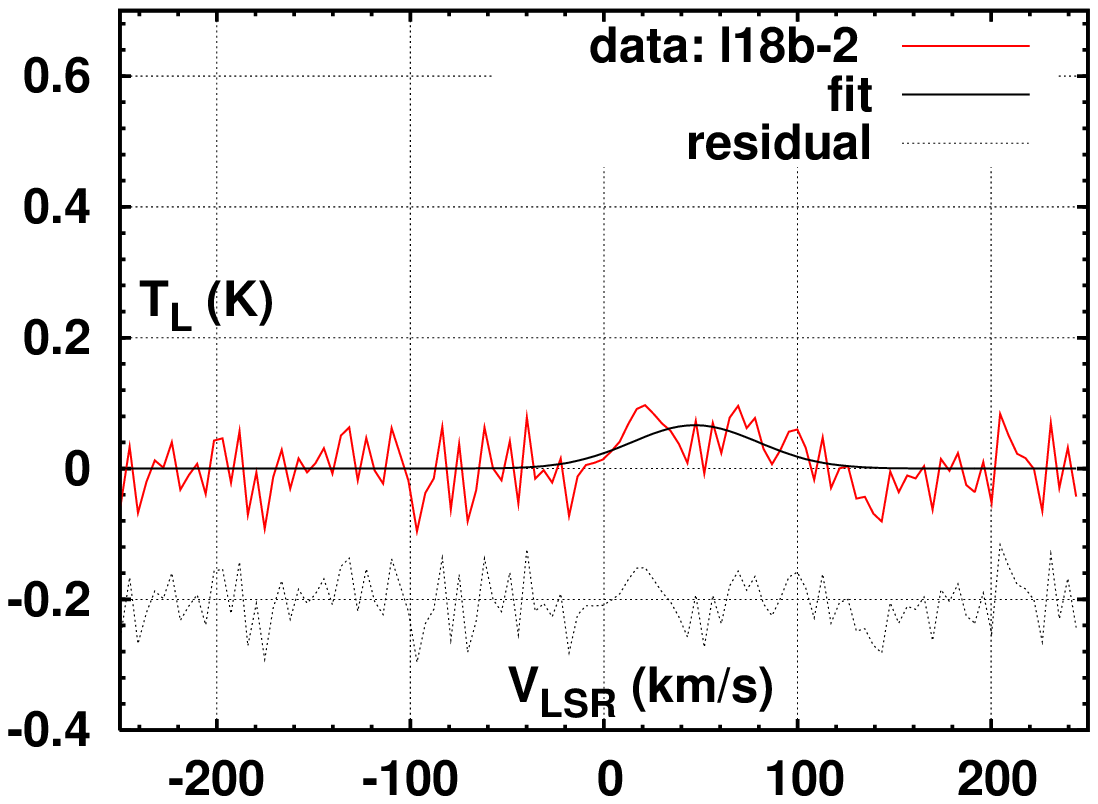}
\includegraphics[width=32mm,height=24mm,angle=0]{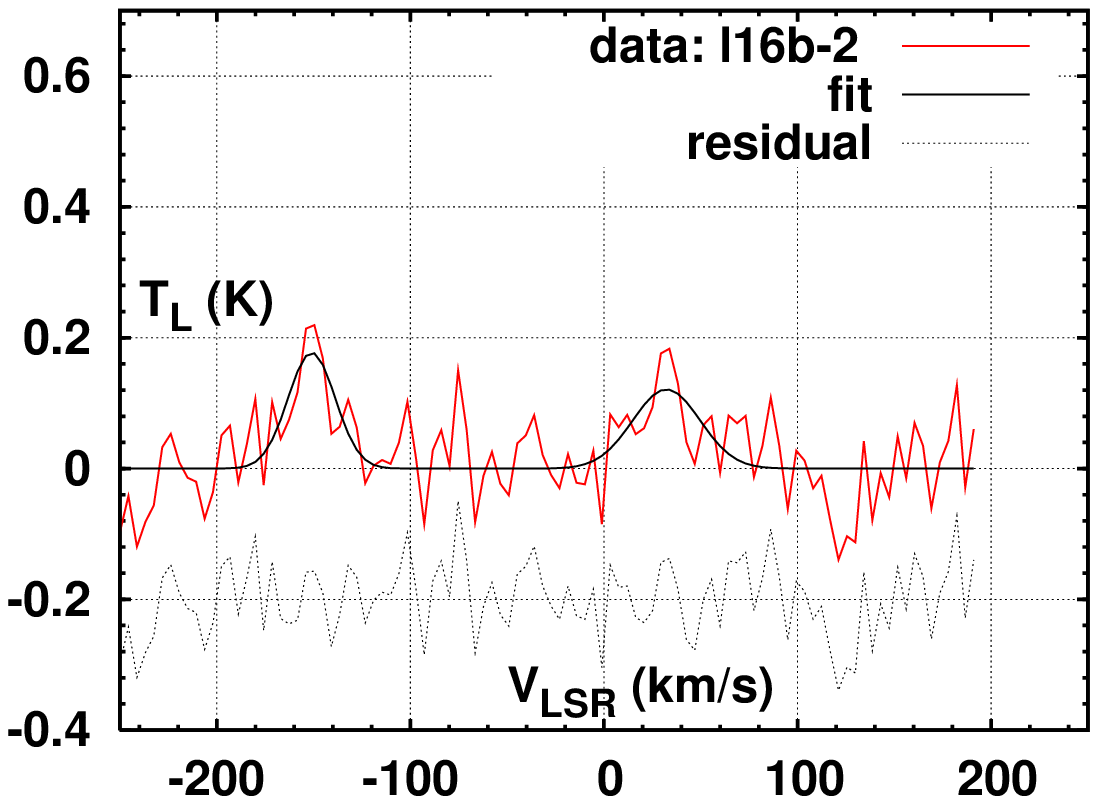}
\includegraphics[width=32mm,height=24mm,angle=0]{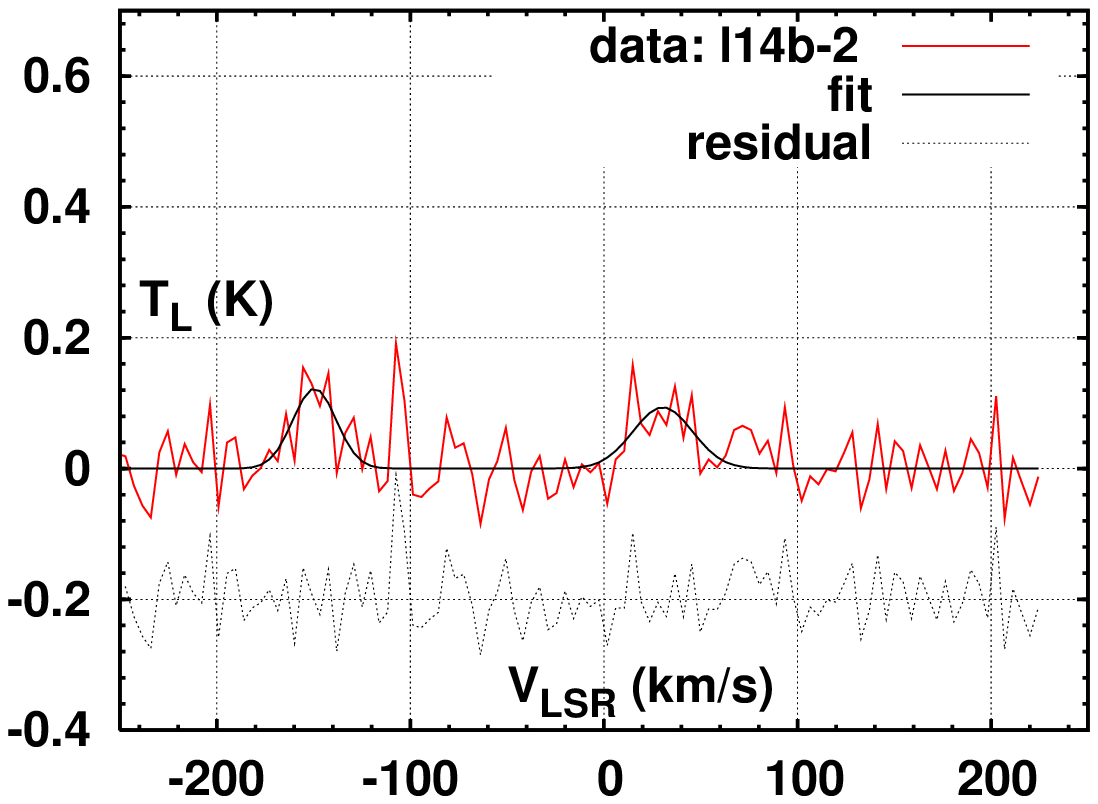}
\includegraphics[width=32mm,height=24mm,angle=0]{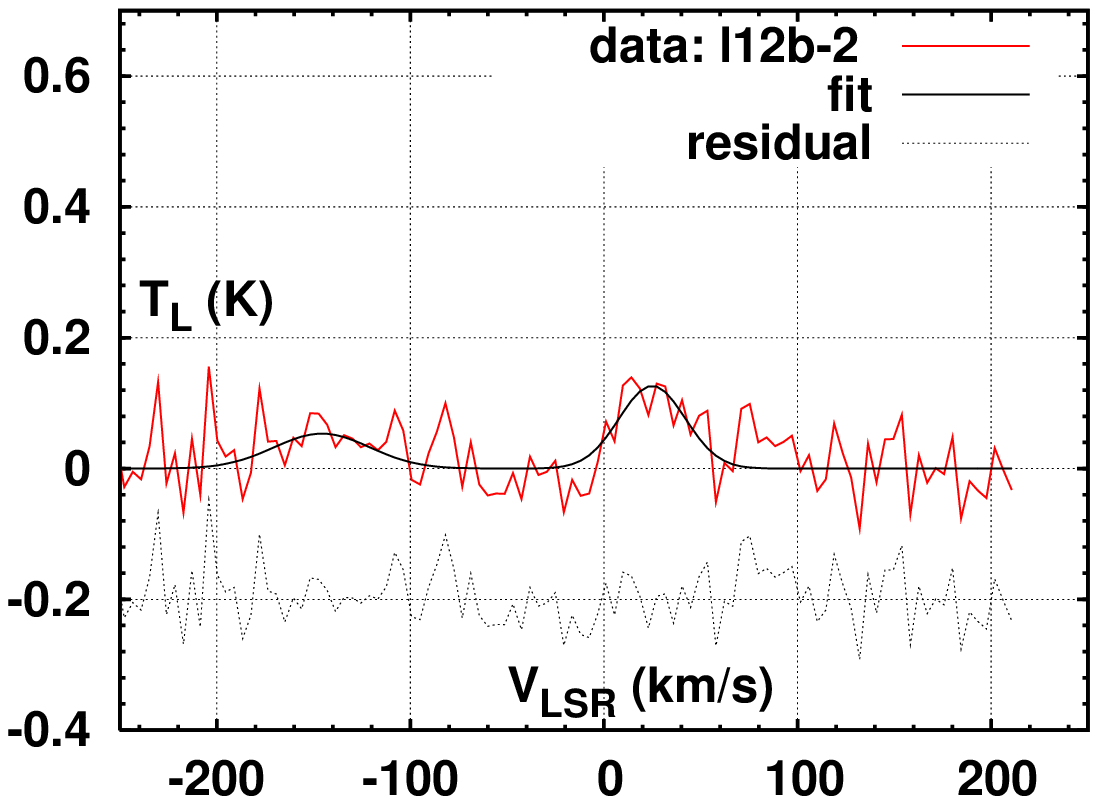}
\includegraphics[width=32mm,height=24mm,angle=0]{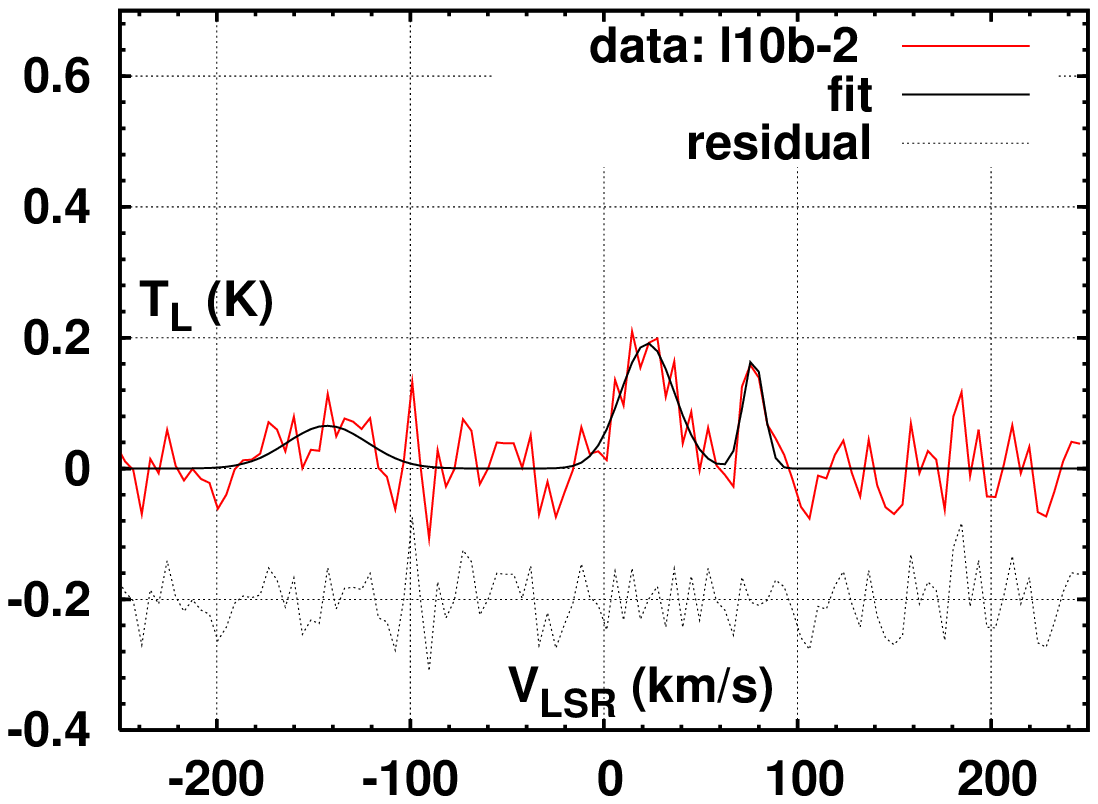}\vspace{1cm}
\includegraphics[width=32mm,height=24mm,angle=0]{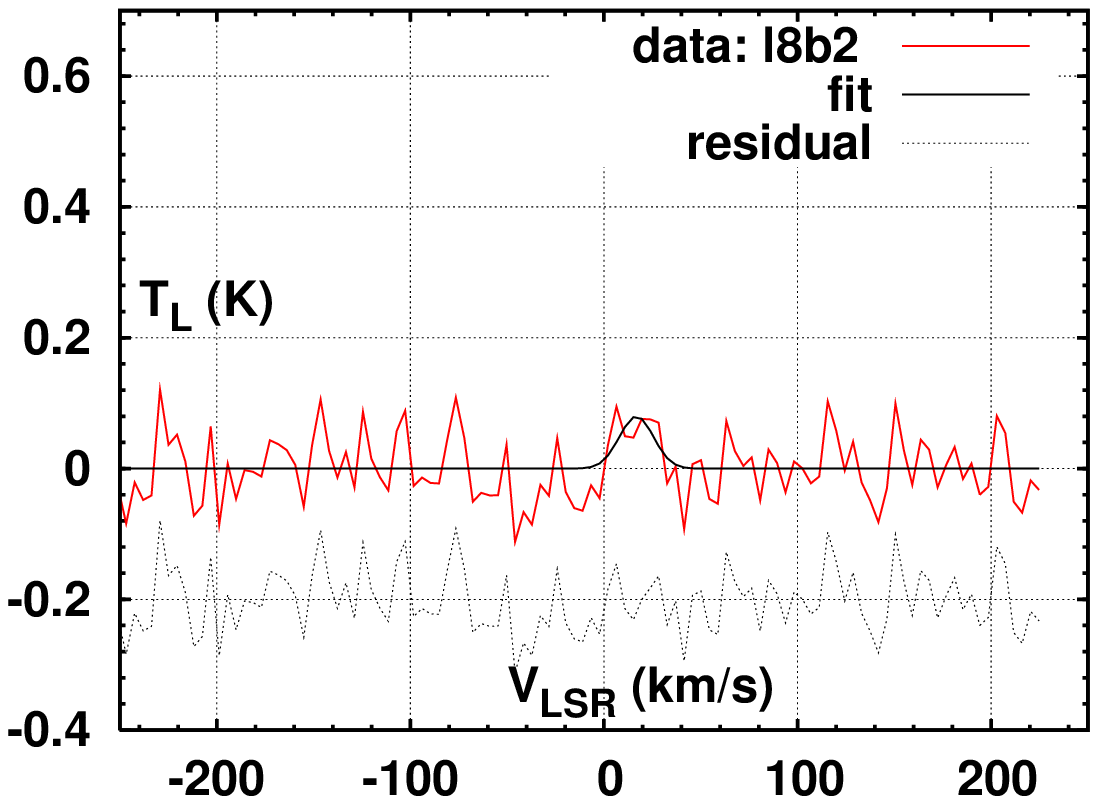}
\includegraphics[width=32mm,height=24mm,angle=0]{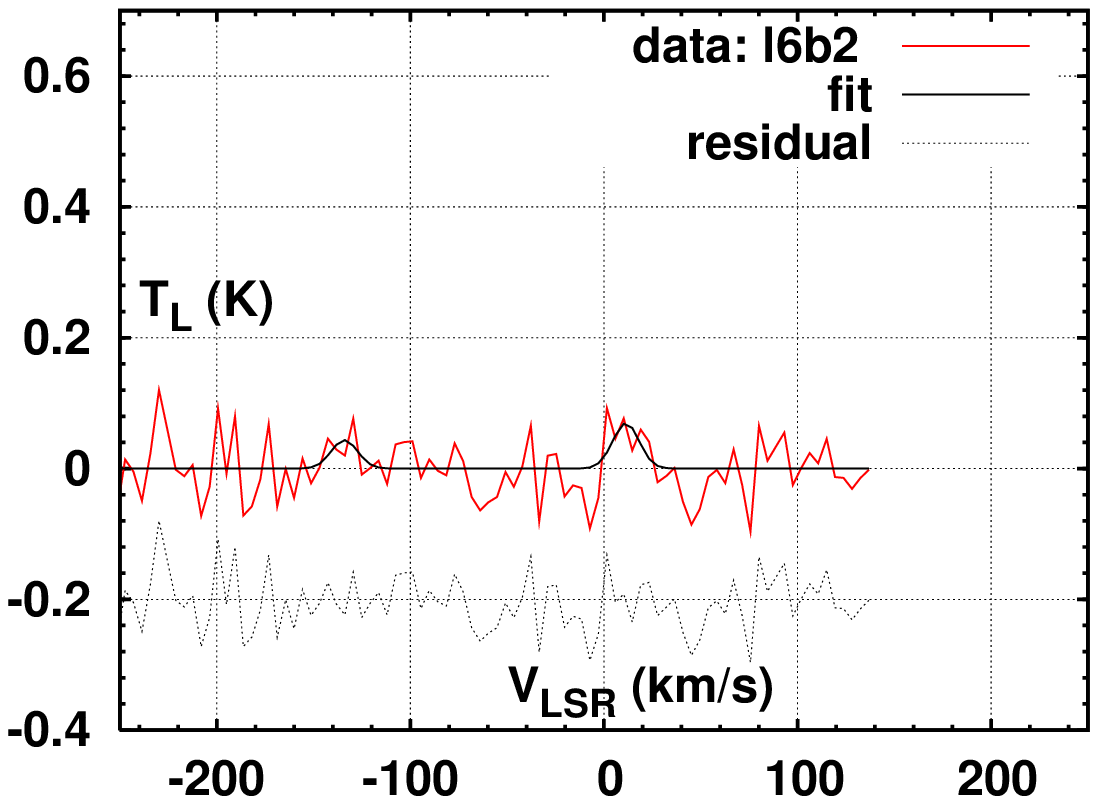}
\includegraphics[width=32mm,height=24mm,angle=0]{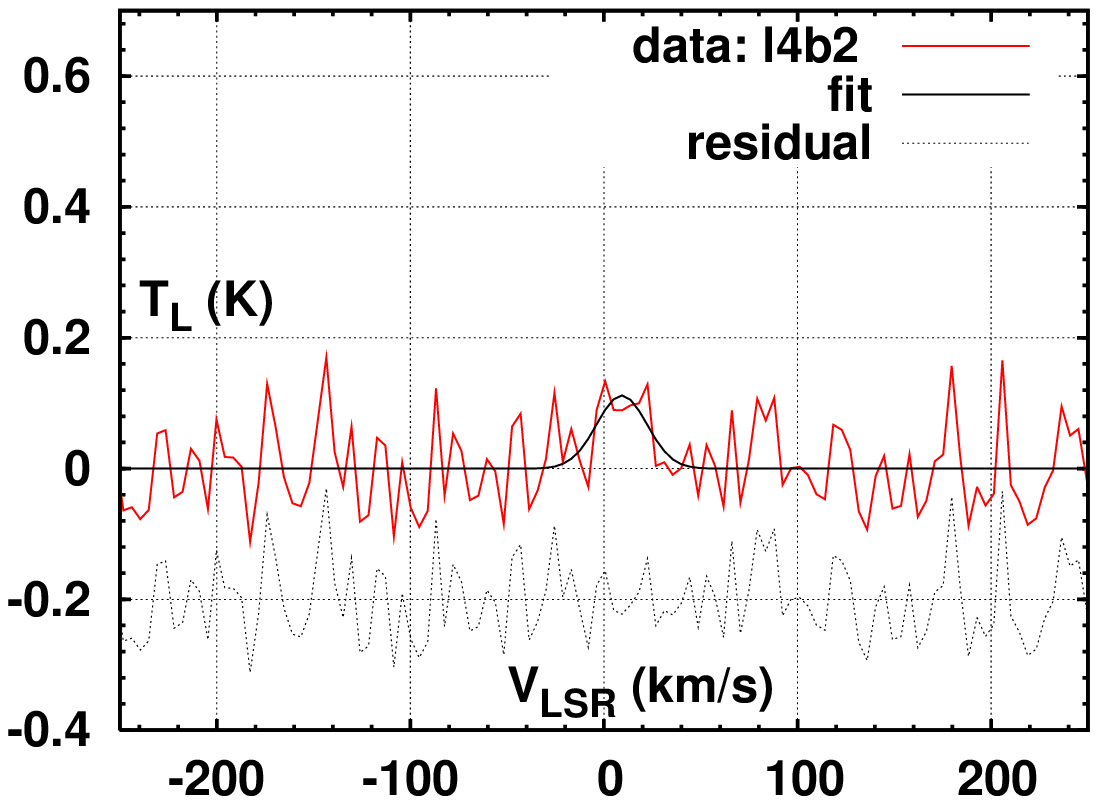}
\includegraphics[width=32mm,height=24mm,angle=0]{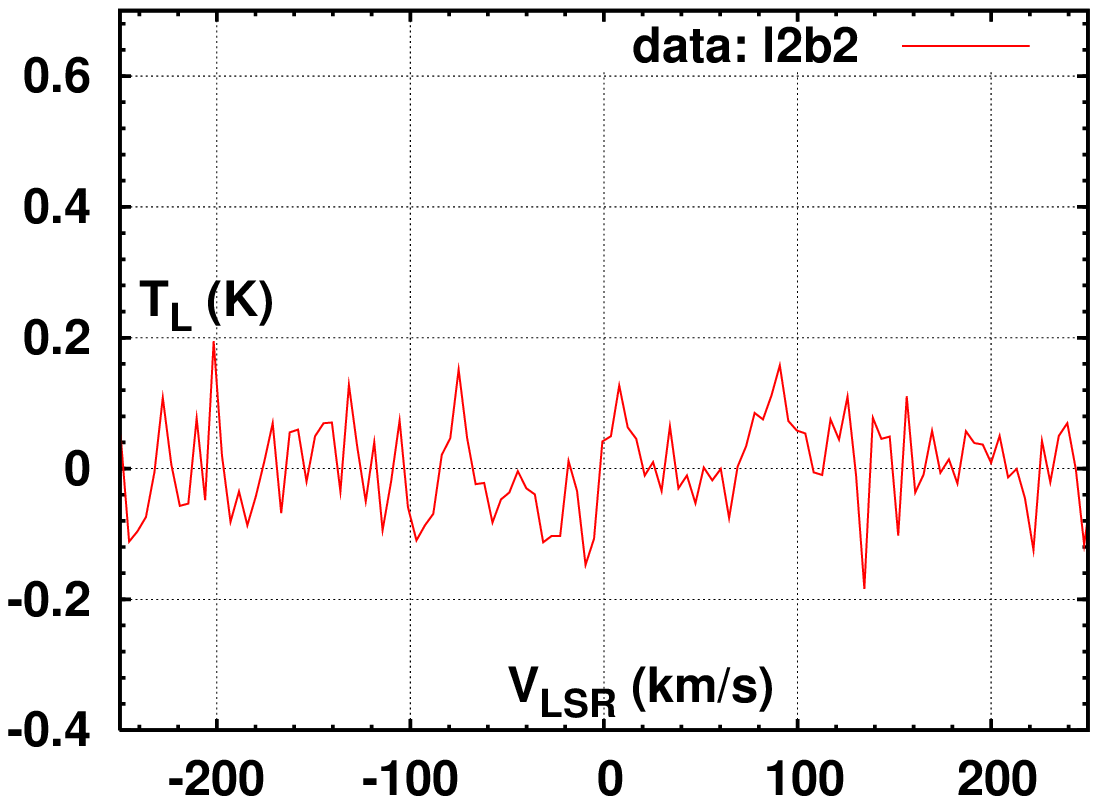}
\includegraphics[width=32mm,height=24mm,angle=0]{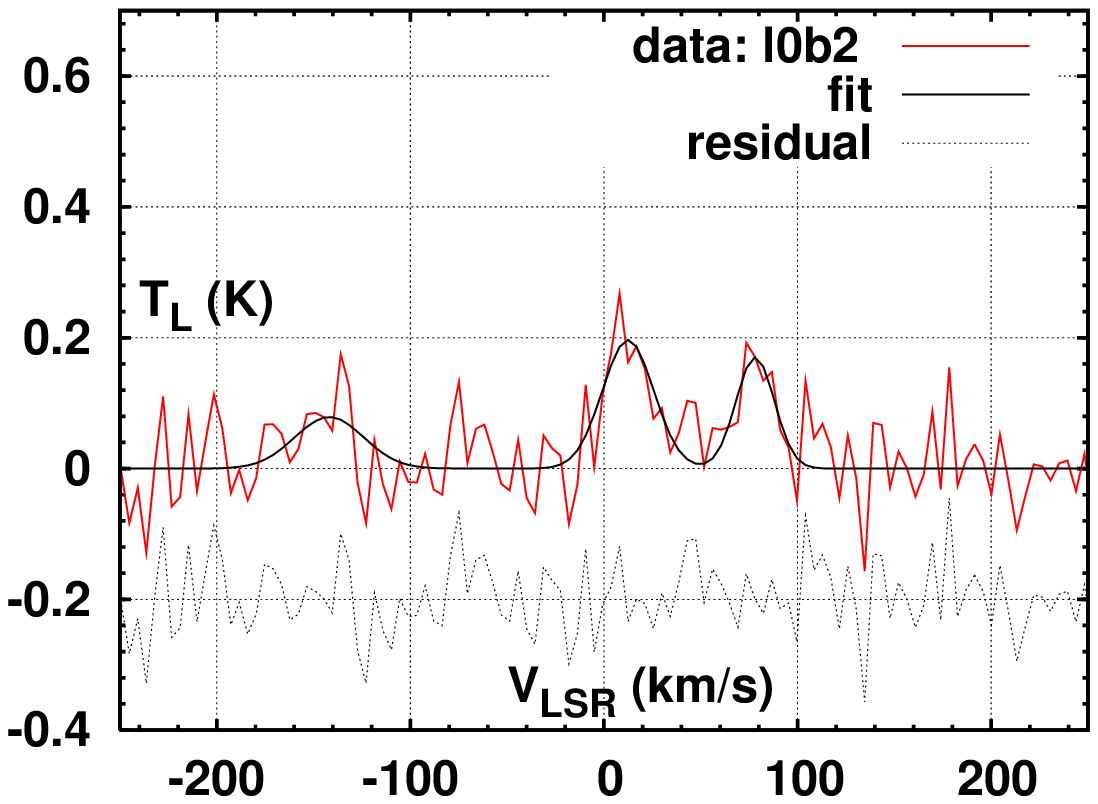}
\includegraphics[width=32mm,height=24mm,angle=0]{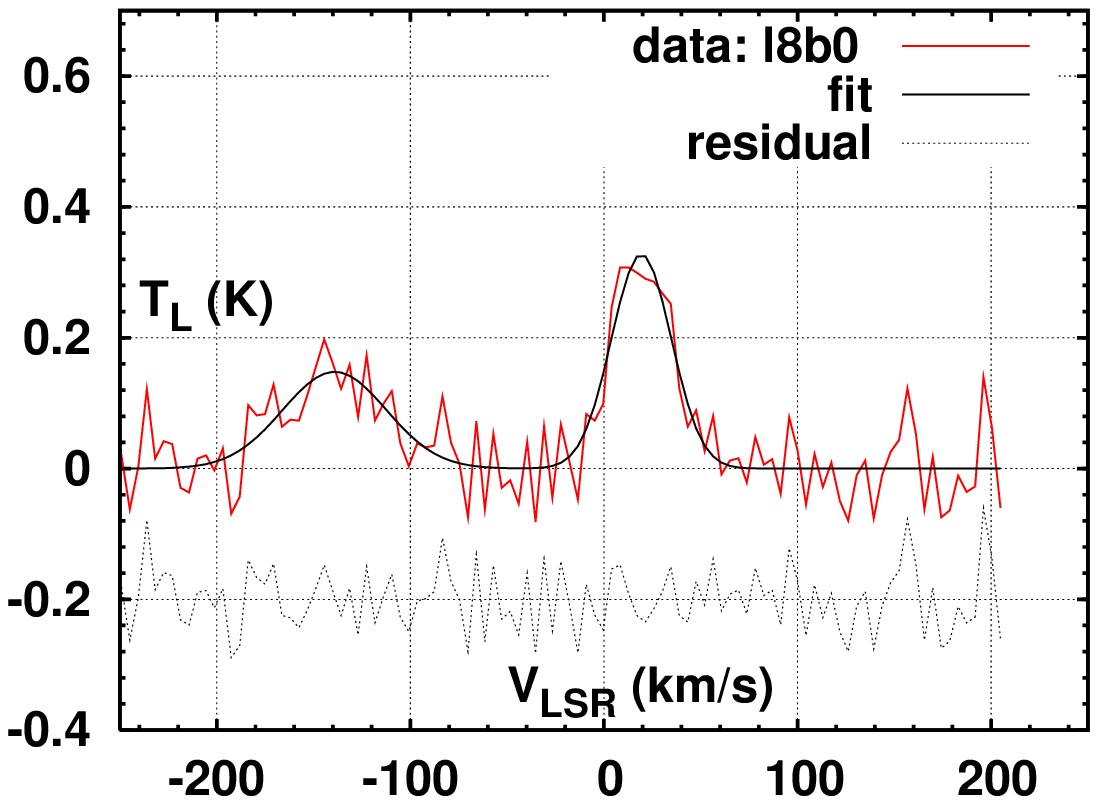}
\includegraphics[width=32mm,height=24mm,angle=0]{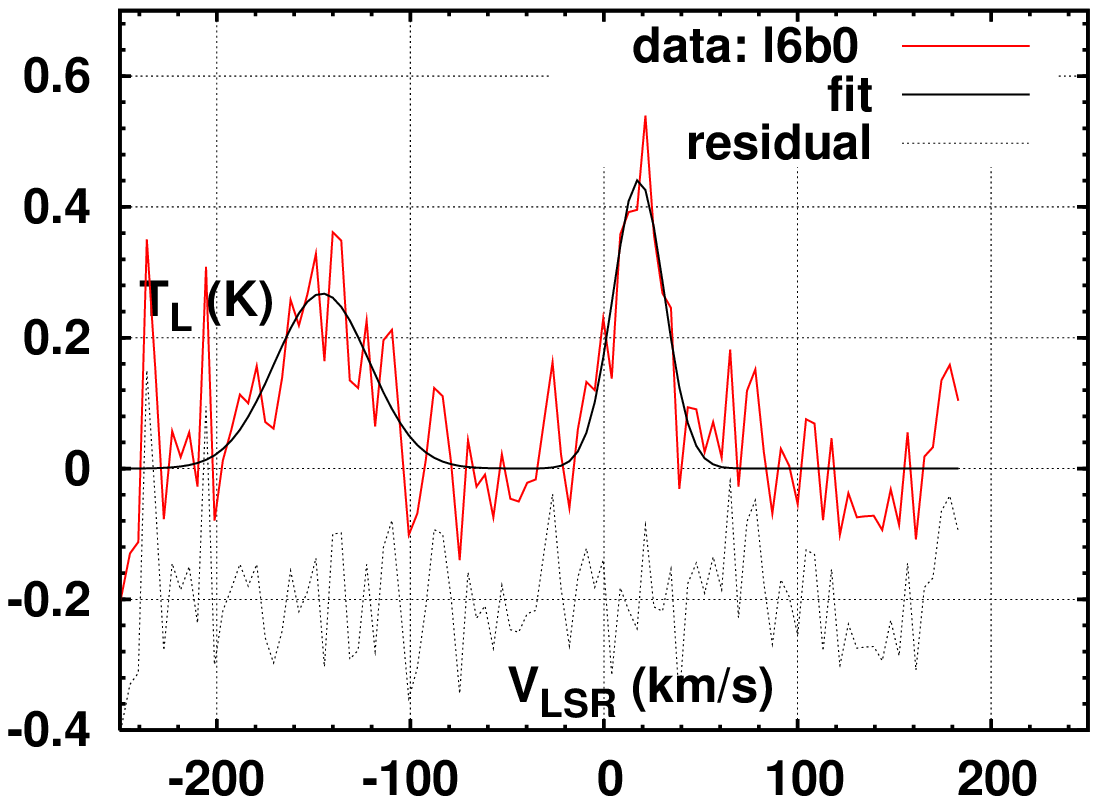}
\includegraphics[width=32mm,height=24mm,angle=0]{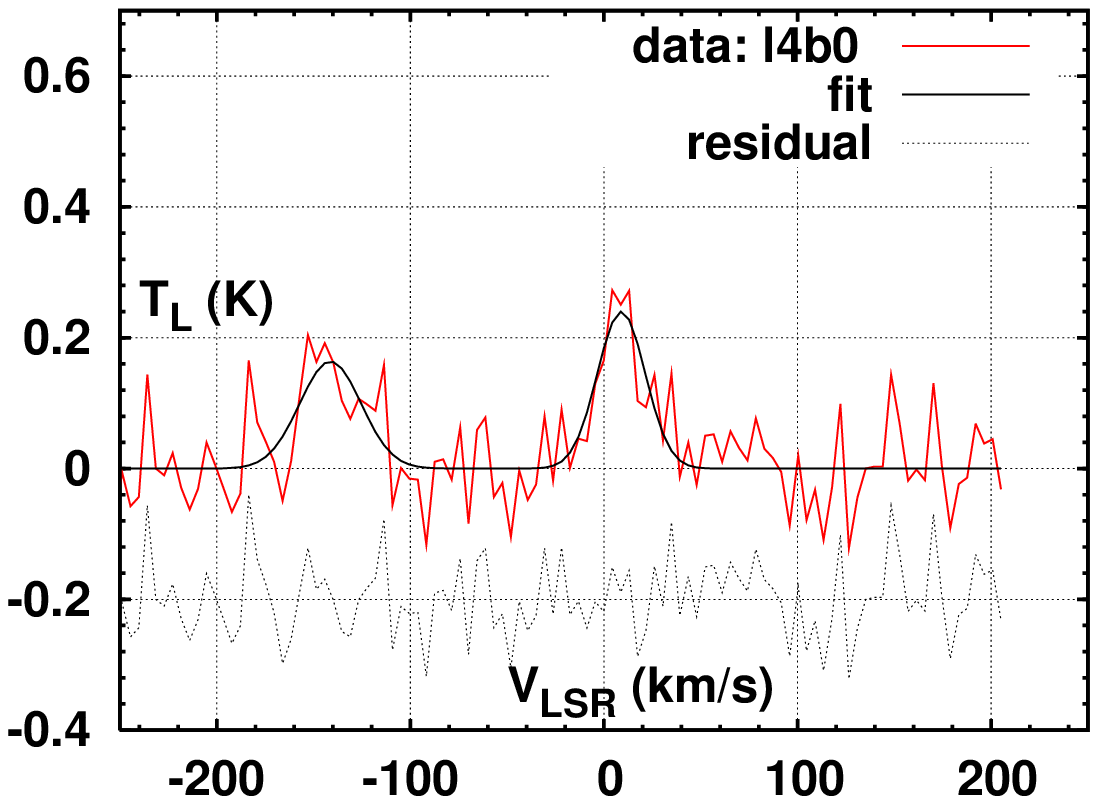}
\includegraphics[width=32mm,height=24mm,angle=0]{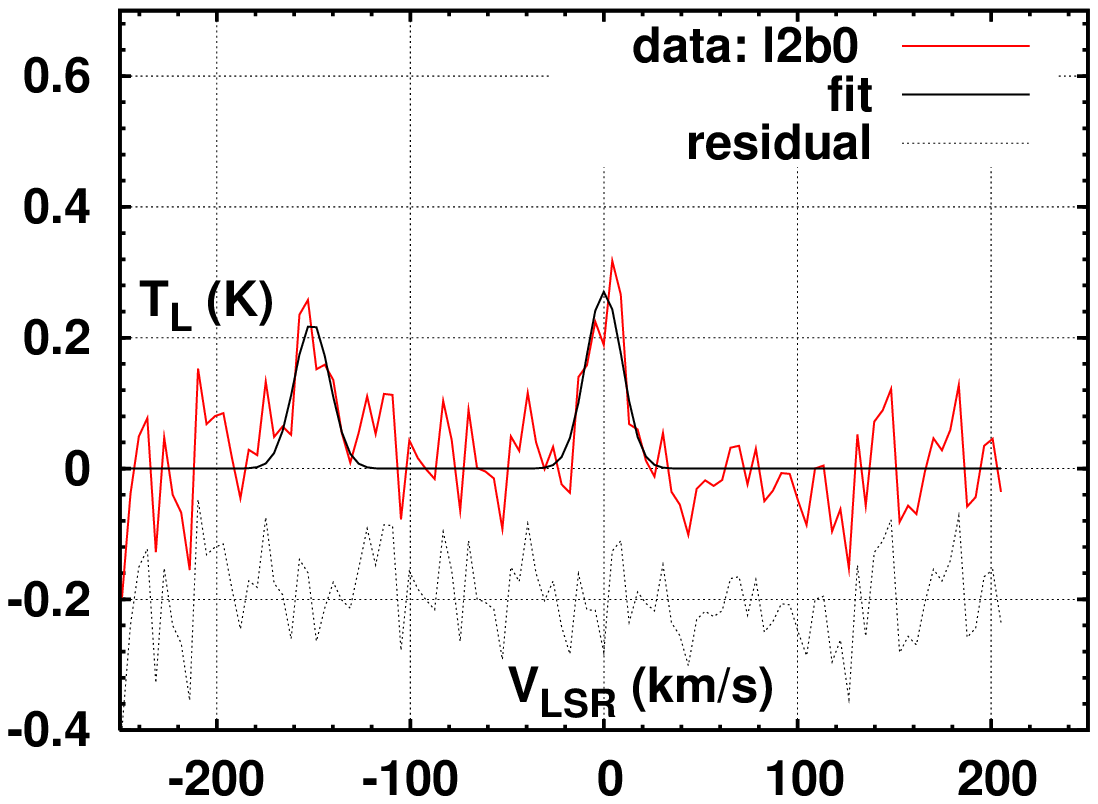}
\includegraphics[width=32mm,height=24mm,angle=0]{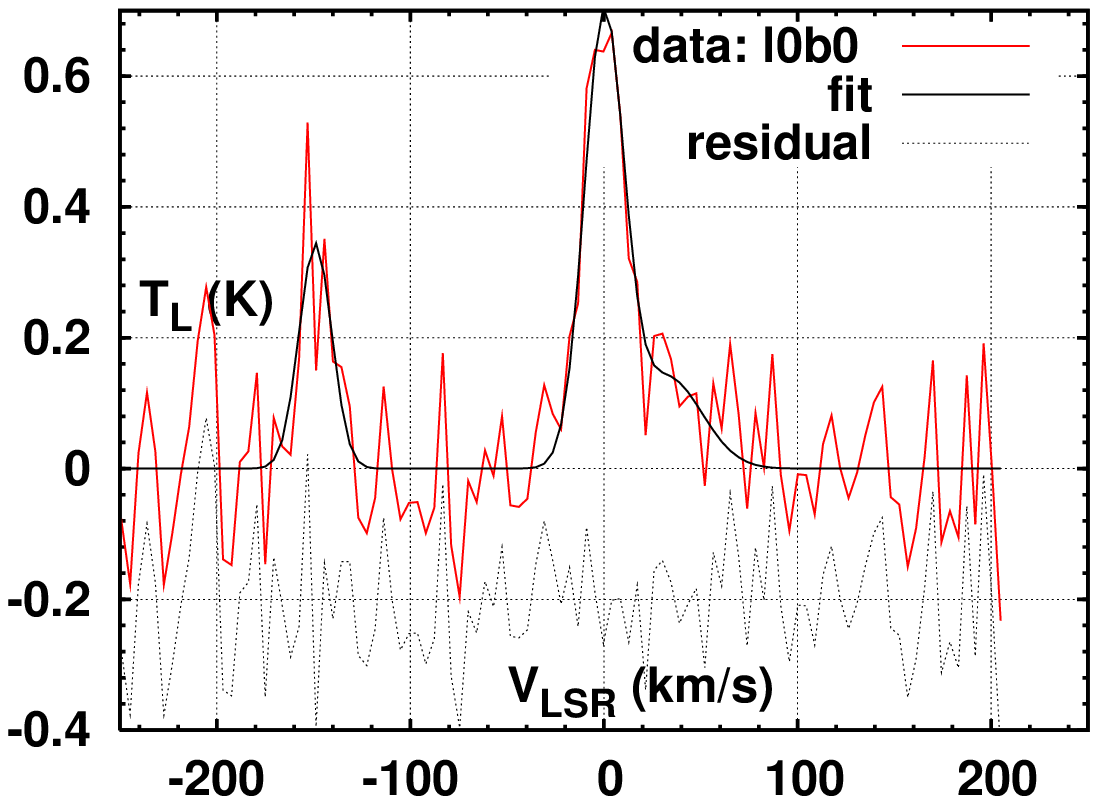}
\includegraphics[width=32mm,height=24mm,angle=0]{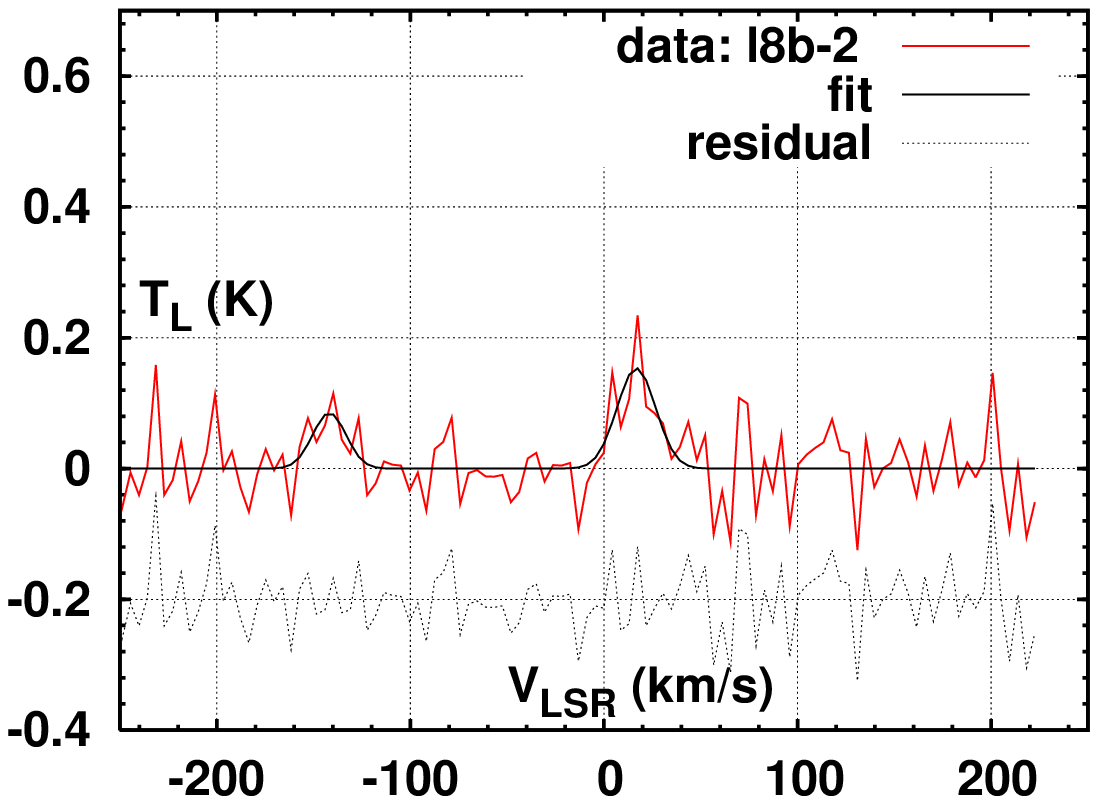}
\includegraphics[width=32mm,height=24mm,angle=0]{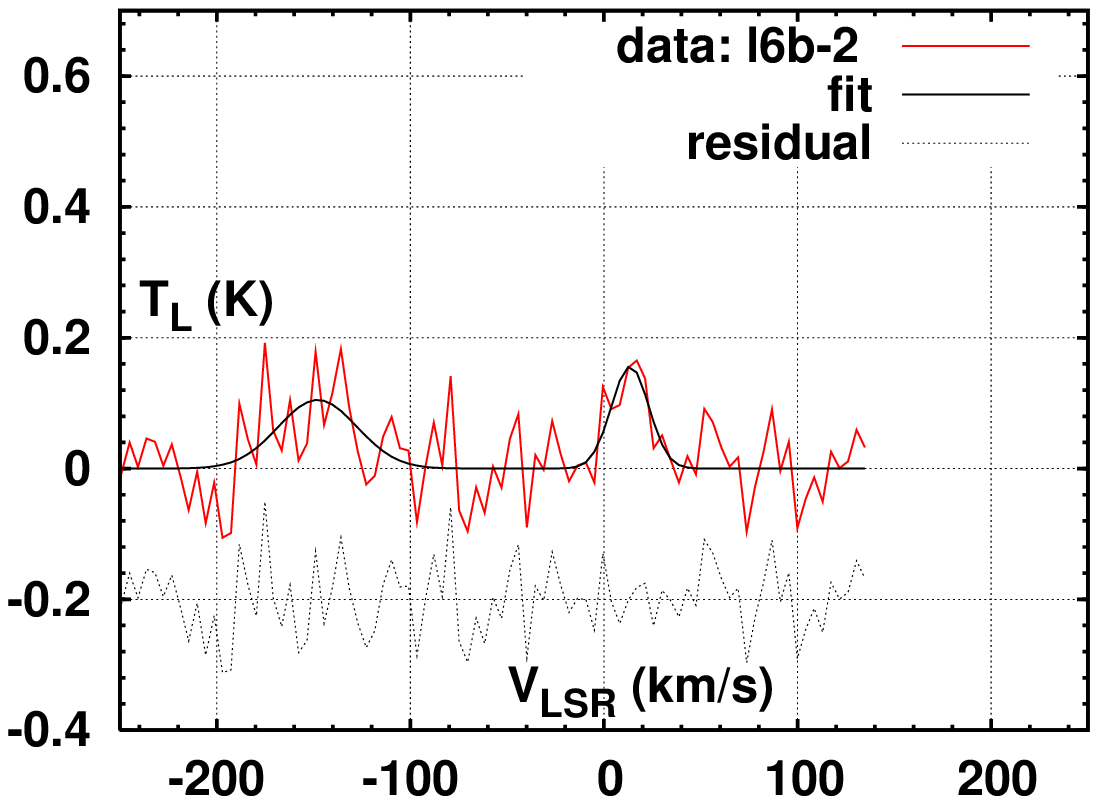}
\includegraphics[width=32mm,height=24mm,angle=0]{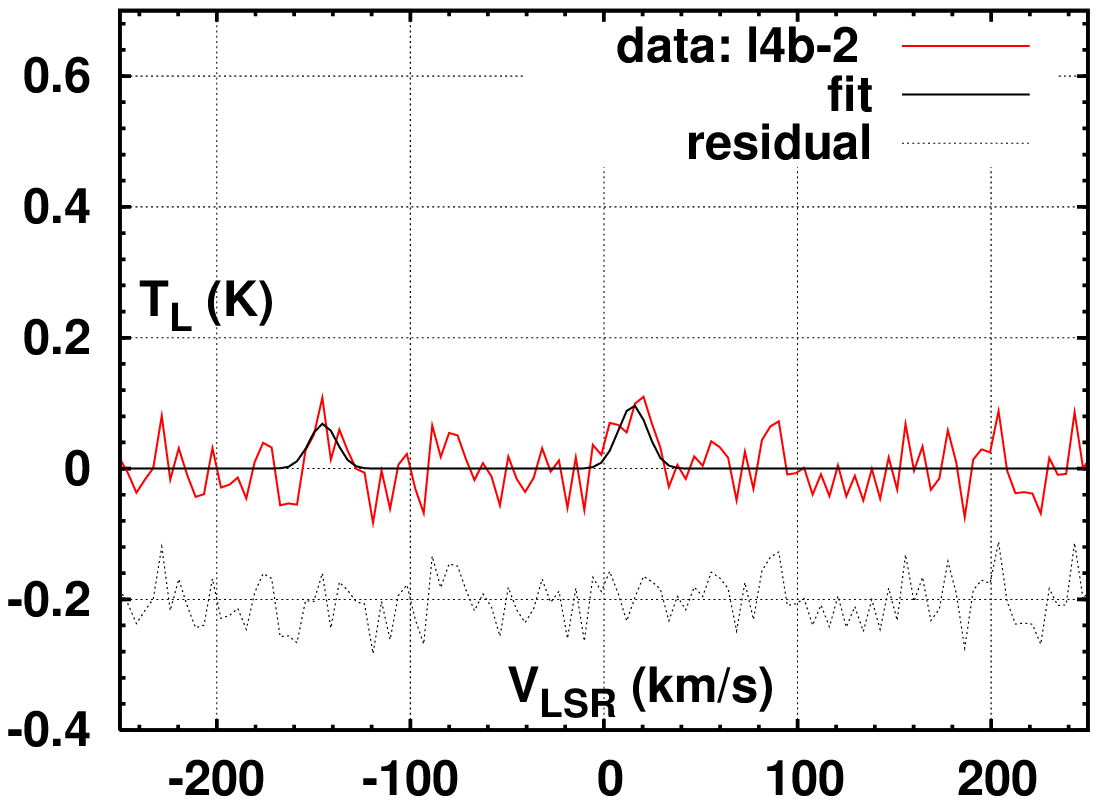}
\includegraphics[width=32mm,height=24mm,angle=0]{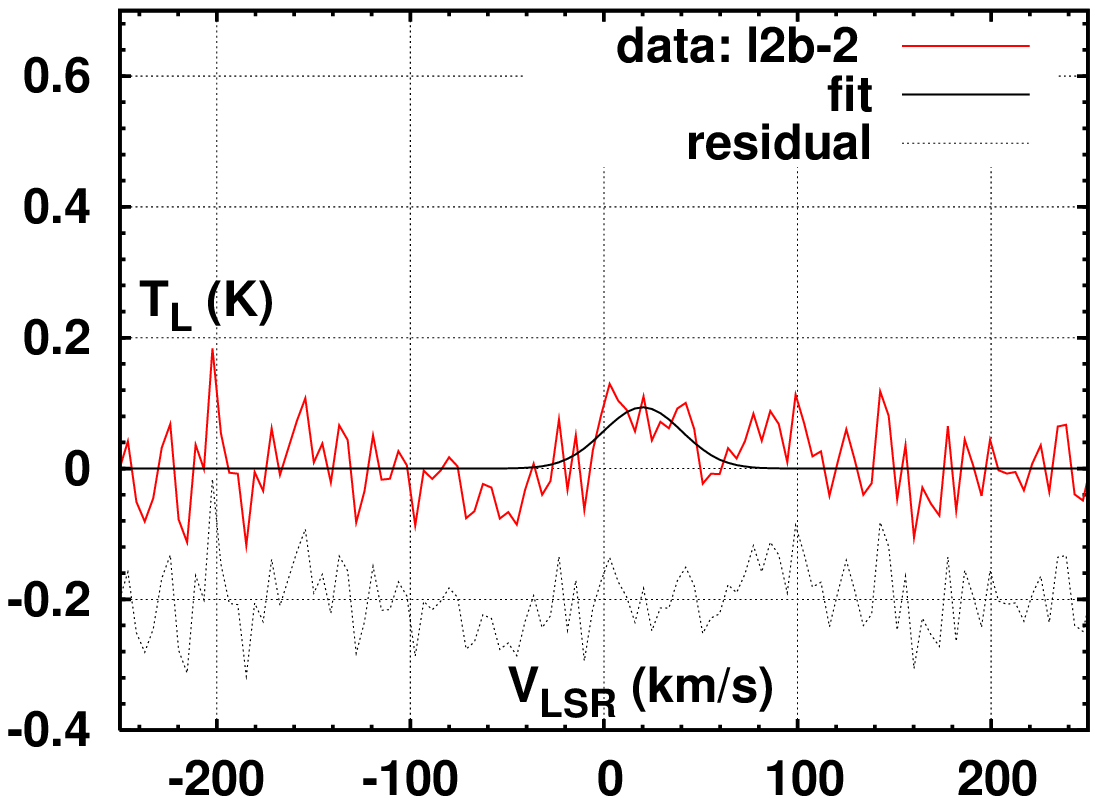}
\includegraphics[width=32mm,height=24mm,angle=0]{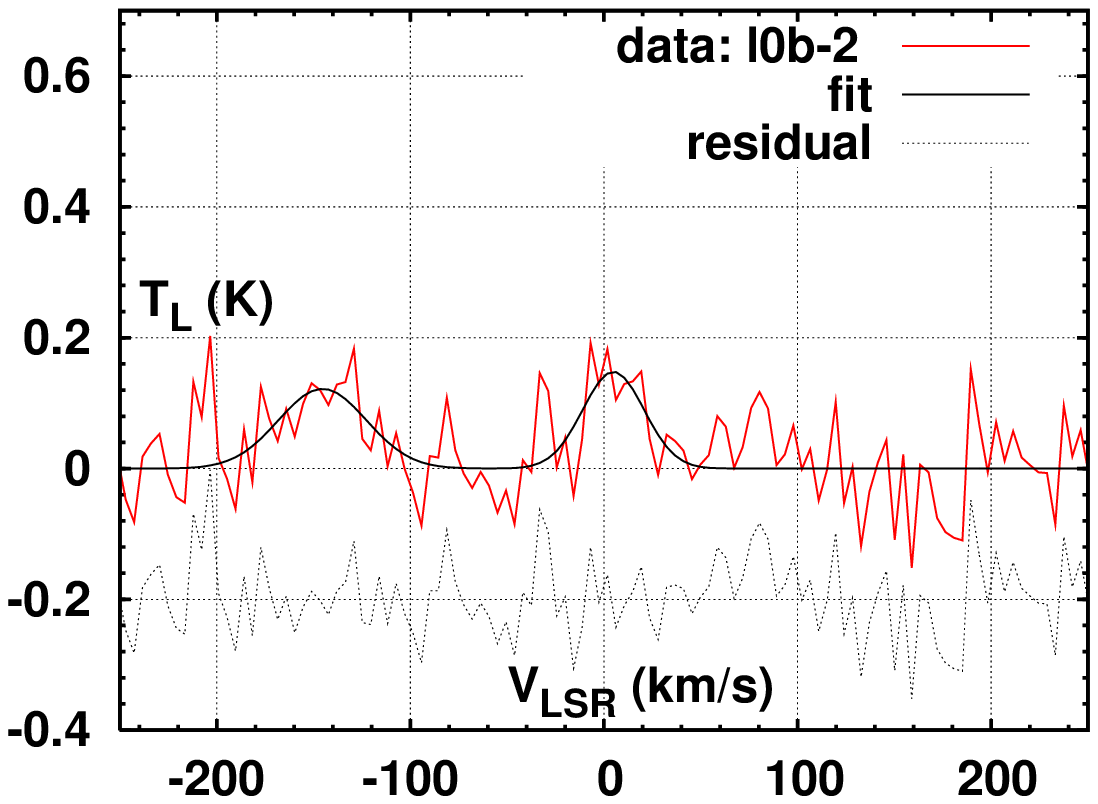}

 \caption{ ORT H271$\alpha$ RL observation.}
\end{center}
\end{figure}

\begin{figure}[ht]
\begin{center}
\includegraphics[width=32mm,height=24mm,angle=0]{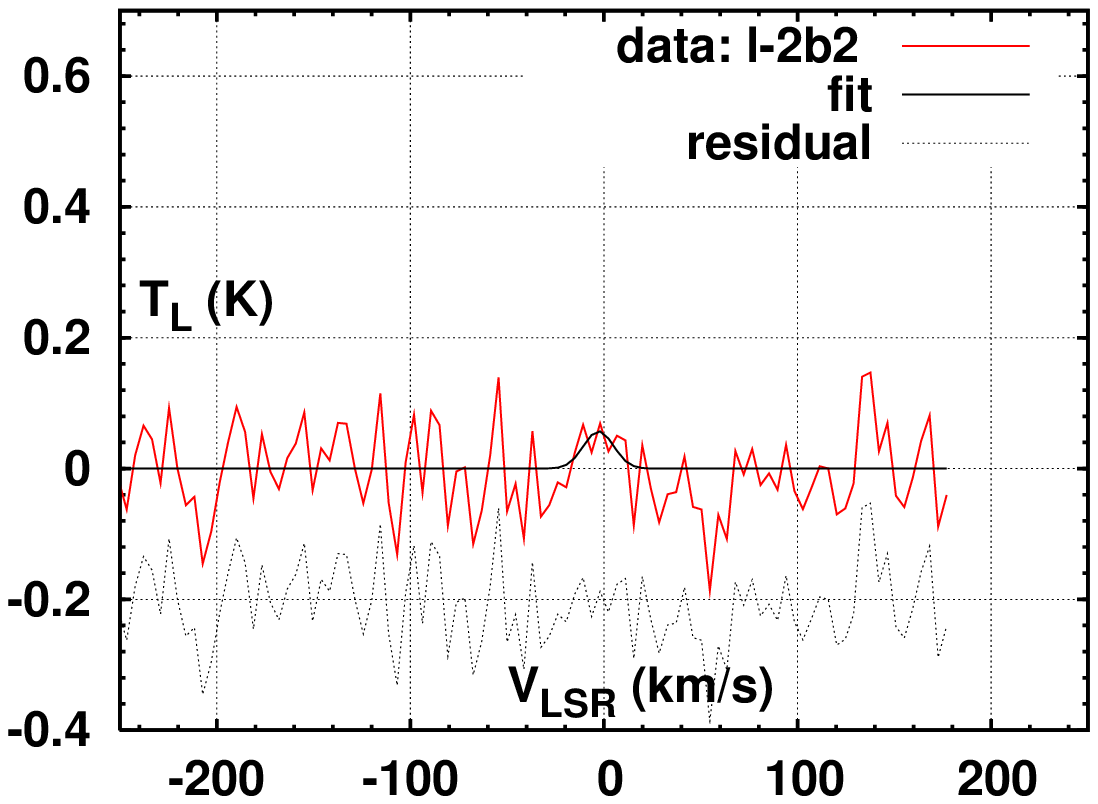}
\includegraphics[width=32mm,height=24mm,angle=0]{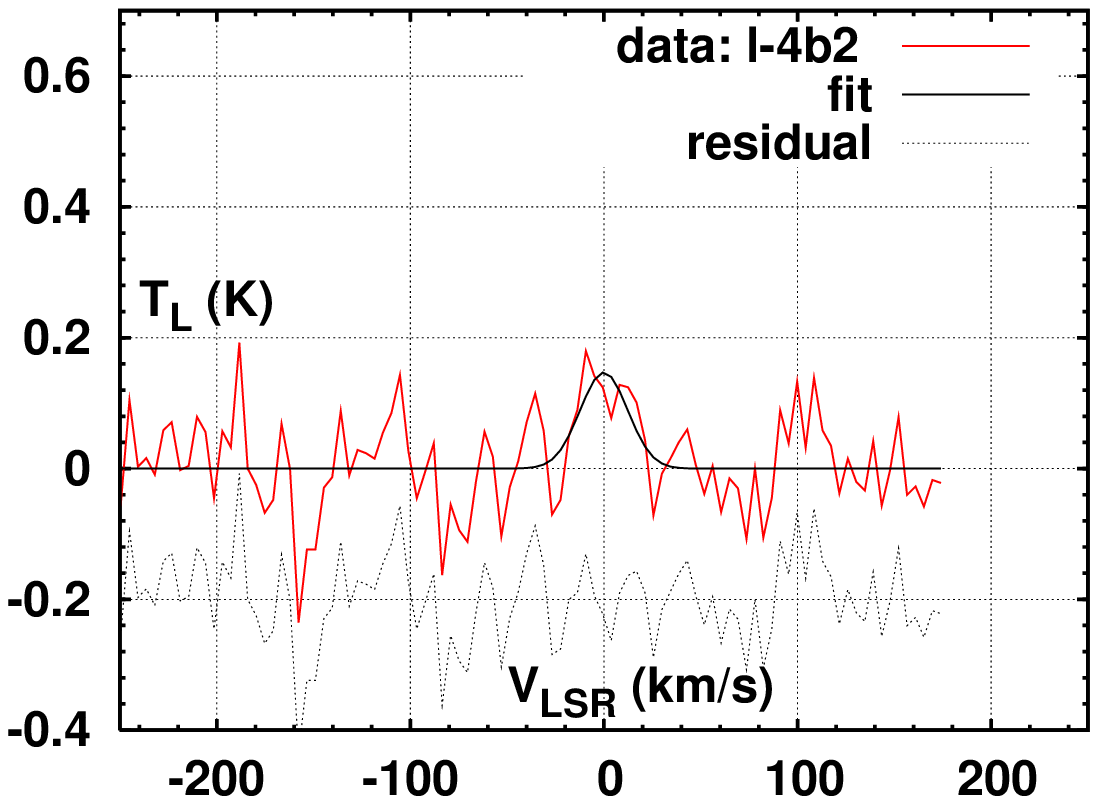}
\includegraphics[width=32mm,height=24mm,angle=0]{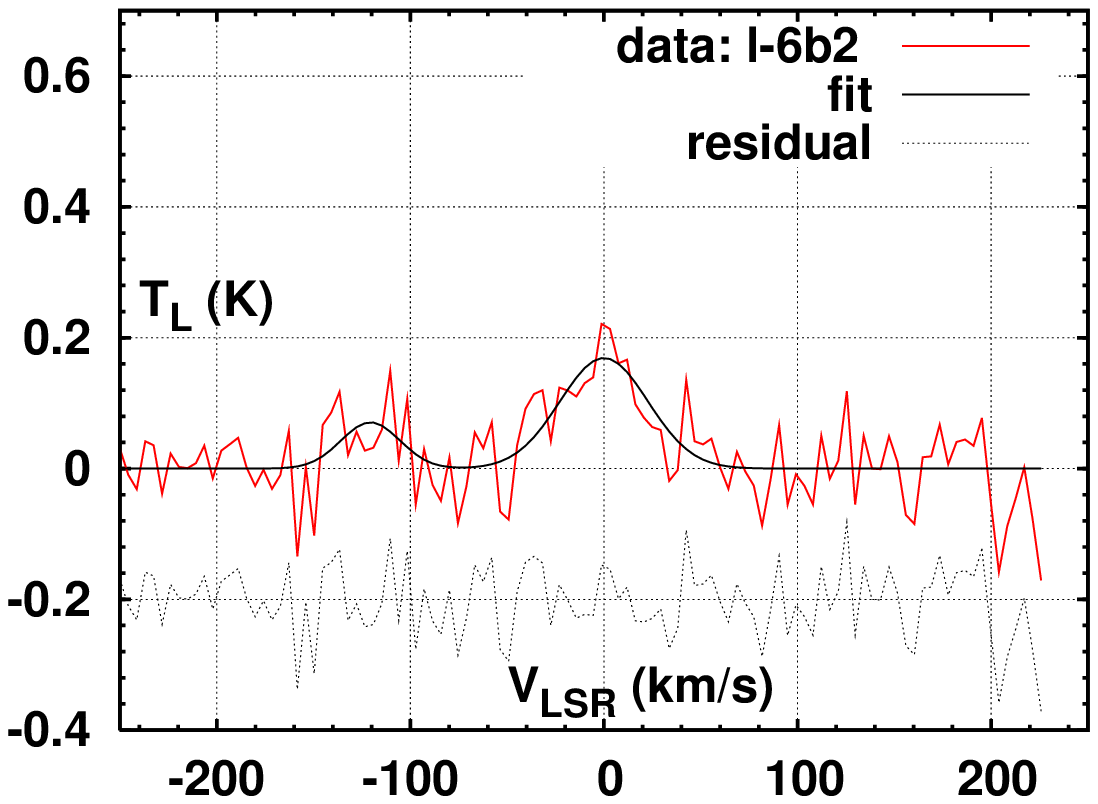}
\includegraphics[width=32mm,height=24mm,angle=0]{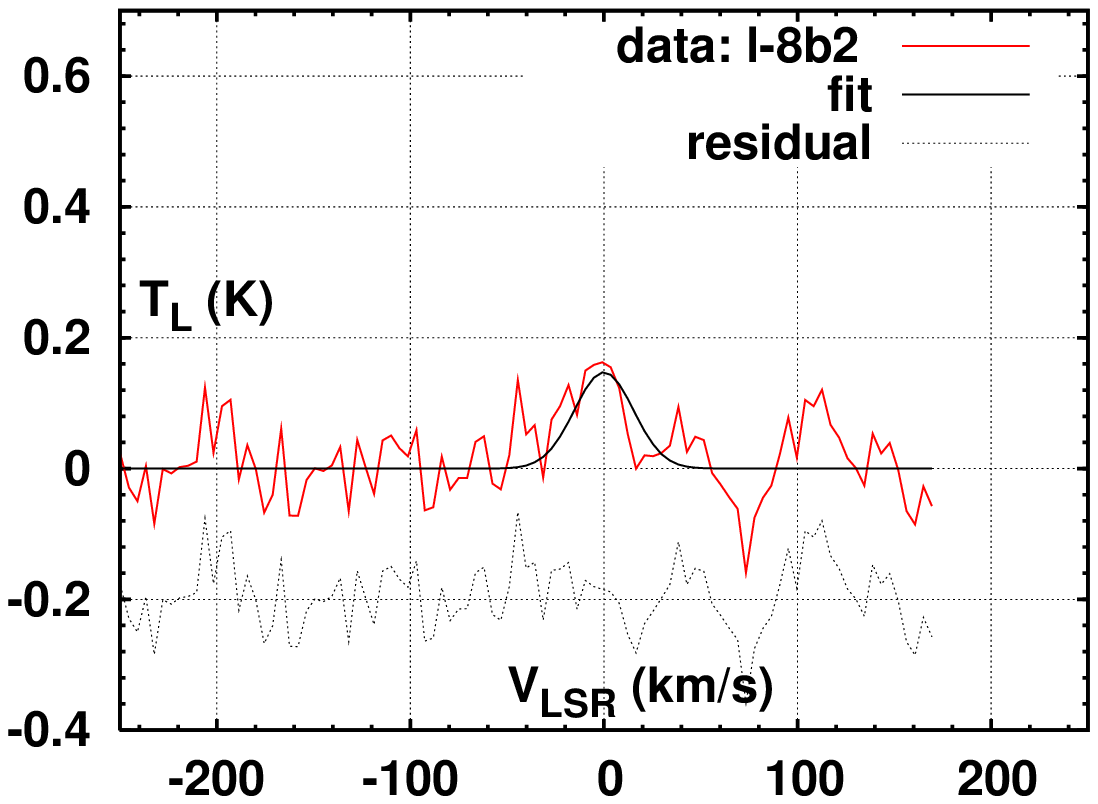}
\includegraphics[width=32mm,height=24mm,angle=0]{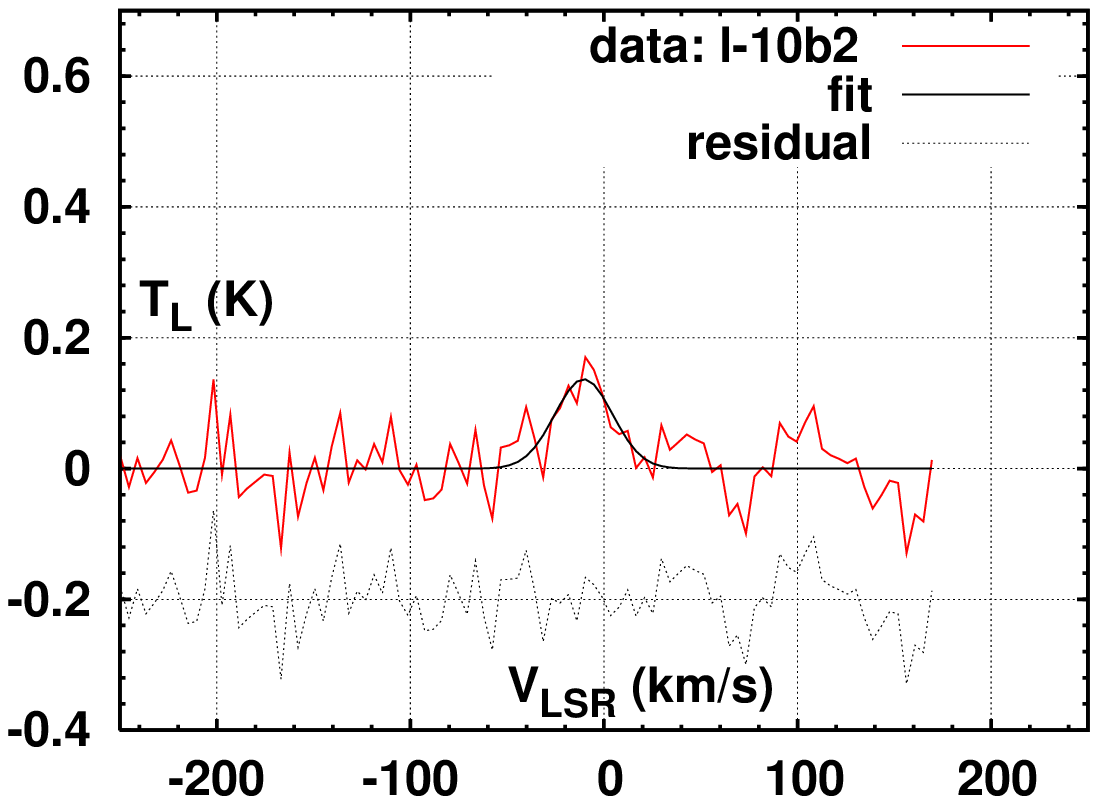}
\includegraphics[width=32mm,height=24mm,angle=0]{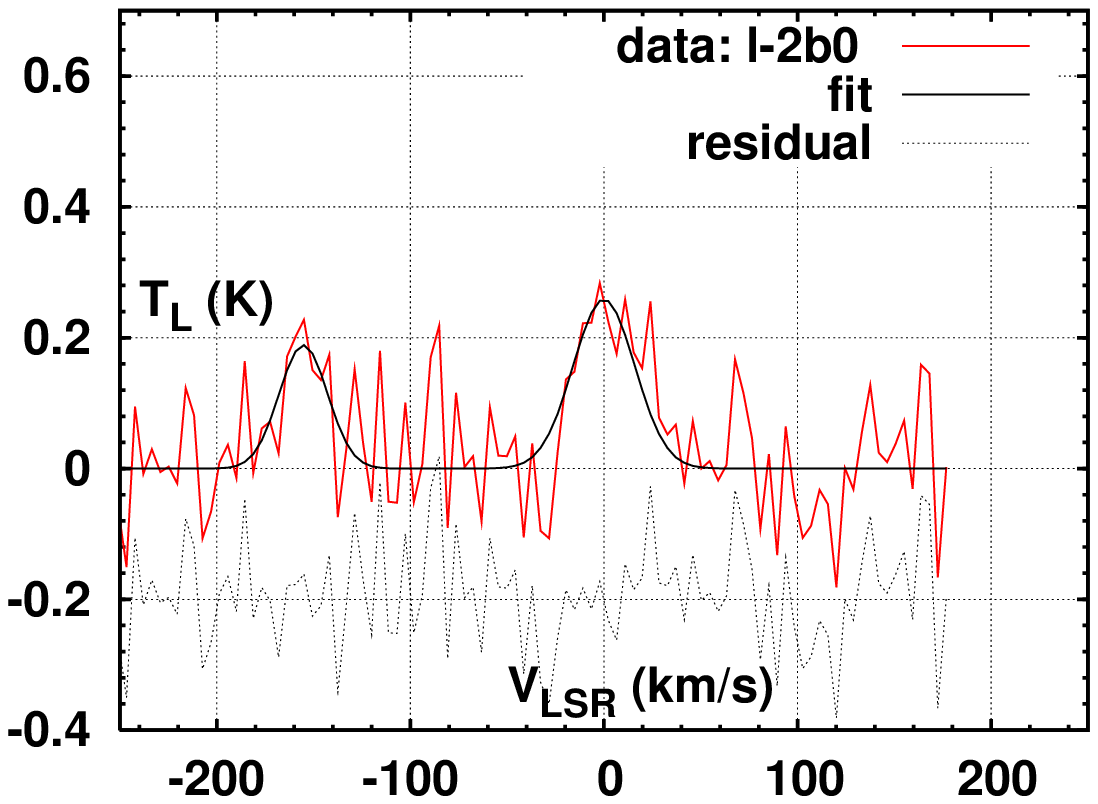}
\includegraphics[width=32mm,height=24mm,angle=0]{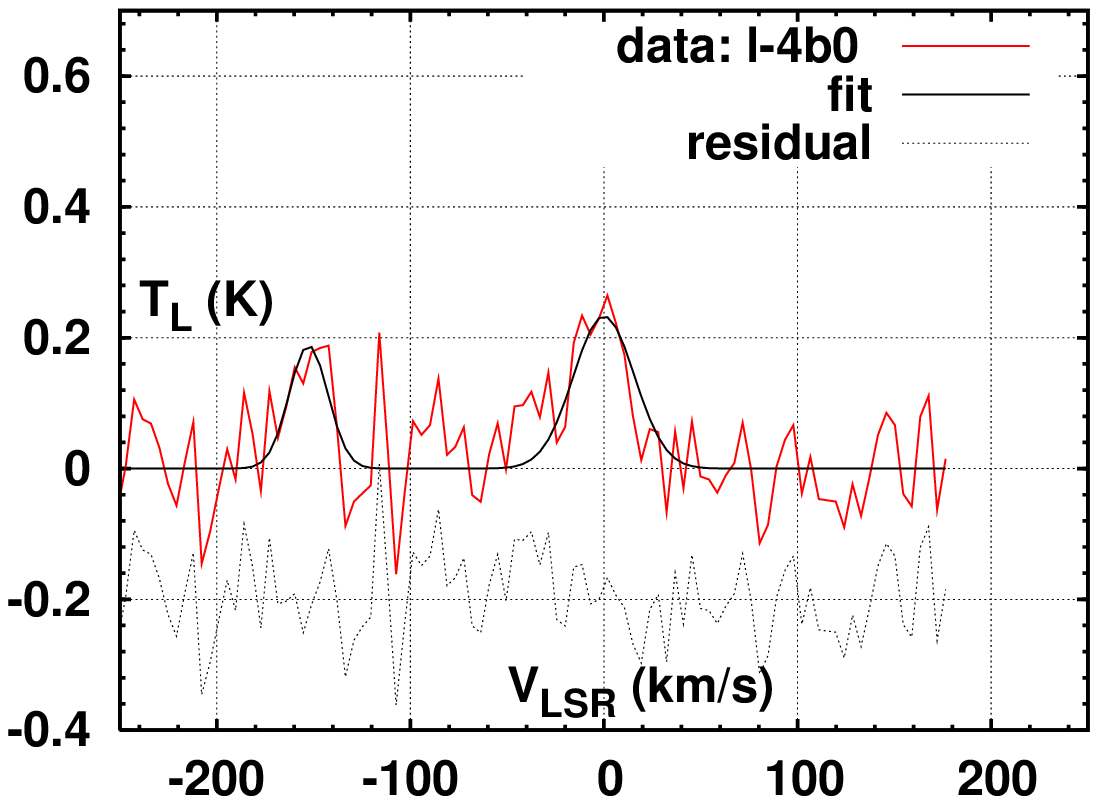}
\includegraphics[width=32mm,height=24mm,angle=0]{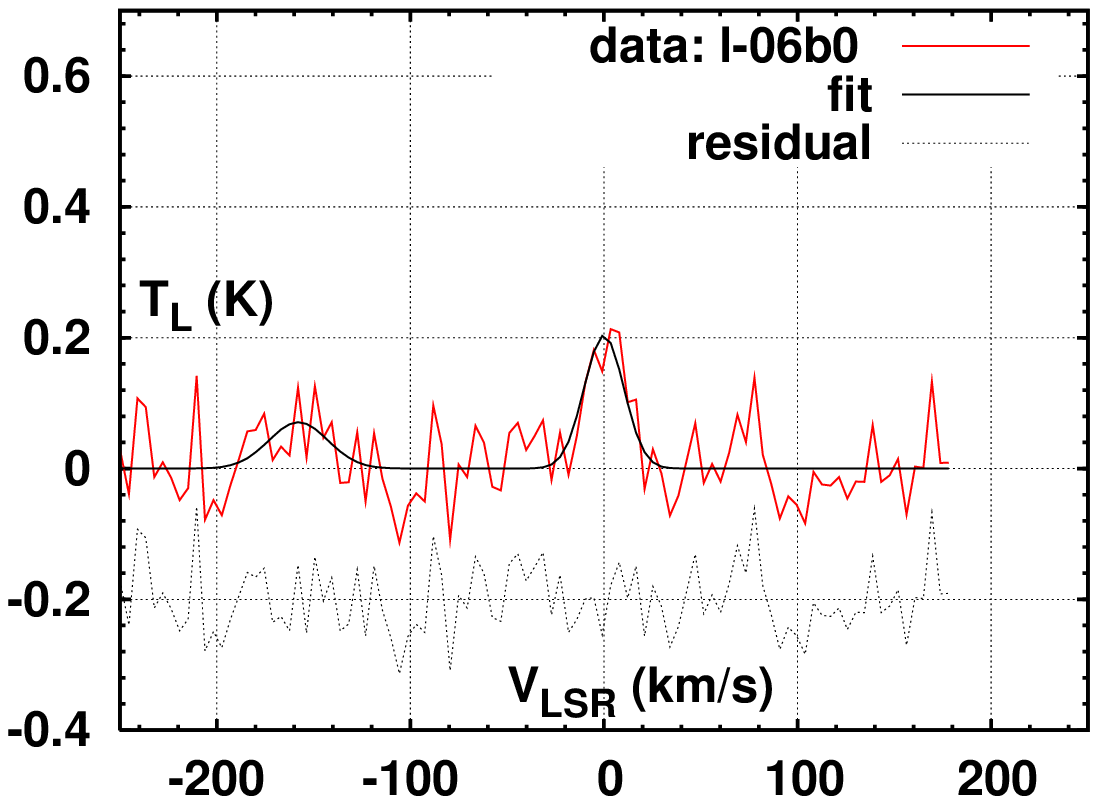}
\includegraphics[width=32mm,height=24mm,angle=0]{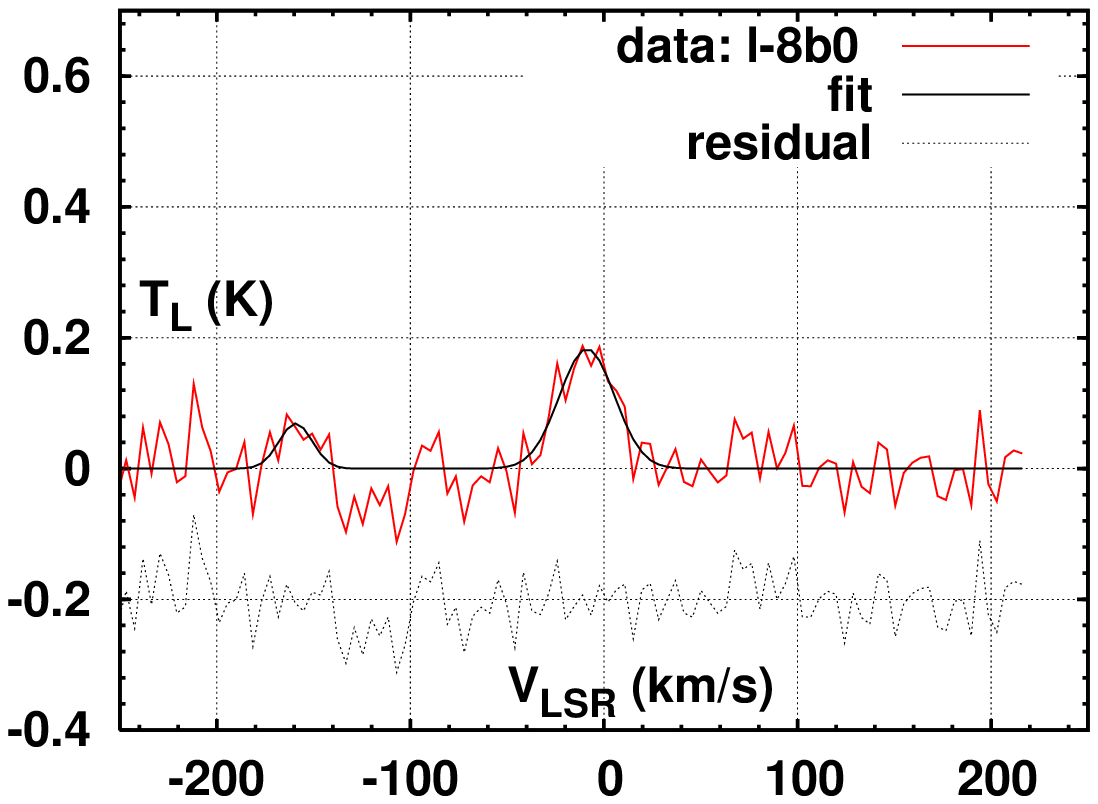}
\includegraphics[width=32mm,height=24mm,angle=0]{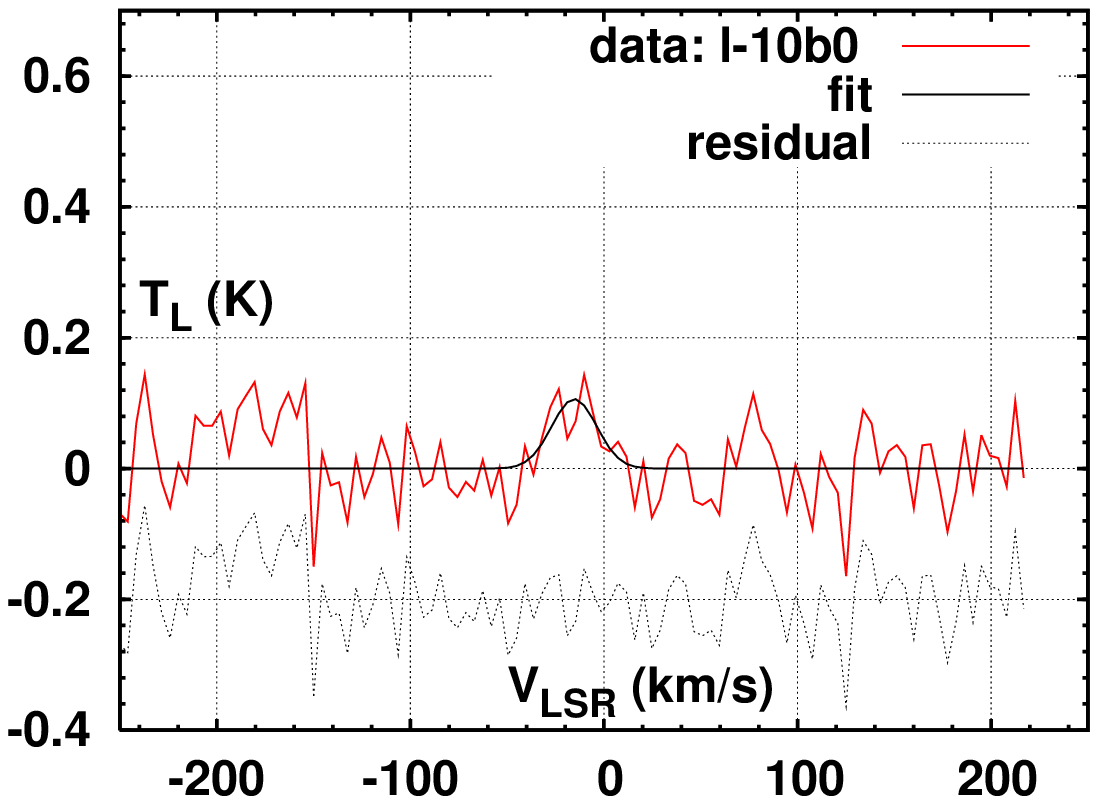}
\includegraphics[width=32mm,height=24mm,angle=0]{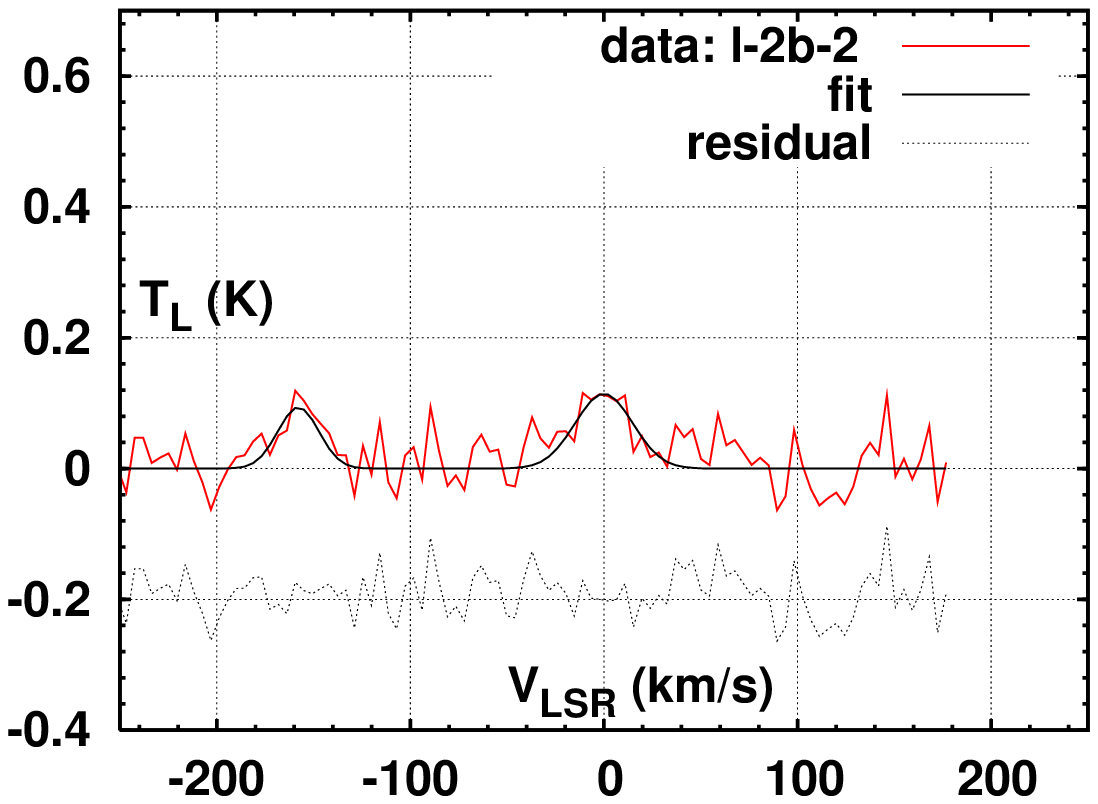}
\includegraphics[width=32mm,height=24mm,angle=0]{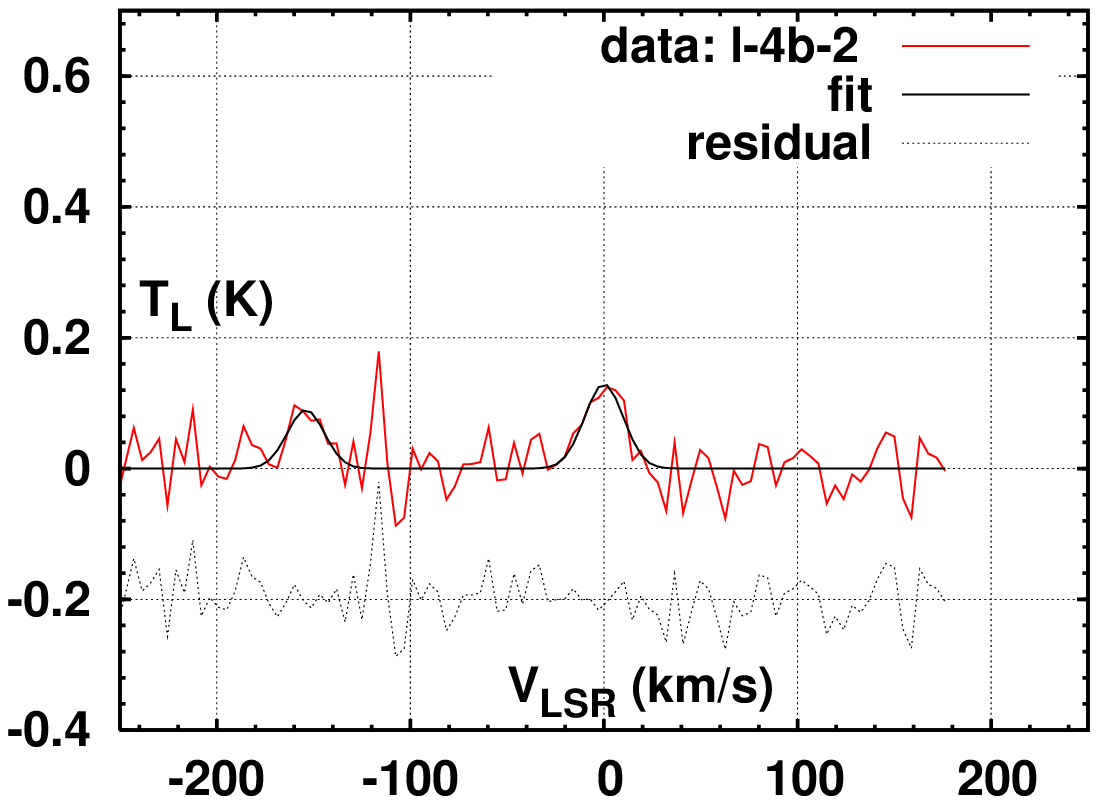}
\includegraphics[width=32mm,height=24mm,angle=0]{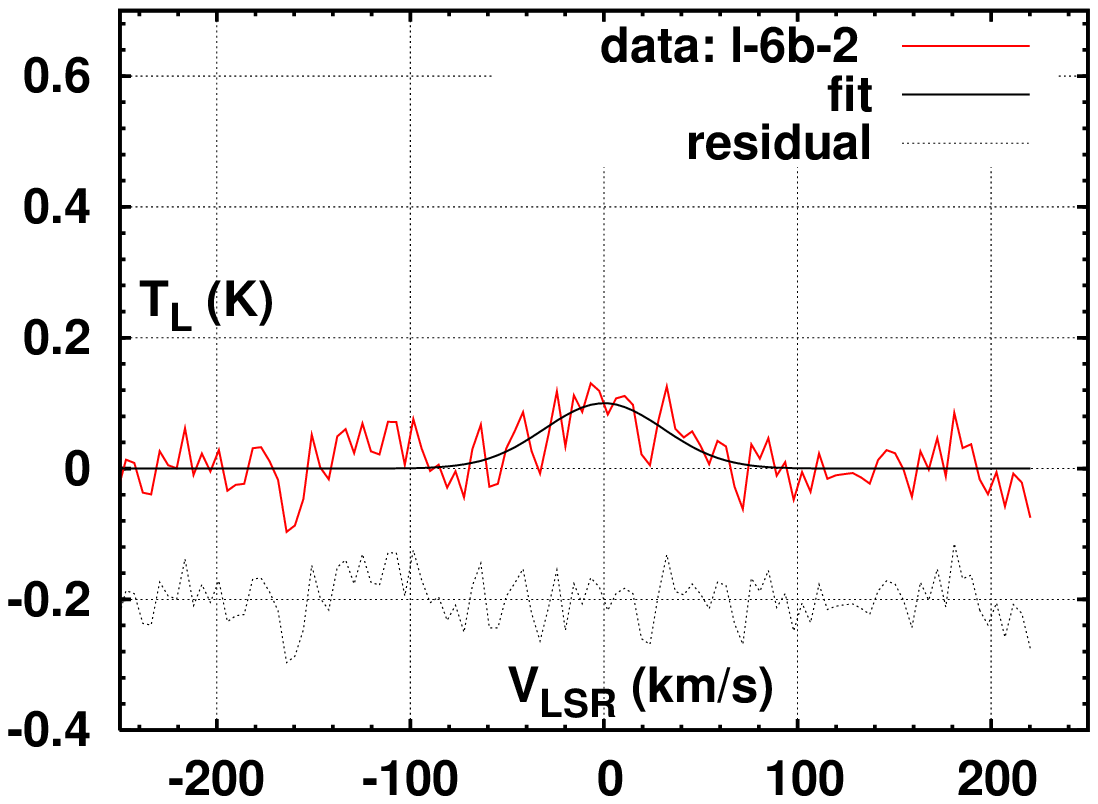}
\includegraphics[width=32mm,height=24mm,angle=0]{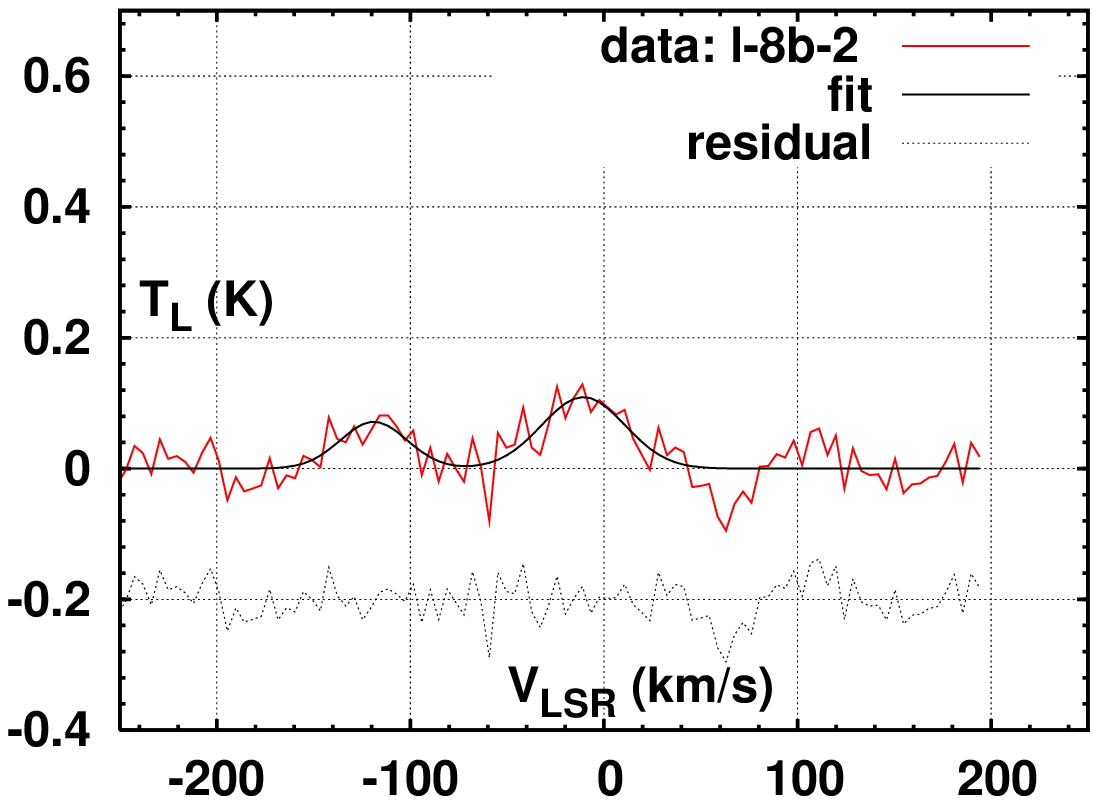}
\includegraphics[width=32mm,height=24mm,angle=0]{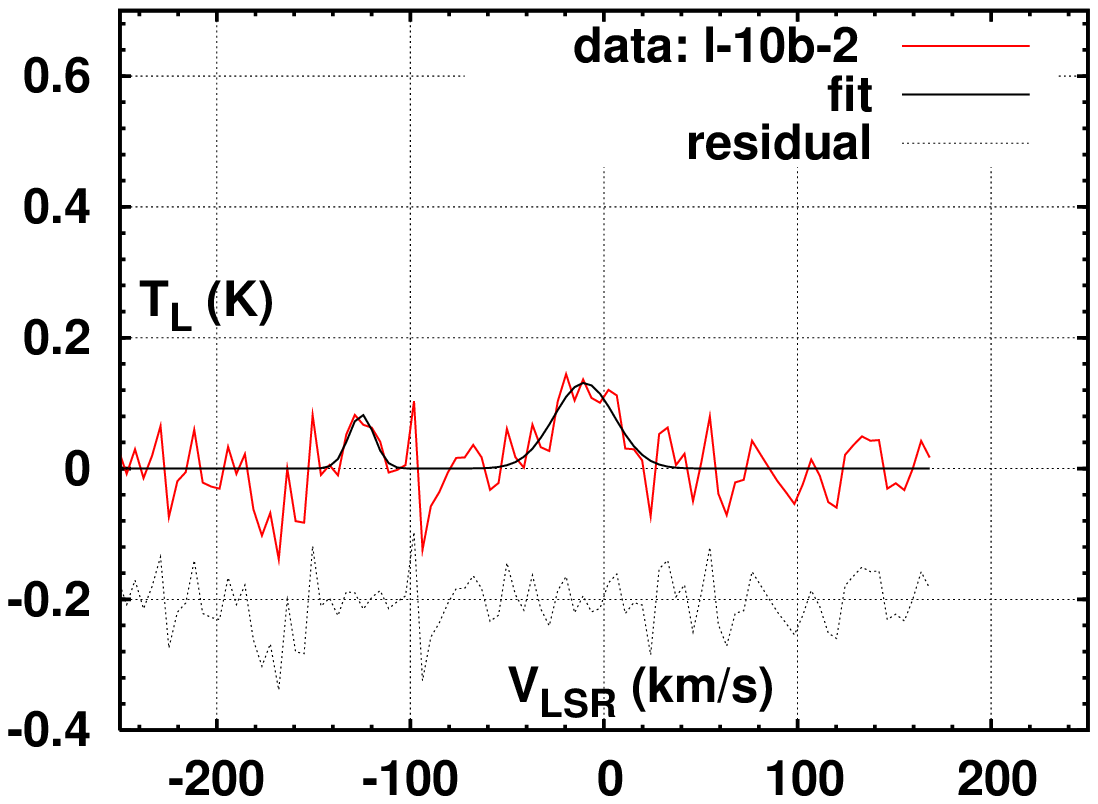}\vspace{1cm}
\includegraphics[width=32mm,height=24mm,angle=0]{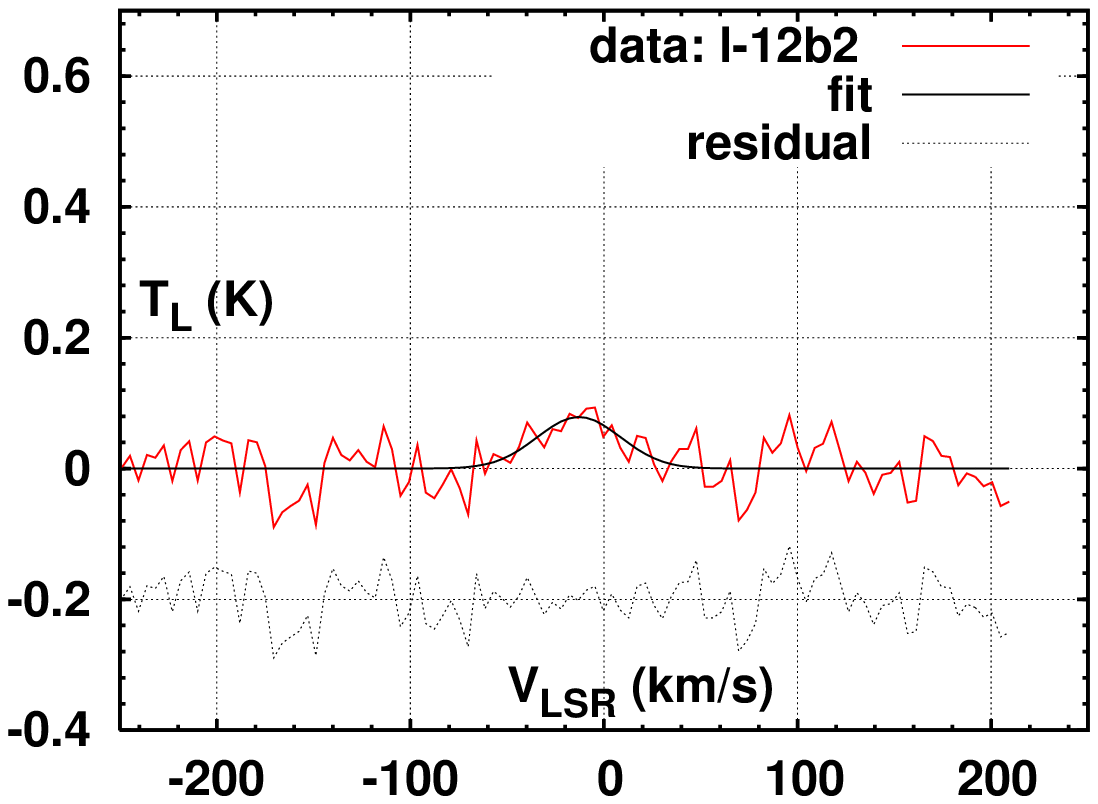}
\includegraphics[width=32mm,height=24mm,angle=0]{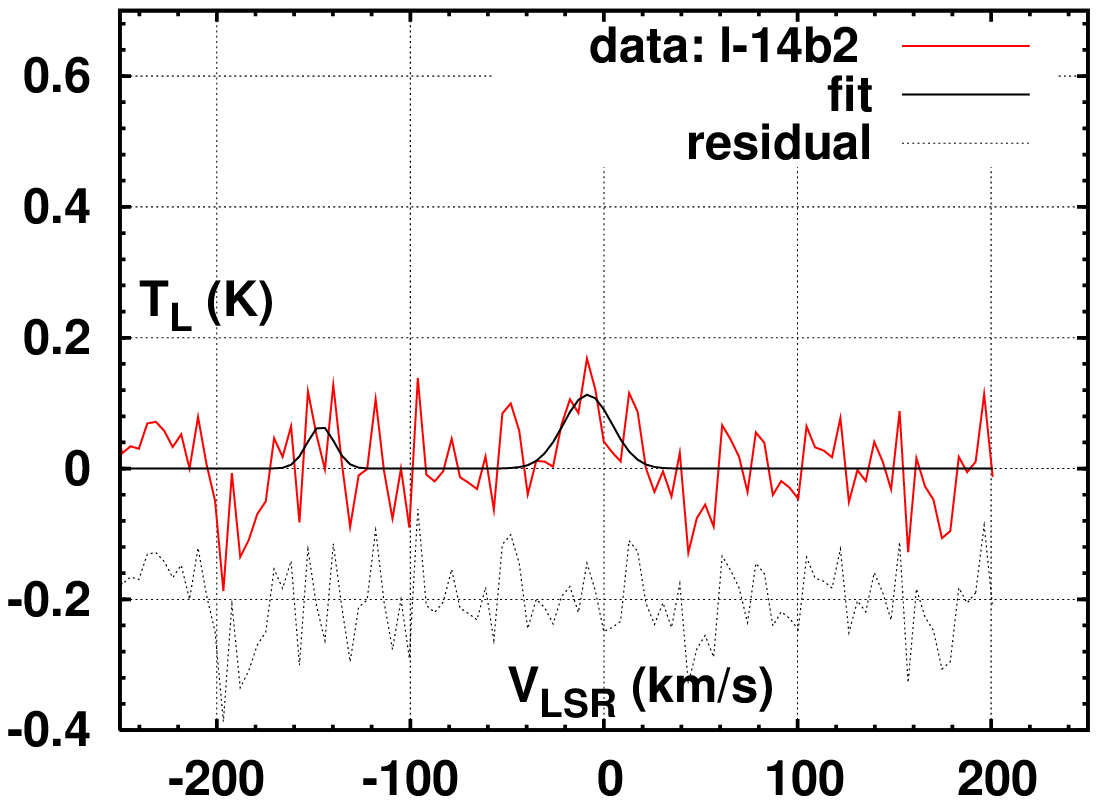}
\includegraphics[width=32mm,height=24mm,angle=0]{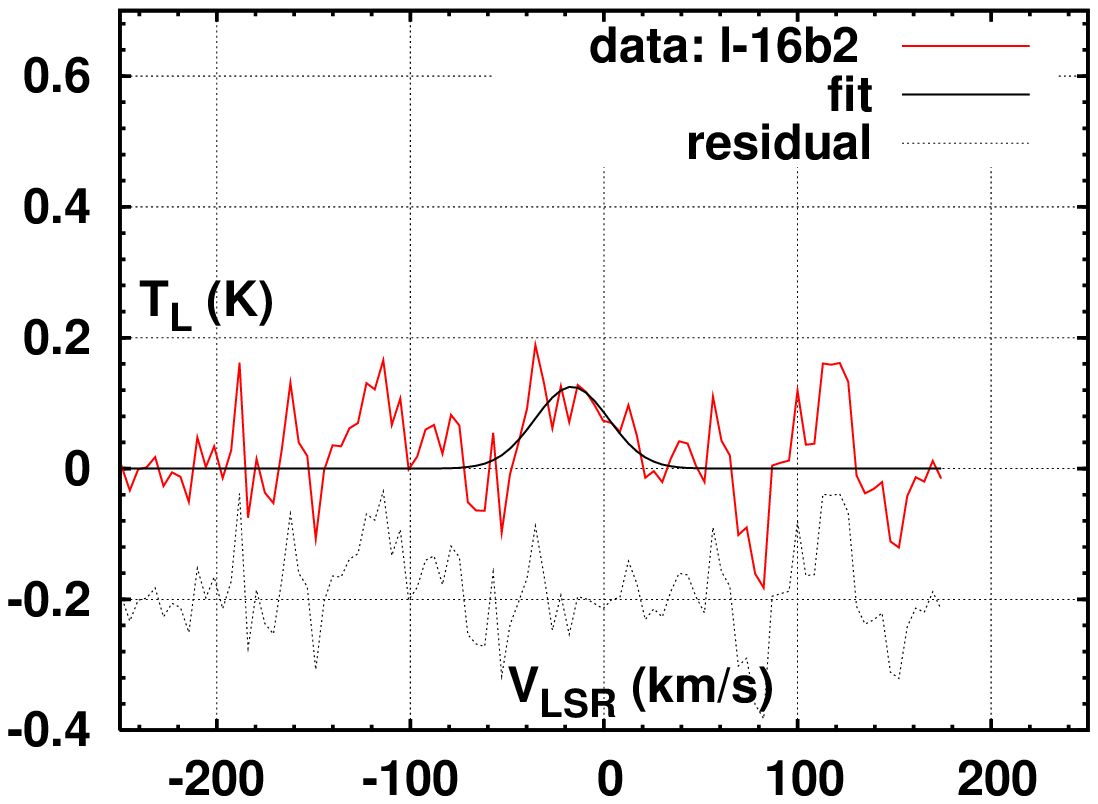}
\includegraphics[width=32mm,height=24mm,angle=0]{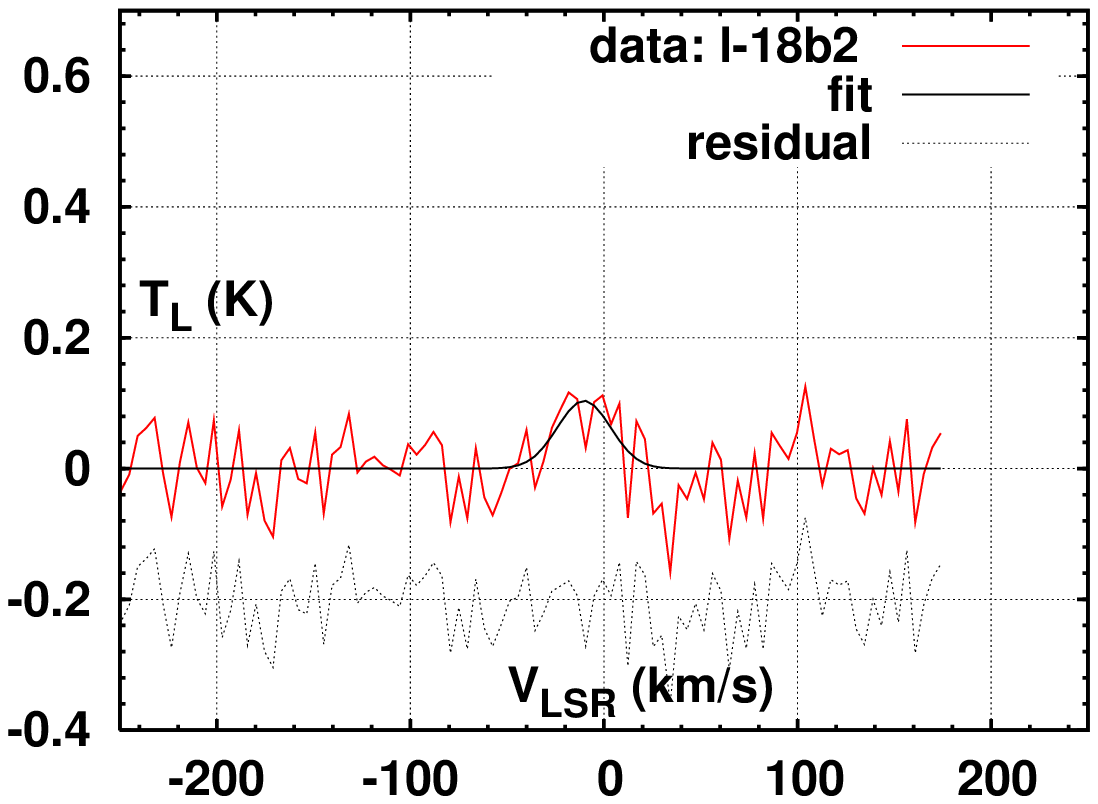}
\includegraphics[width=32mm,height=24mm,angle=0]{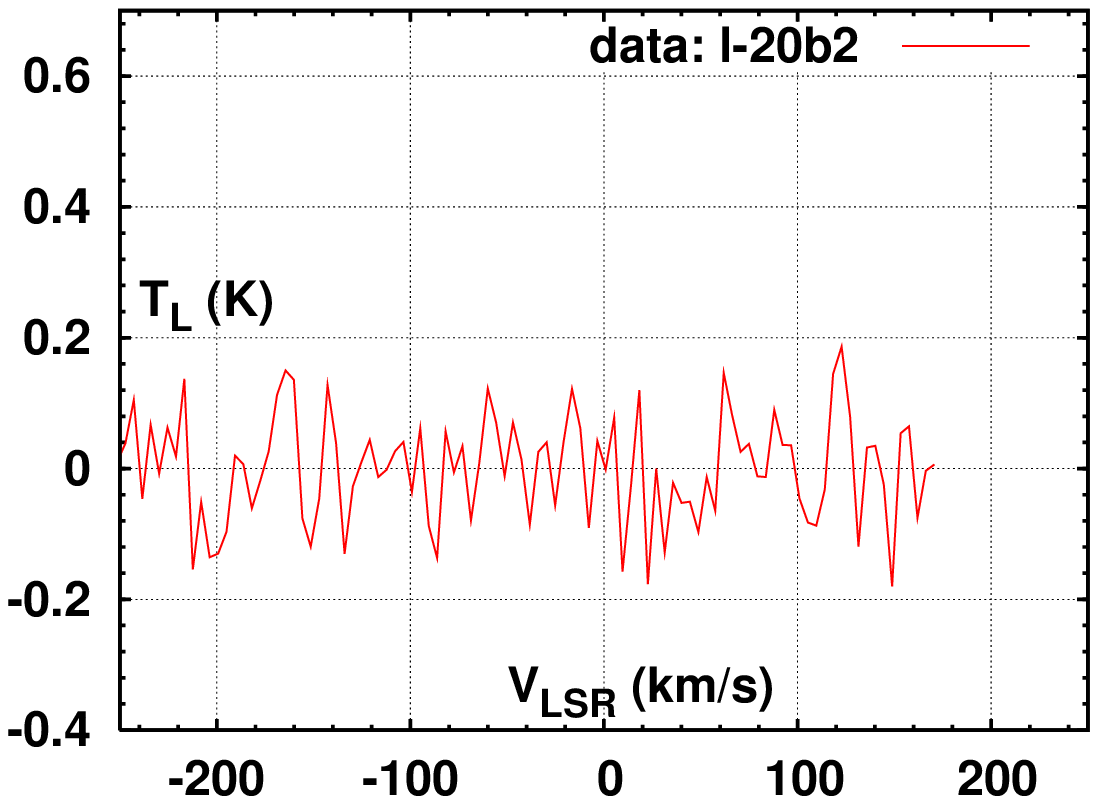}
\includegraphics[width=32mm,height=24mm,angle=0]{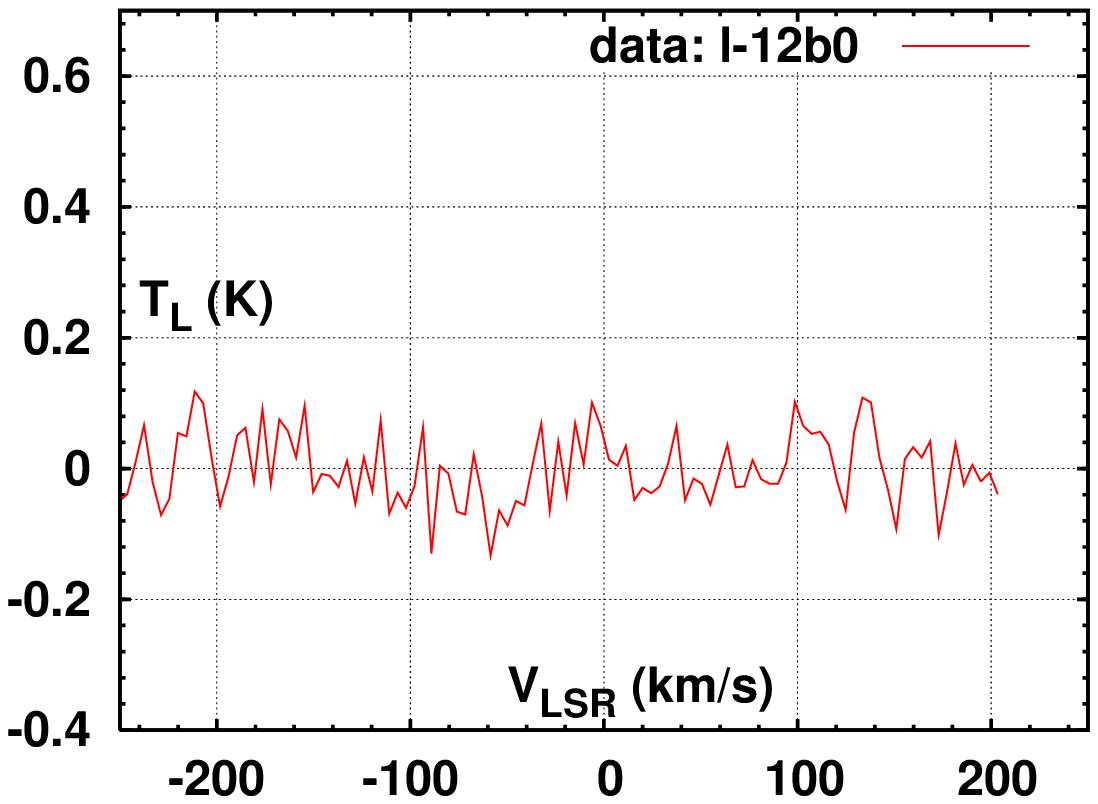}
\includegraphics[width=32mm,height=24mm,angle=0]{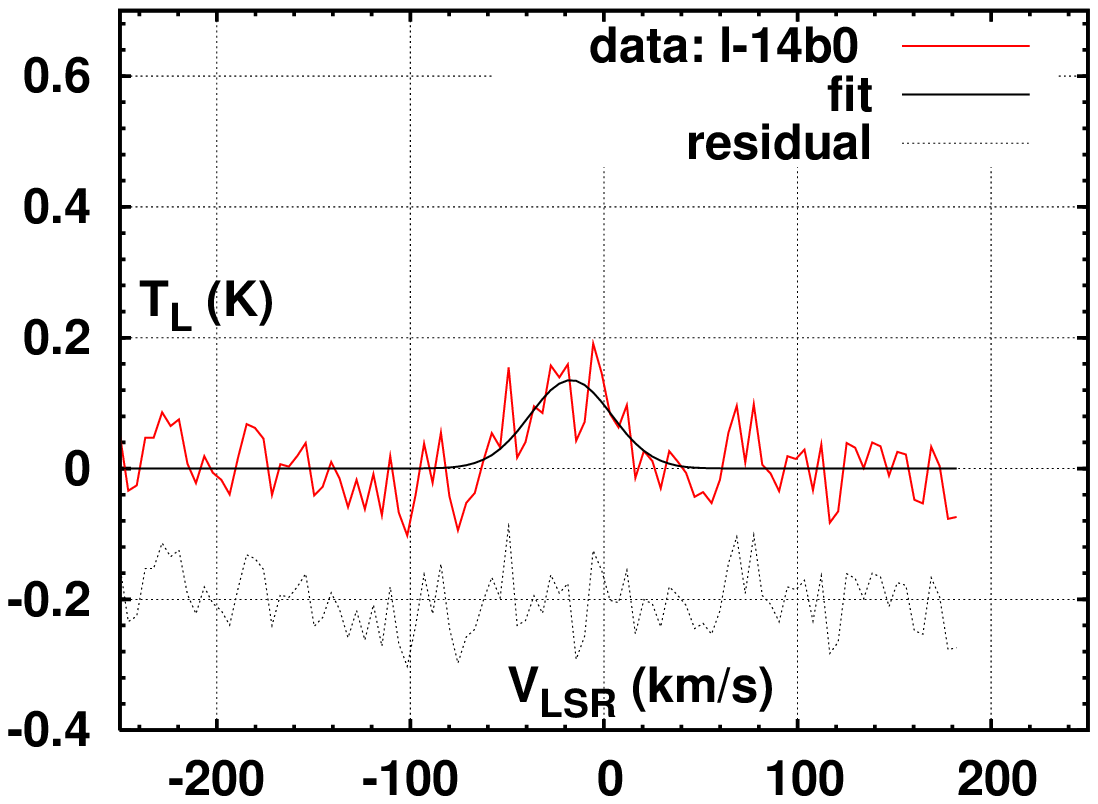}
\includegraphics[width=32mm,height=24mm,angle=0]{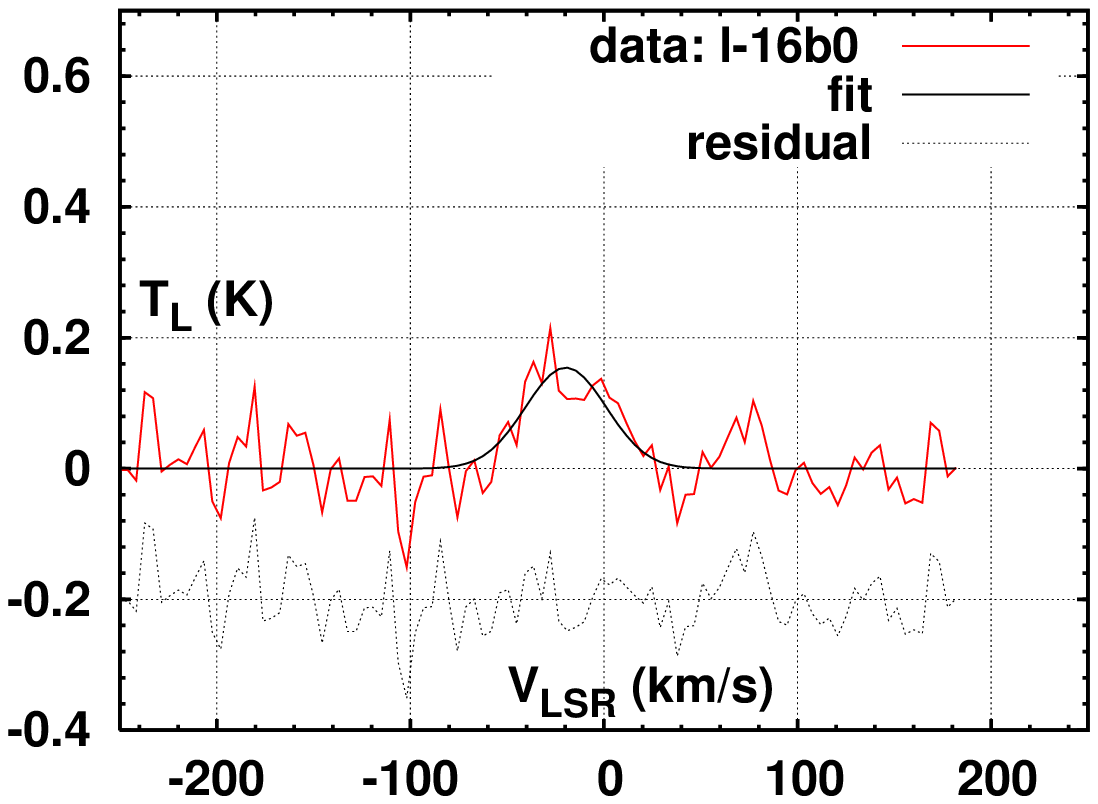}
\includegraphics[width=32mm,height=24mm,angle=0]{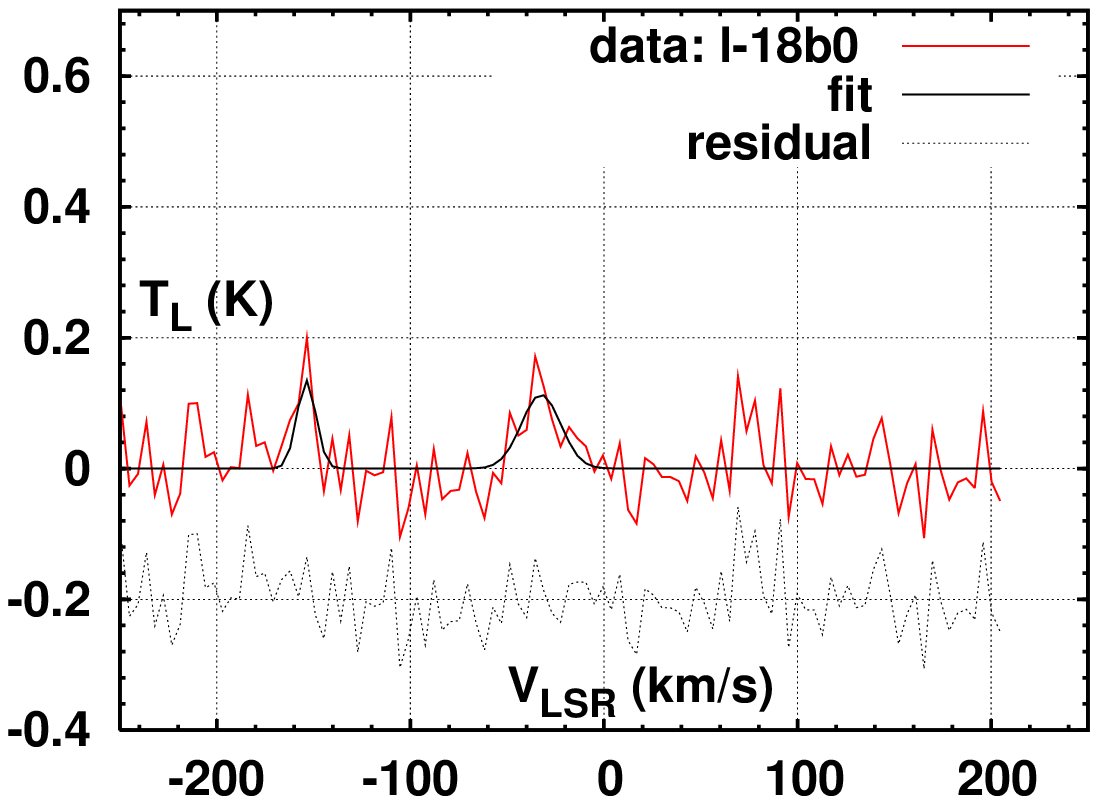}
\includegraphics[width=32mm,height=24mm,angle=0]{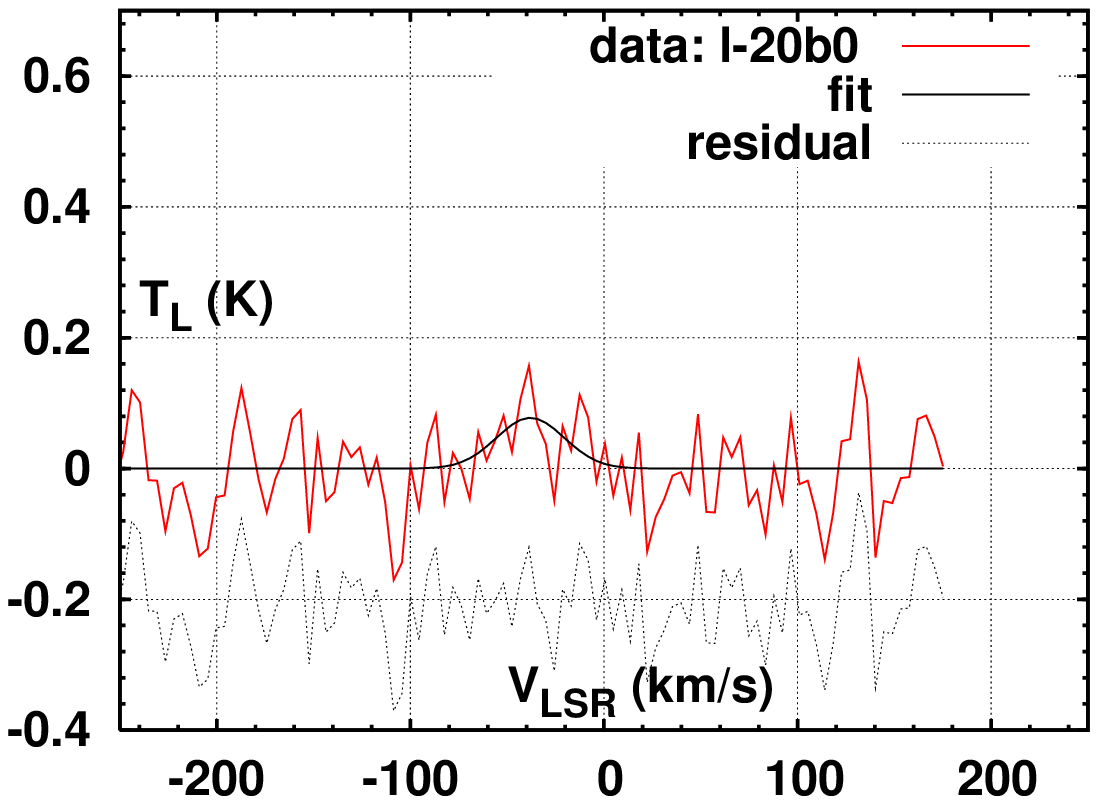}
\includegraphics[width=32mm,height=24mm,angle=0]{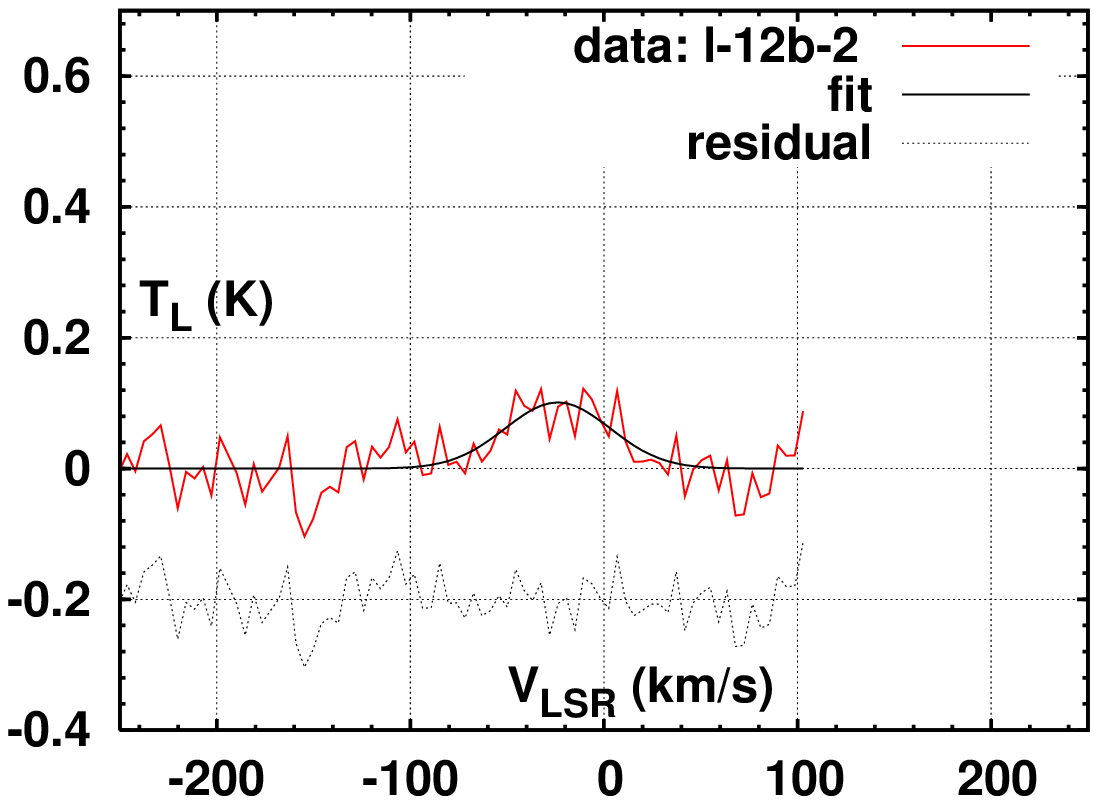}
\includegraphics[width=32mm,height=24mm,angle=0]{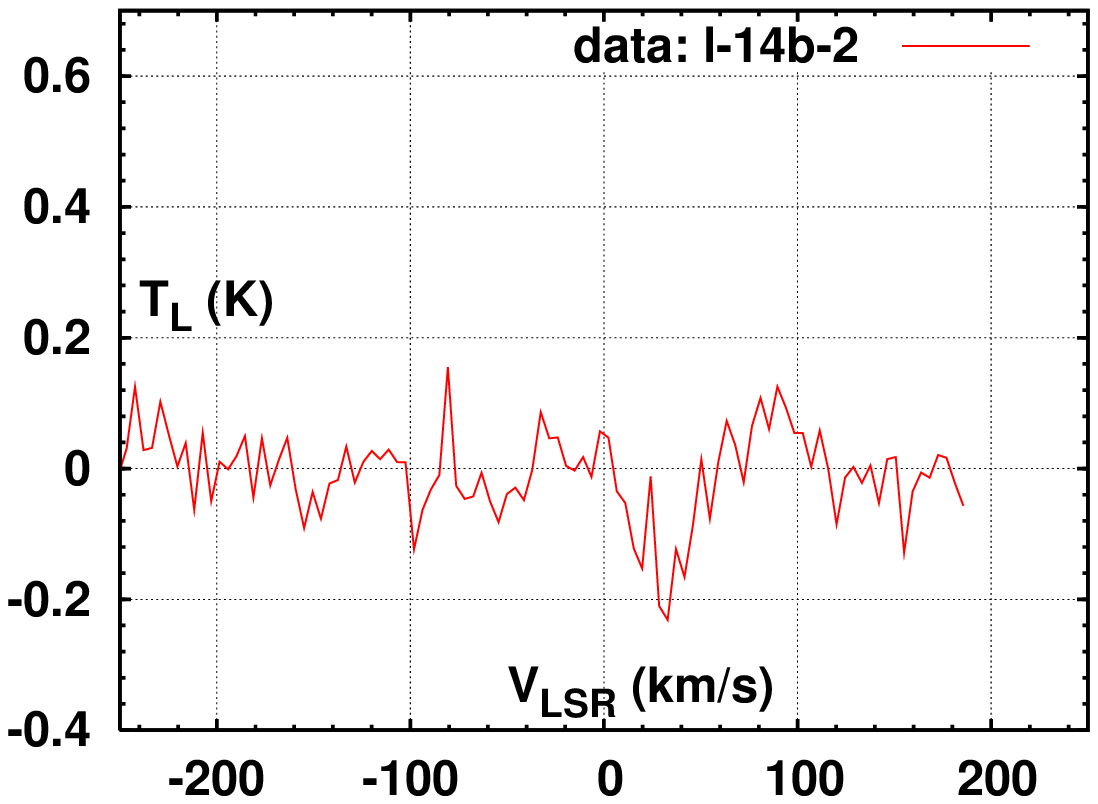}
\includegraphics[width=32mm,height=24mm,angle=0]{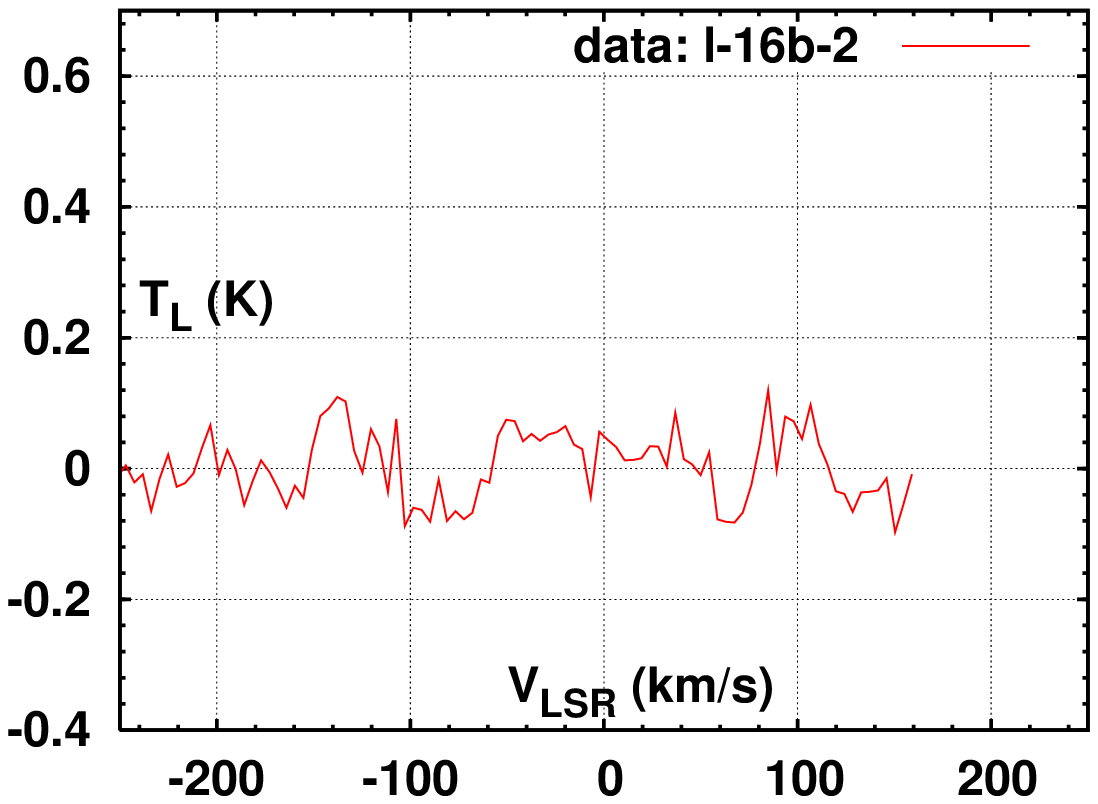}
\includegraphics[width=32mm,height=24mm,angle=0]{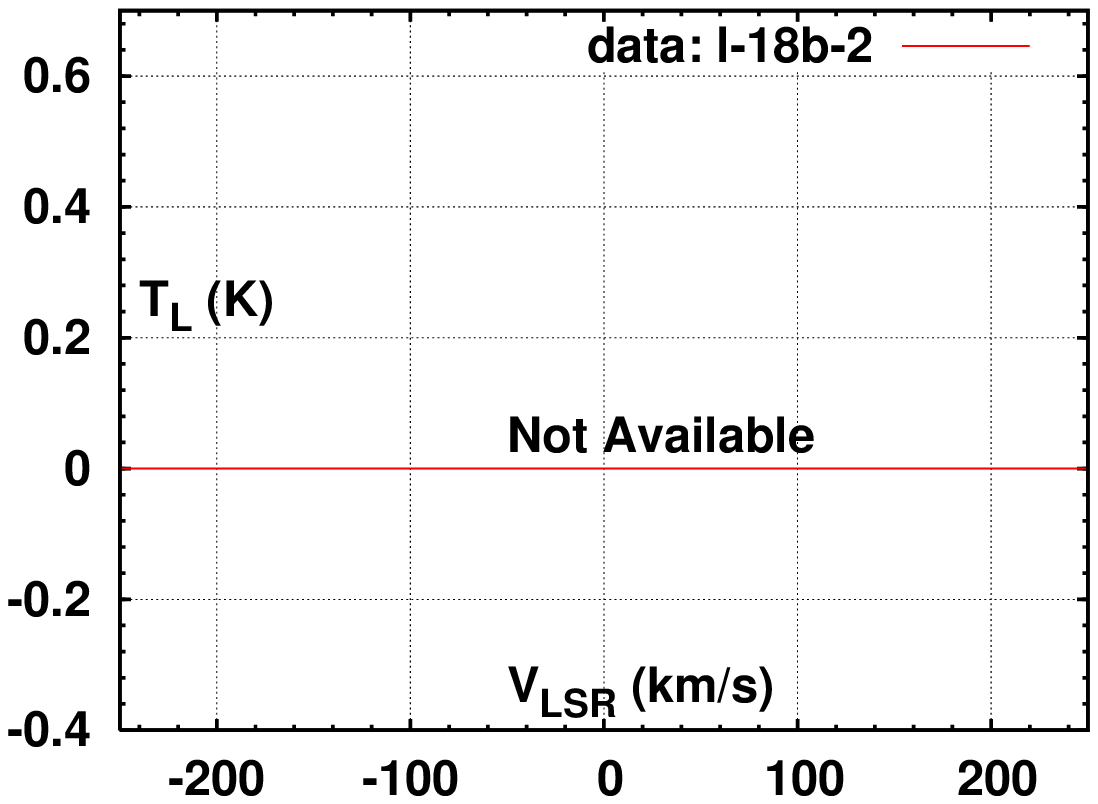}
\includegraphics[width=32mm,height=24mm,angle=0]{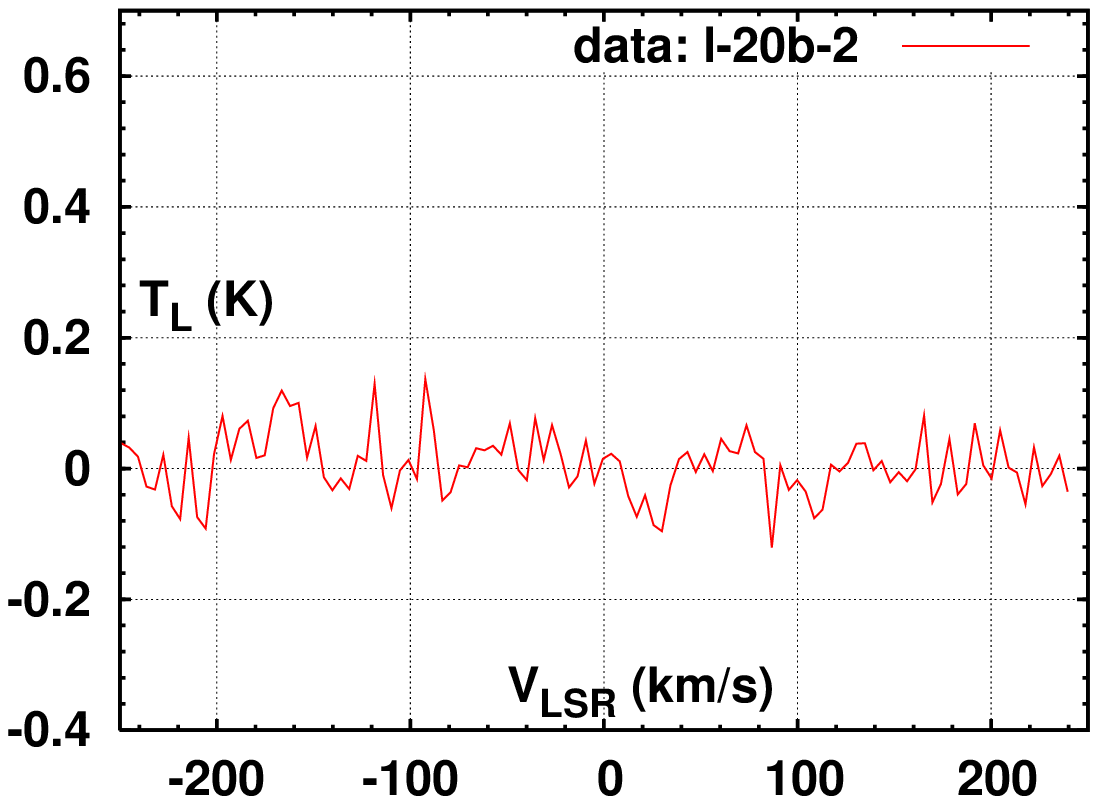}

 \caption{ ORT  H271$\alpha$ RL observation.}
\end{center}
\end{figure}

\begin{figure}[ht]
\begin{center}
\includegraphics[width=32mm,height=24mm,angle=0]{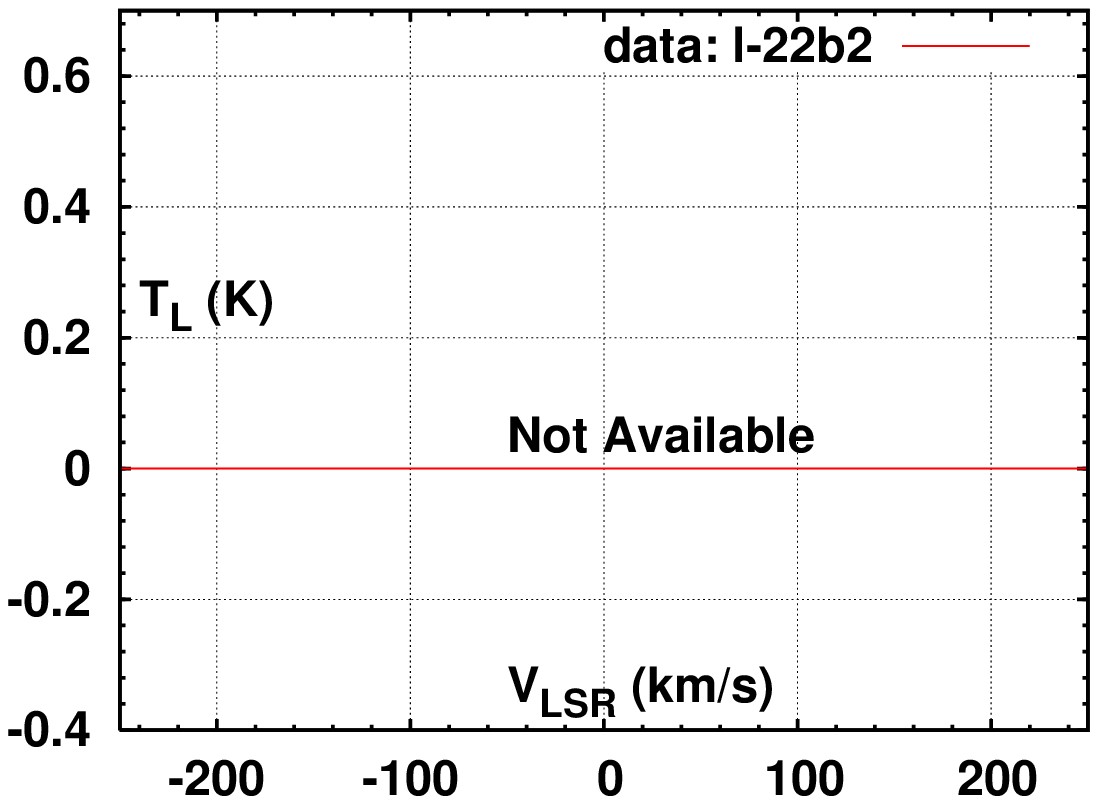}
\includegraphics[width=32mm,height=24mm,angle=0]{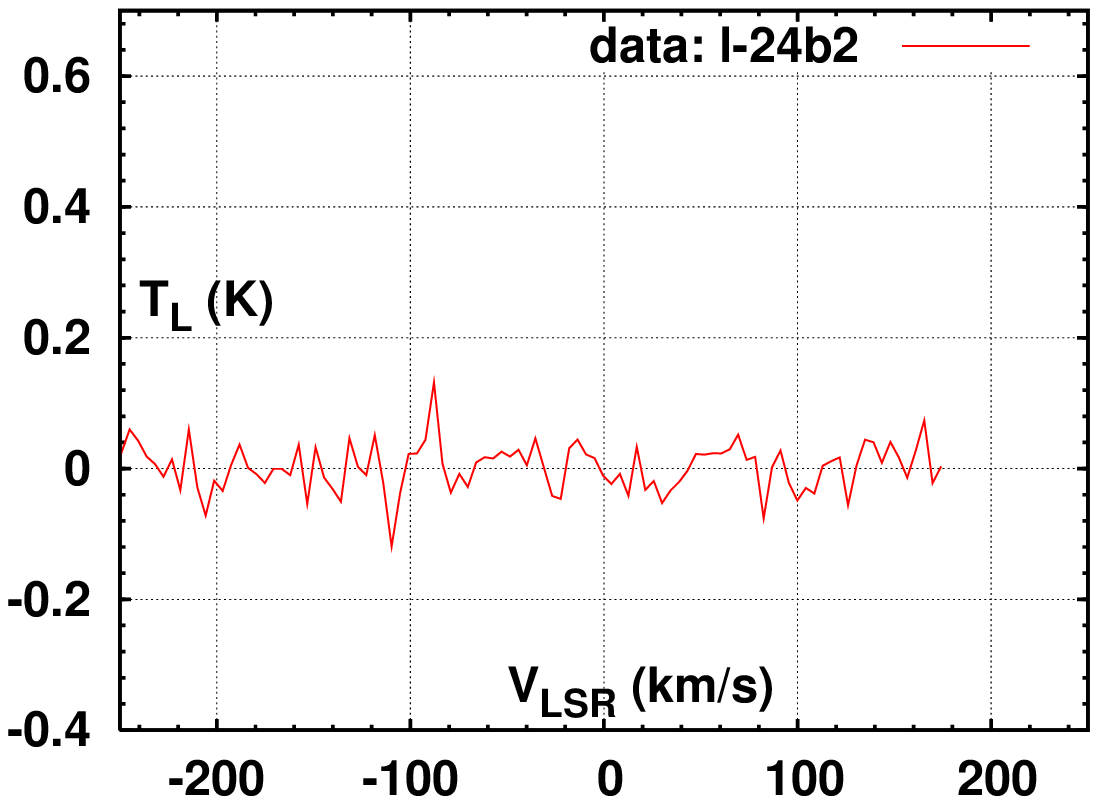}
\includegraphics[width=32mm,height=24mm,angle=0]{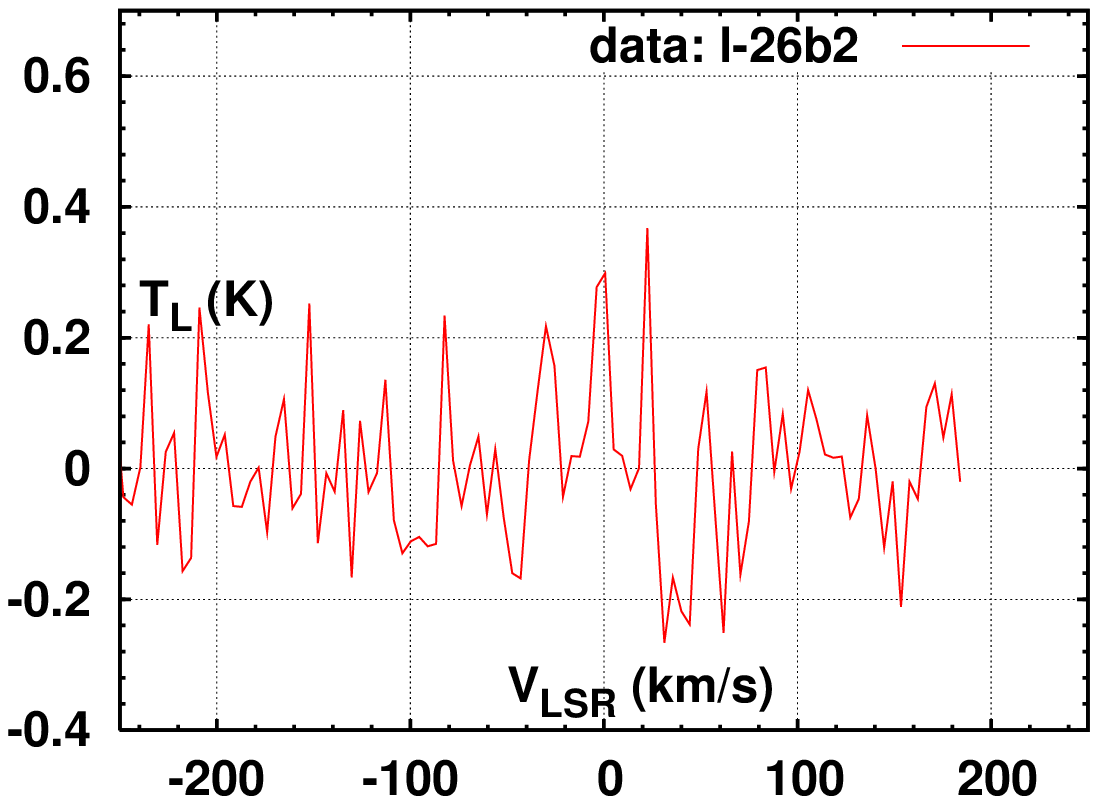}
\includegraphics[width=32mm,height=24mm,angle=0]{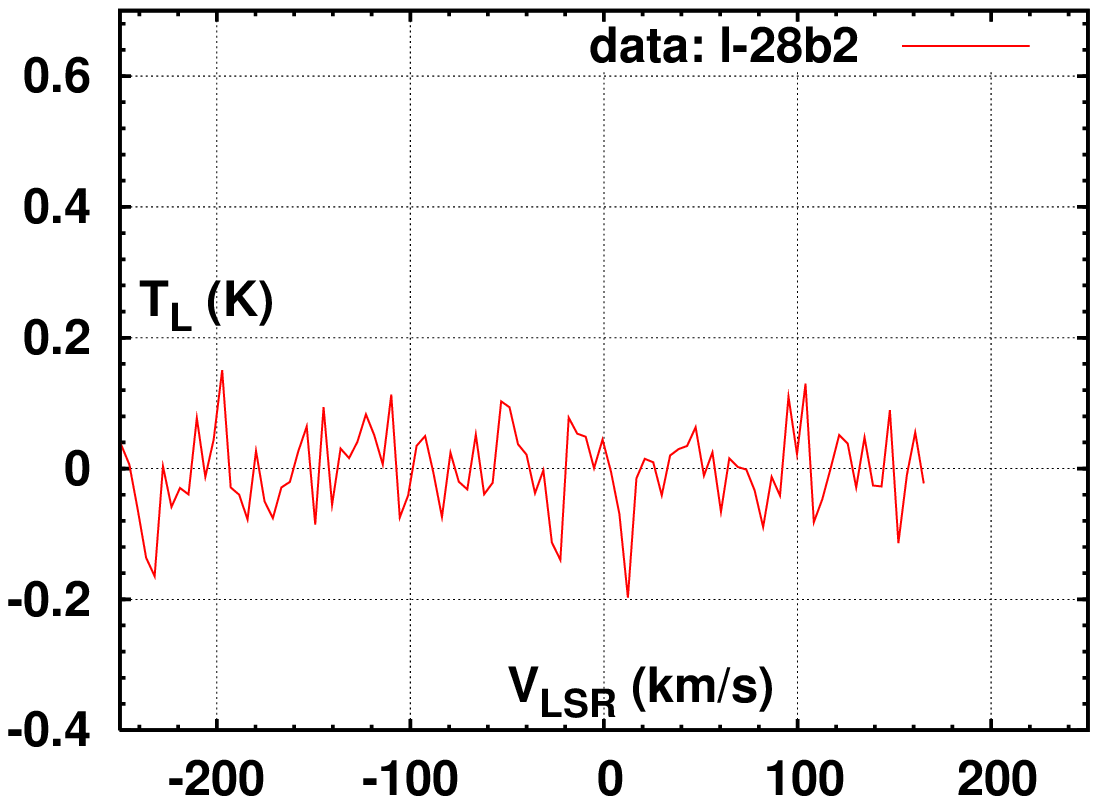}
\includegraphics[width=32mm,height=24mm,angle=0]{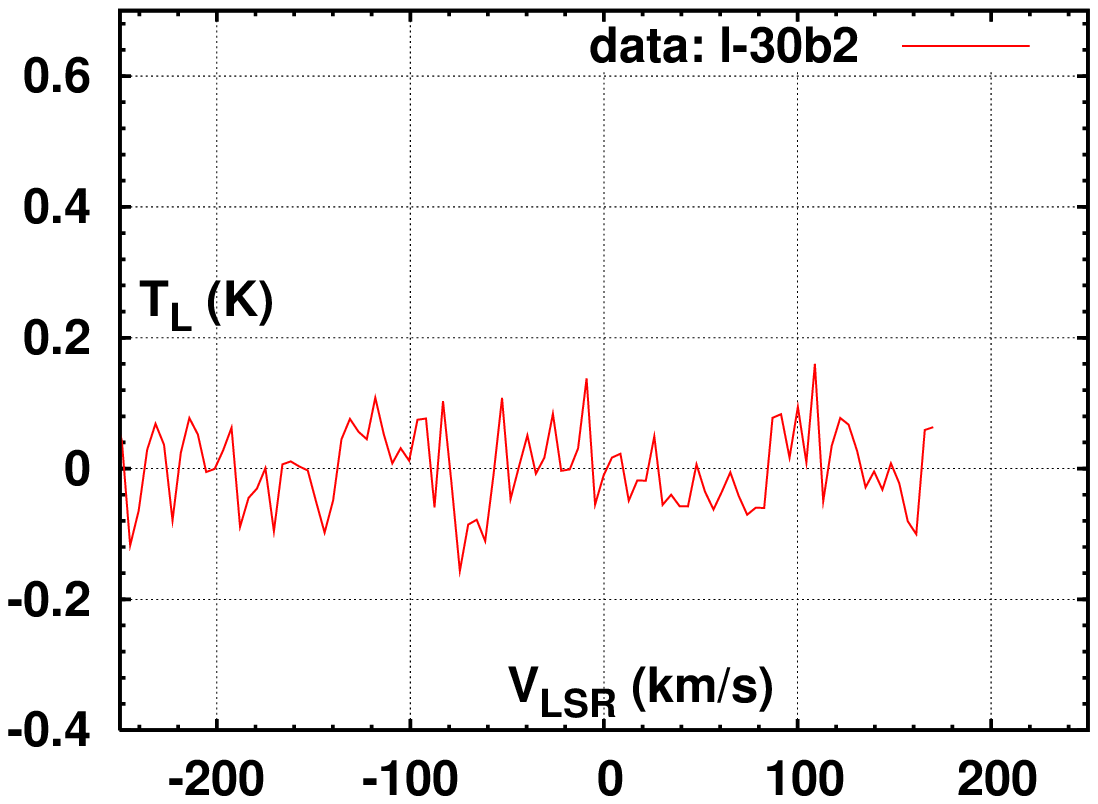}
\includegraphics[width=32mm,height=24mm,angle=0]{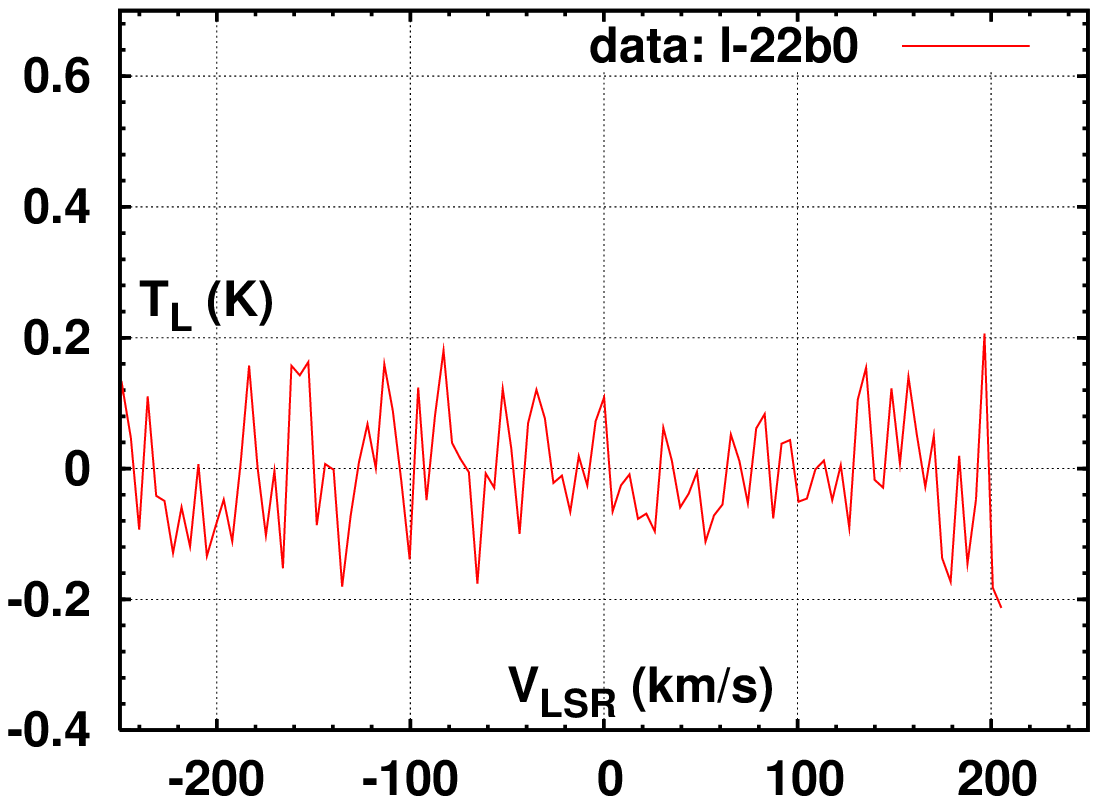}
\includegraphics[width=32mm,height=24mm,angle=0]{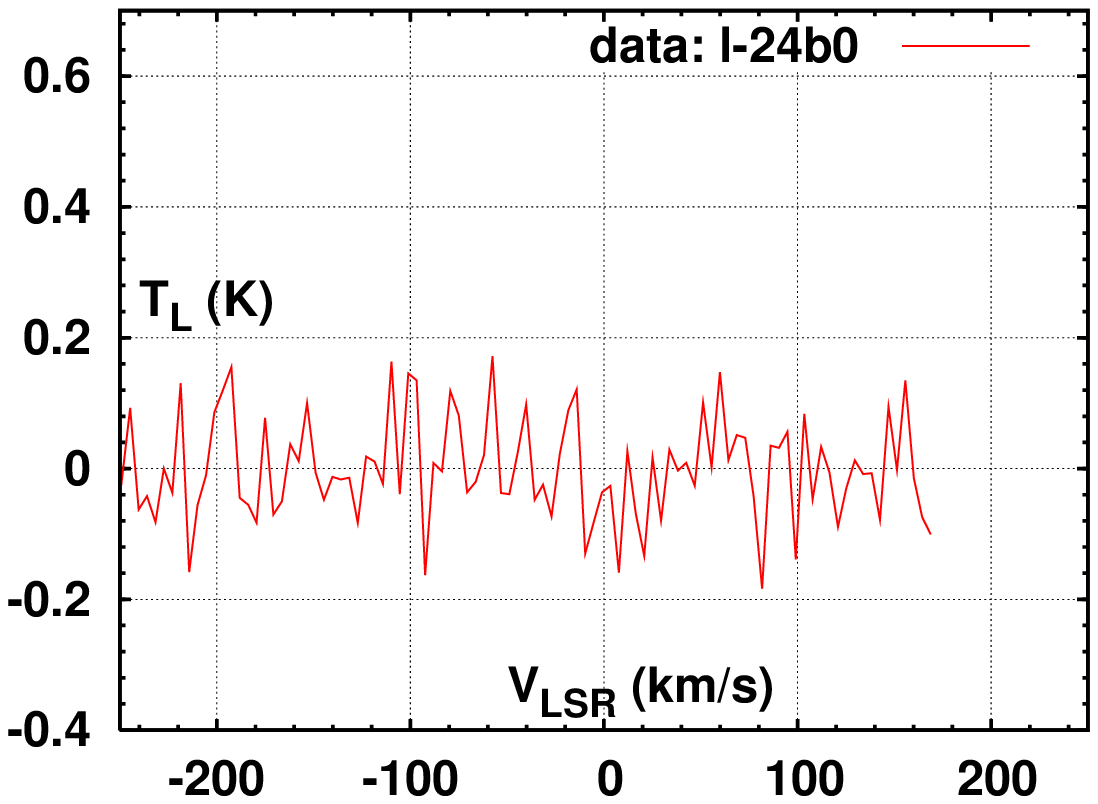}
\includegraphics[width=32mm,height=24mm,angle=0]{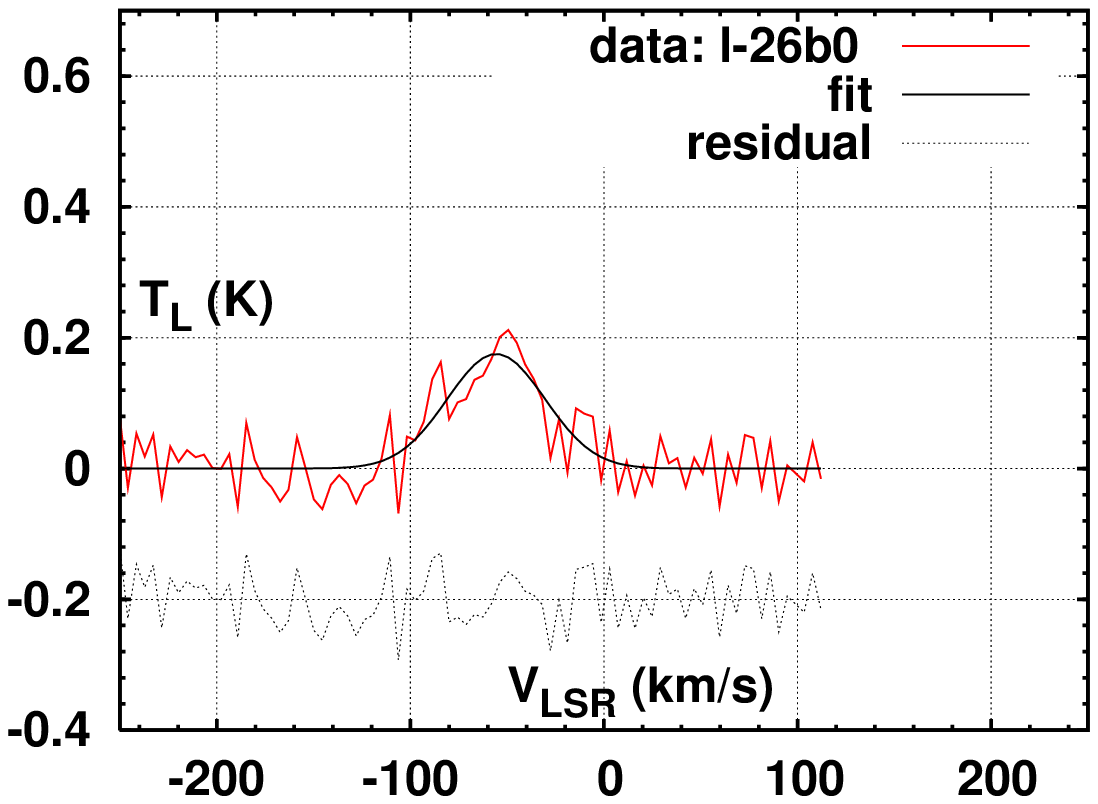}
\includegraphics[width=32mm,height=24mm,angle=0]{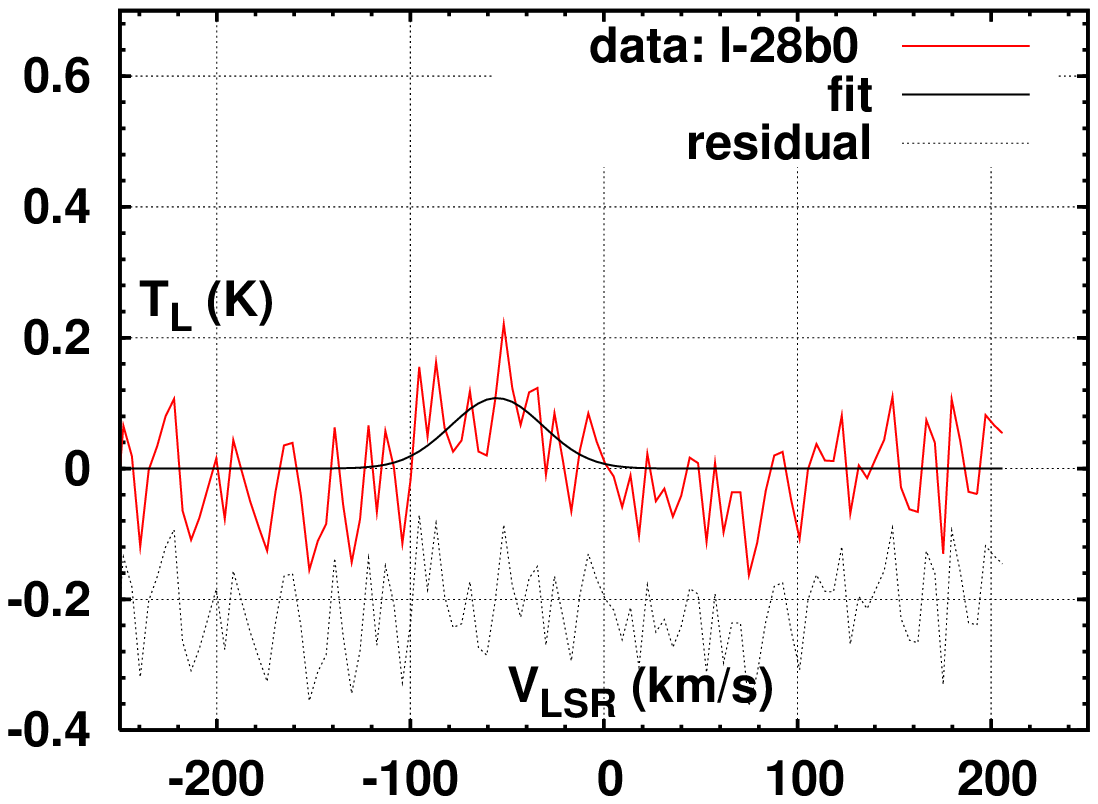}
\includegraphics[width=32mm,height=24mm,angle=0]{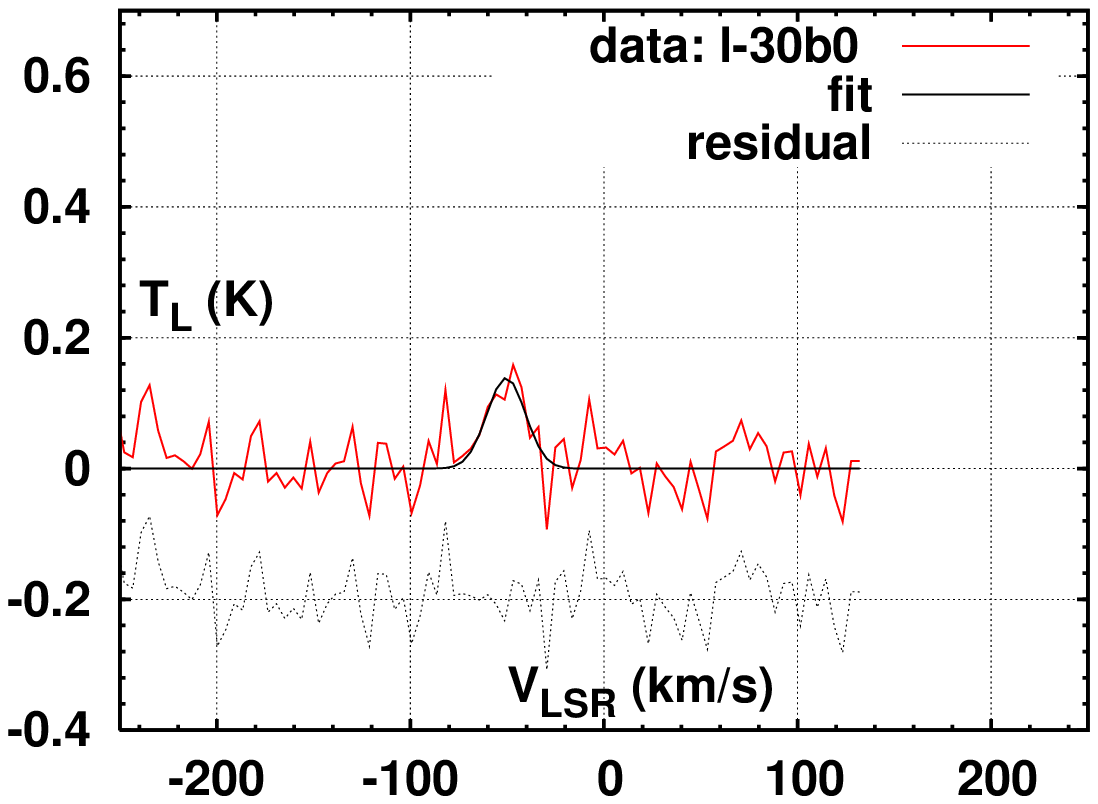}
\includegraphics[width=32mm,height=24mm,angle=0]{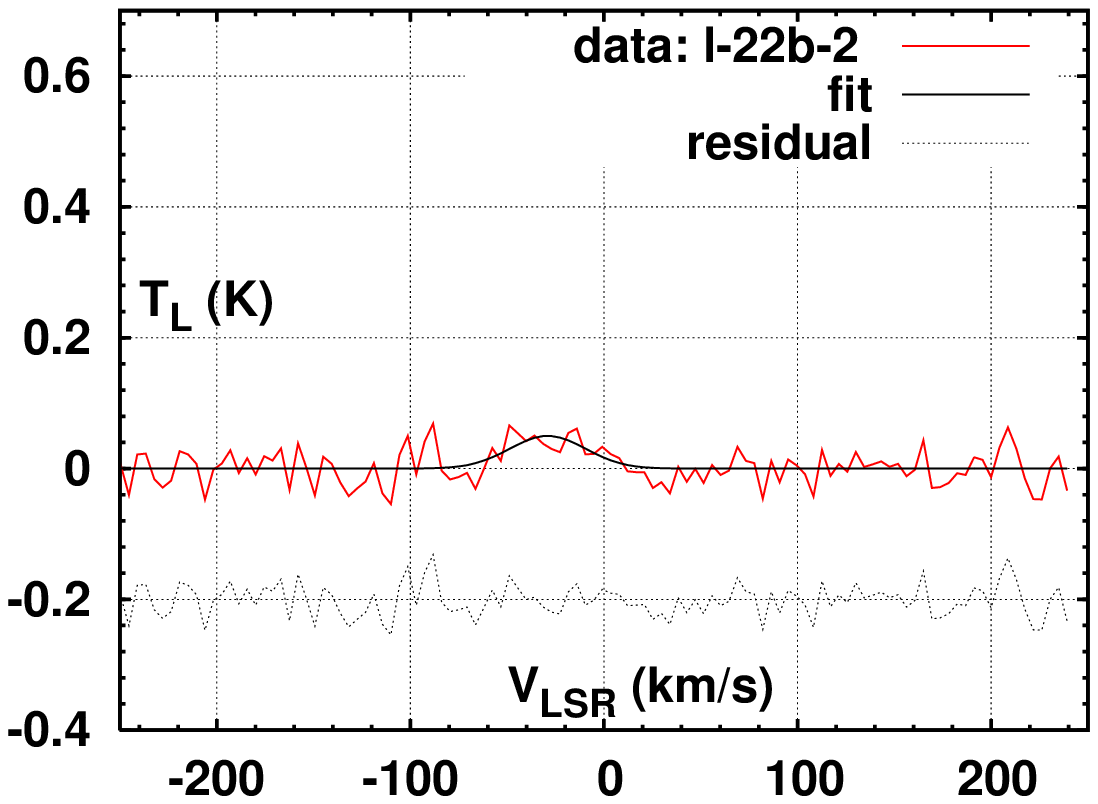}
\includegraphics[width=32mm,height=24mm,angle=0]{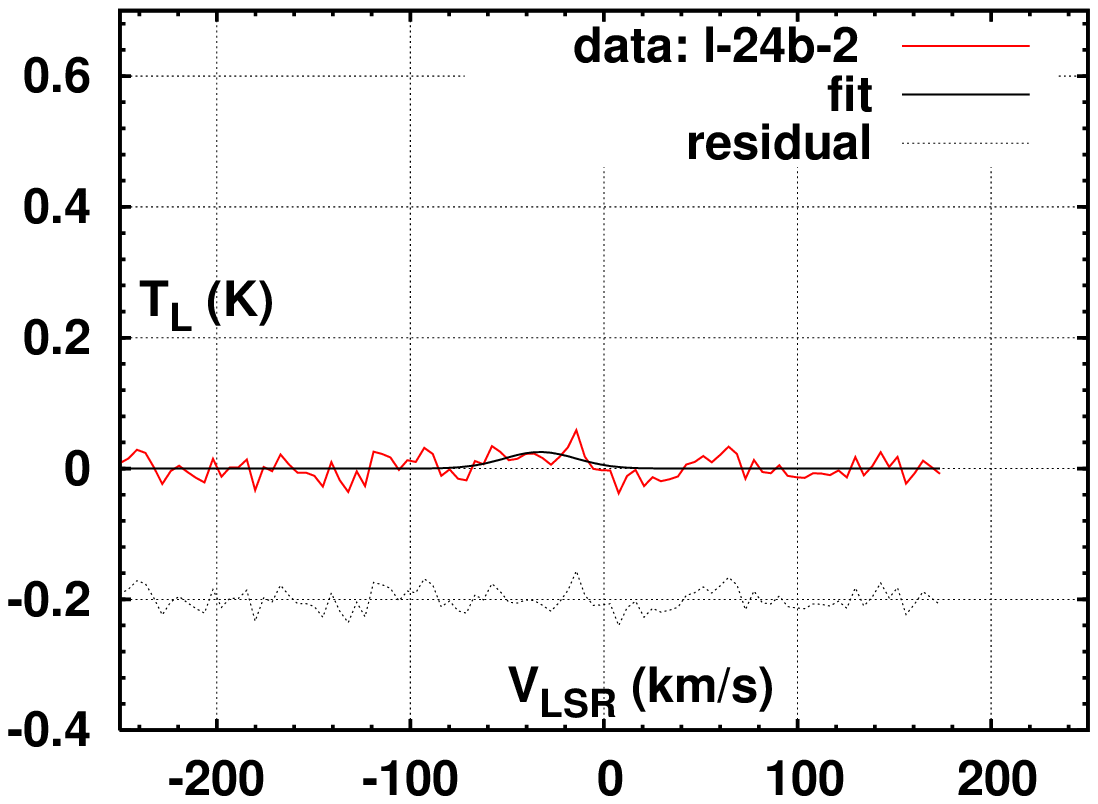}
\includegraphics[width=32mm,height=24mm,angle=0]{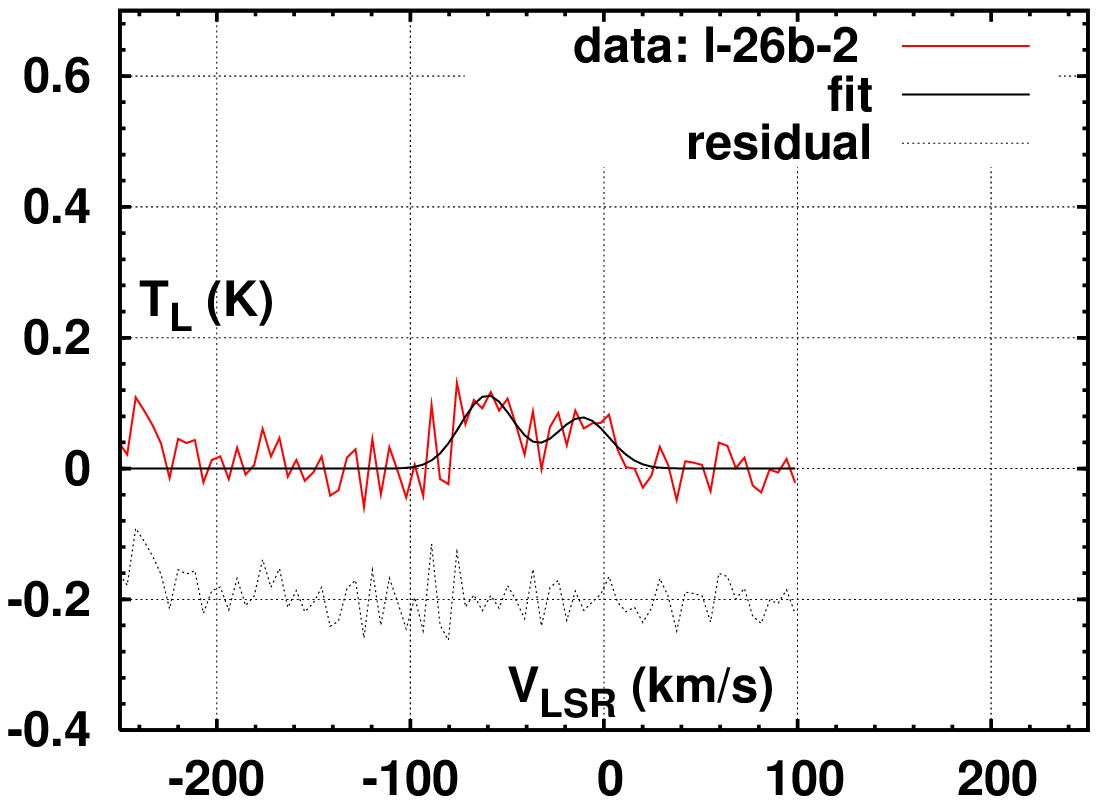}
\includegraphics[width=32mm,height=24mm,angle=0]{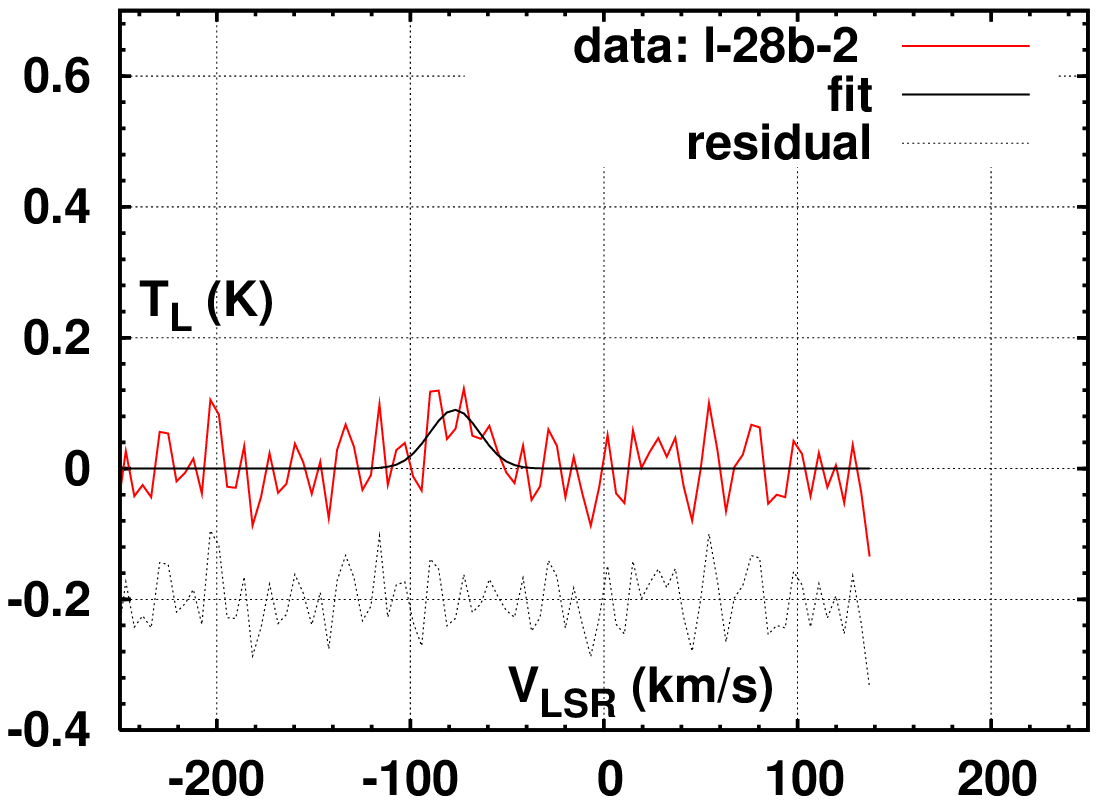}
\includegraphics[width=32mm,height=24mm,angle=0]{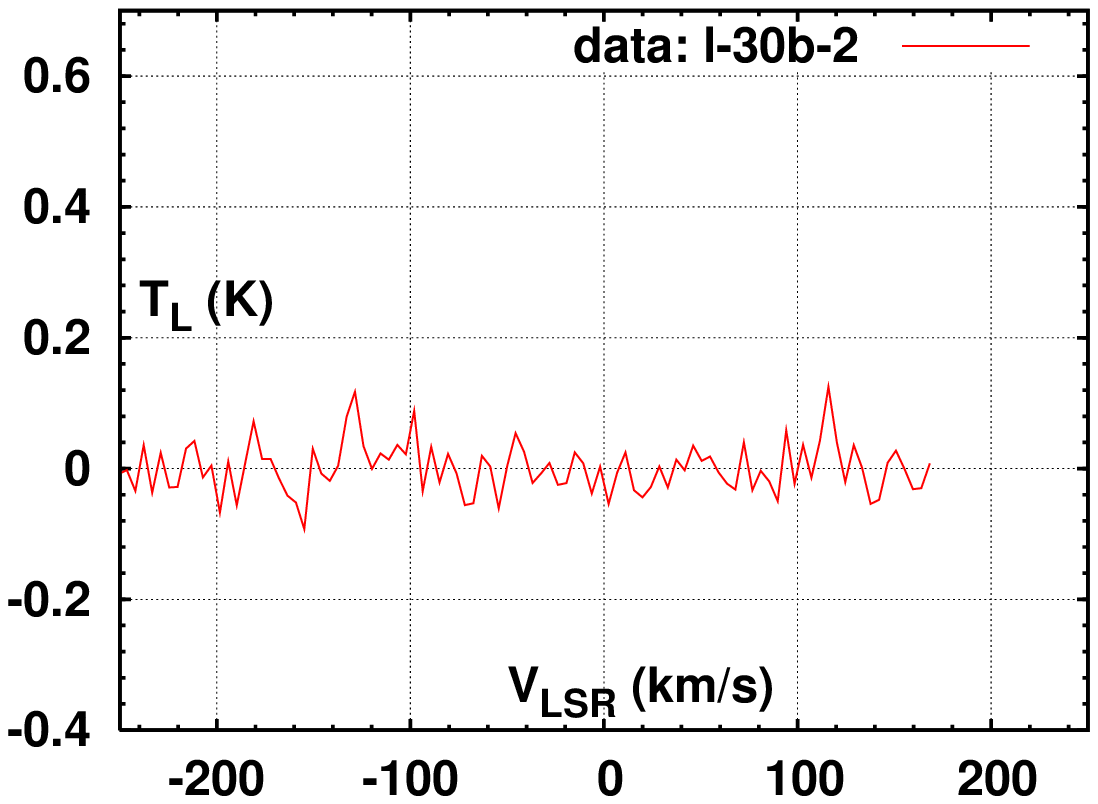}\vspace{1cm}

 \caption{ ORT H271$\alpha$ RL observation.}
\end{center}
\end{figure}

\begin{figure}[ht]
\begin{center}
\includegraphics[trim = 1mm 1mm 30mm 1mm, clip, width=163mm,angle=0]{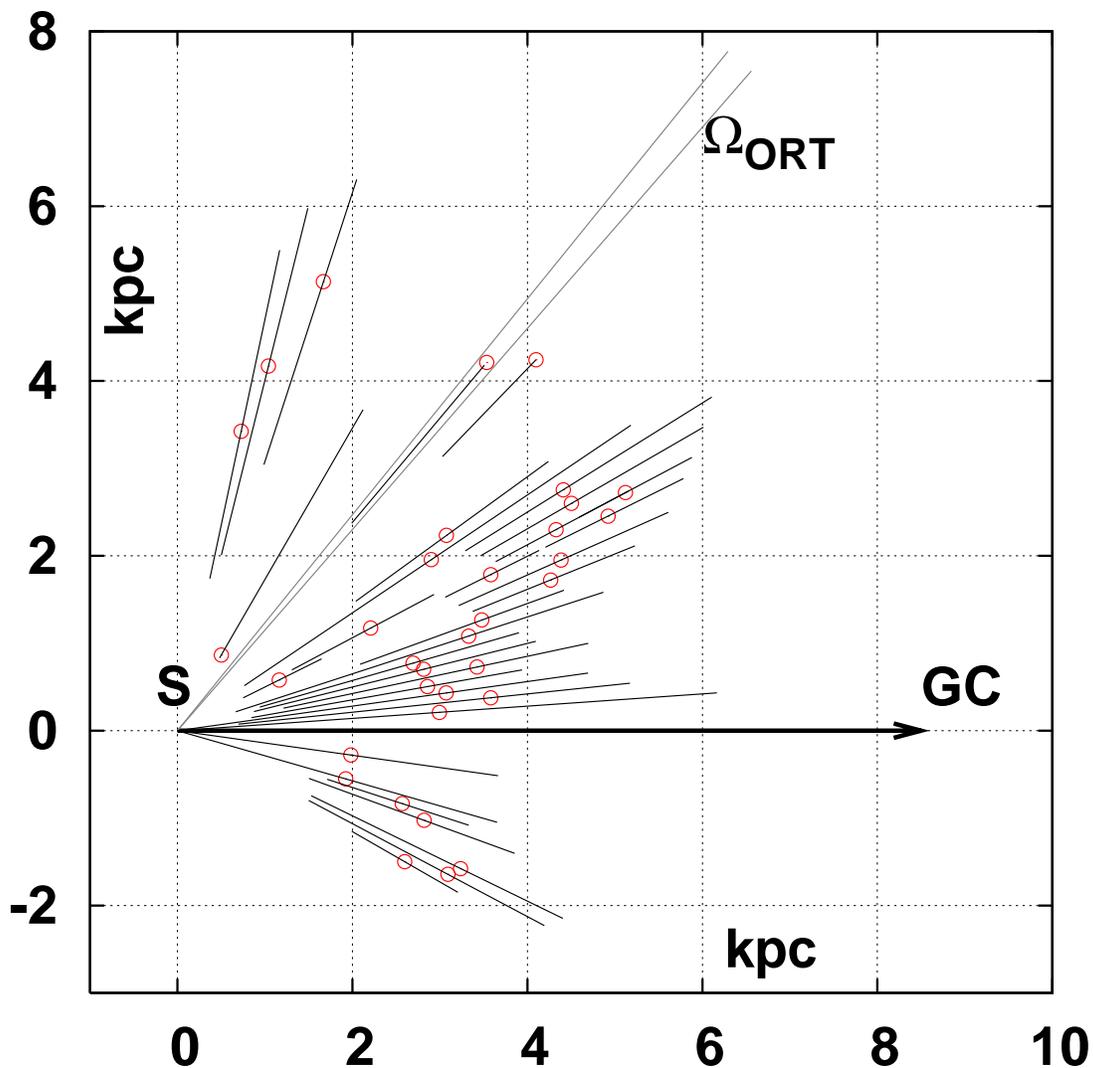}
 \caption{Distribution of ELDWIM clouds in the plane(b=$0^o$) of the 
Galaxy. The origin marks the Solar system S and GC is the Galactic Center. 
$\Omega_{ORT}$ is the beam of ORT. The red circles mark the center of line 
origin obtained from the fitted $V_{LSR}$. The length of the line through 
these circles is the FWHM converted to distance, indicating an upper limit 
on the spread of the gas. Due to observed $V_{LSR}$ the clouds above(b=+$2^o$) 
and below(b=-$2^o$) the plane follow a similar distribution. }
\end{center}
\end{figure}

\label{lastpage}

\end{document}